# Autonomic Management in a Distributed Storage System

PhD Thesis

by

Markus Tauber

School of Computer Science

University of St Andrews

December 2009

# Abstract

This thesis investigates the application of autonomic management to a distributed storage system. Effects on performance and resource consumption were measured in experiments, which were carried out in a local area test-bed. The experiments were conducted with components of one specific distributed storage system, but seek to be applicable to a wide range of such systems, in particular those exposed to varying conditions.

The perceived characteristics of distributed storage systems depend on their configuration parameters and on various dynamic conditions. For a given set of conditions, one specific configuration may be better than another with respect to measures such as resource consumption and performance. Here, configuration parameter values were set dynamically and the results compared with a static configuration. It was hypothesised that under non-changing conditions this would allow the system to converge on a configuration that was more suitable than any that could be set *a priori*. Furthermore, the system could react to a change in conditions by adopting a more appropriate configuration. Autonomic management was applied to the peer-to-peer (P2P) and data retrieval components of ASA, a distributed storage system. The effects were measured experimentally for various work-load and churn patterns. The management policies and mechanisms were implemented using a generic autonomic management framework developed during this work.

The motivation for both groups of experiments was to test management policies with the objective to avoid unsatisfactory situations with respect to resource consumption and performance. Such unsatisfactory situations occur when either the P2P layer or the data retrieval mechanism is configured statically. In a statically configured P2P system two unsatisfactory situations can be identified. The first arises when the frequency with which P2P node states are verified is low and membership churn is high. The P2P node state becomes inaccurate due to a high membership churn, leading to errors during the routing process and a reduction in performance. In this situation it is desirable to increase the frequency to increase P2P state accuracy. The converse situation arises when the frequency is high and churn is low. In this situation network resources are used unnecessarily, which may also reduce performance, making it desirable to decrease the frequency.

In ASA's data retrieval mechanism similar unsatisfactory situations can be identified with respect to the degree of concurrency (DOC). The DOC controls the eagerness with which multiple redundant replicas are retrieved. An unsatisfactory situation arises when the DOC is low and there is a large variation in the times taken to retrieve replicas. In this situation it is desirable to increase the DOC, because by retrieving more replicas in parallel a result can be returned to the user sooner. The converse situation arises when the DOC is high, there is little variation in retrieval time and there is a network bottleneck close to the requesting client. In this situation it is desirable to decrease the DOC, since the low variation removes any benefit in parallel retrieval, and the bottleneck means that decreasing parallelism reduces both bandwidth consumption and elapsed time for the user.

The experimental evaluations of autonomic management show promising results, and suggest several future research topics. These include optimisations of the managed mechanisms, alternative management policies, different evaluation methods, and the application of developed management mechanisms to other facets of a distributed storage system. The findings of this thesis could be exploited in building other distributed storage systems that focus on harnessing storage on user workstations, since these are particularly likely to be exposed to varying, unpredictable conditions.

# Declaration

I, Markus Tauber, hereby certify that this thesis, which is approximately 45,000 words in length, has been written by me, that it is the record of work carried out by me, and that it has not been submitted in any previous application for a higher degree.

*date* ____________________ *signature of candidate* ____________________

I was admitted as a research student in March 2005 and as a candidate for the degree of Doctor of Philosophy in March 2006; the higher study of which this is a record was carried out in the University of St Andrews between 2005 and 2009.

*date* ____________________ *signature of candidate* ____________________

I hereby certify that the candidate has fulfilled the conditions of the Resolution and Regulations appropriate for the degree of Doctor of Philosophy in the University of St Andrews and that the candidate is qualified to submit this thesis in application for that degree.

*date* ____________________ *signature of supervisor* ____________________

# Acknowledgement

I would like to thank Graham Kirby and Al Dearle, my supervisors, for their help, patience and support. They have provided a great introduction to research.

My thanks also go to everyone in the School of Computer Science and in the Internet Club Burgenland (ICB) with whom I had stimulating discussions about my work or about tools that make a PhD student's life easier.

I also would like to express my thanks to Rhona Maclean for her proof-reading.

My deepest thanks go to my family, for their help, encouragement and unwavering belief in me. This also applies to Sandra Vater, my partner, she has been a great support during the demanding process of writing my thesis.

This work was partially funded by EPSRC grant GR/S44501/01: "Secure Location-Independent Autonomic Storage Architectures" (ASA).

# Contents

# Chapter 1

# Introduction

## Outline

This chapter introduces distributed storage systems and the benefits of such systems. It explains the type of distributed storage system that this thesis focuses on and introduces the hypothesis that autonomic management can be used to improve such systems with respect to performance and resource consumption. The chapter finishes with an outline of the thesis.

## 1.1 Distributed Storage Systems

Distributed storage systems are an approach towards the provision of reliable, scalable and fault-tolerant storage. The benefits of such systems include that they may be more scalable than non-distributed storage systems, partly because computers which contribute storage can be added dynamically, and partly because they are able to balance the usage of storage and the computational load of individual machines. Another benefit of distributed storage systems is that data may be replicated on multiple physical machines, which improves the fault-tolerance of such systems. Additionally, distributed storage systems can be built on a decentralised infrastructure which results in the absence of a single point of failure.

Distributed storage systems may operate over a dedicated set of machines. They may, however, also be used to harness storage on user workstations. This thesis investigates distributed storage systems which may be used in both usage scenarios. Potential implementation challenges result from changing conditions to which such systems are exposed. These include changes in the available network performance, the frequency with which users disconnect and reconnect their machines from the distributed storage system, and the frequency with which users access data items. The perceived characteristics of distributed storage systems depend on their configuration parameters and on combinations of various dynamic conditions.

For a given set of conditions, one specific configuration may be better than another with respect to measures such as performance and resource consumption. Performance in this context is the average time it takes to retrieve or store data in the distributed storage system.

When referring to resource consumption, this thesis focuses on network resources which are consumed by the distributed storage system. A motivation for considering network resources rather than, for instance, storage resources is given by reports ([10, 12]) that network resources are generally exhausted faster than storage or computational resources in the considered usage scenarios. In [12] it is concluded that *network bandwidth is a scarce resource in a wide-area distributed storage system*. The overall performance and resource consumption of a distributed storage system depends on the performance of its constituent components and more specifically on individual facets of these components.

One such facet is represented by the scheduling of maintenance operations in Peer-to-Peer (*P2P*) overlays used as a distributed storage system's decentralised infrastructure. P2P overlays provide data item to storage host mappings in a changing network topology. These mappings are resolved by P2P routing operations (lookups). A peer-set is used to make routing decisions, and to adapt the overlay network to new nodes joining and existing nodes leaving or failing. The interval between maintenance operations of this peer-set is a configuration parameter. Two possible unsatisfactory situations can be identified when considering the interval between such maintenance operations. The first arises when both the maintenance interval and the membership churn are high. Peer-sets become inaccurate due to high membership churn, leading to errors during the routing process and a reduction in performance. In this situation it is desirable to decrease the interval in order to increase the accuracy of peer-set. The converse situation arises when the interval is low and churn is low. In this situation network resources are used unnecessarily, which may also reduce performance, making it desirable to increase the interval.

Another facet of a distributed storage system whose performance and resource consumption depends on a specific configuration parameter is the data retrieval mechanism. This configuration parameter is the degree of concurrency (DOC) with which data is retrieved. The DOC controls the eagerness with which multiple redundant replicas are retrieved. Again, two unsatisfactory situations can be identified. The first arises when the DOC is low and there is a large variation in the times taken to retrieve replicas. In this situation it is desirable to increase the DOC, because by retrieving more replicas in parallel a result can be returned to the user sooner. The converse situation arises when the DOC is high, there is little variation in retrieval time and there is a network bottleneck close to the requesting client. In this situation it is desirable to decrease the DOC, since the low variation removes any benefit in parallel retrieval, and the bottleneck means that decreasing parallelism reduces both bandwidth consumption and elapsed time for the user.

## 1.2 Problem Definition and Hypothesis

The above situations represent cases in which a specific configuration is beneficial with respect to performance and resource consumption but ceases to be so as conditions vary. In such situations a dynamic adaptation of the configuration may be more beneficial than a static configuration. Due to its complexity and dynamic nature, the adaptation of configuration parameters of facets of distributed storage systems ideally happens without the need for human interference. An approach for doing this is *autonomic management*.

Autonomic management is a methodology inspired by the autonomic nervous system. In the human body the autonomic nervous system controls the heart rate without the conscious brain being aware of it. The autonomic nervous system increases the heart rate in situations in which a human is highly alerted, but decreases it in moments of relaxation and thus prevents unnecessary resource consumption. The same principle can be applied to, for instance, the rate with which maintenance operations in a distributed storage system's P2P layer are executed or the degree of concurrency with which redundant data items are retrieved in a distributed storage's data layer.

*It is hypothesised that controlling configuration parameter values autonomically in response to changing conditions yields improved results with respect to performance and resource consumption, compared with a static configuration. Furthermore, under non-changing conditions this allows the system to converge on a configuration that is more suitable than any that could be set* a priori.

The evaluation of the hypothesis involves:

- An analysis of the scope for optimisation with respect to performance and resource consumption with autonomic management.
- The development of an autonomic management framework.
- An experimental evaluation of the effects of autonomic management applied to facets of a distributed storage system.

This work is carried out as part of the *Autonomic Storage Architecture (ASA)* project [47]. All investigations are carried out using components of the ASA storage system.

## 1.3 Autonomic Management in Distributed Systems

Related work with respect to autonomic management in distributed systems in general is discussed in this section.

In the *Architectural Artefacts for Autonomic Distributed Systems ($A^3DS$)* project [86] and the project concerned with the development of the *k-component* [20, 19] the focus lies on the development of concepts and tools to add autonomic behaviour to distributed systems. The usage of those tools in the context of this thesis is discussed at a later point (chapter 6). $A^3$DS introduces an approach for specifying high level policies via a *contract specification language*. The k-component provides a similar approach which is evaluated in a distributed file system use case. In this use case a connection manager (specific to *k-components*) and a file forwarding mechanism were autonomically managed. Unlike to the facets considered in this work, those facets do not apply to a wide range of distributed storage systems. On the other hand, both the $A^3$DS and the *k-component* provide interesting starting points with regards to the process of applying autonomic management. Generic conceptual components of those approaches which can also be found in [45] were applied in this thesis.

In [92] facets of a distributed storage system which may benefit from autonomic management are discussed at a high level. The facets comprise replica placement strategies, error detection and recovery, management of file locking mechanisms, and intrusion detection. All of the facets are discussed with respect to DMSuite. DMSuite is a generic storage system designed for use in a Grid environment. Some of the discussed facets can also be found in distributed storage systems.

In [5] policy based autonomic managers are proposed to control various facets of a network in response to a changing environment. The policies are inspired by biological processes like *Chemotaxis* which triggers a positive or negative stimulus depending on the amount and type of monitored information. This is applied to the management of a routing protocol which implements a hop-by-hop route discovery from source to destination. The routing table is managed autonomically. Though the area of investigation is different to that in this thesis, the authors are also concerned with the performance of underlying network protocols in the presence of a changing environment, and address disadvantages with autonomic management.

In work on distributed storage systems like, for instance, PAST [21] and OceanStore [78], the attribute *self-organising* is introduced for the ability to relocate data items in order to compensate for membership churn. This suggests that those distributed storage systems are autonomically managed. In fact, they rely on the maintenance operations in their underlying P2P layer and react to reported changes in the key space. This represents an additional motivation to investigate whether autonomic management can be used to optimise performance and resource consumption in P2P overlays (section 1.1).

## 1.4 Thesis Structure

The structure of the thesis is influenced by it dealing with various areas including autonomic management/computing, P2P computing and distributed storage architecture. Background information and related work is discussed in the relevant chapters.

In chapter 2, general principles of autonomic management are explained and related to computer science. Chapter 3 introduces the ASA storage system, and outlines the connection between the different layers and how the P2P layer is utilised by a higher level storage layer. Chapter 4 investigates the scope for optimisation with autonomic management in a distributed storage system's P2P layer. Chapter 5 explores the scope for optimisation with autonomic management in a distributed storage system's data layer.

In chapter 6, an autonomic management framework developed as part of this work is outlined. Chapter 7 reports experiments in which the effects on resource consumption and performance of autonomic management applied to a P2P overlay are compared with a statically configured overlay under various conditions. Chapter 8 reports experiments in which the effects on resource consumption and performance of autonomic management applied to a distributed storage system's retrieval mechanism are compared with a statically configured system under various conditions. Chapter 9 finishes the thesis by drawing some general conclusions and giving an outline for potential future work, including an outline of how further facets of a distributed storage system can be controlled with an autonomic manager.

Extensive appendices are provided containing preliminary work, as well as detailed views on the data for the experiments reported in chapters 7 and 8. Further information to support wider applications of the framework developed here is also provided.

## 1.5 Conclusions

This chapter outlined potential implementation challenges in distributed storage systems that may be exposed to various changing conditions. Situations were briefly discussed in which autonomic adaptation of specific configuration parameters may improve performance and resource consumption in comparison to a statically configured system.

# Chapter 2

# Autonomic Management

## Outline

To provide an understanding of the terms and concepts used throughout this thesis, the analogy between nature and computer science with respect to autonomic management is briefly outlined. The general concepts of autonomic management are also explained in this chapter.

## 2.1 Introduction

Autonomic computing or autonomic management [45, 39, 32, 34, 36] is a term inspired by the autonomic nervous system [24, 23], which frees the conscious brain from controlling vital parts of the human body in the presence of a changing environment. The hypothesis is made in section 1.2 that autonomic management may be able to dynamically control facets of a distributed storage system in order to improve performance and resource consumption with respect to a statically configured system. Like in nature, such autonomic control is envisioned to operate without human intervention. To provide an understanding of the terms and concepts used throughout this thesis, the autonomic nervous system is briefly outlined in section 2.2, followed by an introduction to autonomic management in computer science in section 2.3. The chapter is concluded in section 2.4.

## 2.2 The Autonomic Nervous System

The autonomic nervous system is a control and monitoring system which exists in all mammals, including humans. In the human body the autonomous nervous system coordinates rapid responses (adjustments of the function of, for instance, organs, muscles or the digestion system) to specific stimuli, without the conscious brain being aware of it. The basic parts of the autonomous nervous system are described as:

- *Receptors* sense changes in the internal or external environments. Sensory input can be in many forms, including, blood pressure, acidity in the stomach or blood pH. Information about any sensory input is either sent via electrical impulses or via different hormone levels to the brain or spinal cord.

- In the *sensory centres* of the brain or in the spinal cord, two separate processes are carried out. Firstly the input is processed. This provides the brain or spinal chord with information about the current situation of the controlled part of the human body or a facet of it. Secondly an appropriate response is generated.

- Via a *signalling system* which interacts with the controlled part of the body, the response is transmitted to organs which convert the signal into some form of action. Examples of such actions include changes in heart rate or release of hormones which trigger further nervous system related actions.

An example of a facet of our body functions controlled by the autonomic nervous system without our conscious brain being aware of it is the heart rate. This is part of a well understood phenomenon in biology, *the fight- or flight-response* [68]. Controlled by the amount of specific hormones, excitement and fear increase the heart rate. On the other hand in situations in which we relax or digest, the heart rate is decreased, which is controlled by antagonistic hormones. The autonomic nervous system controls multiple vital functions simultaneously in order to keep the human body literally going. Constituent components of our body are often controlled in isolation; the autonomous behaviour of our body as a whole, however, is determined by *the autonomous nervous system.*

## 2.3 Autonomic Management in Computer Science

In computer science, a system's behaviour is referred to as autonomic if it is, for instance: self-configuring, self-healing or self-protecting. These behaviours are often referred to as *self-** behaviours and are applied without the need for a human operator. A system's autonomic behaviour as a whole depends on the self-managing capabilities of its constituent components [70] and on facets of those components (at a finer granularity). A system, a component and a facet of a component describe different granularities of entities to which autonomic management can be applied. At the finest granularity a controlled facet and the manager together are referred to as an *autonomic element* [45, 39, 32].

The underlying principles of an autonomic manager are based on a control loop [45, 39, 32] (figure 2.1) consisting of the following:

- A monitoring phase where the manager receives information about the target system or the facet of a controlled component of the target system. This is analogous to, for instance, hormones being sent to the autonomic nervous system in the human body. Such pieces of information are referred to as *events*.
- An analysis phase in which complex situations are modelled by aggregating received information. This is analogous to one category of operations in the sensory centres in nature. In this work a modelled situation is expressed as an abstract *metric*.
- A planning phase during which decisions are made about how to react to the current situation. This is analogous to another category of operations in the sensory systems in nature. In this work such a plan is referred to as *policy*.

- An execution phase during which the planned reactions are carried out. This is analogous to the signalling system in nature. Planned actions are applied to a target system via *effectors*.

Information which is shared between all phases of the autonomic control loop is referred to as shared knowledge.

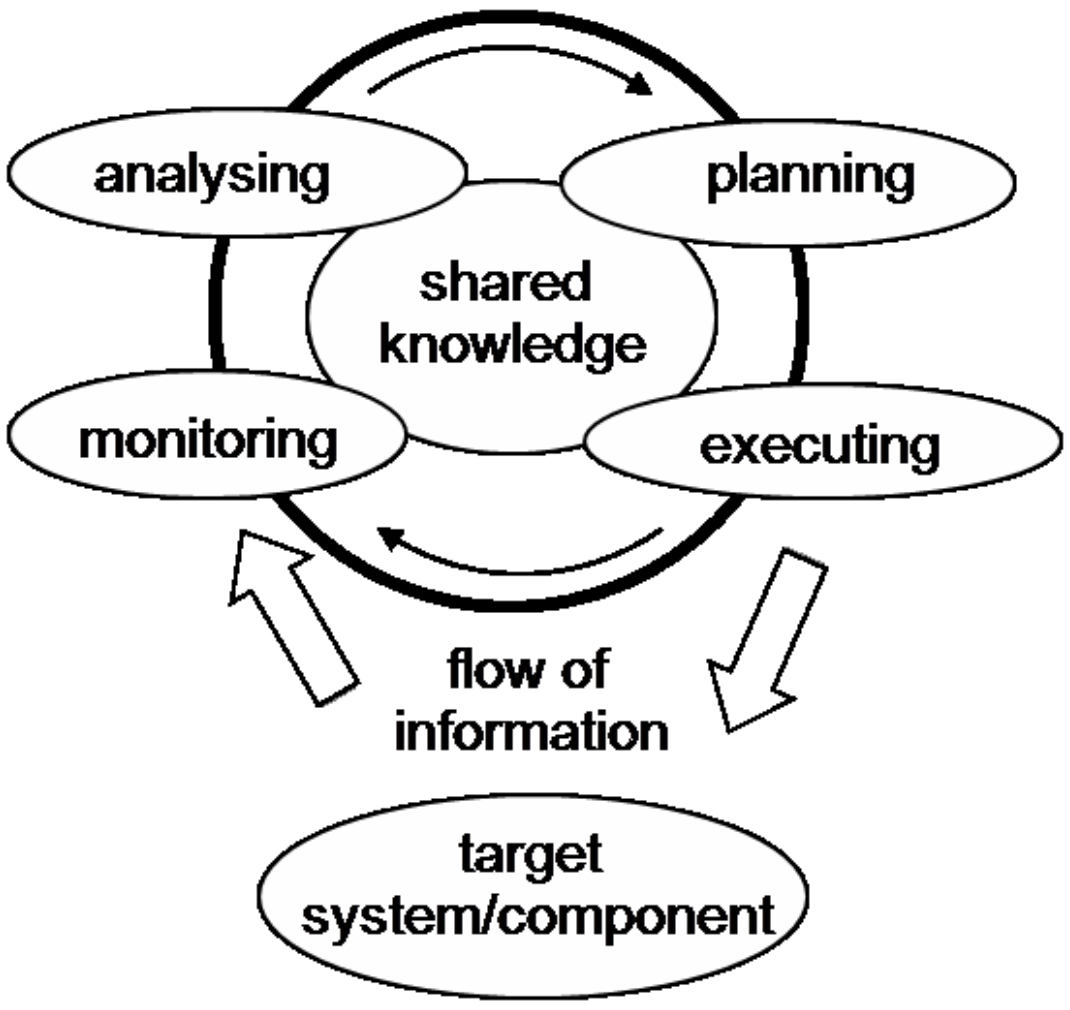

Figure 2.1: The autonomic control loop [45].

## 2.4 Conclusions

This chapter described how autonomic management is inspired by the autonomic nervous system. The concepts described here provide a conceptual framework of operations which can be used to apply autonomic management to facets of target systems.

# Chapter 3

# The ASA Storage System

## Outline

This chapter introduces the *Autonomic Storage Architecture (ASA)*, which is used as experimental platform in the work reported in this thesis. It outlines how several building blocks interact with each other and explains those details which are relevant for this thesis.

## 3.1 Introduction

The work for this thesis is carried out as part of the development of a distributed storage system termed the *Autonomic Storage Architecture (ASA)* [47, 48, 49]. In common with other distributed storage systems like *CFS* [13], *Ivy* [66], *Past* [21], *Oceanstore* [50], *Pond* [77], *Koorde* [42] and *ConChord* [3], the goal of ASA is to develop a resilient ubiquitous distributed storage system with the following attributes:

- Data can be accessed efficiently and securely from any physical location.
- Data is stored resiliently, such that failures of individual machines and network links do not result in data loss.
- An historical record of data is available.

ASA also has the following more specific goals:

- To provide general autonomic tuning mechanisms to allow low-level facets of the system's operation to be managed automatically, controlled by policies that are driven by high-level user preferences.
- To develop specific policies suitable for managing a distributed file storage system, and to investigate their efficiency.

To support the goals stated above, a modular ASA structure is defined. ASA consists of the following layers:

- *File System Adapter:* Provides a file system interface.
- *Abstract File System:* Translates file system operations to storage requests and maps file names to keys.
- *Generic Storage:* Stores and replicates data on multiple P2P nodes; uses keys to identify data; actively maintains replicas as nodes become unavailable.
- *P2P infrastructure:* Returns a node for a given key. The P2P Infrastructure is a key-based routing system; its principal functionality is to map keys to nodes and to maintain the mapping when nodes join or leave the infrastructure.

Rather than providing a complete description of ASA, this chapter focuses on the specific facets which are used for experimental evaluation of autonomic management. Hence, the reminder of this chapter is as follows. After an introduction and a brief outline of ASA in section 3.1 ASA's P2P layer is explained in section 3.2. This is followed by explaining relevant facets of the storage layer in section 3.3. A summary of the chapter is provided in section 3.4.

## 3.2 P2P Layer

Fundamental concepts of ASA's generic storage layer are based on properties of the P2P infrastructure *StAChord*, which is an implementation of the structured P2P overlay *Chord* [85]. Structured P2P overlay network protocols support the *key-based routing (KBR)* abstraction [14]. A KBR implementation allows any given abstract key value to be dynamically mapped (or routed) to a unique live node in the overlay network. This mapping operation is referred to as *lookup*. A P2P overlay actively maintains the key to node mapping despite dynamic changes in network membership. Distributed storage systems can utilise such a mechanism to maintain data to storage host allocations. Additionally correlations between P2P nodes in an overlay are used to determine hosts on which data items are replicated. Besides ASA, such correlations between the P2P infrastructure and the storage layer are used in distributed storage systems like *CFS* [13], *Ivy* [66], *Past* [21], *Oceanstore* [50], *Pond* [77], *Koorde* [42] and *ConChord* [3].

In ASA, data items are identified with keys, derived from the P2P key space, which is organised as a ring. Figure 3.1 illustrates how StAChord is utilised by the generic storage layer for locating the storage host for a data item identified with a specific key. The figure shows a segment of a StAChord key ring in which a data item, identified with key 50, is stored on node 2. StAChord's *lookup* protocol will return node 2 if key 50 is looked up, as all keys from 26 to key 80 belong to node 2.

Another property of StAChord is utilised to determine nodes for replicated data items. Given the fixed size of the key space ($KS$), keys at specific distances around the ring are determined via a so-called *cross algorithm* [15]. For a replication factor of $r$, the data item associated with key $k$ is replicated on $r-1$ nodes associated with keys $k + n \times \frac{KS}{r}$ where $n$ ranges from $1$ to $r-1$. Figure 3.2 illustrates this with a replication factor of four, which results in replica keys for key 5 being identified after each $90°$ along the ring.

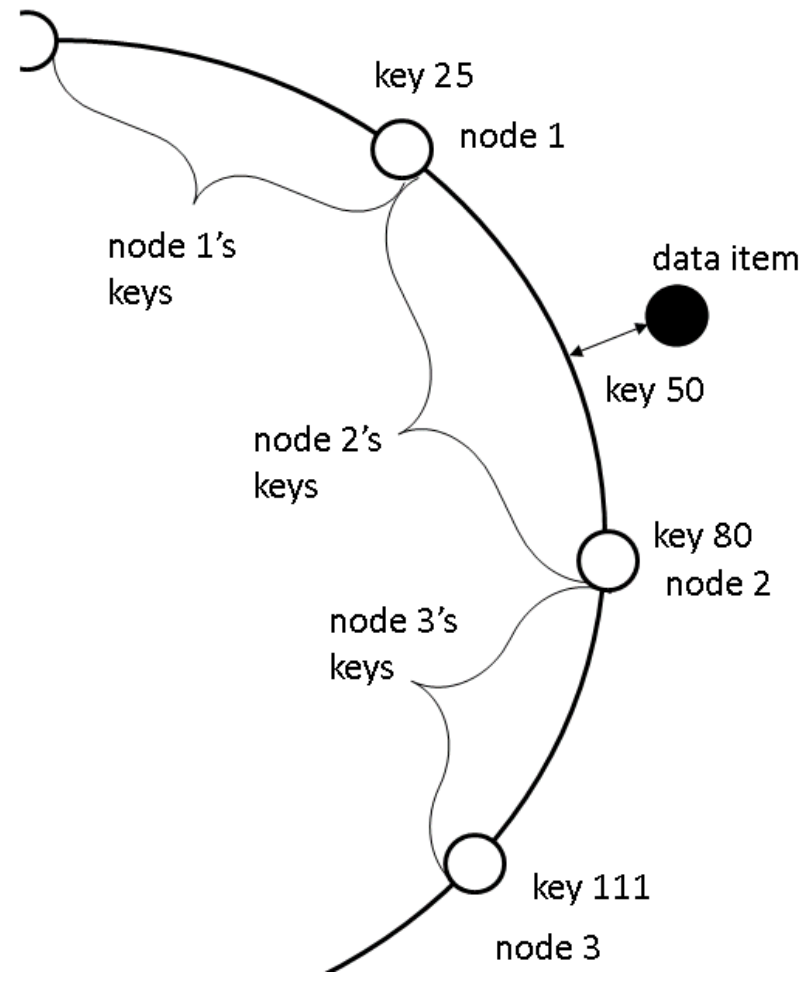

Figure 3.1: Simplified representation of StAChord's key to node and data mapping.

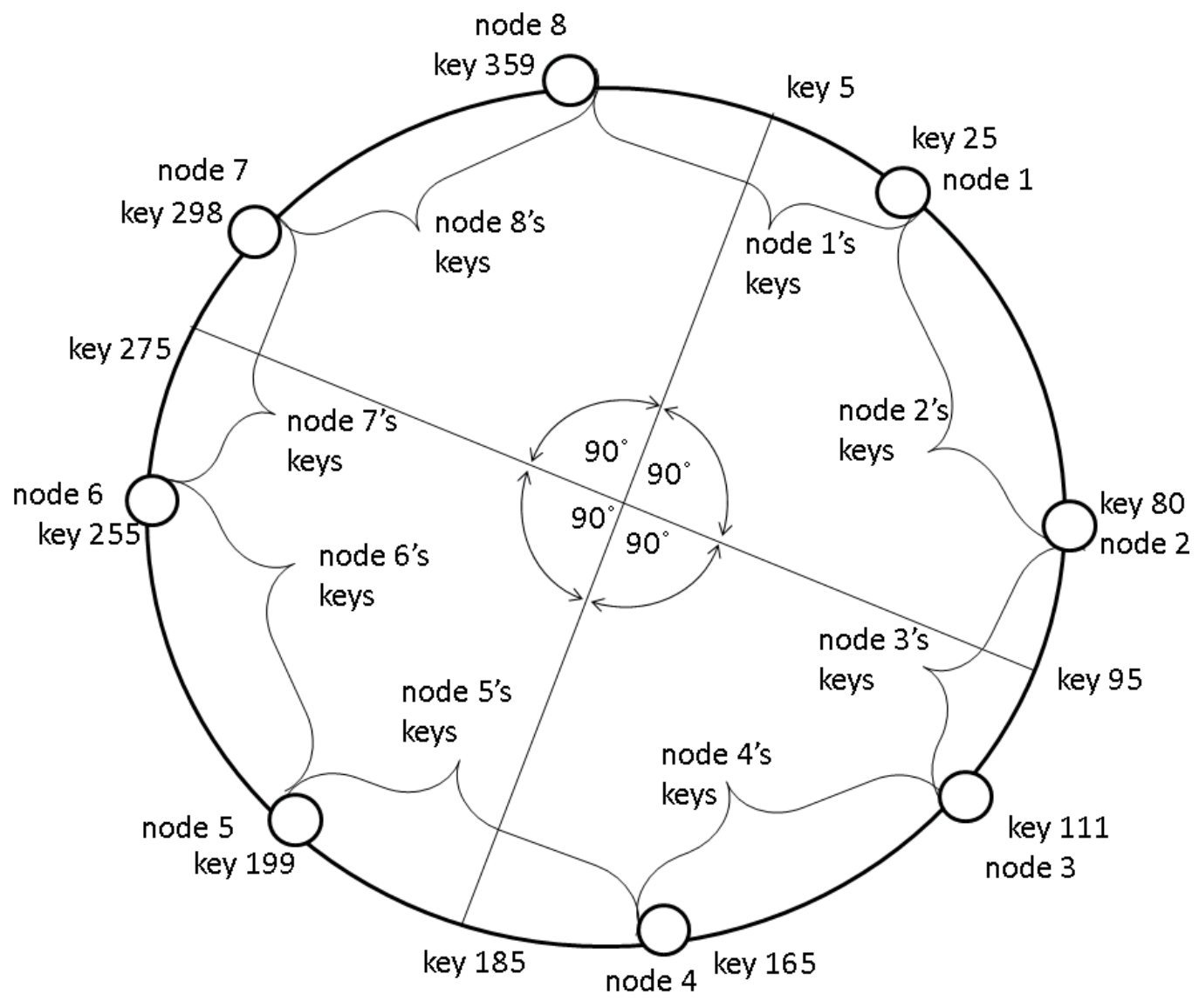

Figure 3.2: Identification of replica nodes in the StAChord key space.

## 3.3 Generic Storage Layer

The generic storage layer provides a ubiquitous resilient mutable storage facility, for unstructured data, with an historical record. To support the historical record, updates are appended rather than being destructive. The main entities supported are data blocks, PIDs and GUIDs. All entities are identified with a unique key from the underlying P2P infrastructure.

- A data block contains unstructured immutable data.
- A PID (Persistent Identifier) is used to denote a particular data block.
- A GUID (Globally Unique Identifier) is used to denote something with identity, such as a file, object or directory (that means meta-data).

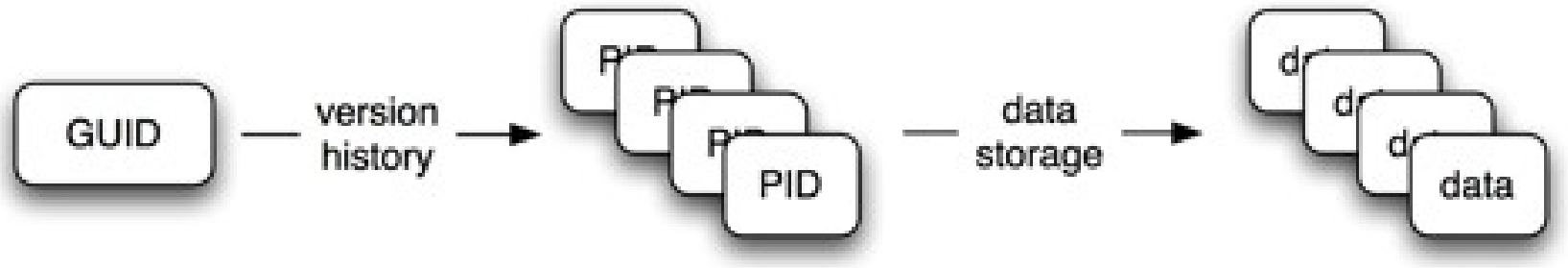

Figure 3.3: ASA Storage Model [49]

Figure 3.3 shows how historical data is associated with PIDs and GUIDs. Additionally to maintaining historical versions of data items, the generic storage layer replicates data and meta-data on multiple nodes, and actively maintains these replicas as nodes fail, misbehave or leave the network. It is insufficient, however, for data to be replicated; the replicas must also be accessible. The placement of replicas is thus organised so as to reduce the probability of a malicious node being able to hinder access to particular replicas. Originally inserted data is referred to with a PID created by hashing its content and replicas of this data item are distributed as determined by the cross algorithm (section 3.2). Storage hosts for meta-data are determined by GUIDs, which are derived from randomly selected P2P keys. Thus the authenticity of a specific data item can be verified with this data item's content and the key with which it is addressed. In the original ASA implementation a replication factor of four is chosen.

Access to any file stored in ASA results in a sequence of operations in which meta-data and data items are fetched. A file is specified with a file path, where each element of the file path represents meta-data identified by a GUID. As meta-data is not self-verifying, at least three meta-data items have to be fetched before the data item can be fetched, in accordance with the protocol outlined in [15]. The meta-data maintains pointers to historical versions of the data. The default configuration ensures that the latest version of a data item is fetched at any request for data. In the original implementation four replicas are available for every data block (if none of the replica holding servers failed). Only one of these data items needs to be received by the requesting client. Its authenticity can be verified by recomputing the PID with the data item's content. After receiving the data item, the meta-data items for the next child directory can be fetched and so forth.

## 3.4 Conclusions

This chapter explained the fundamental building blocks of the Autonomic Storage Architecture ASA relevant for the work carried out in this thesis. It showed how the generic storage layer utilises properties of the P2P infrastructure and provided an understanding of how specific effects on the P2P infrastructure affect the generic storage layer. For instance, incorrect key to host mappings in the P2P infrastructure would result in failed accesses to data items or would cause the replication mechanism to break.

# Chapter 4

# Optimisation of P2P Overlays

## Outline

This chapter investigates the scope for optimisation of P2P overlays with autonomic management in order to improve performance and resource usage. It outlines related work and some background on the maintenance mechanisms of existing structured P2P overlay networks. It introduces an autonomic management mechanism, the goal of which is to detect whether maintenance effort is wasted, or required, in order to adapt the scheduling mechanisms appropriately.

## 4.1 Introduction

As outlined in chapter 3, ASA and other distributed storage systems use structured P2P overlay networks to provide scalable data location facilities even in the presence of churn in the network membership. The wide usage of P2P overlay networks in distributed storage systems is the motivation for this investigation.

In structured P2P overlay networks, each node maintains a set of nodes as its *peer-set*. The peer-set is used to make routing decisions, and to adapt the overlay network to new nodes joining and existing nodes leaving or failing. The validity of the peer-set in existing P2P overlay networks is checked (and if required, repaired) periodically. These periodic operations are referred to as *maintenance* operations in this work. Each maintenance operation involves the exchange of one or more messages with other nodes in the overlay network. This means that each maintenance operation requires the usage of some network resources. The optimal scheduling of such maintenance operations depends both on the workload – that is, the pattern of routing calls applied to the network – and the churn in network membership – that is the temporal pattern with which nodes join or leave the network. For example, if the network membership is completely static, then the optimal behaviour is to perform no maintenance, since it represents pure overhead and network resources are used up unnecessarily. Conversely, under rapid network churn it is beneficial for nodes to expend significant maintenance effort in order to sustain a high success rate for routing operations.

In all current structured P2P overlay protocols, maintenance operations are scheduled at a statically fixed interval. Workload and churn will often vary dynamically. Even if a predetermined fixed interval is appropriate for the initial circumstances, it will cease to be so as conditions vary. An autonomic management mechanism is proposed for dynamically controlling maintenance scheduling, by adapting the interval between maintenance operations, in response to changing conditions. The proposed management mechanism is governed by a policy which has the objective to detect and correct unsatisfactory situations with respect to performance and resource consumption. Autonomic management may balance resource usage and performance better than a statically configured system.

This chapter is structured in the following way: in section 4.2 peer-set maintenance protocols of existing overlay networks are outlined. Related work on optimisation of P2P overlay networks in the presence of membership churn is discussed in section 4.3. This is followed by section 4.4 where the problems of current approaches are summarised and autonomic management is introduced to address the outlined problems. In section 4.5 an autonomic manager for the scheduling of maintenance operations in a P2P overlay is proposed. This is followed by some conclusions in section 4.6.

## 4.2 Background

Existing P2P overlay networks like Tapestry [96, 91], CAN [74], Pastry [81, 57] and Chord [85] define protocols for updating their peer-sets when new nodes join the overlay network. Additionally, individual nodes in each overlay periodically probe for their peers to detect failed nodes, or changes in the key space, in order to trigger peer-set update operations.

None of them provides a mechanism to adapt the interval between maintenance operations dynamically.

Because a Chord implementation is used for the work described in this thesis, Chord is explained here in some detail. Each node is assigned a unique key and the nodes form a logical ring, ordered on key values. Each key value $k$ in the key space is mapped to a node succeeding $k$ in the key space. The peer-set of a node comprises the addresses of:

- Its *predecessor* and *successor* in the ring.
- A list of nodes following its successor (known as the *successor list*).
- A list of nodes (known as the *finger table*) at specific distances around the ring.

Figure 4.1 shows nodes 5 to node 10 which are associated with specific keys in the circular key space. Pointers originating from node 6, indicate its fingers (node 9 and node 10), successors (node 7 and node 8) and the predecessor (node 5).

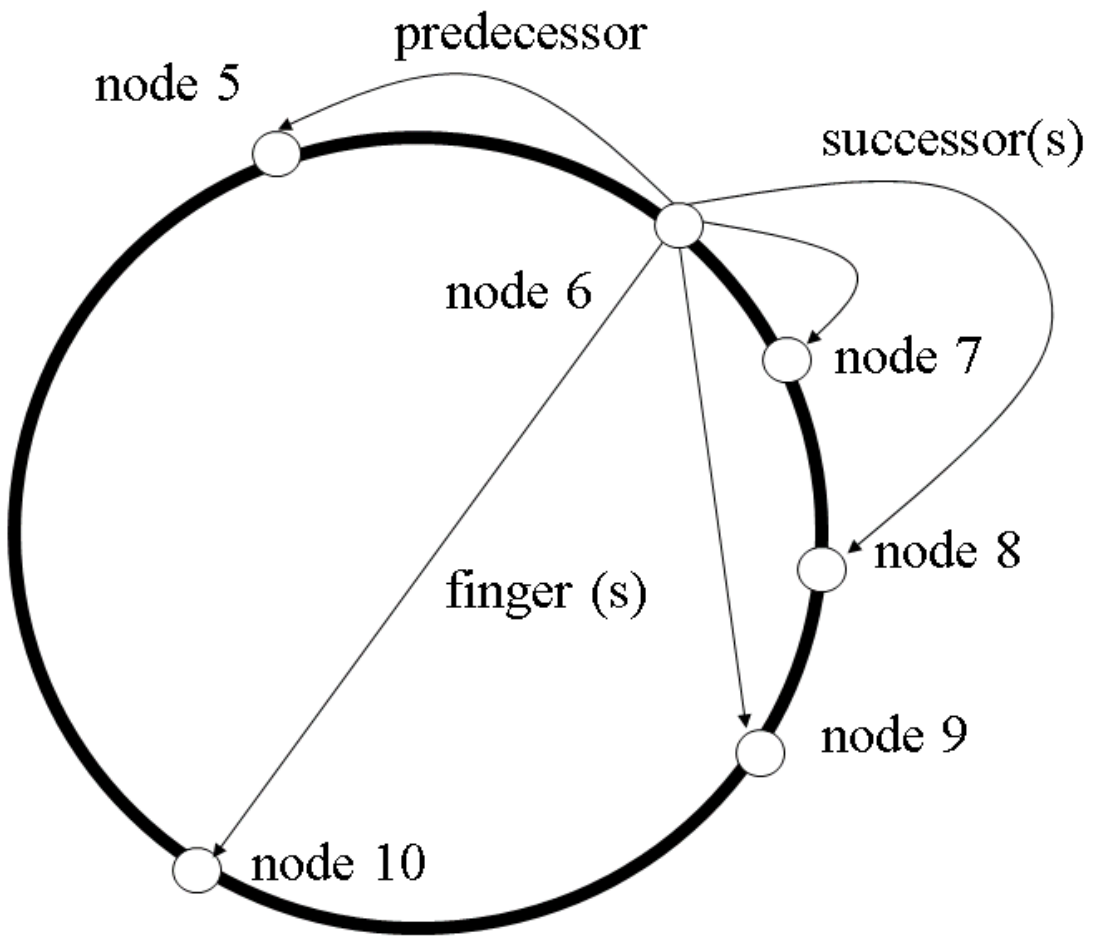

Figure 4.1: A Chord node's peer-set, in a simplified representation of the key space.

The predecessor and successor are used to form the ring. The successor list allows repair of the ring if the successor fails. The finger table is used to support scalable routing. To lookup a given key $k$, a node $n$ determines whether $n$'s successor is associated with $k$. If that is not the case, $n$'s fingers are used to query the finger identified by the closest preceding key to the given key $k$. Applied to figure 4.1, this means that, if node 6 looks up a key $key$ which is associated with node 10, it first checks if its successor (node 7) is associated with $key$. As this is not the case the closest preceding finger, with respect to the querried key, (node 9) is queried which identifies node 10 as being associated with $key$ and thus successfully resolves the lookup. In a network with $N$ nodes this means that in average $log(N)$ steps are required to resolve a lookup.

In order to successfully resolve lookups the peer-set needs to be valid, even if the network membership changes. In the original Chord maintenance protocol, the fingers are periodically verified by an operation termed *fixFinger*. This is done by each node by looking up the node associated with the key at a specific distance from the executing node. Additionally each node executes *stabilize* periodically to maintain its successor list and predecessor. When an error in the current peer-set is detected – due to a change in network membership – a more appropriate reference is established. During the *stabilize* operation a node $n$ verifies whether a new node has joined between $n$ and its successor. If the predecessor of $n$'s current successor is not $n$, a new node may have joined between $n$ and $n$'s successor and $n$ updates its successors. After that $n$ calls *notify* to tell the new node that $n$ may be its predecessor.

## 4.3 Related Work

This work investigates how performance and resource consumption can be improved in structured P2P overlays. The problem this chapter addresses is that an optimal maintenance interval in P2P overlays cannot be predicted and thus has to be adapted dynamically. Some related work was identified which shares aspects of this work's high-level objective of optimisation with respect to performance and resource consumption and provides similar approaches (section 4.3.1). Other related work was identified which shares the high-level objectives but addresses them via different approaches, retaining statically configured maintenance intervals (section 4.3.2). All of the latter potentially face the same problem which is that an ideal static interval cannot be predicted.

### 4.3.1 Dynamic Control of P2P Overlays

The work which is most closely related to this is based on the Pastry overlay [57]. The authors propose to improve both performance and resource consumption in structured P2P overlays by dynamically adapting the interval between maintenance operations. The optimisation focus however lies only on the performance, because they propose to adapt the maintenance interval in order to achieve a specific minimum performance. Thus, their manager will not change the maintenance interval once this minimum performance is achieved. That means that in a situation in which churn keeps decreasing and the minimum performance is reached, their manager ceases to increase the maintenance interval. In such a situation an increasing maintenance interval correlates directly with decreasing unneces-sary use of network resources. Thus resource usage is only improved up to a fixed point by their manager. Further comparison of [57] and the approach of this thesis is provided in chapter 7 which reports on the experimental evaluation of the approach introduced here.

[7] focuses on Chord's *stabilize* operation and is based on a *churn estimation mechanism*. Although this is introduced as a first step towards a self-tuned maintenance mechanism, the dynamic adaptation of maintenance intervals is not specified nor evaluated. Their churn estimation mechanism is based on an analytical model into which data gathered during *stabilize* executions is fed. A potential problem of this approach is the scope of the input data for the churn estimator. As only data which is gathered during maintenance operations is processed, a high interval may result in out-dated information being used for the churn estimator. Thus, this approach potentially results in estimations of the degree of churn which do not represent the current situation in the case illustrated above.

### 4.3.2 Static Control of P2P Overlays

In [8] an adaptation of structured P2P overlays based on the Kademlia [59] overlay is proposed. The maintenance interval which governs Kademlia's maintenance operations is not adapted dynamically. Instead an additional maintenance mechanism is developed which improves performance, and increases or decreases maintenance overhead on top of the original maintenance overhead depending on the exhibited churn. In more detail: Kademlia is a combination of a P2P overlay network and a Distributed Hash Table (DHT). Kademlia's P2P routing protocol maintains a node's routing state lazily. Its DHT component, however, executes a periodic replica maintenance operation, during which potential routing-state errors are repaired. In [8] a modification is proposed based on a distributed data structure for maintaining failed node addresses. As the number of failed node addresses in this data structure corresponds with the membership churn, the maintenance overhead for keeping this list up-to-date correlates with the membership churn. Thus the resulting network usage due to maintenance varies dynamically but the maintenance interval is not adapted.

In [94] an additional layer to a Chord overlay is proposed. Here again the maintenance intervals are not adapted but an additional maintenance mechanism is introduced. This is based on a second overlay layer for finger table maintenance. This layer consists of a manually selected set of *super nodes* with better stability characteristics than the average P2P participants. P2P nodes which are not in the maintenance layer do not maintain their finger tables. However, they periodically maintain their immediate neighbours in the key-space. The *super nodes* are informed about failed nodes, which triggers a distributed finger-table repair process.

In the FS-Chord project [43] a two-step joining protocol for Chord is proposed to reduce maintenance overhead and to improve stability. The modified joining protocol is based on the following: a node that has sent a join request is only accepted as an overlay node after a fixed time during which the new node's availability is monitored. If the new node does not fail during this time it is granted permission to fully join the overlay network. This may stop unstable nodes from joining the overlay network. Subsequently any network usage caused by maintenance work due to the unstable node is saved. This approach does, however, not enable Chord to adapt to a change in the environmental conditions after the monitoring period has passed.

In [9] a model of a Chord network is developed which shows that increasing the size of the successor list in a Chord overlay network improves stability. The authors suggest the dynamic adaptation of the successor list length. This mechanism is however not further specified or evaluated. Even though their proposal can be considered as a dynamic adaptation of the peer-set it is not based on the same principles as the approach introduced here. It may increase the stability of a Chord ring up to a certain level, but may also reach the point where maintenance is not executed frequently enough.

In [51] a number of modifications of Chord are introduced to improve stability in the presence of membership churn. None of the modifications is however evaluated. They comprise periodic rejoins to maintain the structure of the Chord overlay network in the event of node failures. Additionally, a modified lookup algorithm which improves the stability in the event of failed successors is proposed. This involves nodes discovered during lookup operations being used for peer-set maintenance. It also involves the suggestion to decrease the *stabilize* interval when errors are detected, but no rule is specified for increasing the interval.

In [52] it is experimentally evaluated whether a modified *stabilize* algorithm can improve stability in a Chord network in the presence of high churn. A modification of the Chord protocol is made which maintains a list of predecessors and successors. The interval however is kept at a fixed value. It is suggested that a small interval is desirable in networks with high membership churn and thus stated that there is a correlation between the interval and the degree of churn. However, it is not proposed to dynamically adapt the maintenance interval.

In [53] Chord is compared with an overlay network based on a hierarchical grouping schema. The maintenance mechanism of the hierarchical groups overlay network is not further specified. The authors experimentally evaluate how different *stabilize* intervals affect the lookup error rate in Chord but do not evaluate or propose dynamic adaptations.

In [55] Chord's maintenance mechanism is analysed and it is concluded that its execution rate is an important configuration factor. The authors analyse the correlation between maintenance rate and performance, stress the importance of conservative network usage and ask the question whether an optimum maintenance rate can be learned. An analytic model is developed for computing a lower bound maintenance rate given the time interval during which 50% of the P2P participants leave an overlay network (this time is referred to as half-life).

In [54] Chord joining and leaving protocols are modified to execute faster. A positive aspect of this approach is that nodes may be able to fully establish a valid peer-set sooner than in the original Chord protocol. Its applicability may be limited as the authors assume a fault-free overlay network in which nodes leave only voluntarily.

## 4.4 Problem Definition

Unsatisfactory situations with respect to performance and resource consumption can be identified in a statically configured P2P overlay. An optimum interval between peer-set maintenance operations depends on dynamic conditions and cannot be predicted and configured statically. Such a static configuration is used in the P2P overlays outlined in section 4.2 and in most of the modified overlays as outlined in section 4.3. In the following section the problems involved in statically configuring peer-set maintenance intervals are discussed.

### 4.4.1 Evaluation Criteria

A user of an application like a decentralised distributed storage system may never interact directly with the P2P overlay network, but nevertheless the P2P overlay may have a large impact on the application's perceived performance and its resource consumption. The fundamental operation (lookup) of a P2P overlay, which implements KBR, is to return a host for a given key. In decentralised distributed storage systems, using KBR, at least one lookup operation will be executed every time a data item is accessed. The time it takes to resolve a lookup request (*lookup time*) is therefore a performance related metric. In the case where the lookup fails it is assumed here that it is repeated after a specific time (*lookup error time*) by some lookup fall back mechanism. From a user's perspective lookup failures thus increase the time until a lookup is successfully resolved. Therefore the lookup time, the lookup error rate and lookup error time determine the performance of a P2P overlay. Addi-tionally network usage may be of interest for a user if the user is billed for used resources. It is also of interest for a developer who may be concerned with enough resources being available for other applications [55] or for reasons outlined in chapter 1. Performance and network usage are referred to as *user-level metrics* in the rest of this work. Both user-level metrics are considered as average measurements over a specific observation period.

### 4.4.2 Discussion and Hypothesis

Following the definition from section 4.4.1, the performance of a P2P overlay improves as the proportion of valid entries in the peer-set of individual nodes increases. Invalid entries may result in wrong routing decisions or in lookups failing completely due to nodes which have left the overlay. Thus invalid entries may require repetitions of lookup operations. The objective of maintenance operations is to provide valid peer-sets, which happens at the cost of some network usage.

In a situation in which no membership churn is exhibited, a fixed short interval between maintenance operations may be disadvantageous with respect to network usage. Here the network usage represents pure overhead and thus wasted maintenance effort as the key to node association does not change. A fixed long interval between maintenance operations instead may be beneficial because of reduced network usage. Additionally a long interval potentially frees computational resources for lookup operations and is thus also beneficial with respect to performance, especially if a workload with frequent lookups is executed.

Conversely if high network membership churn is exhibited it is beneficial for maintenance intervals to be short. In such situations the node to key association changes frequently. This requires the maintenance operations to be executed more often in order to provide a valid peer-set. If the maintenance interval is long the peer-set may become out of date and is thus disadvantageous with respect to performance. Here maintenance effort is required and network usage becomes justifiable. It however, ceases to be justifiable if no workload is executed as this does not require good performance from a user's perspective.

The optimum maintenance interval depends thus on the workload and the membership churn which cannot be predicted. It is hypothesised that autonomic management (as explained in chapter 2) may be able to detect whether network usage is wasted, due to unnecessarily executed maintenance operations, and to increase maintenance intervals as an appropriate reaction. Autonomic management may also be able to detect if maintenance operations are required and to decrease intervals as an appropriate response. Thus it may be able to discover an optimum maintenance interval that cannot be predicted statically, and it may also be able to adjust the interval dynamically to keep it near an optimal value in the presence of dynamic variations in workload and membership churn.

## 4.5 Autonomic Management in P2P Overlays

### 4.5.1 Overview

To achieve the behaviour envisioned in section 4.4.2 an autonomic manager is proposed whose aim it is to dynamically adapt the maintenance interval in response to varying con-ditions. A single manager is added to each individual node in a P2P overlay. The manager is based on the autonomic control loop outlined in section 2.3. The current situation is modelled with *metrics* which are extracted from monitored *events*. The response to a specific situation is specified by *policies* and applied via *effectors*.

The manager's underlying principle is to detect when maintenance effort is wasted and to increase the current maintenance interval as an appropriate response. The manager can also detect when more maintenance is required and the interval should be decreased as an appropriate response. It balances out these two competing priorities by averaging all responses. Each of a set of *sub-policies* determines a new maintenance interval with respect to a specific metric. All intervals determined by sub-policies are averaged by an *aggregation-policy*. Policy evaluations trigger the extraction of metrics values from the monitoring data. Only events which have not been considered since the last policy evaluation are extracted. The policy evaluation interval thus determines the *observation period* with respect to events and metrics. Only locally generated events are processed and thus no additional network traffic is generated when adding this manager to the P2P node.

### 4.5.2 Autonomic Management Components

#### Metrics

**Non-Effective Maintenance Operation (NEMO) Metric:** The NEMO metric models the amount of effort invested in maintenance operations without effect. A metric value (denoted as $METRIC.value$, wherever appropriate) represents the number of events which indicate that a maintenance operation did not update the maintained peer(s) during the analysed observation period. A high value suggests that a lot of network usage took place unnecessarily as either little change in the key to node allocation was detected, or maintenance was executed frequently enough to compensate for such changes. Conversely, a small value implies that the network usage due to maintenance operations was effective and could therefore be regarded as justifiable.

**Error Rate (ER) Metric:** An ER metric value represents the number of failed accesses to maintained peers during the analysed observation period. A high value suggests that a large proportion of the peer-set was not valid (meaning that it was out-of-date) when being accessed during recent lookup operations. Conversely, a low value suggests that the peer-set was valid due to frequent maintenance. Additionally a low value could also mean that no lookups (workload) were executed, or that the churn was low.

**Locally Issued Lookup Time (LILT) Metric:** The LILT metric models the speed of locally issued lookup requests. A LILT value is represented by the mean of all lookup times issued during the analysed observation period. A high value suggests that poor routing decisions, due to an invalid peer-set, increased the hop count and thus the time until lookups were successfully resolved. Conversely a low value suggests that routing decisions and thus the peer-set were valid and resulted in lookups completing in short times.

**Policies**

An aggregation-policy combines the responses of sub-policies by computing the mean of the intervals specified by the sub-policies. Each sub-policy determines a new interval with respect to a single metric in isolation. The general structure of the management process, with respect to a sub-policy is based on a negative feedback loop [30]. The goals of the various sub-policies may be different and may even conflict with each other; however, the aim is that they balance each other out after some time. Each sub-policy determines a new interval which differs from the current interval, by a magnitude ($P$) which is proportional to the difference of the metric value under consideration from its ideal value. As each metric models a specific aspect of the current situation the goal of each sub-policy is to put the system in an ideal situation, with respect to this aspect. Additionally, the aggregation-policy specifies the immediate execution of a maintenance operation if an error is detected.

- In an ideal situation no non-effective maintenance operations are monitored ($NEMO.value = 0$). The sub-policy using the NEMO metric always specifies an *increase* of the controlled interval as an appropriate response to non-ideal metric values.
- In an ideal situation no errors are monitored ($ER.value = 0$). The sub-policy using the ER metric always specifies a *decrease* of the controlled interval as an appropriate response to non-ideal metric values.
- In an ideal situation monitored lookup times are 0 ($LILT.value = 0$). The sub-policy using the LILT metric always specifies a *decrease* of the controlled interval as an appropriate response to non-ideal metric values.

A new value for an interval is determined by each sub-policy as defined in formula 4.1, where $\pm$ reflects that the NEMO sub-policy determines an increase of the controlled interval and the ER and LILT sub-policies a decrease.

$$new\ interval = current\ interval \pm current\ interval \times P \tag{4.1}$$

Factor $P$ (formula 4.2) determines the proportion by which the interval is changed. Its underlying principle is to specify that the magnitude of change correlates with the difference between the observed metric value and the metric's value in an ideal situation.

$$P = 1 - \frac{1}{\frac{metric.value - t}{k} + 1} \tag{4.2}$$

Formula 4.2 assumes $metric.value$ to be greater than $t$, otherwise $P$ is defined to be $0$. Factor $k$ is a constant non-zero factor which controls the rate of change of $P$ with respect to the metric value. This allows to specify how rapidly an interval is changed depending on the difference between the current metric value and its ideal value. $t$ specifies a threshold for metric values below which no change to the interval results. When applying the underlying principle to the the error rate, that means that a small error rate will result in a small decrease, and a large error rate in a large decrease of the controlled interval. To do so, the current metric value is mapped to a value between 0 and 1 by the term $\frac{1}{\frac{metric.value-t}{k}+1}$ in formula 4.2.

To illustrate the effect of various $k$ values, figure 4.2 shows the progression of the proportion $P$ by which an interval is changed with respect to values of the ER metric.

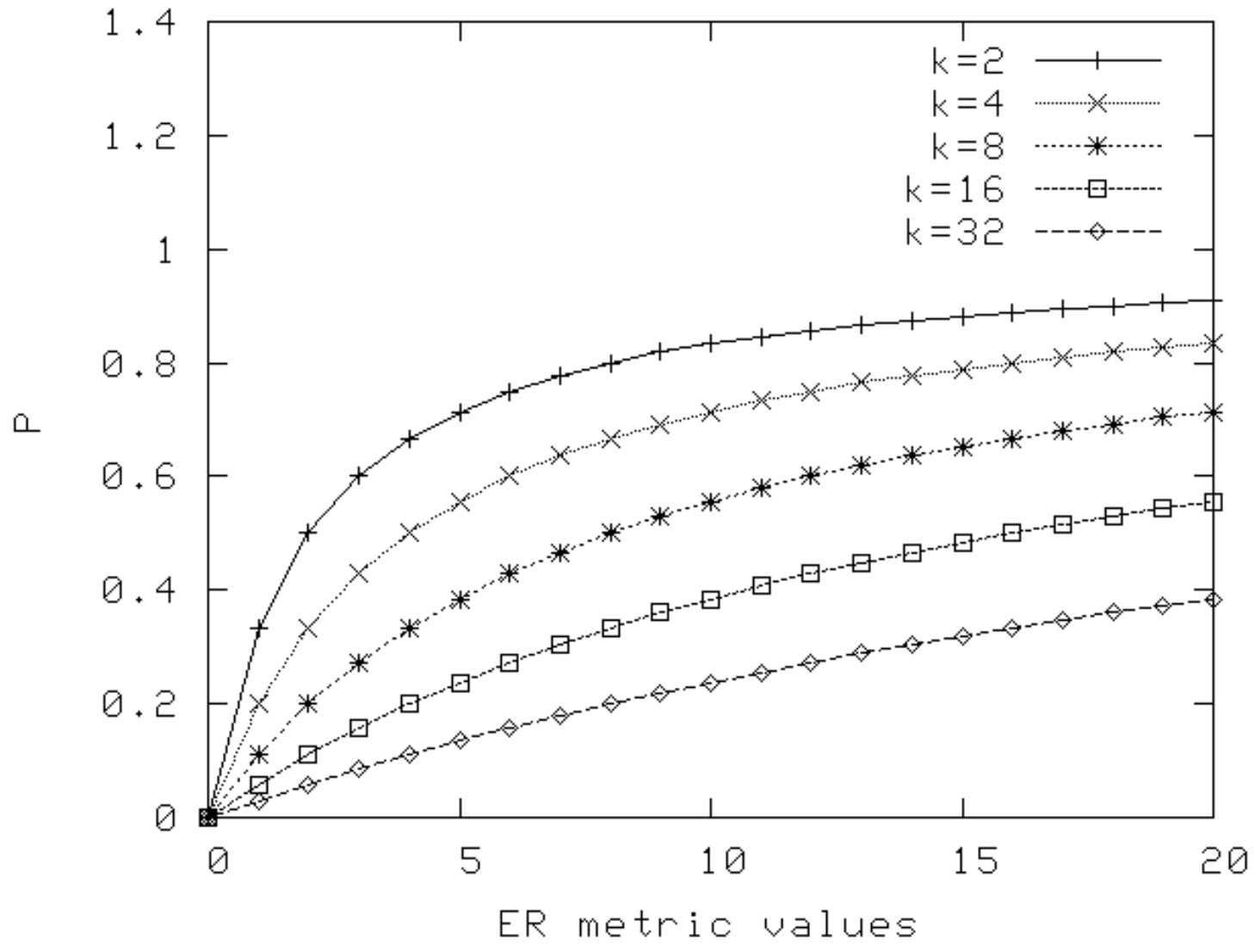

Figure 4.2: Relationship between $P$ and ER values for various values of $k$.

As illustrated, lower $k$ values yield more rapid policy responses. The policy configuration parameters $t$ and $k$ allow this approach to autonomic management to be used to build various differently behaving autonomic managers.

## 4.6 Conclusions

This chapter investigated the scope for optimising a P2P overlay with respect to performance and resource consumption by autonomically controlling the maintenance scheduling. Related projects identified in section 4.3 aim to improve a P2P overlay's performance and resource consumption by modifications of the peer-set maintenance protocols. The work on Pastry is based on dynamic adaptation of the maintenance intervals, but this work fo-cuses on performance and does not improve resource consumption beyond a specific point. All other related projects aim to improve a P2P overlay's performance and resource consumption by maintenance protocol adaptations in which the intervals remain statically configured. This makes all of them prone to the problems discussed in section 4.4 which are addressed by autonomic management as introduced in section 4.5. This autonomic manager could be applied to a wide range of P2P overlays. Furthermore it has the potential to be easily adapted to manage the scheduling of any periodic maintenance operation in response to varying demand.

# Chapter 5

# Optimisation of Data Retrieval Mechanisms in Distributed Storage Systems

## Outline

In this chapter the scope for optimisation via autonomic management of a distributed storage system is investigated. The effects on performance and resource consumption of various degrees of concurrency (DOC) in a distributed store client's data retrieval mechanism are analysed with the help of an analytical model. An autonomic management mechanism is proposed with the aim of identifying and correcting disadvantageous situations by dynamically adapting the DOC.

## 5.1 Introduction

As outlined in chapter 3, ASA and other distributed storage systems such as *CFS* [13], *Ivy* [66], *Past* [21], *Koorde* [42], *ConChord* [3] and *Google-FS (GFS)* [27] replicate data items on some number, $R$, of storage hosts in order to improve availability and resilience. This enables a requesting client to either fetch one replica from a specific host, or to concurrently fetch up to $R$ replicas from different hosts. For the rest of this document, the number of concurrently initiated fetch operations is referred to as the *Degree of Concurrency (DOC)*, which ranges from $1$ to $R$.

In this chapter the scope for optimisation with autonomic management of the DOC with respect to performance and resource consumption is investigated. It is hypothesised that, depending on various conditions, either a high or low DOC is more advantageous. Thus dynamic adaptation may yield an overall benefit in comparison to a statically configured DOC.

Performance, in this context, is defined as the average time for completing users' *get requests*[1]. Resource consumption is defined as the amount of data sent to the network by all storage hosts involved in any *get* request. Both are of interest for a user and are therefore also referred to as *user-level metrics (ULM)* for the rest of this thesis. Environment conditions are specified by the failure rate of *fetch* operations, by the degree of variation between *fetch times* and by the available network speeds. The state of a specific condition or combinations of various states cannot be predicted and thus an optimum DOC cannot be statically configured. Even if a DOC is initially optimal, it may cease to be so as conditions vary. Autonomic management (see chapter 2) may be able to adapt the DOC dynamically in the presence of changing conditions, or may learn an optimum DOC under unchanging conditions in order to improve performance and resource consumption.

[1] An individual *get* request initiates up to $R$ concurrent *fetch* operations of which the fastest determines the time for completing the *get* request – if no failures occur.

This chapter is structured in the following way. In section 5.2 background information is provided on the data retrieval mechanism under consideration. This is followed in section 5.3 by an outline of other approaches to the optimisation of distributed storage systems. In section 5.4 the problems this work addresses are discussed in order to support the development of an analytical model which shows the effects of various DOC settings and specific network conditions on performance and resource consumption. The model is used to illustrate different use cases and to argue for dynamic adaptation of the DOC. Following that, an autonomic management mechanism for dynamically controlling the DOC in a distributed store client is proposed in section 5.5. The chapter is summed up in section 5.6.

## 5.2 Background

In the previous section it is hypothesised that dynamically adapting the DOC in a distributed storage system's data retrieval mechanism in response to various conditions improves performance and resource consumption. To allow the DOC to be varied at all, a distributed storage system must exhibit the following properties:

- The store holds up to $R$ identical copies of the requested data item on different phys-

ical server hosts. A client is aware of which replica resides on which host.

- The client only needs to retrieve one out of *R* replicas successfully. This in turn implies that replicas must be self-verifying. That means that a client can itself verify that a retrieved replica is valid.

- The self-verifying capability of replicas involved in individual *get* requests allows all remaining fetch operations to be terminated after the first out of $R$ replicas is successfully retrieved. This results in no further network resources being consumed (unnecessarily) after the requested item is retrieved.

- A fall-back mechanism retries fetching a data item if the previous attempt failed.

The above properties allow the adaptation of the DOC in ASA, which shares these properties with a range of other distributed storage systems. Data retrieval mechanics in existing decentralised systems like *CFS, Ivy, Koorde*, and *ConChord* exhibit such properties[2]. Additionally the same properties can be found in centralised distributed storage systems like *Google FS*. A subset of the above properties can be found in *PAST*. This sub set could however be extended to the full set of properties as listed above by some simple modifications.

## 5.3 Related Work

All of the distributed storage systems outlined in section 5.2 exhibit properties which enable them to fetch up to $R$ replicas concurrently. However, in none of them is the DOC adapted to address disadvantageous conditions. For instance, in *CFS* [13] and *GFS* [27] any client determines, for an individual *get* request, the server from which it fetches one out of $R$ replicas based on a performance measure. On the other hand, in *Past [21]* the underlying P2P overlay *Pastry* [81] transparently prioritises well performing hosts during routing operations. That means that Pastry first routes to well performing servers which store a specific replica when the corresponding data item is requested. The approach towards improvement of performance taken in PAST allows various adaptations to be made. It can easily be modified to move the decision about which host to fetch a replica from, to the client (see CFS and GFS). Thus this section focuses on the former approach, which is referred to as a *Server Ranking Mechanism (SRM)* for the rest of this thesis.

[2]All are built on the P2P overlay Chord (see section 4.2).

The objective of such a SRM is to improve data retrieval performance by ranking servers based on predictions about which host will result in the shortest fetch time[3]. This is based on the assumption that historical monitoring data can be used to predict future performance of specific hosts. A SRM provides good performance as long as conditions are not changing in an unpredictable manner, which may then result in the selection of a potentially badly performing host. Another (obvious) disadvantage of a SRM is that specific network conditions may not be taken into account by the considered performance metric. This may result in nominating a server that is performing badly as a well performing server – for example, if latency is used as a performance metric and the network connection exhibits a good performance with respect to the latency but a bad one with respect to network bandwidth.

[3]Fastest servers are ranked first, failed ones at the end.

*DHash* is a generic block storage layer which is built using Chord. It is used in various distributed storage projects including CFS. The measurements utilised by DHash, as used, for instance in CFS, to determine the performance of a specific host, are based on latency. GFS decides on the performance of a host based on *IP* addresses. It is assumed in GFS that server hosts in the same subnet as the client are connected to the same network switch and thus potentially perform better than hosts outside this network segment. In Pastry a so-called *proximity metric*, which is based on the number of IP hops between two nodes, is used to determine the performance of a specific host.

## 5.4 Problem Definition

In section 5.1 the hypothesis was introduced that an autonomically adapted DOC yields benefits with respect to performance and resource consumption in comparison to a statically configured one. In section 5.2 and 5.3 it was outlined how existing distributed storage systems which exhibit properties that would allow an adaptation of the DOC, instead address disadvantageous situations with *server ranking mechanisms (SRM).*

The following unsatisfactory situations can be identified with respect to a static DOC in combination with a SRM. An unsatisfactory situation arises when the DOC is low and there is a large error rate for fetch operations or a large variation in the times taken to fetch replicas. The latter may cause a SRM to be unable to make accurate predictions. In this situation it is desirable to increase the DOC, because by retrieving more replicas in parallel a result can be returned to the user sooner. The converse situation arises when the DOC is high, no fetch failures are observed, there is little variation in the time it takes to fetch replicas and there is a network bottleneck close to the requesting client. In this situation it is desirable to decrease the DOC, since the low variation or error rate removes any benefit in parallel retrieval, and the bottleneck means that decreasing parallelism reduces both bandwidth consumption and elapsed time for the user.

In the following section both approaches towards the optimisation of distributed storage systems (SRM and DOC adaptation) are compared. An analytic model is developed to support this comparison. The analytical model shows how the time it takes to complete a *get* request (*get time*) depends on available network speeds, fetch failure rates and on the degree of variation between fetch times. Three cases are considered:

- A *low DOC*, in which the DOC is statically configured at a minimum (1) and no SRM is considered.
- A *high DOC*, in which the DOC is statically configured at a maximum ($R$).
- A *perfect SRM*, whose performance predictions are always correct; here the DOC is also statically configured at a minimum (1).

The time it takes to execute individual *fetch* operations, in order to serve a specific *get* request, determines the *get time* in each case. In cases in which a high DOC is statically configured, the degree by which parallel fetch operations have to share bandwidth increases individual fetch times. A statically configured low DOC also represents cases in which a high degree of variation in fetch times removes any benefit from a SRM as it is not able to predict performance accurately. A perfect SRM always selects the best performing host to fetch a specific replica.

All three cases are considered by the analytical model which is developed in the following section by: defining the parameter space; developing a formula to compute the fetch time; modelling the three cases above; and illustrating the differences between them in specific use cases.

### 5.4.1 Analytical Model

The analytic model is based on a simplified distributed storage system as shown in figure 5.1; it represents one distributed store client which is connected to $R$ servers (one per replica). Client and servers of the simplified system in figure 5.1 are referred to as participants

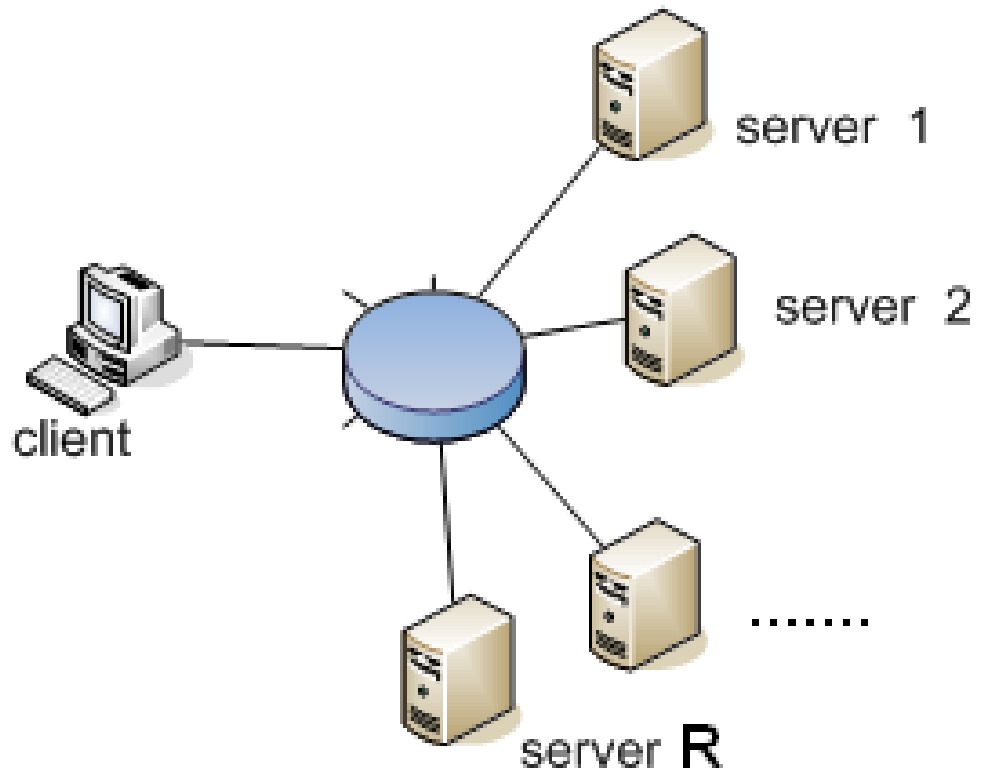

Figure 5.1: A simplified distributed storage system.

where appropriate. They are connected via some interconnection[4], whose internal network links are assumed to exhibit significantly higher bandwidth and lower latency in compar-

[4]the disc in the centre of figure 5.1

ison to any link[5] from a participant to the interconnection. Thus, the effect on fetch times when data is transferred across this interconnection is assumed to be negligible.

Model parameter overview:

- Replication factor: $R$ is the maximum number of replicas available (when no server has failed).
- Data item size: $S$ is the size of a replica. All replicas are assumed to have the same size.
- Degree of concurrency: $DOC$ is the number of concurrently initiated fetch operations. The maximum value for DOC is $R$ and the minimum value is $1$.
- Bandwidth: $BW_i$ is the available bandwidth between *participant i* and the interconnection.
- Latency: $L_i$ is the latency of the connection between *participant i* and the interconnection.
- Probability of failure: $P_{failure}$ is the probability that any individual fetch operation fails.

In each of the considered cases the *get time ($t_{get}$)* depends on the times taken by fetch operations (fetch time), which is the elapsed time from initiating a *fetch* operation until the replica is received. In cases in which a low DOC is statically configured, $P_{failure}$ represents the probability with which the fetch operation has to be repeated and thus increases $t_{get}$.

[5]the individual lines from a participant to the disc in figure 5.1

In any other case failures are either compensated for by perfect performance predictions or redundant parallel fetch operations. In cases in which a perfect SRM is used, the smallest fetch time (involved in a single get request) determines $t_{get}$. In cases with a statically configured high DOC the fetch time (and subsequently $t_{get}$) depends on the bandwidth available for individual fetch operations ($BW_{client}/DOC$) on the client link.

### Fetch Time

In each case the fetch time can be expressed with a formula which is developed in the following section by analysing the steps involved in a fetch operation, in temporal order[6]. The fetch time for server $i$ ($t_{fetch\ i}$) has three components:

- the time for requesting a replica from server $i$, $t_{request\ server\ i}$
- the time for the response from server $i$ to the interconnection, $t_{response\ server\ i\ link}$
- and the time for the response from the interconnection to the client, $t_{response\ client\ link}$

The fetch operation is initiated by a request for a replica to server $i$. The request itself has negligible size therefore only latencies are significant.

$$t_{request\ server\ i} = L_{client} + L_{server\ i} \tag{5.1}$$

The server then sends the replica back to the client via the interconnection. The time $t_{response\ server\ i\ link}$ which it takes the server to send the replica to the interconnection is

[6]in consideration of the simplified system illustrated in figure 5.1

determined by the size of the replica and the available bandwidth and latency on the server side, as shown in formula 5.2.

$$t_{response\ server\ i\ link} = \frac{S}{BW_{server\ i}} + L_{server\ i} \tag{5.2}$$

The last component of the fetch time is the time it takes to transfer the replica from the interconnection to the client $t_{response\ client\ link}$. The DOC specifies how many concurrent fetch operations share the available bandwidth between client and interconnection.

$$t_{response\ client\ link} = \frac{S}{\left(\frac{BW_{client}}{DOC}\right)} + L_{client} = \frac{S \times DOC}{BW_{client}} + L_{client} \tag{5.3}$$

The combination of formulas 5.1, 5.2 and 5.3 define the overall fetch time for server $i$, $t_{fetch\ i}$ as shown in formula 5.4.

$$t_{fetch\ i} = 2 \times L_{client} + 2 \times L_{server\ i} + S \times \left(\frac{1}{BW_{server\ i}} + \frac{DOC}{BW_{client}}\right) \tag{5.4}$$

**Get Time for the Static Low DOC Case**

In the case that a low DOC (1) is configured[7], the *get* request is initiated by fetching a replica from a randomly selected server $i$. If the replica is retrieved successfully the *get* request is completed. This fetch operation may however fail with a probability $P_{failure}$. If the fetch operation fails it is repeated with another randomly selected server. The fetch time resulting from the random selection is denoted as $t_{fetch\ rnd}$. For simplicity, it is assumed

[7] no SRM is considered

that a server fails after 100% of the replica is fetched. For a successful *get* request, fetch operations can be repeated up to $R - 1$ times. Thus the sum of all probabilistic cases determines the *get time* as shown in formula 5.5 (derived from formula 5.4).

$$t_{get} = t_{fetch\ rnd} + \sum_{k=1}^{R-1} (k \times t_{fetch\ rnd} \times P_{failure}^{k}) \tag{5.5}$$

### Get Time for the Static High DOC Case

In the case that a high DOC ($R$) is configured, the *get* request initiates $R$ concurrent fetch operations. Failures are compensated for by redundant concurrent fetch operations and thus the fastest successful fetch operation determines the *get time*. In accordance with formula 5.4, $t_{fetch\ i}$ is increased by the high DOC which reduces the bandwidth for an individual fetch operation at the client side.

Thus the fetch time is denoted as $t_{fetch\ min,\ DOC=R}$, and determines $t_{get}$ as shown in 5.6.

$$t_{get} = t_{fetch\ min,\ DOC=R} \tag{5.6}$$

### Get Time for the Perfect SRM Case

In the case that a perfect SRM is used and a low DOC (1) is configured, the *get* request initiates 1 fetch operation from the fastest server. Failures are compensated for by a perfect prediction and thus, again, the fastest successful fetch operation determines the *get time*. Note, that this takes the low DOC into account which does not reduce the bandwidth on the client side. Thus the fetch time is denoted as $t_{fetch\ min,\ DOC=1}$, and determines $t_{get}$ as shown in 5.7.

$$t_{get} = t_{fetch\ min,\ DOC=1} \tag{5.7}$$

### Use Case Scenarios

In the following the *get time* is analysed with all three of the above considered cases, with an increasing $P_{failure}$, and two different scenarios with respect to network speeds. In both scenarios a data item size $S$ of *500 [KByte]* and a latency $L$ on the server and client side of *100 [ms]* is defined. The replication factor $R$ is $4$ and for simplicity, all bandwidth and latencies on the server side $BW_{server\ i}$, $L_{server\ i}$ are equal.

In scenario 1 (figure 5.2) the bandwidth of each server ($BW_{server\ i}$) link is specified as *100 [Mbps]* and the bandwidth on the client side ($BW_{client}$) as *1 [Mbps]*. This represents a network bottleneck at the client side.

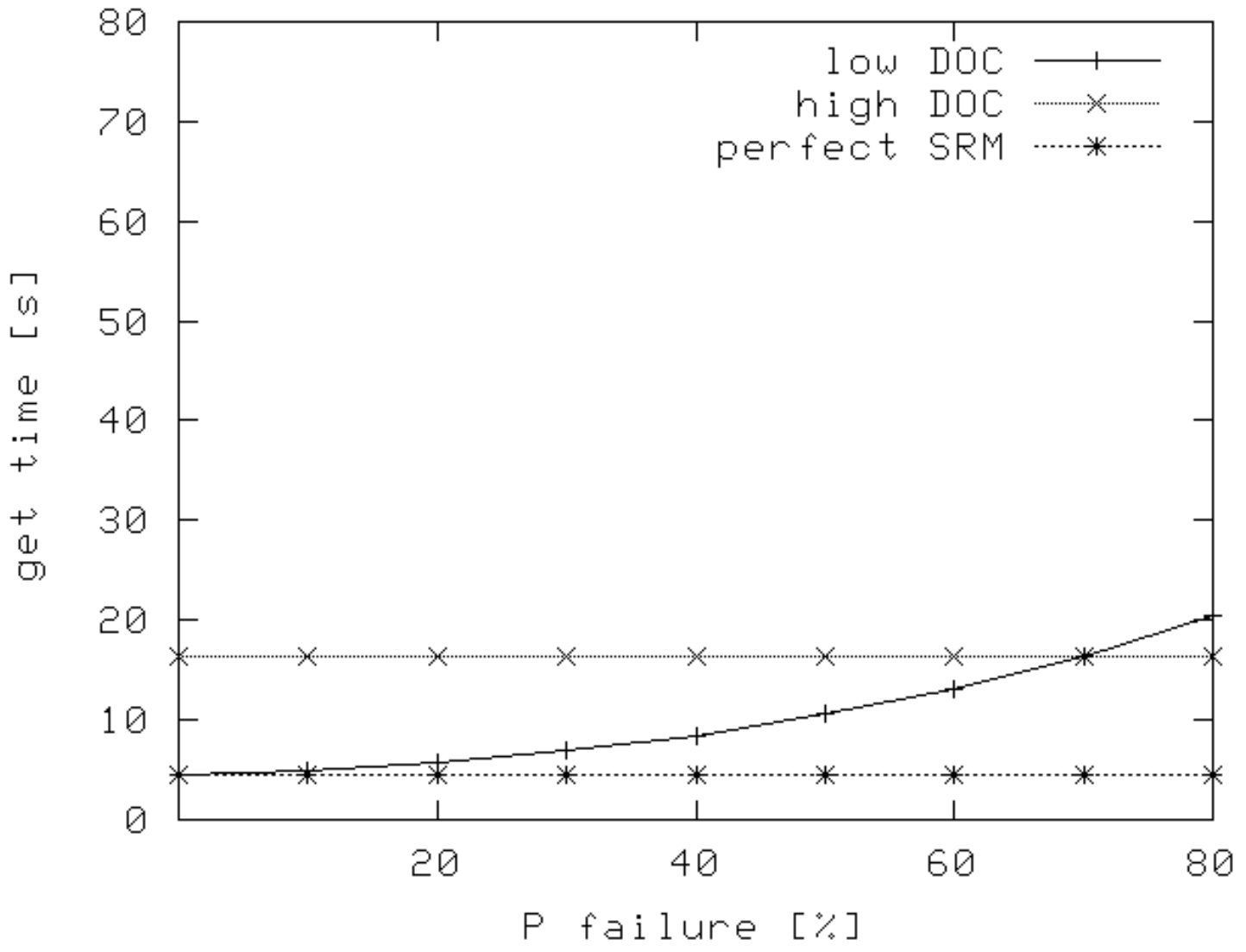

Figure 5.2: The effects of failures on the get time when a bottleneck exists at the client side.

In figure 5.2 it is shown that a low DOC results in shorter $t_{get}$ than a high DOC, for failure rates below 70 %. $t_{get}$ with a perfect SRM yields the same value as a statically low DOC, even with increasing $P_{failure}$ as it is assumed that it also detects failed hosts.

In scenario 2 (figure 5.3) the bandwidth of each server ($BW_{server\ i}$) link is specified as *1 [Mbps]* and the bandwidth on the client side ($BW_{client}$) as *100 [Mbps]*. This represents a network bottleneck at the the server side.

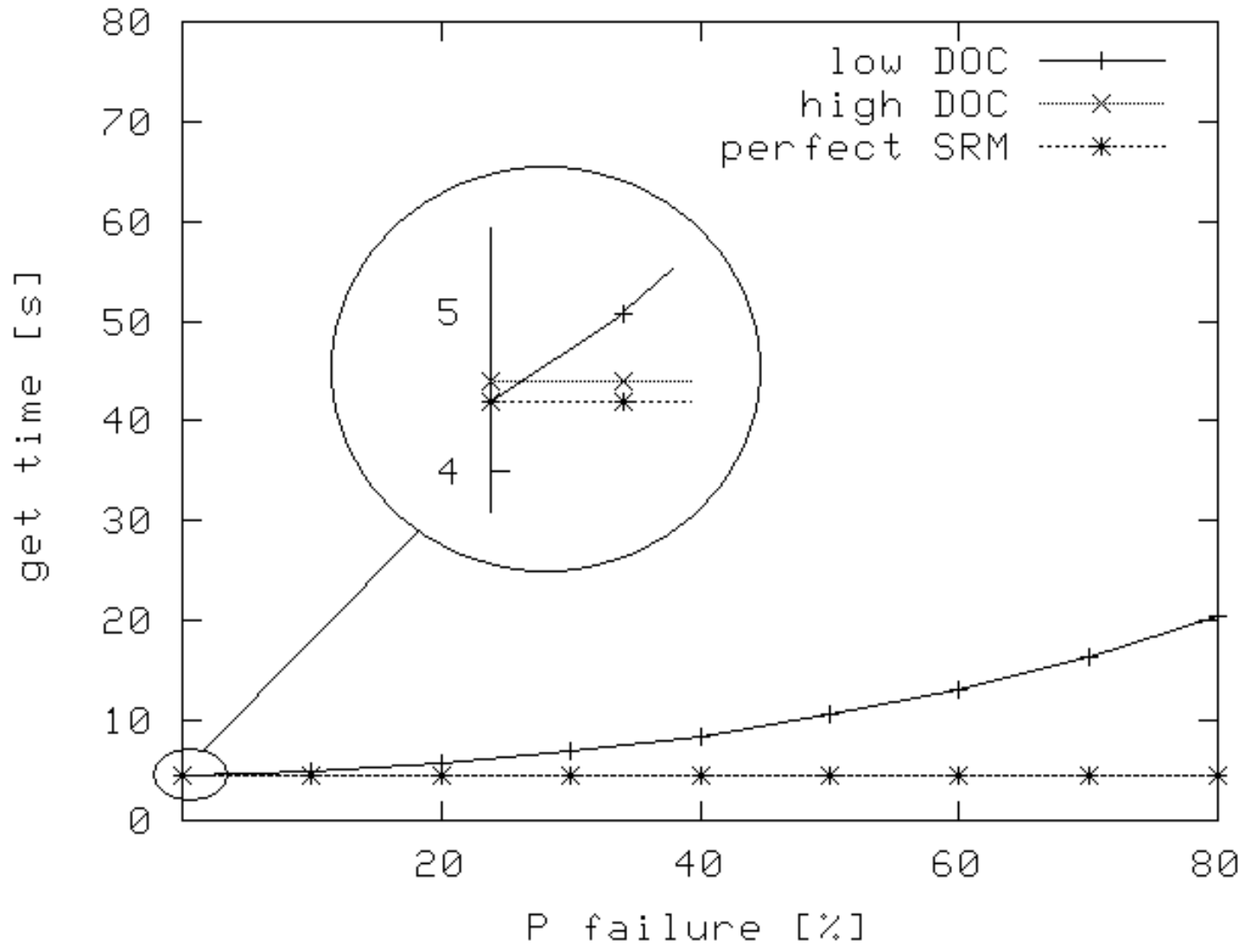

Figure 5.3: The effects of failures on the get time when a bottleneck exists at the server side.

In figure 5.3 it is shown that a high DOC results in shorter $t_{get}$ than a low DOC, for high error rates. $t_{get}$ with a perfect SRM yields the same value as a statically low DOC, even with increasing $P_{failure}$ as it is assumed that it also detects failed hosts.

### 5.4.2 Discussion and Hypothesis

In the previous section it is shown that situations can be identified in which a high DOC is beneficial with respect to the *get time* and others where a low DOC is beneficial. The effects of a SRM on the *get time* are analysed under the assumption that the SRM makes perfectly accurate predictions. Such predictions are however only possible if the available data on which the SRM's predictions are based is up-to-date and future conditions are similar to past ones. In the case of a high degree of variation of fetch times due to, for instance, a variation in the available network speeds, this data may cease to be valid. In this case the resulting *get time* will be similar to that for a client with a statically configured low DOC. Furthermore the degree of variation may require a high DOC to compensate for invalid predictions.

Even though situations were identified in which a high DOC yields benefits to $t_{get}$ it has disadvantageous effects on the network usage, which is proportional to the DOC. This represents an argument to decrease the DOC if it does not generate a disadvantageous situation with respect to $t_{get}$.

Thus performance and network usage of a distributed storage system depend on the DOC and on environmental conditions such as the degree of variation in fetch times, the failure rate of fetch operations and the available network speeds. None of those can be predicted statically, thus it is hypothesised that an autonomic manager may be able to adapt the DOC in the presence of changing conditions in order to yield benefits with respect to performance and network usage in comparison to a statically configured DOC.

## 5.5 Autonomic Management Applied to a Distributed Store Client

### 5.5.1 Overview

As hypothesised in section 5.4.2, an autonomic management mechanism may be able to detect various environmental conditions and to set an appropriate DOC in a distributed store client. An autonomic manager is proposed here for a distributed store client which works in combination with a SRM.

The autonomic manager is based on a generic autonomic control loop (see section 2.3) and reuses information gathered by the SRM in order to predict the performance of specific hosts. As the information provided by the SRM is considered as important for the manager, the scope of this information is now briefly sketched. As part of SRM processes, every *participant* in a distributed storage system periodically evaluates latency and bandwidth on its link to an interconnection (see figure 5.1). Since a client cannot measure server-side conditions directly, the SRM on a *client* periodically gathers latency and bandwidth information from servers. In order to combine bandwidth and latency measurements, the gathered data is used to compute an *expected data transfer time*. This is representative of how long it would take to transfer a replica of a given size across links with such bandwidth and latency. The expected data transfer time serves as a ranking metric for each server's performance with respect to the SRM.

The autonomic manager runs locally on each client. It extracts timestamped monitoring data from the SRM and other locally monitored events. These events are used to model specific aspects of the situation during each specific observation period via metrics. Metrics values are extracted to represent:

- the currently used DOC;
- the fetch operation's failure rate;
- the variation between recent fetch times;
- and the location of any network bottleneck[8];

The manager's policy decides on the magnitude by which the DOC should be varied depending on the currently used DOC, the variation between fetch operations, the failure rate and the location of the bottleneck.

- If the DOC is currently low, and the failure rate or the variation between recent fetch operations are high, the policy determines an increase of the DOC (to the maximum if there is a bottleneck on the client side).
- If the DOC is currently high, and the failure rate or the variation between recent fetch operations are low, the policy determines a decrease of the DOC.

[8]A bottleneck on the client side, for instance, is specified by significantly less bandwidth being exhibited on the client link than on the server link (see figure 5.1).

### 5.5.2 Autonomic Management Components

Metric values represent aspects of the state of the system in a specific observation period. Beside the monitoring information reused from the SRM, locally monitored data is used to compute the specific metrics. This means for the autonomic manager itself no monitoring data is gathered which would require the use of network resources. Here follows a brief description of how the specific metrics are defined and how they are extracted from the monitoring data.

#### Metrics

**Current DOC Metric:** This metric represents the value of the DOC. To be consistent with all other inputs to the policy this is modelled by a metric.

**Fetch Failure Rate (FFR) Metric:** The fetch failure rate metric is the ratio of failed fetch operations to the total number of initiated fetch operations.

$$FFR = \frac{number\ of\ failed\ fetch\ operations}{total\ number\ of\ fetch\ operations}$$

**Fetch Time Variation (FTV) Metric:** The FTV metric represents the degree of variation between all recently computed expected data transfer times, EDTT (generated by the SRM). It is specified by the standard deviation ($\sigma$) normalised by the mean ($\mu$) of EDTT in the specific observation period.

$$FTV = \frac{\sigma_{EDTT}}{\mu_{EDTT}}$$

A high FTV suggests that there is a large variation between fetch times.

**Bottleneck (BN) Metric:** As for the FTV metric, SRM monitoring data is used to compute the BN metric. Bandwidth is the limiting factor when sharing links between concurrent fetch operations. Therefore SRM server and client bandwidth measurements are used to compute the BN metric. The BN metric is the ratio of the mean ($\mu$) of all bandwidths monitored on the client side to the mean of all bandwidths monitored on server side links.

$$BN = \frac{\mu_{client\ bandwidths}}{\mu_{server\ bandwidths}}$$

A BN value smaller than one suggests that there is a bottleneck on the client side.

**Policy**

A policy controls the DOC depending on the current DOC, the FFR, the FTV and the BN metric values. The policy is defined as follows:

- If the DOC is currently low, and FFR and FTV are high, increase the DOC to the highest possible value.
- If the DOC is currently low, and either FFR or FTV is high, incrementally increase the DOC.

- If the DOC is currently high, and FFR and FTV are low, incrementally decrease DOC.
- If the DOC is currently high, and FFR and FTV are low and the BN determines a bottleneck on the client side, set DOC to the minimum value (1).

**Server Ranking Mechanism (SRM)**

The objective of a SRM is to rank specific servers, based on a prediction of how fast each server, involved in a specific *get* request, will transfer data to a client. The prediction is based on periodically gathered monitoring data. A SRM is proposed here based on monitored network bandwidth and latency of the specific server links (as illustrated in figure 5.1). The address of the host in the interconnection which lies closest to each participant is used to determine the participant's latency ($L$) and bandwidth ($BW$), which are gathered periodically. In order to combine bandwidth and latency measurements the gathered data is used to compute an expected data transfer time (*EDTT*) for a data item of size $S$. This EDTT serves as ranking metric for each server's performance.

$$EDTT = L + \frac{S}{BW}$$

The lower the EDTT associated with a specific server, the better its ranking. At each *get* request the SRM compares expected data transfer times for all involved servers.

## 5.6 Conclusions

This chapter investigated the scope for optimising a distributed storage system's data retrieval mechanism by autonomically controlling the DOC. Related work on the optimisation of data retrieval mechanisms and their shortcomings was identified and analysed in section 5.3 and 5.4. Existing systems use Server Ranking Mechanisms (SRM) to select a specific host for fetching a replica. To address identified problems of this approach an autonomic manager is introduced in section 5.5, which dynamically adapts the DOC and works in combination with a SRM. This autonomic manager is envisioned to work for any system which exhibits the properties outlined in section 5.2.

# Chapter 6

# A Generic Autonomic Management Framework

## Outline

This chapter introduces the requirements for an autonomic management framework for the experimental evaluation carried out in this thesis, and concludes that existing tools and frameworks are inadequate for the task. It introduces a new framework which has been developed as part of this thesis. The framework is designed to be generic to allow it to be used outwith this thesis.

## 6.1 Introduction

This thesis reports on the investigation of whether distributed storage systems can be improved with autonomic management. A tool was needed for implementing an autonomic manager for the distributed storage system ($ASA$) considered in this thesis. The tool was envisioned to fulfil the following requirements:

- ASA and other distributed storage systems (chapter 3) consist of several layers of abstraction. Therefore the tool was required to be generic enough to be used for all layers.
- It was required to reduce complexity and to ease the process of developing an autonomic manager. This means that it should allow a developer to focus on developing system-specific management components rather then generic management processes, without any complex configuration of the tool itself.
- It was required to be lightweight, to have little impact on the runtime performance of the target system and to allow management components to operate in the same address space as the target system.
- With regards to this research project it was required to allow development of components in *Java*.

Available tools do not meet all these requirements, therefore a *generic autonomic management framework (GAMF)* was developed (in *Java*) and introduced in this chapter.

In this chapter shortcomings of available tools for applying autonomic management to a target system are briefly outlined with respect to the above requirements, in section 6.2. The design of the GAMF is explained in section 6.3. The chapter finishes with a summary and information about features beyond the scope of this thesis in section 6.4.

## 6.2 Related Work

The J2EE server manager [95], *vGrid* [46], *AutoMate* [1], *k-component* [20, 19], *IBM autonomic computing toolkit* [11, 60] and *Accord* [56] were all considered as either domain-specific, heavyweight or unnecessarily complex with respect to this work. These tools only offer interfaces to manage specific resources in a grid system or for a J2EE server, therefore they were considered as not sufficiently generic for the work in this thesis.

AutoMate and Accord were considered as too heavyweight and complex for the work carried out as part of this thesis. Both are expected to have a significant impact on the target system's runtime-performance as they contain P2P overlay networks for discovering components in a distributed system, and mechanisms which extract policies from *XML* formatted configuration files. These additional mechanisms also make the use of those frameworks overly complex.

The *IBM autonomic computing toolkit* allows the application of autonomic management using a wide variety of built-in interfaces which are however limited in their scope. Customisation of such interfaces could allow this framework to be considered as generic. These interfaces however require the manager to operate in a different address space to the target system. This and the complex machinery which comes with this tool impose a significant load on the target system. Thus the IBM autonomic computing toolkit was considered as too heavyweight for this work. Additionally it was considered as too complex for the work carried out as part of this research as it requires the use of a customised IDE, as well as the configuration of the entire machinery that comes with it.

For similar reasons, the *k-component* architecture was not considered in this work. It provides a *C++* library which defines an *Adaptation Contract Description Language (ACDL)* to specify how a controlled component is adapted by some control mechanism. A *Collaborative Reinforcement Learning* methodology is used to gather information from remote components via *CORBA*. Component meta-information, used for the ACDL, is stored in separate files in XML format. This meta-information is generated via the use of custom configuration tools.

A novel approach to describe and specify controlled systems and high level policies via *contracts* is given in the *Architectural Artefacts for Autonomic Distributed Systems (A$^3$DS)* project [86]. It provides concepts and tools to specify controllable properties of target systems and contracts for how these properties may be controlled. It also provides a specification language and a code generation application which transforms the high level policy or property description into Java code. A$^3$DS does not provide an autonomic manager implementation and is thus not considered in this work.

It was decided to address the above limitations by developing a *Generic Autonomic Man-*

*agement Framework (GAMF)*. This should incorporate the requirements stated earlier (see section 6.1).

## 6.3 Design

### 6.3.1 Overview

The *GAMF* provides a generic control mechanism based on an autonomic control loop (see figure 2.1) and a set of interfaces to allow interaction between the control mechanism and system-specific management components (*system adapters*). System adapters are: *event generators* and *effectors* which allow interaction of the control mechanism with the target system; *metric extractors* and *policy evaluators* which provide the means for computing a specific response, determined by policies, to an observed situation, modelled by metrics (see chapter 2).

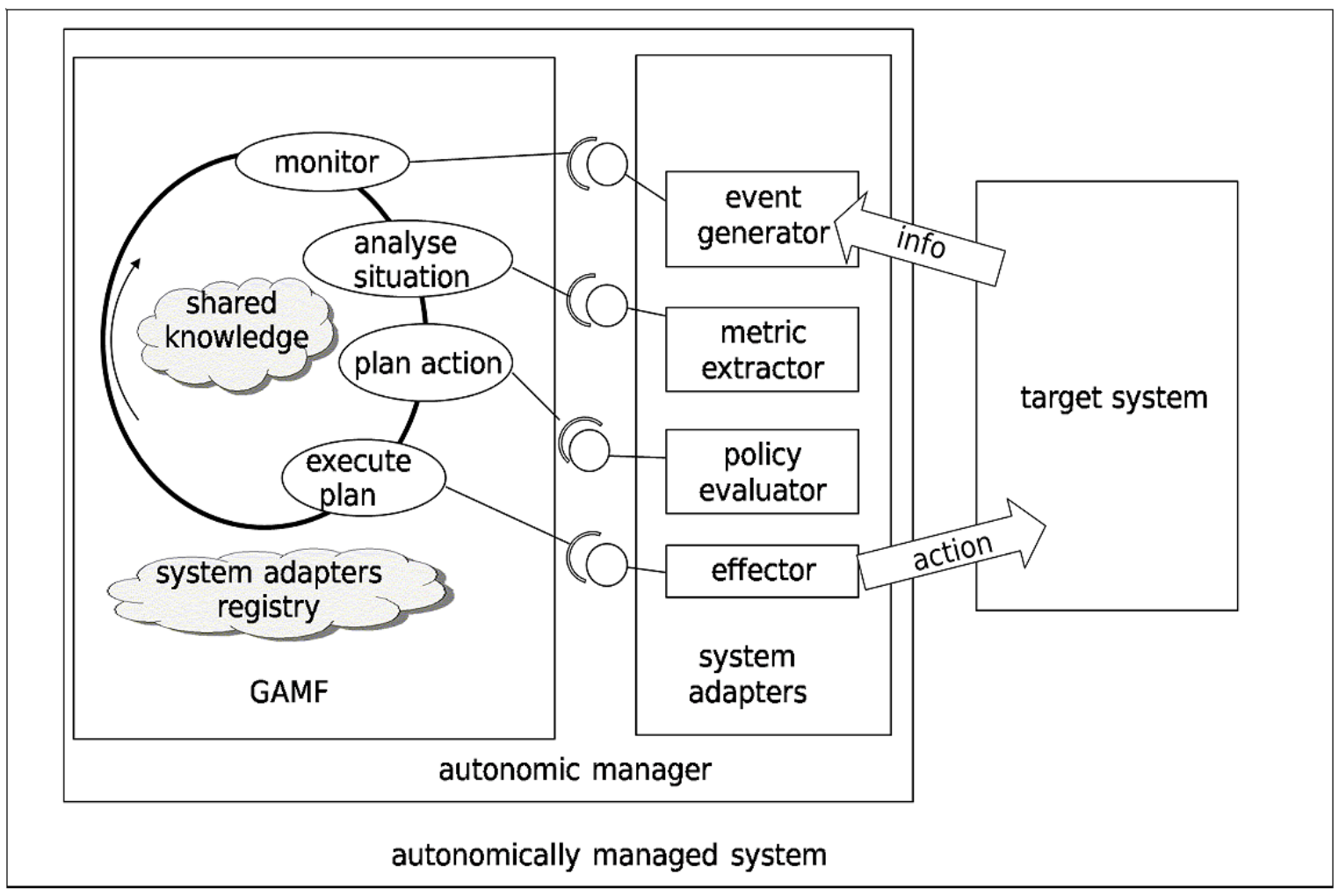

Figure 6.1: A target system autonomically managed using GAMF.

Figure 6.1 illustrates the general architecture by showing the components used for managing a target system. Additionally the flow of information between management components and target system is illustrated. An autonomically managed system consists of the originally unmanaged target system and an autonomic manager (which itself comprises system adapters and the GAMF). All system adapters are registered in the GAMF's *system adapters registry*. Amongst other features, this grants them permission to access the GAMF's *shared knowledge* database in which events and metric values are stored.

### 6.3.2 Detail

System adapters provide operations that correspond to a specific phase of the autonomic control cycle. Their execution is triggered by the GAMF, as configured by a system adapter developer (for instance, periodically). The information about how a specific system adapter is triggered is held in the system adapters registry along with access permissions to the shared knowledge database. Access to any data stored in the shared knowledge data base is concurrency-safe.

The following types of system adapters are defined:

- *Event generators* provide the GAMF with time-stamped information about specific *events* in the target system. An event comprises an event type, a time-stamp specifying the generation time of the event and a field for additional information, specified by the system adapter developer. In order to allow unambiguous usage of event types, the types of events generated by an individual event generator can be registered with GAMF to prevent them being used by other event generators.

- *Metric extractors* are used to extract monitoring data from the shared knowledge database in order to represent a specific situation, modelled by the *metric*. The metric is specified by the system adapter developer with a metric type, a time-stamp specifying the computation time of the metric value, a field for additional information and one for a numerical metric value.

- *Policy evaluators* evaluate the policy specified by the system adapter programmer. A policy is used for determining which action has to be carried out in response to the target system's current situation (represented by specific metric values).

- *Effectors* carry out the specific action in the target system when triggered. An effector is envisioned to be triggered by the policy evaluator to change a controlled system configuration parameter and to make the system aware of the change.

GAMF includes a flexible mechanism to filter for specific events or metrics in the shared knowledge database. Filter options include filtering for events or metrics of a specific type, and metrics or events recorded within a specific time window. To allow reuse of system adapters or simply to organise them, a system adapter can be categorised as being used for managing a specific *facet* of the target system. Additionally different options are provided to control when metric extractors and policy evaluators are triggered:

- scheduled at regular intervals;

- triggered by the arrival of a specific event type;

- triggered on an arbitrary schedule.

## 6.4 Conclusions and Future Work

The framework introduced here allows developers to focus on the control logic rather than on implementation details of generic autonomic control processes and components by factoring out generic parts of the autonomic control loop. It is applicable to a wide scope of applications which are not covered by this thesis. Examples are listed in appendix C. The GAMF is implemented in *Java*; its source code and additional information, including an API, can be obtained from http://www-systems.cs.st-andrews.ac.uk/gamf.

# Chapter 7

# Experimental Evaluation of the Management of P2P Nodes

## Outline

The effects on performance and resource consumption of autonomic management of the maintenance scheduling in P2P nodes were experimentally evaluated. P2P nodes were configured with autonomic management, as proposed in chapter 4, and deployed in a local area test-bed. The P2P nodes were exposed to various churn patterns and workloads. In the majority of the experiments autonomically managed nodes yielded benefits with respect to both performance and resource consumption compared to statically configured nodes.

## 7.1 Introduction

In chapter 4 it was hypothesised that an autonomic manager which dynamically adapts the interval between maintenance operations in P2P overlays (section 4.5) may yield improved performance and network usage in comparison to a statically configured system. Performance was defined in section 4.4.1 as a combination of lookup time, error rate and lookup error time. Performance and network usage are referred to as user-level metrics (ULM) wherever appropriate for the rest of this chapter. The hypothesis was experimentally evaluated with a local area deployment of autonomically managed and statically configured P2P nodes which were exposed to various conditions, and the effects on individual ULMs were compared. The static and autonomic scheduling of maintenance operations was specified by *policies*. One policy determined a fixed maintenance interval in order to represent an unmanaged system (the baseline). Two other policies determined an autonomic adaptation of maintenance intervals as specified in chapter 4, with different configurations.

The effects of each policy on performance and network usage were determined in a series of experiments, each specified by a different combination of:

- churn patterns (temporal patterns of nodes joining and leaving the P2P overlay);
- workloads (temporal patterns of P2P lookups);
- policies which controlled the maintenance scheduling mechanism.

Performance, in terms of lookup time, error rate and lookup error time, was measured for each experiment by executing the workload. The network usage was measured in terms of the amount of data each node sent to the network. To verify the reproducibility of the measurements, every experiment was repeated three times.

The motivation for this research was the wide usage of P2P overlays in distributed storage systems. Thus, representative conditions for specific scenarios were derived from distributed storage use cases. Experiments were carried out with components used in the ASA storage system, which is outlined in chapter 3. ASA uses a P2P overlay which is based on Chord [85] and is named StAChord [47]. StAChord is built in *Java* and uses the *Rafda Run Time (RRT)* [90, 89] library as middleware. The RRT was extended with a network traffic monitor which recorded the number of bytes sent to the network, in order to monitor the network usage. The autonomic manager was implemented using a framework (*GAMF*) developed as part of this work, which is described in chapter 6.

This chapter is structured in the following way: the implementation of autonomic management for StAChord is outlined in 7.2. The conceptual principles of the experimental parameters, such as workload and churn pattern, as well as the machinery used to apply them are explained in section 7.3. The experimental setup is explained in terms of the actual configuration of policies, churn patterns and workloads; additionally the extraction and aggregation of performance and network usage measurements are outlined in 7.4. Following that, a numerical breakdown of the measurements is provided in several levels of detail in section 7.5. The obtained results are compared with results from related work in section 7.6. In section 7.7 this chapter is concluded by discussing the results, revisiting initial hypotheses and by looking at future work.

## 7.2 Implementation of Autonomic Management for StAChord

The original maintenance mechanism of StAChord consists of the periodically executed operations *stabilize, fixNextFinger* and *checkPredecessor* (see chapters 3 and 4). When autonomic management was applied, the adaptation of the intervals between executions of each operation was managed individually by a single autonomic manager. In accordance with the used framework ($GAMF$), *event generators* and *effectors* were developed to allow interaction of the autonomic manager with the target system, and *metric extractors* and *policy evaluators* were developed to determine the interval.

The objective of the manager was to balance the requirement for potentially conflicting maintenance intervals. The following basic principles, introduced in section 4.5.2, were shared between the management of each maintenance operation. Each metric represented extracted time-stamped information, originally produced by event generators. For any metric value, only information monitored in a specific observation period was considered. Metrics were defined to represent non-effective maintenance operations (*NEMO*), the error rate when accessing peers (*ER*), and the time it took a node to locally execute a lookup (*LILT*). Sub-policies determined new intervals considering each metric in isolation by using the metric values to compute the magnitude of the change of the specific interval, in response to the value of the specific metric. An aggregation-policy averaged all sub-policy responses. The magnitude of change was determined by the metric value, a threshold $t$, and a factor $k$ (see equation 4.2).

### 7.2.1 Event Generation

The operations monitored were: *stabilize, checkPredecessor, fixNextFinger, lookup* and *findWorkingSuccessor*. The maintenance operations *stabilize, checkPredecessor* and *fixNextFinger* verify specific peer-sets and repair them if necessary. Each maintenance operation generated an operation-specific *event* whenever it was executed unnecessarily. The routing operation *lookup* was called when a key to node mapping needed to be resolved. An *event* was generated at each *lookup* execution to monitor the time it took to execute anindividual routing operation. If any remote access to a peer (fingers, successors, predecessor) failed, a peer-specific failure-event was generated.

### 7.2.2 Management of the *stabilize()* Interval

The *stabilize* operation maintains the successor and the successor list, both accessed during *stabilize* and *lookup* operations. Thus the following metrics were used to determine a new interval between *stabilize* executions, in order to assess whether maintenance-effort was wasted or if more effort was required (section 4.5.2):

- $NEMO_{stabilize}$: The *stabilize* operation had an effect if a new successor was found and a new successor list item was installed. The number of any other outcomes of *stabilize* was reflected by the $NEMO_{stabilize}$ metric value.
- $ER_{stabilize}$: Any failed access to a successor triggered a call of *findWorkingSuccessor* which installed a new successor from the successor list. Thus the number of *find-*

*WorkingSuccessor* calls determined the $ER_{stabilize}$ metric value.

- $LILT_{stabilize}$: The mean time it took *lookup* calls to complete (averaged over an observation period) was used as the $LILT_{stabilize}$ metric value.

Sub-policies determined new intervals with respect to each individual metric in isolation, which were then averaged by an aggregation-policy. The sub-policies which determined a new interval with respect to $ER_{stabilize}$ and $LILT_{stabilize}$ metrics determined a decrease. The $NEMO_{stabilize}$ metric specific sub-policy determined an increase. For the extraction of any metric only monitoring data within a specific observation period was considered. This applies also to the management of any of the following maintenance operations.

### 7.2.3 Management of the *fixNextFinger()* Interval

The *fixNextFinger* operation maintains the items in the *finger table* (fingers), accessed during *lookup* and *fixNextFinger* execution. Thus the following metrics were used to determine a new interval between *fixNextFinger* executions, in order to assess whether maintenance-effort was wasted or if more effort was required (section 4.5.2):

- $NEMO_{fixNextFinger}$: The *fixNextFinger* operation had an effect if a new finger table item was installed. The number of any other outcomes of *fixNextFinger* was reflected by the $NEMO_{fixNextFinger}$ metric value.

- $ER_{fixNextFinger}$: The number of events indicating failed calls to any finger determined the $ER_{fixNextFinger}$ metric value.

- $LILT_{fixNextFinger}$: The mean time it took *lookup* calls to complete (averaged over an observation period) was used as the $LILT_{fixNextFinger}$ metric value.

### 7.2.4 Management of the *checkPredecessor()* Interval

In contrast to the other maintenance operations, *checkPredecessor* only verifies the predecessor and sets its address to *null* if an access to the predecessor has failed, whereas the peer-set maintenance operations *stabilize* and *fixNextFinger* verify peers and install new ones if required. In contrast to fingers or successors, a predecessor access is not initiated by any local P2P operation other than by *checkPredecessor.* Thus the following metrics were used to determine a new interval between *checkPredecessor* executions, in order to assess whether maintenance-effort was wasted or if more effort was required (section 4.5.2):

- $NEMO_{checkPredecessor}$: *checkPredecessor* had an effect if the predecessor was changed, $NEMO_{checkPredecessor}$ was defined as the total number of times that *checkPredecessor* did not have an effect.

- $ER_{checkPredecessor}$: The number of failed connections to the predecessor node was used to compute $ER_{checkPredecessor}$ metric value.

Note: $LILT$ was not considered when computing the interval between *checkPredecessor* calls, as the predecessor was not accessed during a call to the *lookup* operation.

## 7.3 Experimental Parameters

The effects of the autonomic manager, outlined in section 7.2, on StAChord's performance and network usage were measured in various experiments. Each experiment involved a particular workload (a temporal pattern of lookup requests) and membership churn (a temporal pattern of nodes joining and leaving the overlay). As the motivation for this research is the use of P2P overlays in distributed storage systems, workloads and churns were derived from distributed storage use cases. In each of the use cases a P2P overlay as used in ASA was considered. This means that the nodes of which the P2P overlay was comprised represent individual storage servers. One of these storage servers acted as a dedicated *gateway* to issue lookups during the application of workloads. Here the general concepts and motivations for both workload and churn patterns are explained as well as the machinery used for applying them. The configurations used in the experiments can be found in section 7.4.

### 7.3.1 Churn Pattern

Each churn pattern modelled the behaviour of a set of nodes, in terms of a sequence of alternating on-line and off-line phases for each node.

- During on-line phases a node can be routed to by its key.
- During off-line phases a node cannot be routed to by its key.

The durations of these phases were pseudo-randomly generated according to two normal distributions, one for on-line and one for off-line phases. Thus the churn pattern was defined by the two distributions.

**Example Churn Pattern**

The following example shows how a churn pattern was applied to an individual node by the machinery. The churn pattern was given by the on-line and off-line phase-durations. Each phase-duration was selected from a normal distribution of values which was specified by its mean and standard deviation ($\mu_{t_{on/off-line}} \pm \sigma_{t_{on/off-line}}$). A *duration* defined the minimum overall length of all alternating phases. All durations are given in seconds, *[s]*.

- *duration:* $500[s]$
- $t_{on-line}$ : $102 \pm 3[s]$
- $t_{off-line}$ : $106 \pm 10[s]$

Figure 7.1 illustrates the behaviour of an individual node due to the above churn pattern, which resulted in alternating on-line and off-line phases for an overall duration of *518 [s]*.

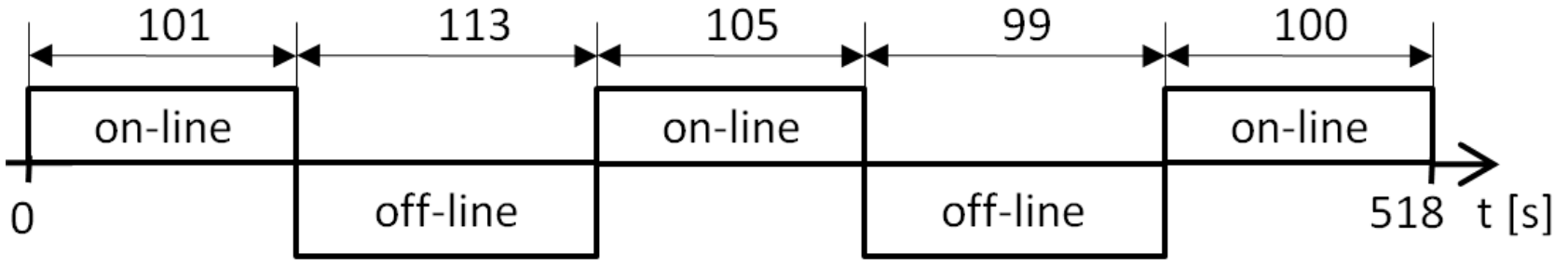

Figure 7.1: Temporal pattern of on-line/off-line phases of an individual node.

All participating nodes in an overlay, as used in the experiments, exhibited similar behaviour to that illustrated in figure 7.1 as a result of the churn pattern just described. The particular length of each phase was pseudo-randomly selected from the corresponding distribution. Additionally the type of the first phase (on-line or off-line) was also selected pseudo-randomly. The motivation for the pseudo-randomness was that churn patterns were required to exhibit variation during the course of an experiment and between nodes but not between repetitions of experiments with the same churn pattern configuration.

**Distributed Storage Usage Scenarios/Churn Patterns**

The conceptual approach towards the simulation of specific churn patterns as described above was used to support the following four usage scenarios. The four scenarios represent edge cases and combinations of edge cases in order to have significant variations between the tested scenarios. In each scenario one of the nodes was a dedicated gateway through which a workload was executed. All other nodes exhibited behaviour corresponding to the scenarios. The gateway was permanently on-line because it was required by any node to join the overlay at the initiation of any node's on-line phase and by the machinery which applied the workload, for issuing lookups. The scenarios evaluated were:

- A *low membership churn* in which the overlay is composed of nodes representative of dedicated servers that join the network and rarely leave it. In this scenario a high interval between maintenance operations is desired.
- A *high membership churn* in which the overlay is composed of nodes representative

of workstations that join and leave with a high frequency. In this scenario a short interval between maintenance operations is desired.

- A *locally varying membership churn* in which the overlay is composed of nodes representative of either servers or workstations with different joining and leaving patterns. In this scenario it is desired that nodes which are exposed to high churn maintain links to their peers with a higher frequency than those exposed to low churn.

- A *temporally varying membership* in which the overlay is composed of nodes which change their behaviour over time, from behaving like workstations to behaving like servers and vice versa. Workstations exhibit high churn whereas servers exhibit low churn. In this scenario small intervals are desirable when nodes exhibit high churn and vice versa in periods of low churn.

### 7.3.2 Workload

Each workload was specified as a temporal pattern of P2P lookups, which were executed via the previously introduced gateway. Machinery was developed which allowed the expression of various temporal lookup patterns representing four scenarios which starkly differ from each other. These can be categorised as synthetic or file system specific. In each type of workload the keys were pseudo-randomly generated, to allow a variation of keys within an individual experiment but not between repetitions.

### Synthetic Workloads

The evaluated synthetic workloads were:

- To represent scenarios in which no or very few lookups were executed, a *synthetic light weight workload* was defined. Such a workload was specified by a number of $l$ lookups which were spread evenly over the experimental duration. The interval between lookups was $d$ seconds.
- To represent scenarios in which lookups were executed at a high rate a *synthetic heavy weight workload* was defined. Such a workload was specified by $l$ lookups which were executed sequentially without any delay between them.
- To represent scenarios in which a workload exhibits alternating phases of heavy and light weight workloads, a *synthetic variable weight workload* was defined. Such a workload was specified by $l$ sequential lookups which were spread over the experimental duration. The spreading factor was specified by $s$ and $d$. $s$ sequential lookups were executed without any delay between them, and after every $s^{th}$ lookup a delay of $d$ seconds was exhibited.

### File System Specific Workload

To represent a distributed storage usage scenario, a workload specific for a *file system workload* was defined. File system operations were extracted from existing “real world” file system traces and the corresponding ASA operations were generated. These ASA operations corresponded with temporal patterns of lookup operations in ASA's P2P layer. This transformation was based on ASA semantics as described in chapter 3. More details are available in the appendix A.1.4.

## 7.4 Experiment Setup

### 7.4.1 The Test-Bed

The experiments reported here were conducted on a local area test-bed consisting of 16 dedicated hosts each with a 3.00GHz Intel®Pentium®4 CPU and 1GB of RAM. The hosts were connected to a dedicated switch and isolated from the rest of the network. A single overlay node was executed on each network host to ensure that the performance of the overlay network was not skewed by multiple overlay nodes competing for resources (CPU-time, memory and network bandwidth) within a host. A separate host, the *workload-executor*, ran the workload and conducted performance measurements. Each participating node monitored the number of bytes it sent and received as well as autonomic management details. To avoid measurements being skewed by collecting data from the individual hosts during an experimental run, monitoring data was kept locally and collected after each experiment finished. As monitoring data was time-stamped the system clocks on all hosts were synchronised using $NTP$ [63]. Information about the motivation for choosing this specific test-bed can be found in appendix A.1.5.

### 7.4.2 Derivation of User-Level Metrics

The experiments[1] were carried out to measure the effects of the various policies on the user-level metrics (ULM) performance and network usage. Single values for both performance and network usage was computed by aggregating measurements for each experiment in order to compare effects of the specific policies. To verify reproducibility each experiment was repeated three times.

Measurements were aggregated over observation periods of *5 minutes*. Performance measurements were derived from the execution of workload lookups. Performance was previously defined in section 4.4.1 as a combination of *lookup time, lookup error rate* and *lookup error time*[2] collected during individual observation periods. Network usage was measured as the amount of data all nodes sent to the network during each individual observation period. The time during which monitoring data was gathered in one experimental run (*experiment run time*) was the time interval from the first lookup until the last, whether it was successful or not.

To highlight the fact that a performance measurement was computed for each observation period, the performance is referred to as *expected lookup time* for the rest of this thesis. The motivation for this notion was that a lookup would have been expected to complete after some time if a fall-back mechanism retried failed lookups until they succeeded. Thus for modelling the expected lookup time, $t_{expected}$, it is assumed that any given lookup succeeds, after a lookup time $t_{lt}$, with a probability $p_{success}$. Conversely any given lookup may fail with a probability $p_{failure}$ after a lookup error time, $t_{let}$ has passed. Every failure is followed by a retry, which is repeated $n$ times. Thus, the expected lookup time is given by the weighted sum of all possible cases as shown in formula 7.1.

[1]An experiment is specified by the combination of a specific churn pattern, workload and policy for managing maintenance scheduling.

[2]All are referred to as secondary ULM wherever appropriate for the rest of this chapter.

$$t_{expected} = t_{lt} \times p_{success} + \sum_{i=0}^{n} \left( (t_{lt} + i \times t_{let}) \times p_{success} \times p_{failure}^{i} \right) \quad (7.1)$$

This resulted in (derived) ULM monitoring data being available in form of progressions of expected lookup times and network usages, each for individual observation periods. Including the repetitions, three progressions were available for each ULM. The individual expected lookup times and network usages for each observation period, including their repetitions, were used to create distributions of expected lookup times and network usages. In order to compare the effects of specific policies on an individual ULM, the arithmetic mean of the distribution was calculated.

### 7.4.3 Churn Pattern Configurations

The four churn patterns were specified by pseudo-randomly selected values[3], as:

- Low membership churn:
  - $t_{on-line} >> 2[h]$
  - $t_{off-line} : 157 \pm 20[s]$

[3]Specified as explained in section 7.3.1.

- High membership churn[4]:
  - $t_{on-line}$ : $200 \pm 40[s]$
  - $t_{off-line}$ : $100 \pm 20[s]$
- Locally varying membership churn: 25% of all P2P nodes were representative of dedicated servers which exhibited low churn. 75% of the P2P nodes were representative of user workstations which exhibited high churn.
- Temporally varying membership churn: a phase in which all nodes exhibited low churn, with a duration of 1000 [s], followed by a phase in which all nodes exhibited high churn, again with a duration of 1000 [s] and so forth.

The churn patterns were held constant between experiment repetitions.

### 7.4.4 Workload Configurations

The following workload specifications[5] were derived from preliminary work, as reported in appendix A.1:

- Synthetic light weight workload: 10 lookups were issued in total; between two lookups a period of 300 seconds of inactivity was configured.
- Synthetic heavy weight workload: 6000 successive lookups were issued.

[4] This was a random churn pattern amongst the highest churn patterns the experimental platform supported.
[5] See section 7.3.2 for definitions and examples.

- Synthetic variable weight workload: 1000 lookups were issued in total; 100 successive lookups were followed by 300 seconds of inactivity.
- File system specific workload: a temporal sequence of 14576 P2P lookups derived from a file system trace. In order to represent original ASA semantics, the lookups were organised in sets of keys, those representative of keys for meta-data were looked up in parallel and those representative of keys for data in sequence. The lookups were spread over the experimental duration in accordance with the file system workload. More details are available in the appendix A.1.4.

### 7.4.5 Policy Parameter Configurations

Three different policies for scheduling maintenance operations were defined using the autonomic management mechanism[6]. The management mechanism consisted of an aggregation policy which balanced out interval recommendations of sub-policies which analysed individual metrics in isolation. Each sub-policy determined an increased or decrease of the current interval proportional to the difference of the analysed metric value to its corresponding ideal value[7]. A threshold $t$ determined which metric values were ignored and a constant factor $k$ determined the rate of change. Values for $t$ and $k$ were configured specifically for the individual sub-policies but then had the same values for each of the individual maintenance operations[8]. $NEMO_t$, $NEMO_k$, $ER_t$, $ER_k$, $LILT_t$, $LILT_k$ were referred to as policy parameters.

[6]See section 7.2 and 4.5 for more details.

[7]The NEMO specific sub-polices determined an increase, ER and LILT a decrease.

[8]*stabilize, fixNextFinger, checkPredecessor*

- Policy 0: This policy left nodes unmanaged but still incurred the overhead of the management processes in order to allow comparison.

- Policy 1: This policy determined a new interval between any maintenance operation based on the operation-specific metrics outlined above. The policy parameters were derived from preliminary experiments (appendix A.1.3) with the objective of finding the most suitable parameter set.

- Policy 2: Like policy 1, this policy determined a new interval between any maintenance operation based on the operation-specific metrics outlined above. This policy was configured to ignore *LILT* metrics and to aggressively react to the other metrics.

All policies were evaluated every two seconds. Two seconds was also used as an initial maintenance interval. Thus the configuration of policy 0 resulted in a statically configured interval for each maintenance operation of two seconds. This maintenance interval was derived from the preliminary experiments reported in appendix A.1.1 as the most suitable static interval.

## 7.5 Experimental Results

### 7.5.1 Overview

Autonomic management yielded a significant improvement of the observed user-level metrics (*expected lookup time* and *network usage*, see section 7.4.2) in the majority of the conducted experiments, in comparison to a static configuration of P2P nodes. Each experiment was specified by the combination of a specific churn pattern, a workload and the policy which specified the management of the peer-set maintenance scheduling in deployed nodes. Every experiment was repeated three times to verify the reproducibility of the observed effects. In the following, the experiments are organised by churn pattern and workload for ease of comparison of the effects of a specific policy on an individual ULM. Four different churn patterns were specified in section 7.4.3 and four different workloads in 7.4.4. Sixteen different groups of experiments were conducted in each the effects of three policies were measured separately. Within each group the churn pattern and workload was the same while the policy varied. *Policy 0* represented an unmanaged system and *policies 1 and 2* autonomically managed systems (see section 7.4.5). When comparing the effects of policies in one group of experiments, the smallest ULM value denoted the most beneficial policy.

Table 7.1 shows the number of experiments in which each policy yielded the greatest benefit, with respect to the expected lookup time ($ELT$) and the network usage ($NU$) individually and to both in combination. Policy 0 gave the best results in fewer experiments than policy 1 and policy 2.

| policy | description | $ELT$ | $NU$ | $ELT$ & $NU$ |
|---|---|---|---|---|
| 2 | $autonomic_2$ | 6 | 12 | 5 |
| 1 | $autonomic_1$ | 8 | 3 | 0 |
| 0 | $static$ | 2 | 1 | 0 |

Table 7.1: The number of experiment groups (out of a total of 16) in which each policy yielded the greatest benefits.

Table 7.2 provides an holistic view of the effects of autonomic management. It shows the mean ULMs in managed systems normalised to an unmanaged system. Thus every normalised ULM less than 1 represents a benefit of the specific autonomic management policy with respect to the unmanaged system. More details are provided in appendix A.2.

| experiment specification | | policy 1 | | policy 2 | |
|---|---|---|---|---|---|
| workload | churn pattern | expected lookup time | network usage | expected lookup time | network usage |
| synthetic light weight | low churn | 0.725 | 0.09 | 0.705 | 0.028 |
| | high churn | **0.814** | 0.548 | **0.817** | 0.353 |
| | locally varying churn | **0.835** | 0.454 | **1.207** | 0.438 |
| | temporally varying churn | 0.807 | 0.293 | **0.979** | 0.178 |
| synthetic heavy weight | low churn | 0.727 | 0.314 | 0.698 | 0.23 |
| | high churn | 0.605 | **1.333** | 0.693 | **0.983** |
| | locally varying churn | 0.085 | **1.267**[9] | 0.183 | **1.082** |
| | temporally varying churn | 0.562 | 0.541 | 0.672 | 0.4 |
| synthetic variable weight | low churn | 0.714 | 0.111 | 0.7 | 0.054 |
| | high churn | 0.362 | **1.202** | 0.364 | **0.781** |
| | locally varying churn | **3.239** | 0.416 | 2.954 | 0.421 |
| | temporally varying churn | **1.559** | 0.258 | **0.974** | 0.245 |
| file system specific | low churn | 0.804 | 0.341 | 0.787 | 0.293 |
| | high churn | 5.142 | 0.47 | **1.089** | 0.523 |
| | locally varying churn | 0.6 | 0.453 | 0.592 | 0.882 |
| | temporally varying churn | 0.862 | 0.595 | **0.932** | 0.409 |

Table 7.2: A summary of all normalised ULMs.

Bold and underlined values in table 7.2 represent results which were not statistically significantly different from the baseline (policy 0), according to a *visual approximation test for significance*. This test was conducted following guidelines from [40]. It involved the computation of the *90% confidence interval of the mean ($ci_\mu$)* of the corresponding user level metrics and evaluating whether the confidence intervals overlapped. 25% of the comparisons shown in table 7.2 were not significantly different. In addition *t-tests* were carried out to evaluate the probabilities of the compared data sets being statistically significantly different from each other. All *t-test* results are available in appendix A.2.18. The comparison of expected lookup times abstracts over the variation of the underlying raw data (*lookup time*, *lookup error time* and *lookup error rate*) for simplicity. Samples of the raw data as provided in appendix A.2 for each measurement show that statistically significant differences can be identified in the raw data used to derive the expected lookup time. As an example the figures 7.2a,7.2b and 7.2c show the $ci_\mu$ (with 90% confidence) for all performance-related raw measurements in experiments with heavy weight workload and high churn.

[9] In contrast to the visual approximation, a t-test results in a significant difference in this case.

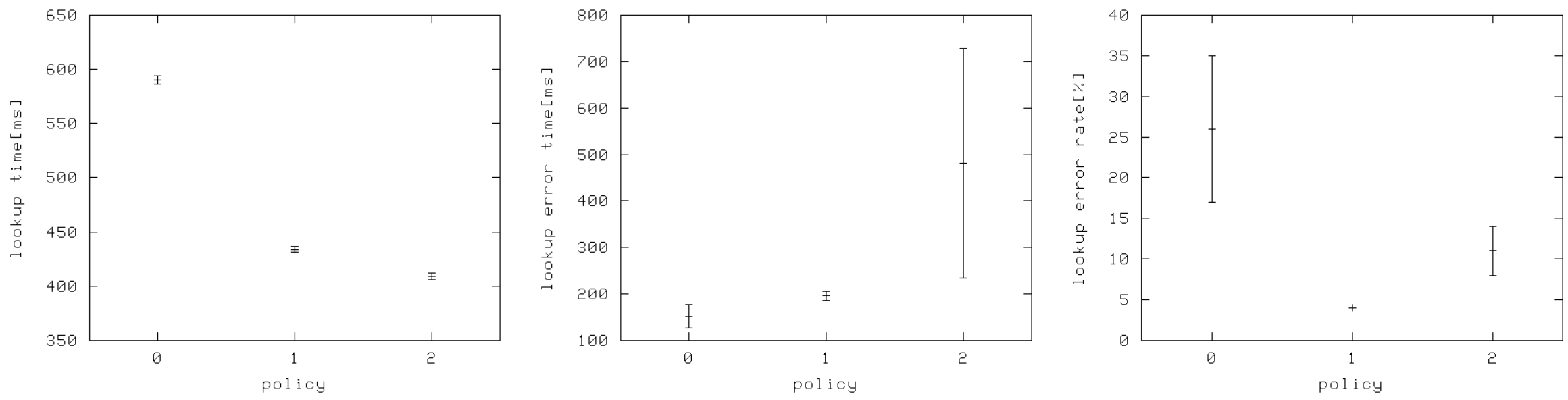

(a) Lookup Time (b) Lookup Error Time (c) Lookup Error Rate

Figure 7.2: Visual Approximation for Statistical Significance of secondary ULMs

### 7.5.2 Detailed Analysis

Benefit was achieved by the autonomic manager, by successfully detecting unsatisfactory situations and adapting the interval of the controlled maintenance operation accordingly. In order to explore the potential of autonomic management, the focus here lies on analysing the experiments in which autonomic management did best. These were experiments in which a synthetic heavy weight workload was executed (table 7.2).

### Experiments with Low Churn

Here the progressions of the expected lookup time, the network usage and the maintenance-interval progressions are portrayed. As it was observed that intervals of the individual maintenance operations progressed at a similar rate, only the finger table maintenance interval progression is shown. All progressions show the mean values of the portrayed observable averaged over three experimental runs in each specific observation time window; the observation time window was specified as five minutes.

The progressions are also specified this way in all of the following sub-sections in which the effects of autonomic management are analysed in detail. All progressions stop after the first sixty minutes of the experimental run time and the same scale is used in all figures showing the same observable to ease comparison between the individual plots.

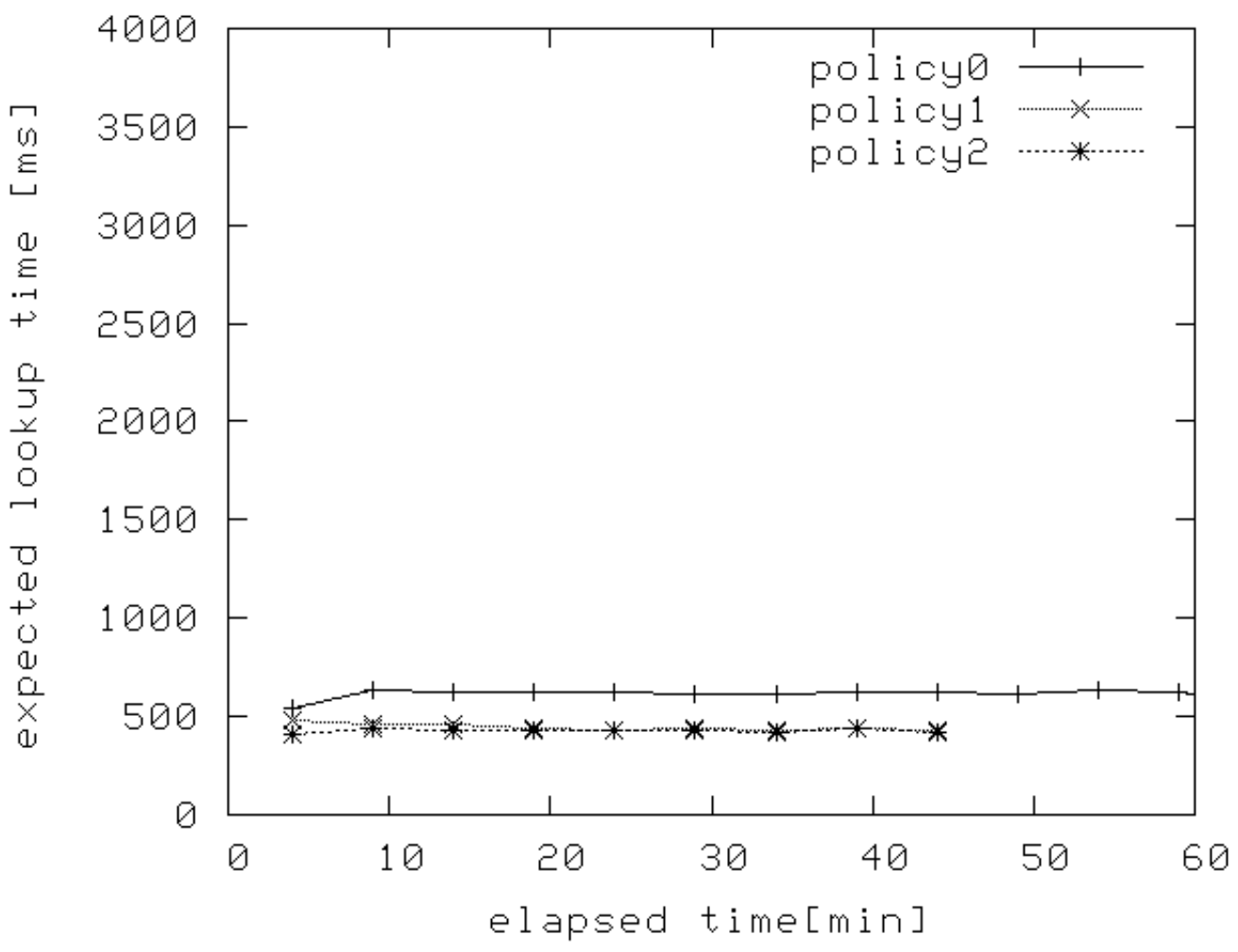

Figure 7.3: Expected lookup time progression with synthetic heavy weight workload and low churn.

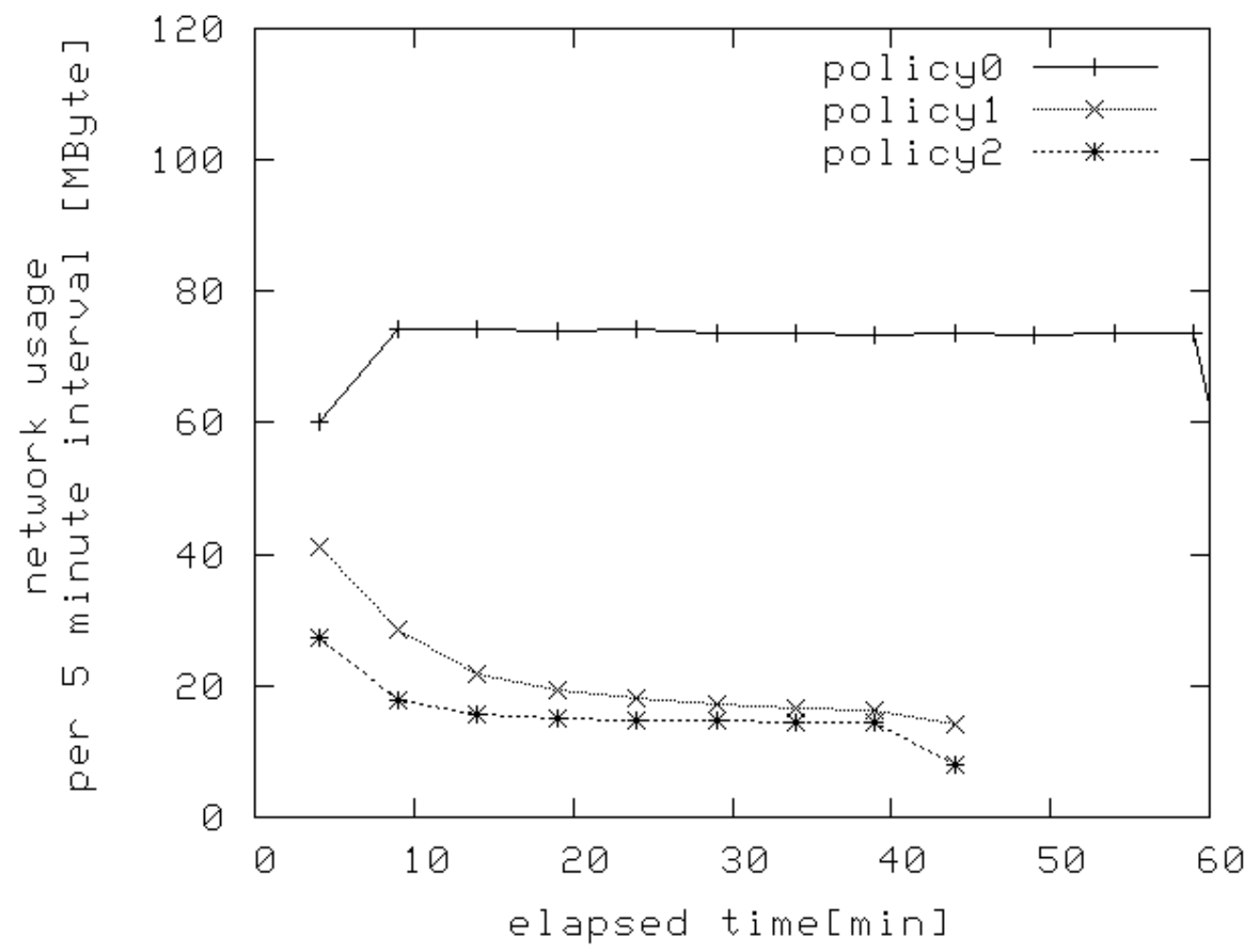

Figure 7.4: Network usage progression with
synthetic heavy weight workload and low churn.

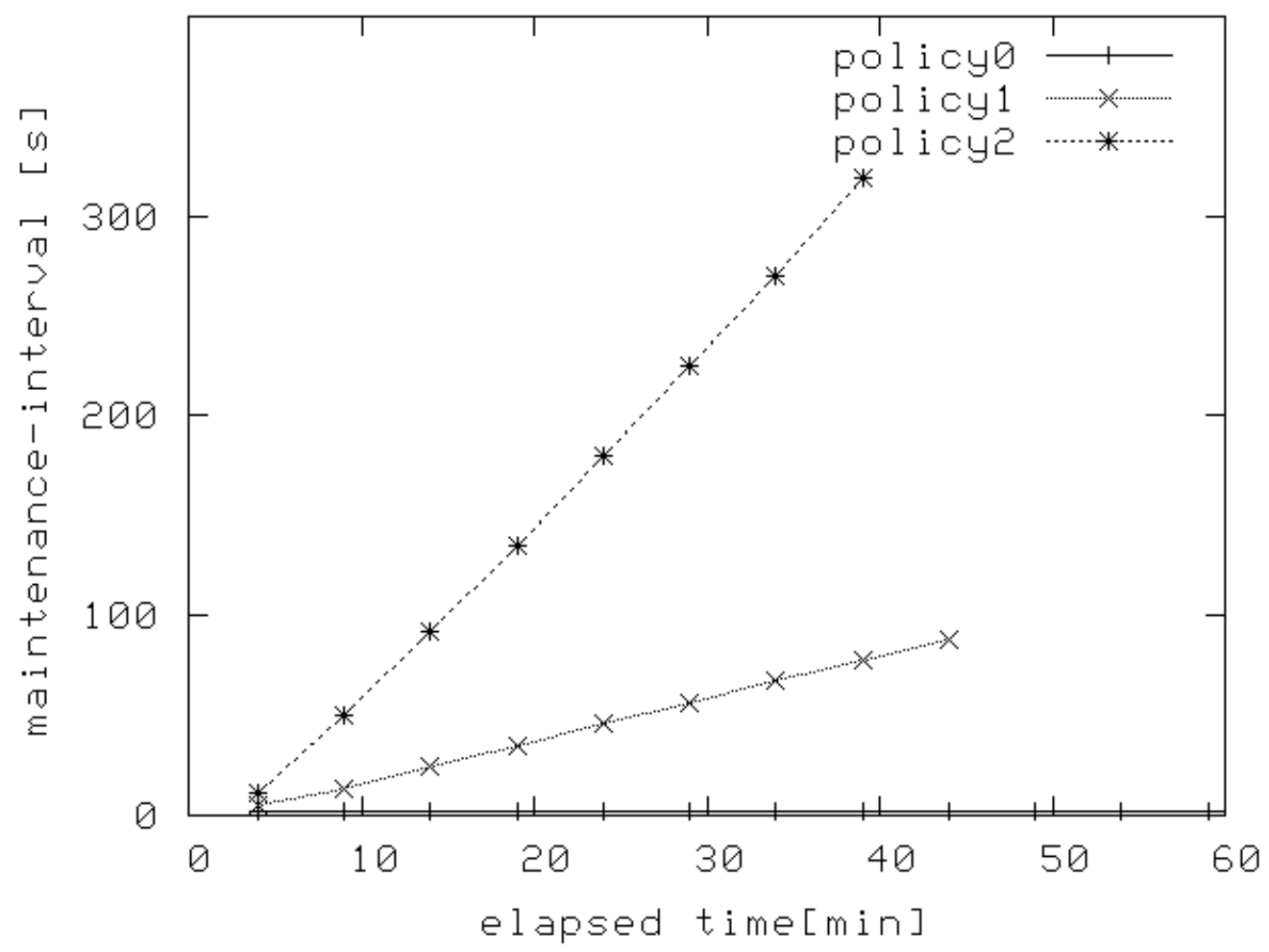

Figure 7.5: Maintenance-interval progression with
synthetic heavy weight workload and low churn.

Figure 7.3 shows that, under low churn, the expected lookup time of nodes which were managed with policies 1 and 2 improved within the first few minutes with respect to unmanaged nodes. After this time no significant improvement was observed. This corresponds with the network usage progression shown in figure 7.4 - after a few minutes the network usage of managed nodes did not further improve significantly versus unmanaged nodes. Further analysis showed that the network usage with managed nodes is mainly due to the execution of lookups in the workload, because hardly any maintenance traffic occurred after a few minutes experimental run time. The network usage progression shown in figure 7.4 exhibits a drop at the end of the experimental run time. This drop is an artefact of the data representation method, which (potentially) applies to any of the following plotted network usage progressions. Data sent during the observation period within which the experiment finished is averaged over that entire observation period.

The effects on the individual ULMs are a result of the maintenance interval adaptation shown in figure 7.5. In experiments with low churn, autonomic management detected an unsatisfactory situation with respect to network usage which was corrected with an increase of the maintenance interval. This decreased the amount of work each node spent (unnecessarily) maintaining its peer-set, and subsequently reduced the amount of data sent to the network in comparison with unmanaged nodes. Additionally, a reduction in the work spent on maintenance operations left more computational capacity for dealing with lookup operations. This decreased the monitored lookup time and subsequently the expected lookup time as no errors occurred in such situations. Policy 2 reacted, as specified, more aggressively to observed metrics and increased the maintenance interval at a higher rate than policy 1. More details are available in appendix A.2.6.

### Experiments with High Churn

Figure 7.6 shows that after the first few minutes the expected lookup time of unmanaged nodes progressed to much higher values than with managed nodes. The expected lookup

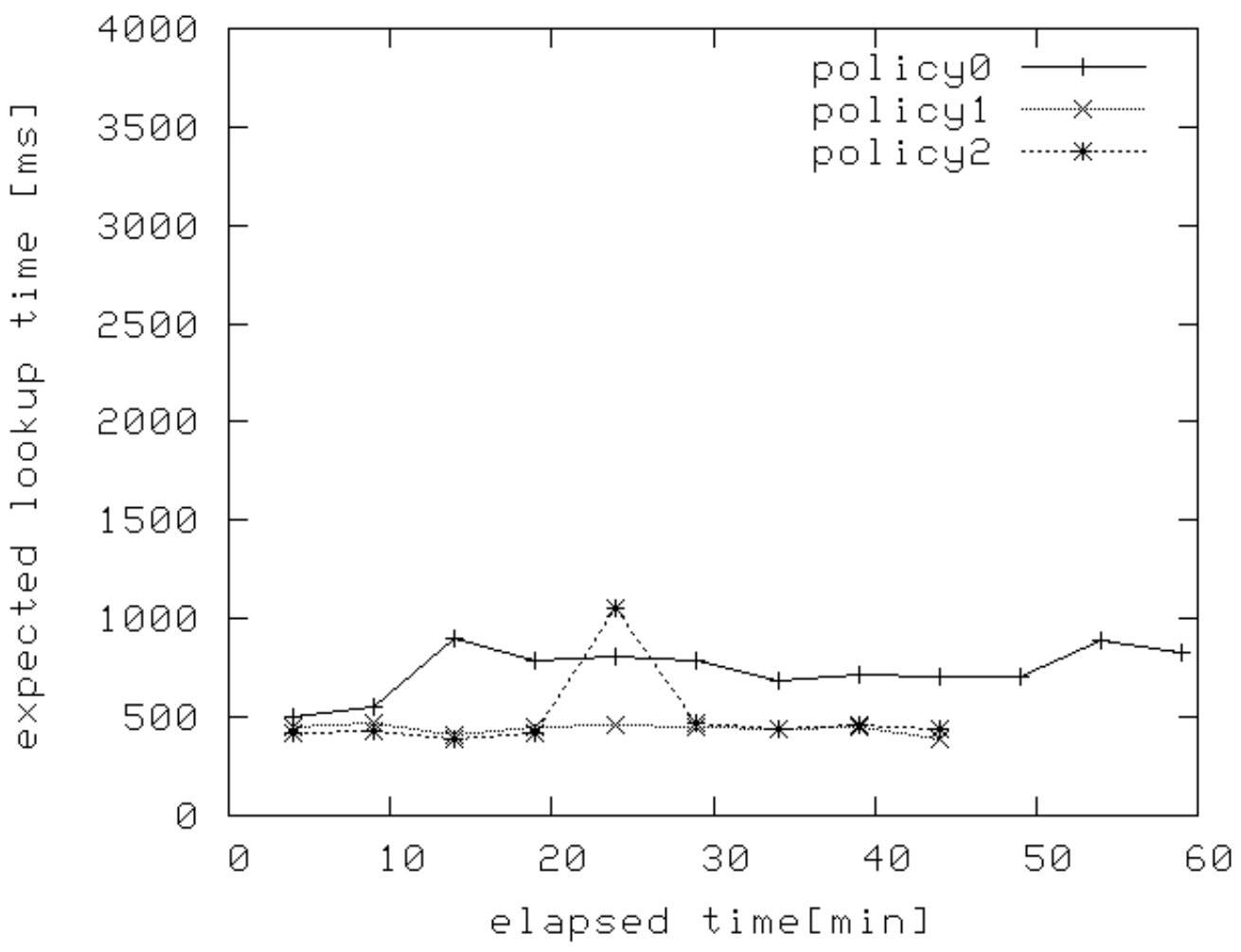

Figure 7.6: Expected lookup time progression with synthetic heavy weight workload and high churn.

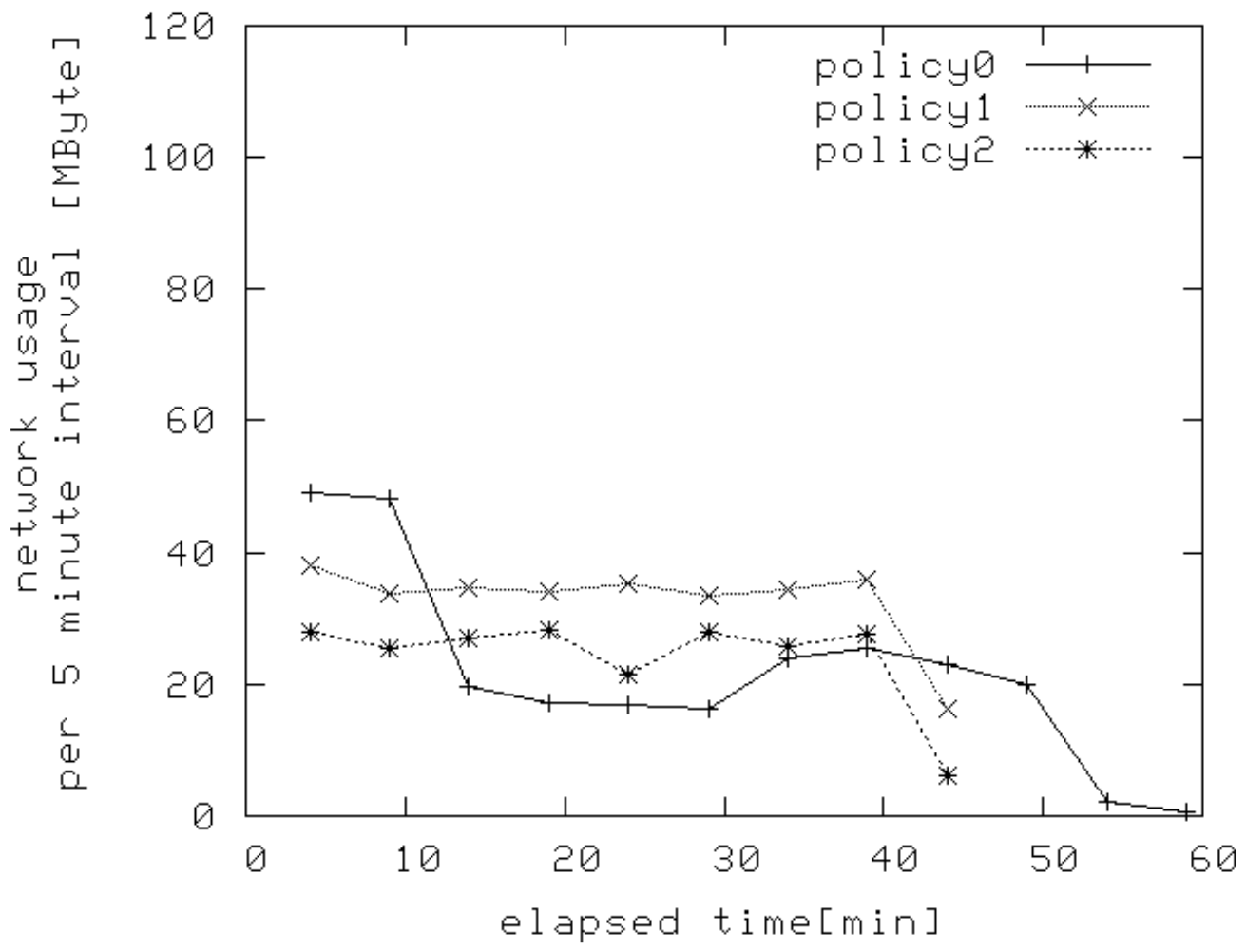

Figure 7.7: Network usage progression with synthetic heavy weight workload and high churn.

time measured in an experiment with nodes managed by policy 2 exhibited a peak after twenty-five minutes run time. Additional analysis showed that this peak and the relatively high values for the expected lookup time of unmanaged nodes were due to an increased error rate in both cases. This increase in the error rate, of nodes managed by policy 2, was compensated for by a decrease of the maintenance interval as illustrated in the magnified

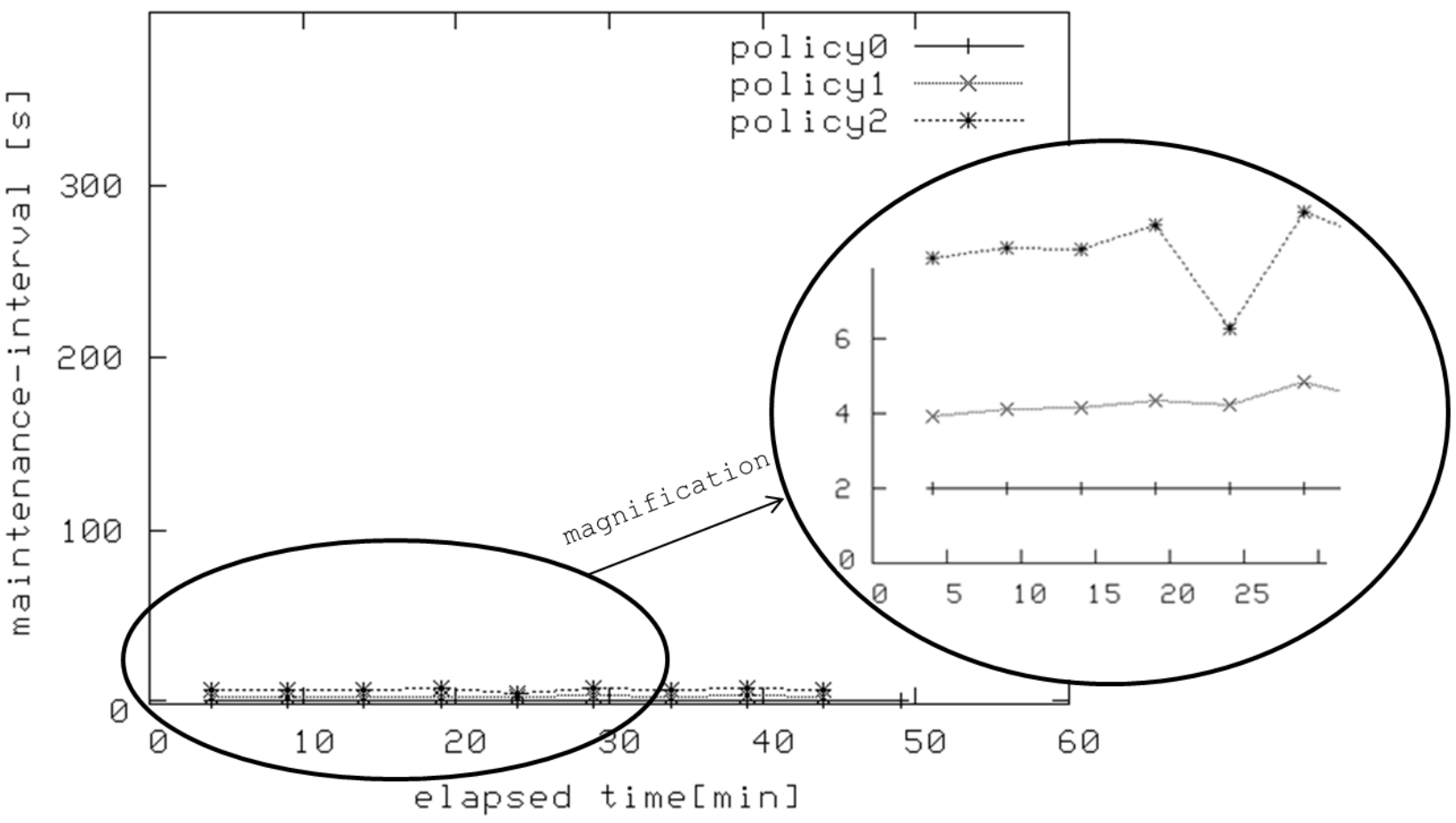

Figure 7.8: Maintenance-interval progression with synthetic heavy weight workload and high churn.

interval progression in figure 7.8.

The relatively high values for network usage in networks with unmanaged nodes in the first few minutes, shown in figure 7.7, result from the low expected lookup times during this time. Low expected lookup times result in more lookups having been issued in the same time than with higher expected lookup times, which subsequently caused more data to be sent to the network during such a phase.

Figure 7.8 shows that autonomic management detected an unsatisfactory situation with respect to the experienced churn and reacted by reducing the maintenance intervals. That resulted in a decrease in the number of failed lookups and subsequently the expected lookup time of managed nodes. Even though high churn was simulated, nodes experienced phases in which the churn allowed autonomic management to temporarily increase the maintenance intervals. This resulted in a better trade-off between work spent for maintenance and expected lookup time than in unmanaged nodes. Here again, policy 2 reacted more aggress-ively to observed metrics and kept the maintenance interval at higher levels than policy 1. More details are available in appendix A.2.7.

### Experiments with Locally Varying Churn

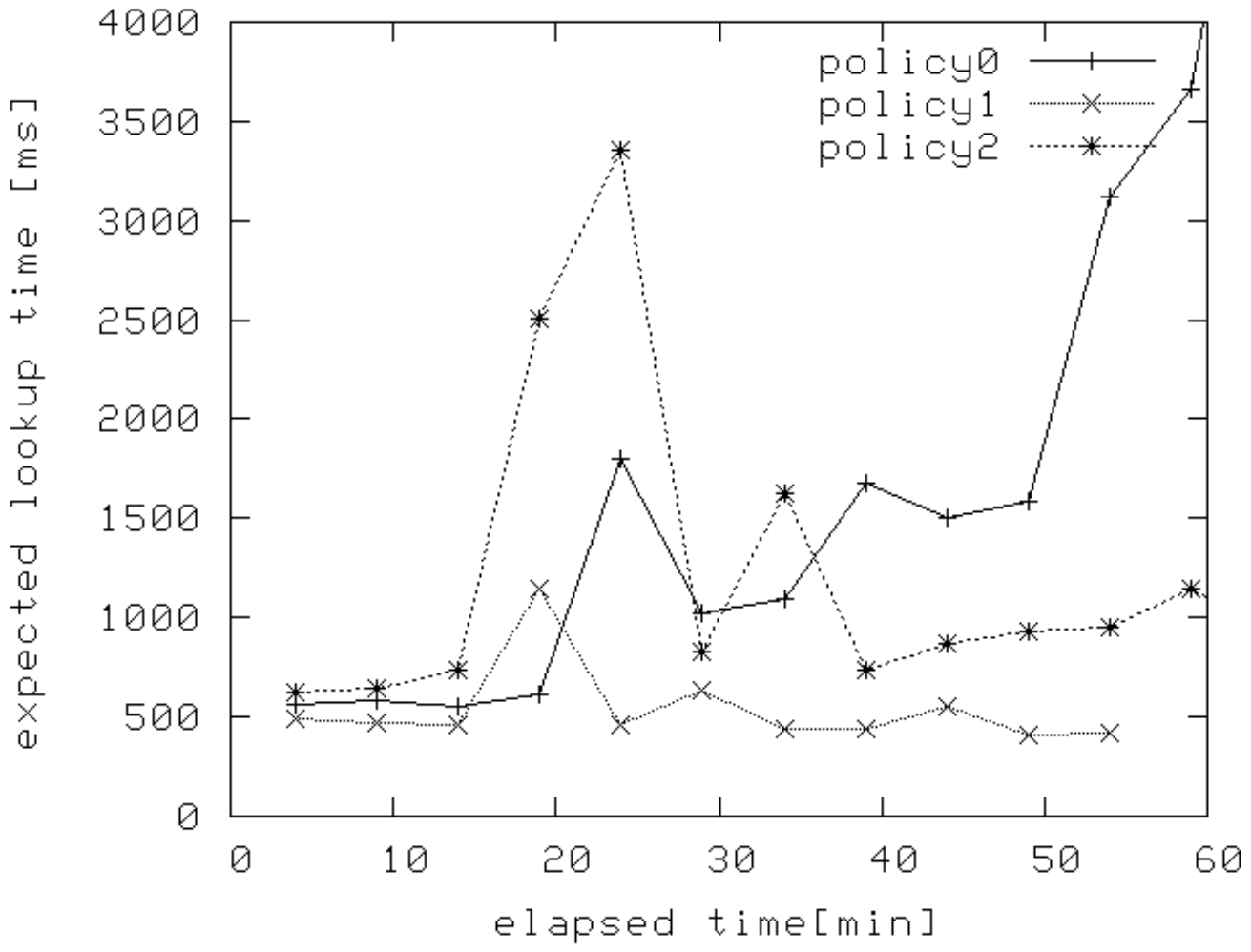

Figure 7.9: Expected lookup time progression with synthetic heavy weight workload and locally varying churn.

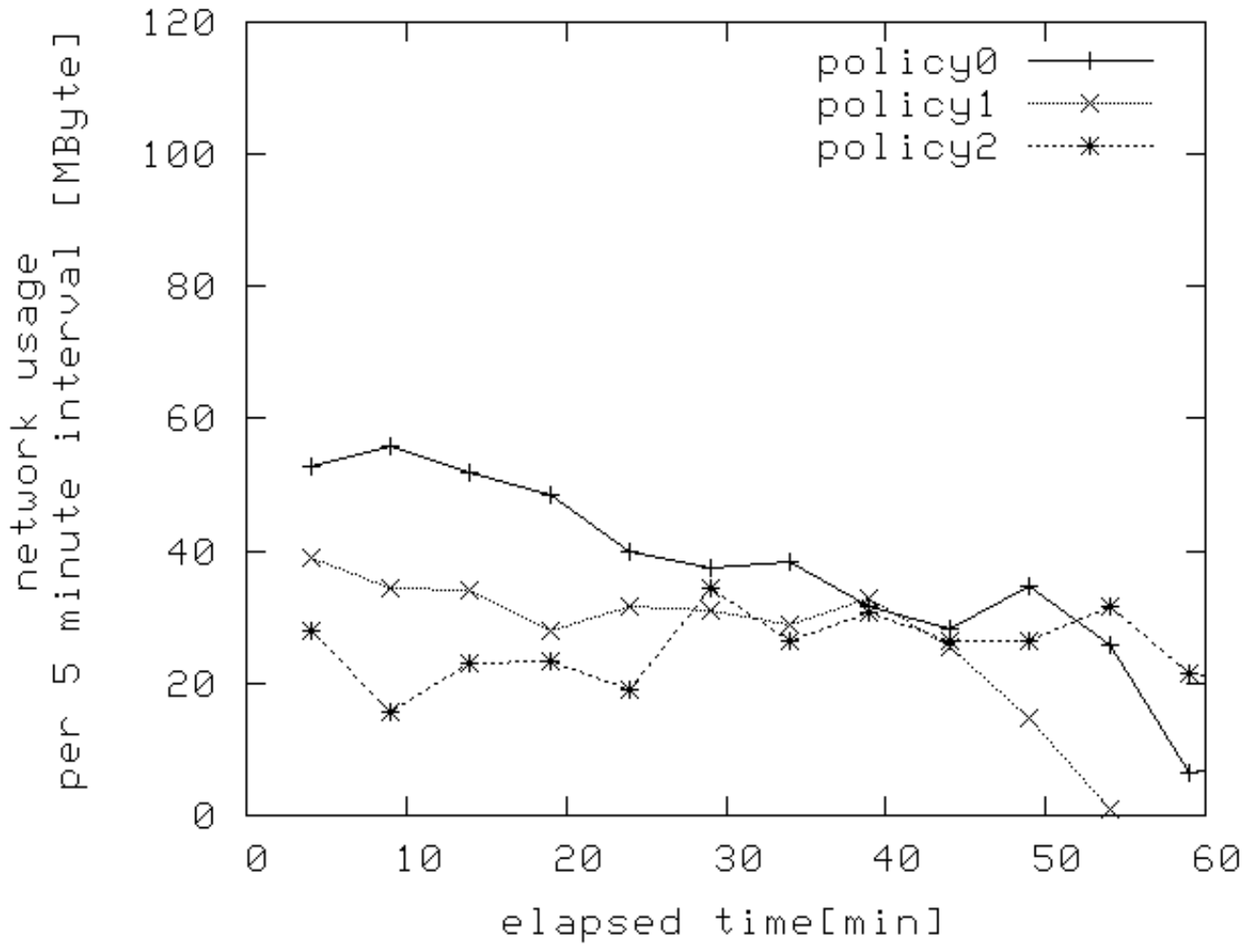

Figure 7.10: Network usage progression with synthetic heavy weight workload and locally varying churn.

In experiments with locally varying churn 25% of the nodes exhibited low churn (referred to as low churn nodes), while the rest exhibited high churn (referred to as high churn nodes). Figure 7.9 shows a peak in the expected lookup time of nodes managed by policy 2, after twenty-five minutes. This peak was identified as a result of an increased error rate after additional analysis. Figure 7.9 also shows an increase of the expected lookup time in unmanaged nodes after the first twenty-five minutes. Additional analysis showed that this increase (and the further increase of policy 0) was due to a combination of an increase in the error rate and the time it took to successfully resolve lookups. The increase of the expected lookup time is reflected by the network usage progression in figure 7.10. As lookups took longer the network was less congested in an individual observation period. Thus, it took longer to finish the entire workload in experiments with unmanaged nodes.

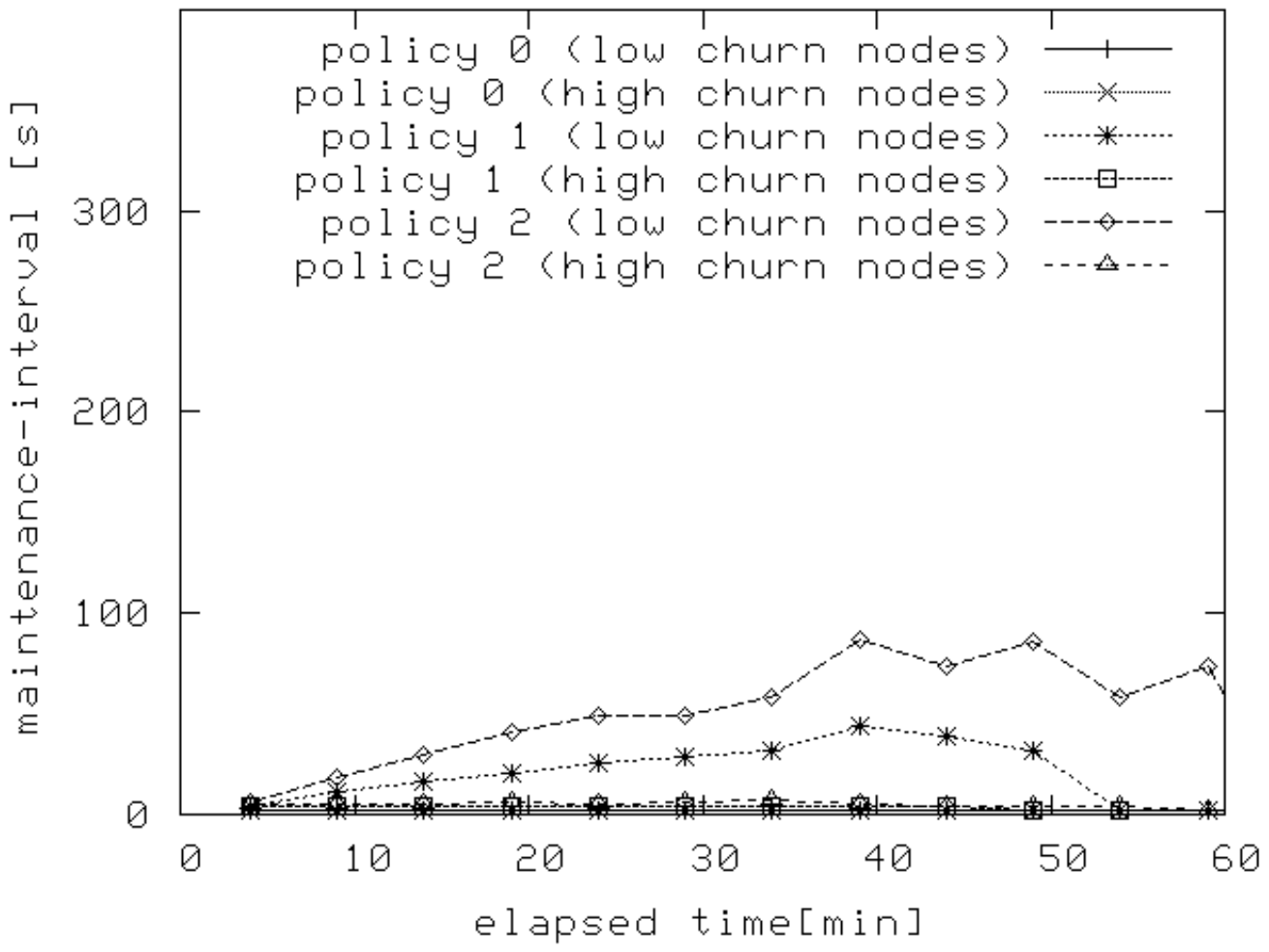

Figure 7.11: Maintenance-interval progression with synthetic heavy weight workload and locally varying churn.

In this experiment, heterogeneous behaviour was exhibited. This caused autonomic management to apply two categories of actions. Maintenance intervals were set to a low value for high churn nodes, and to a high value for low churn nodes. The driver for the adapta-tion in low churn nodes was the detected degree of churn whereas the interval for the high churn nodes was reset to the initial value every time they (re)joined the network. Therefore management actions are plotted separately for both categories in figure 7.11. A network consisting only of low churn nodes would have been expected to exhibit management actions similar to the ones shown in figure 7.5, whereas a network consisting only of high churn nodes would have been expected to exhibit management actions similar to the ones shown in figure 7.8. Even though low churn nodes never left the network, autonomic management never set any interval at such high values as it did for nodes in a network with a homogeneous low churn (see figure 7.5). In fact, the progression shows that management even reduced the maintenance intervals on low churn nodes to a value close to the initial value, when appropriate. By decreasing intervals on nodes exposed to high churn and increasing intervals on nodes exposed to low churn, the management resulted in an overall decrease of the error rate and subsequently a decrease of the expected lookup time and an overall decrease of the network usage with respect to a network with unmanaged nodes. More details are available in appendix A.2.8.

### Experiments with Temporally Varying Churn

Figure 7.12 shows the progression of the expected lookup time with a temporally varying churn. Here nodes exhibited phases of low churn alternating with phases with high churn, each phase lasted about 1000 seconds (17 minutes). When comparing figures 7.12, 7.13 and 7.14 it can be seen that the autonomic manager successfully detected unsatisfactory situations with respect to wasted effort in low churn phases and increased maintenance in-tervals. Conversely, in high churn phases it detected that maintenance effort is required and decreased the intervals. The peak of the expected lookup time (for policy 1) at minute twenty in figure 7.12 corresponds with the start of the high churn phase. It shows a high expected lookup time due to errors which were caused by peer-sets being out of date. Unmanaged nodes exhibit a relatively high error rate towards the end of the experimental runtime which corresponds with an increase of the expected lookup time with unmanaged nodes. The network usage progression plotted in figure 7.13 shows that autonomic management caused more resources to be spent in high churn phases than with unmanaged nodes. The effects on the individual ULM were the result of the maintenance interval progression as shown in figure 7.14. It shows that the interval adaptation carried out by the autonomic manager corresponds with the degree of churn in the specific phases. Autonomic management yielded an adaptation of the controlled intervals in response to the changing situations quickly enough to result in an overall benefit. More details are available in section A.2.9. Here it can be again identified that policy 2 reacted more aggressively than policy 1 and thus resulted in a greater benefit.

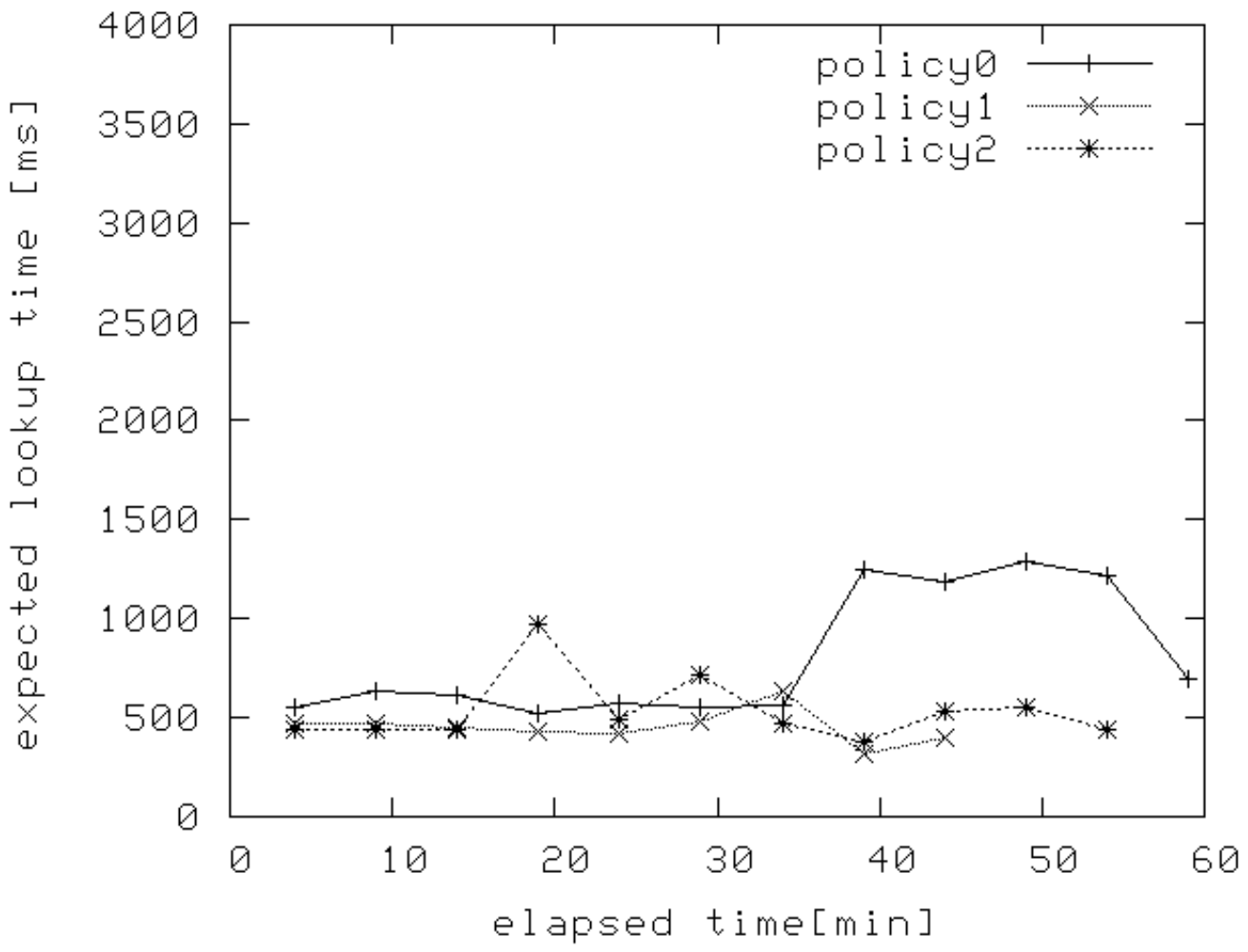

Figure 7.12: Expected lookup time progression with
synthetic heavy weight workload and temporally varying churn.

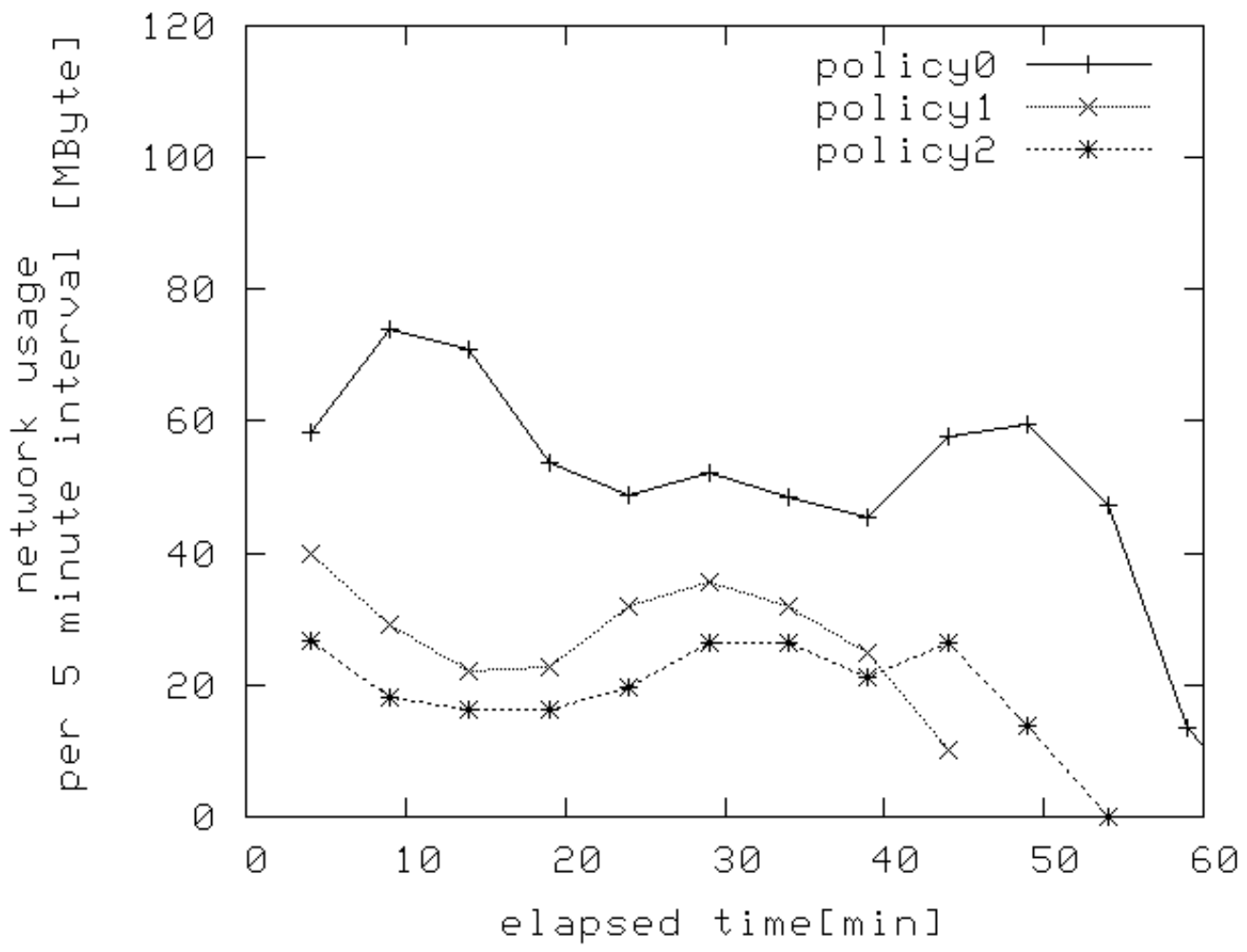

Figure 7.13: Network usage progression with
synthetic heavy weight workload and temporally varying churn.

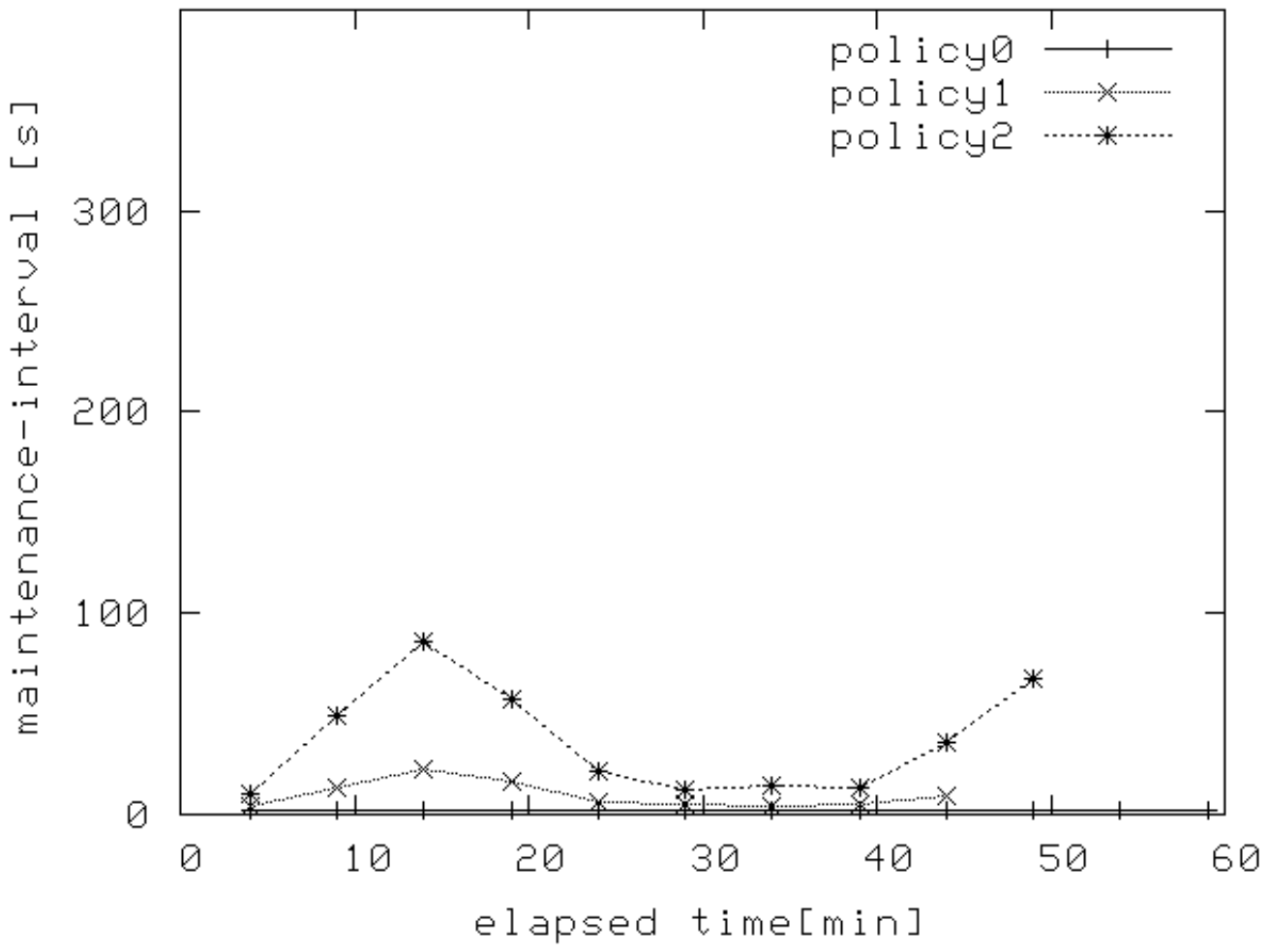

Figure 7.14: Maintenance-interval progression with synthetic heavy weight workload and temporally varying churn.

**Disadvantages of Autonomic Management**

Autonomic management was beneficial in most of the experiments in which it was compared with a static configuration. Performance was improved in some scenarios at the cost of the network usage, which seemed reasonable in those specific scenarios. However, about 19 % of experiments were identified where the tested autonomic policy had a negative effect on performance (table 7.2). Those were in most cases due to the frequency of lookup requests, the resulting amount of available data used as input for the manager, and the resulting management actions. Interesting negative effects were identified with *synthetic light weight workload* and *high churn*; they are representative of a general potential shortcoming of this manager with respect to these specific situations.

During some observation periods in these experiments, more lookup errors occurred with managed than with unmanaged nodes, which resulted in longer expected lookup time. The lookup error rate had approximately the same trend as the progression of the gateway's finger table error rates ($ER_{fixNextFinger}$, see section 7.2.3). This means that, because of too high intervals, out-dated finger table entries at the gateway were accessed during workload execution in such situations.

The reason for autonomically managed nodes with synthetic light weight workload ending up in such a situation was the limited amount of monitoring/input data available to the manager due to the frequency of lookups. A relatively small number of infrequent lookup requests correlated with relatively few failures, even though high churn was exhibited. This was subsequently interpreted by the manager as a low requirement for more maintenance operations. As defined by the autonomic manager's policy, a high error rate was not only interpreted as a strong requirement for a decrease of maintenance intervals, but additionally it initiated immediate maintenance operations. Thus, autonomic management did not apply the desired action in such a scenario with a light weight workload. With a heavy weight workload a lookup failure resulted in the immediately following lookup having an updated peer-set available. With high churn and synthetic heavy weight workload, more lookups also increased the probability of failures which resulted in a greater requirement for decreasing the interval. This is illustrated by comparing the progression of maintenance intervals on the gateway in experiments with synthetic heavy versus light weight workload execution and where a high churn is exhibited in figure 7.15.

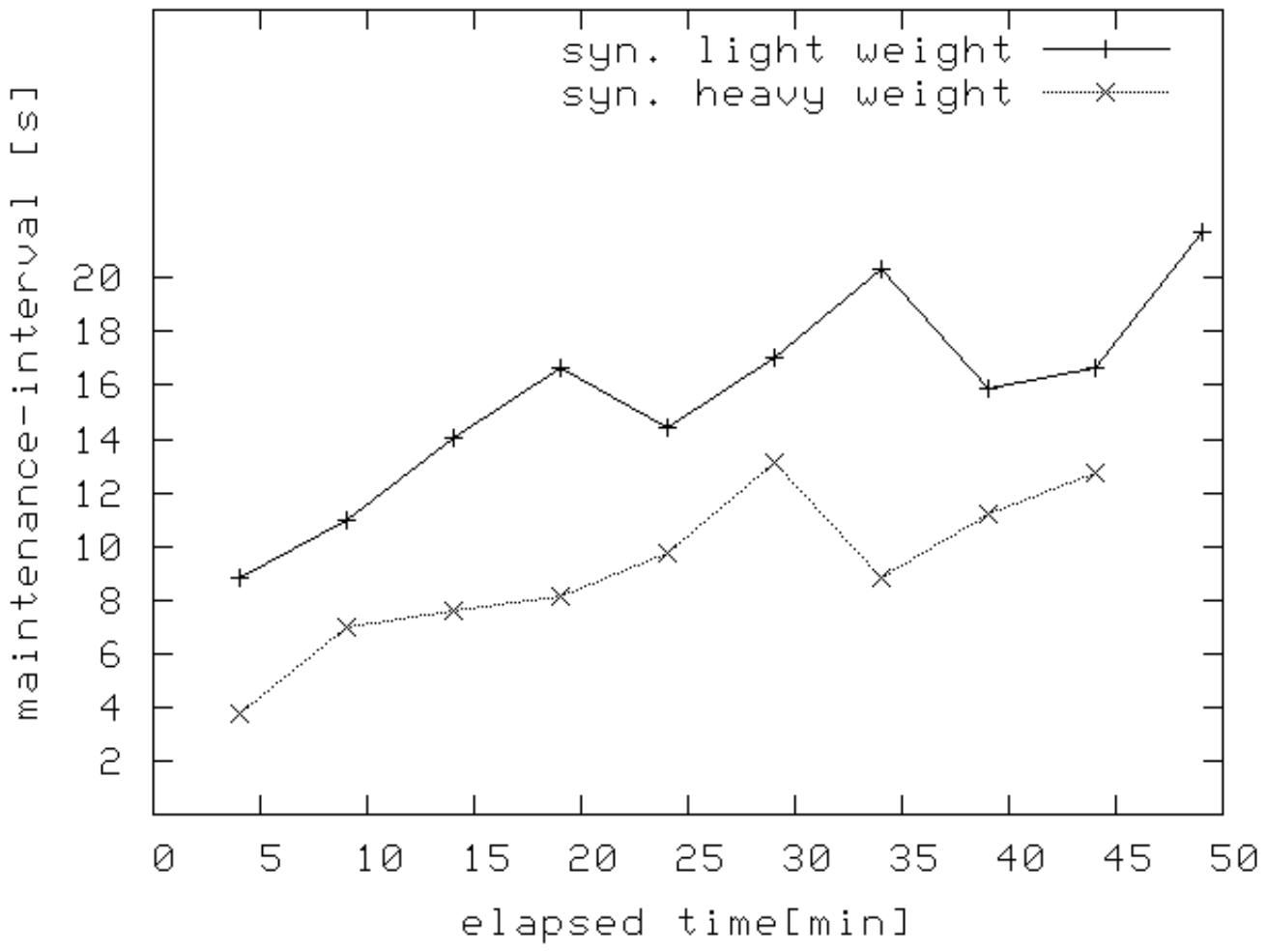

Figure 7.15: Interval progressions with high churn and synthetic heavy versus light weight workload (policy 1).

### 7.5.3 Analysis of Reproducibility

Three repetitions (referred to as runs 1, 2 and 3) of each experiment were executed to investigate reproducibility. A *similarity metric* was defined as the standard deviation normalised ($NSD$) with the mean of the expected lookup times in the corresponding monitoring periods of the three repetitions of an experiment. This provided a set of $NSD$ values for 3 repetitions of an experimental run. $NSD$ values were aggregated by computing the mean, in order to produce a single $NSD$ measure for each experiment.

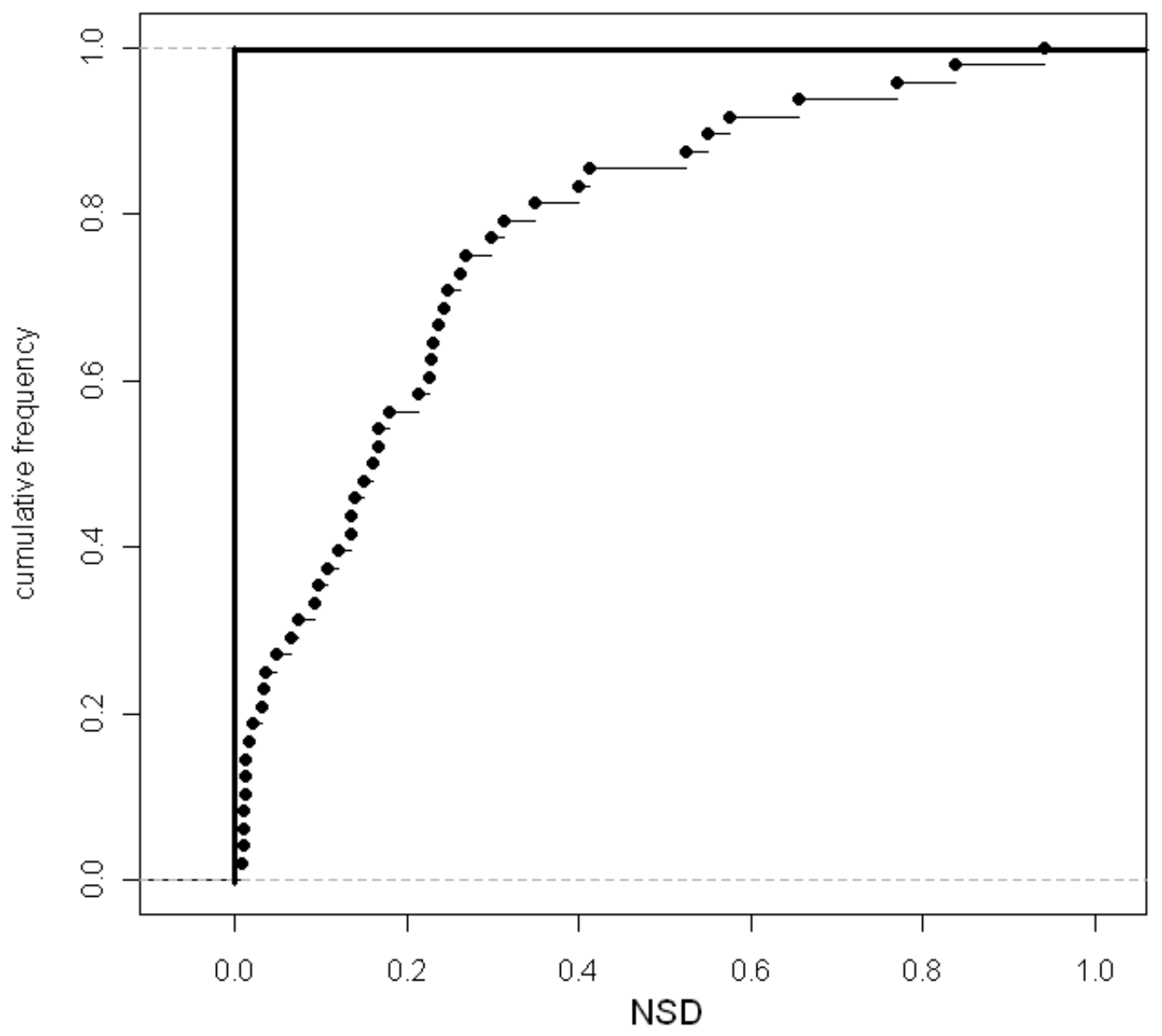

Figure 7.16: NSD cumulative frequency plot.

Figure 7.16 shows the cumulative frequency distribution of all similarity metric values, computed for all three policies in all sixteen groups of experiments. Perfect reproducibility would have been represented by a $NSD$ of $0.0$, which is illustrated as a line in figure 7.17.

To show how the expected lookup times progressed in individual experimental runs with a

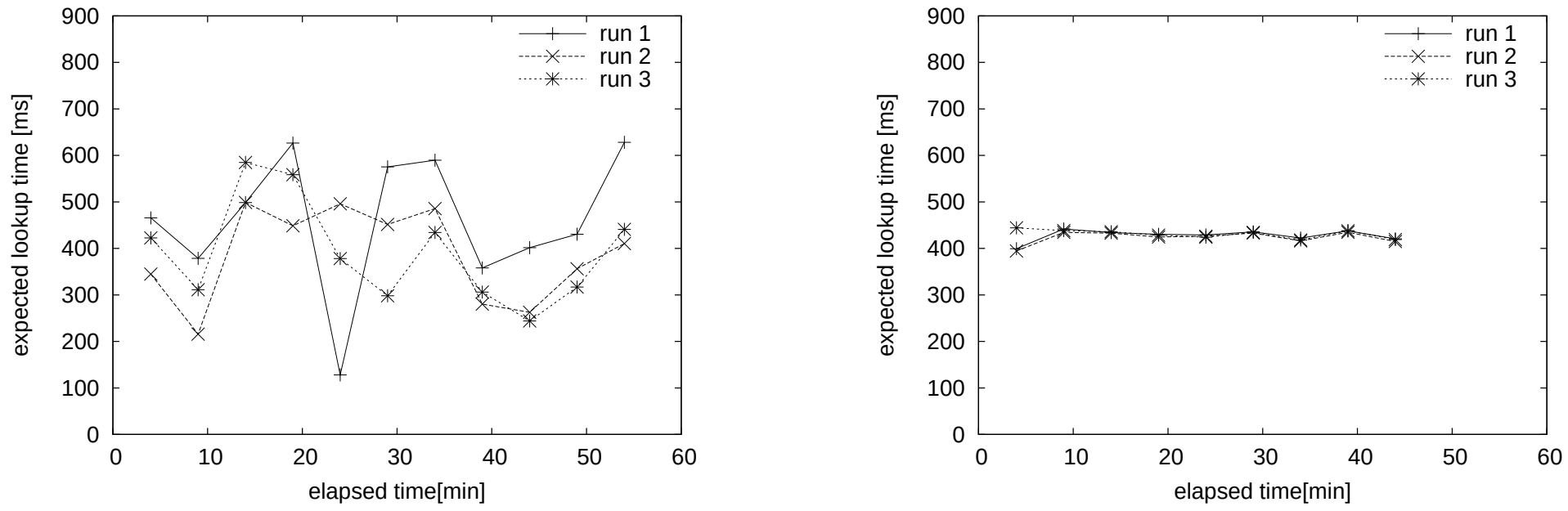

(a) Average $NSD$ value of 0.23 (variable-weight workload, high churn, policy 2)

(b) Average $NSD$ value of 0.01 (heavy-weight workload, low churn, policy 2)

Figure 7.17: Expected lookup time progressions for individual runs of selected experiments

specific $NSD$ value, some selected experiments are portrayed in figures 7.17a and 7.17b. They do not show much variation of the progressions of the expected lookup times and thus on the effects of autonomic management. To summarise, about 50% (the median) of the experiments resulted in an $NSD < 0.2$ and therefore the reproducibility was considered to be sufficient.

## 7.6 Comparison with Related Work

In section 4.3 it is outlined how Ratul Mahajan et. al. [57] evaluated the dynamic adaptation of Pastry maintenance intervals in order to improve performance and resource consumption. Even though no direct experimental comparison was made, some additional data is presented which supports the hypothesis that more benefit may be achieved with the manager introduced here.

The Pastry optimisation focuses on performance because it adapts the maintenance interval in order to achieve a specific minimum performance. In a situation in which churn decreases, this will result in their manager stopping increasing the maintenance interval once the minimum performance is reached, even though churn keeps decreasing. In such a situation an increasing maintenance interval correlates directly with decreasing unnecessarily-used network resources. Thus resource usage is not improved beyond a fixed point by their manager.

The approach introduced in this thesis will increase the maintenance interval as long as no increase in churn is detected. This behaviour was illustrated in the detailed analysis in section 7.5. Figure 7.18 and 7.19 illustrate this here again in different experiments with a synthetic light weight workload.

If the churn level changes and causes an error, maintenance operations will be executed immediately to reduce the possibility of having an outdated peer-set and the interval is decreased. Thus the network usage reduction will not stop at a fixed point if the manager introduced here is used, but still compensates for increased churn by decreasing the in-terval. It is shown in the maintenance-interval progression in figure 7.19 in experiments with temporally varying churn in which nodes exhibited alternating phases of high and low churn, each for a duration of about fifteen minutes.

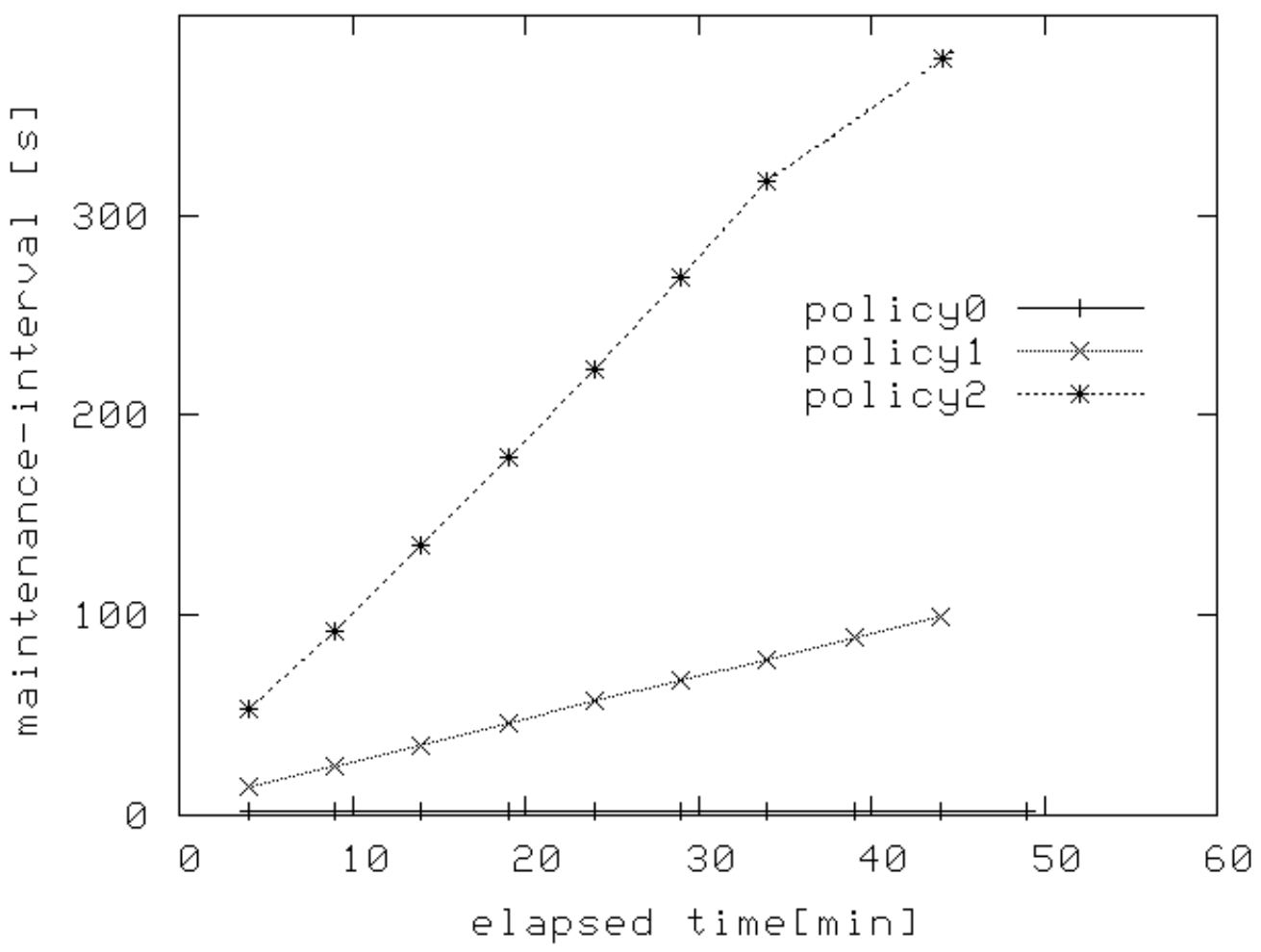

Figure 7.18: Maintenance-interval progression with synthetic light weight workload and low churn.

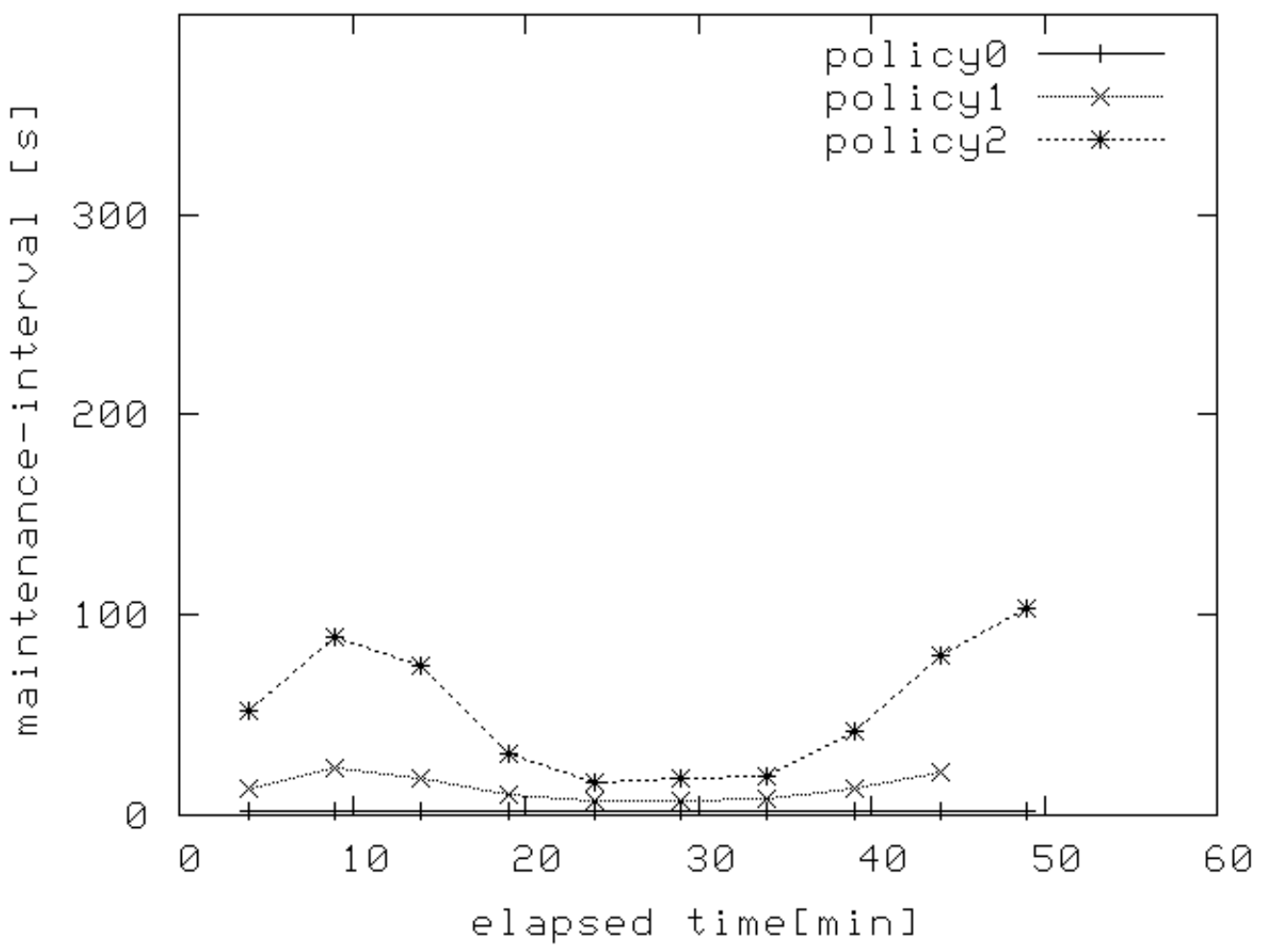

Figure 7.19: Maintenance-interval progression with synthetic light weight workload and temporally varying churn.

## 7.7 Conclusions and Future Work

In the experiments reported in this chapter it was demonstrated that significant benefit can be achieved by controlling a P2P overlay's maintenance scheduling mechanism autonomically, as shown by the example of ASA's underlying P2P overlay *StAChord*. It was hypothesised that autonomic management may yield benefits under dynamic conditions by adapting the intervals appropriately. A further hypothesis stated that under static conditions autonomic management may yield benefits by causing the individual ULMs to converge at better values than an unmanaged system. Both hypotheses are considered to be successfully evaluated. It has been demonstrated that autonomic management is able to detect situations in a dynamic environment in which maintenance effort is wasted or when more maintenance work is required (see figures 7.12 and 7.13). It was also demonstrated that in the case of homogeneously uniform environments, the autonomic manager caused the individual ULMs to converge over time to values much better than in an unmanaged system (see figure 7.3, 7.4, 7.6 and 7.7). However, situations have been identified in which this approach to autonomic management made things worse and thus bears some potential for improvement, which is discussed in the following.

### 7.7.1 Suggestions for Adaptations of Chord

A fall-back mechanism when accessing faulty fingers could improve the benefit which can be gained from autonomic management of the maintenance scheduling.

In the Chord implementation used for these experiments a lookup request fails if the lookup involves an access to a finger which points to a host which is not available anymore. This means the information in the finger table is out-dated. Increasing churn correlates with the probability for a finger to be out-dated. This suggestion for improvement is based on an existing fall-back mechanism which selects the next node from the successor list in the event of an out-dated successor being accessed. The fall-back mechanism for finger table accesses suggested is to select the next node from a list of nodes in the event of an access to a specific finger failing. Such a list could be equivalent to a specific finger's successor list. This adaptation would be an improvement in combination with autonomic management because accesses to failed fingers would not result in an error, since the fall-back mechanism would provide an alternative finger and the autonomic manager would initiate a maintenance operation immediately after an error was detected. This may allow an autonomic manager to set the interval to a high value after a long period of low churn but also keeps the finger table reasonably up-to-date if a sudden high churn is exhibited. A trade-off resulting from this improvement is that more information needs to be transferred over the network when maintaining a finger list than when maintaining a single finger.

Another option which may address this issue would be to modify Chord's lookup operation to choose another finger from the finger table if the accessed finger points to a node which has already left the overlay. This may increase the numbers of hops required for a lookup, but on the other hand no additional overhead would be introduced when transferring information about a node across the network when maintaining a node's finger table.

### 7.7.2 Encountered Nodes in the Network

In addition to the benefits determined by *expected lookup time* and *network usage*, experiments showed that in an overlay consisting of autonomically managed nodes, more nodes stayed connected with their peers than in experiments with unmanaged nodes. This was considered to be a benefit of autonomic management with respect to performance for the following reasons.

In the experiments reported here, the number of nodes connected to the rest of the network was periodically monitored and averaged over an experimental run. The number of encountered nodes available in the overlay network depended on the accuracy of the gateway's finger table. This was due to the network-*join* protocol, in which a joining node executes a lookup via a *known node* to locate its successor for initiating its peer-set. In this work, the gateway was used as the *known node* for joining nodes. It was managed with the same management mechanism as the rest of the P2P nodes. If the management caused the gateway to have out-dated finger table entries, joining nodes could not find their successor and consequently were left out of the overlay network. Additionally, any participating peer with an out-dated peer-set could cause the joining or discovery process to break. Thus the greater the number of discovered nodes in experiments with a high churn, the more up-to-date the peer-set, and the greater the benefit of a specific policy with respect to the number of participants and subsequently to performance.

In more detail: the node discovery process was carried out in every experimental run reported here at every 50 seconds. By making use of the circular key order, the gateway's successor was asked for its successor and so forth until the successor was reached again. The number of detected successors indicated how many nodes were present in the overlay network. The mean number of detected nodes was computed over an entire experimental run. As a representative example, the numbers of encountered nodes in experiments with networks with high membership churn are analysed in the following. The mean number of detected nodes was normalised with respect to the number detected in experiments in which nodes managed with policy 0 were used. When a synthetic heavy weight workload (under which autonomic management gained the highest benefit) was executed, policy 1 resulted in 65% and policy 2 in 33% more node encounters than in networks with unmanaged nodes. When all the normalized node encounters were averaged over experiments in which a high churn was exhibited and every workload pattern was executed, policy 1 resulted in 3.4% and policy 2 in 2.1% more node encounters than in networks with unmanaged nodes. Thus it can be said that autonomic management improved the performance with respect to nodes that stayed connected to the overlay even under high churn.

### 7.7.3 Alternative Validation Approaches

In the evaluation reported here the effects of the specific policies on individual ULMs were quantified with a single value for each ULM. For instance, the single value for the expected lookup time for an individual experiment was represented by the mean of all expected lookup times, computed for individual observation periods. An individual expected lookup time was only computed if monitoring values for lookup times and lookup error times were available for the observation period under consideration. Thus, no expected lookup time was computed if all lookups failed in this observation period. It was however the case that in a small number of cases no expected lookup times were computed as all lookups in some observation periods failed (for instance in experiments with light weight workload and high churn).

An alternative method for quantifying the effects on both ULMs, which captured such situations, showed that autonomic management also yields benefits if ULMs in such observation periods were not missed out. Appendix A.2.19 provides results from a holistic quantification method in which one single expected lookup time was computed for an entire experiment, using the means of all monitored lookup error times and lookup times as well as the lookup error rate computed over the entire experimental run time. However, this holistic approach towards quantifying the effects of the specific policies does not allow to use one set of numbers for showing progressions of the specific user-level metrics and for extracting a single number for each policy to quantify its effect on a specific ULM. Additionally, spikes in the monitoring data could skew the results. Thus the holistic quantification method was not used for this chapter but is however provided in appendix A.2.19 for comparison reasons.

### 7.7.4 Wider Contribution

As the motivation for this research is the application of P2P overlays in distributed storage systems (see chapter 1 and 3), this work is a contribution to the area of distributed storage systems. Additionally this chapter makes a contribution to autonomic management of peer-set maintenance in P2P overlays in general by evaluating the autonomic management approach outlined in section 7.2. Current P2P overlays execute periodic maintenance operations as outlined in section 4.2. Any such overlay could be adapted to use a similar manager to the one introduced here.

# Chapter 8

# Experimental Evaluation of the Management of Distributed Store Clients

## Outline

The autonomic manager introduced in chapter 5 was used with a local area deployment of the ASA distributed storage system. A range of management policies were tested under conditions specified by churn pattern, file access workload, variation of network speed and data item size. The experiments reported here show that autonomic management leads to improved data retrieval times and network usage in the presence of various combinations of conditions.

## 8.1 Introduction

This chapter reports on the experimental evaluation of the approach to autonomic management of distributed storage systems proposed in chapter 5. In chapter 5 it was hypothesised that a distributed storage system in which an autonomic manager is used for dynamically adapting the number of concurrent fetch operations ($DOC$) may yield better performance and resource consumption in various network conditions than a statically configured one.

In an unmanaged system, two unsatisfactory situations can be identified with respect to the data retrieval time (*get time*) and the network usage. These are also referred to as *user-level metrics, ULM*. The first problem arises when the DOC is low and there is a large variation in the times taken to retrieve replicas from various servers. In this situation it is desirable to increase the DOC, because by retrieving more replicas in parallel, a result can be returned to the user sooner. The converse situation arises when the DOC is high, there is little variation in retrieval time and there is a network bottleneck close to the requesting client. In this situation it is desirable to decrease the DOC, since the low variation removes any benefit from parallel retrieval, and the bottleneck means that decreasing parallelism reduces both bandwidth consumption and elapsed time for the user. As reported in section 5.3, some distributed storage systems use a server ranking mechanism ($SRM$) to optimise the distributed storage system.

In section 5.4 situations were identified in which either a statically low DOC, a statically high DOC or a low DOC in combination with a SRM configuration yields the greatest benefit. It was found that in situations where a specific configuration is beneficial the others may not be. It was experimentally evaluated whether autonomic management is able to set an appropriate DOC in each situation. The effects of an autonomic manager and a SRM on both ULMs were measured in such situations and compared with static configurations. This managed client was deployed in a local area test-bed and exposed to various conditions.

This chapter is structured as follows. Section 8.2 reports how autonomic management was implemented and applied to an ASA client. In section 8.3 the experimental parameters: churn pattern, workload, variation of available network speed, and data item size are specified. Section 8.4 outlines the configuration of experimental parameters and the test-bed, including an explanation of how the ULMs *get time* and *network usage* were monitored and how they are summarised in order to determine the benefit of autonomic management. Experimental results are shown in section 8.5. The chapter finishes with some conclusions about the reasons for the monitored effects and potential future work in section 8.6.

## 8.2 Implementation of Autonomic Management for an ASA Store Client

An autonomic manager and a SRM were developed based on the proposal made in section 5.5, and implemented using the generic autonomic management framework GAMF, which was introduced in chapter 6. This section reports on the implementation of the managed distributed storage client as it was used for the experiments reported in this chapter.

A manager was added to an ASA client to adapt the DOC used for all *get* requests. Policies periodically increased or decreased the DOC depending on some metrics. Each metric value specified different aspects of the system's state during a specific time window, extracted from local events. Only events which were not analysed during a previous policy evaluation were considered. The events included reused SRM monitoring data. The metrics used were: the current DOC, the failure rate of fetch operations $FFR$, the variation in fetch times $FTV$, and the bottleneck $BN$ which specified the ratio of bandwidths available on the server and client links. The ASA client and its modifications are outlined below, followed by information about the SRM and some details about the autonomic manager.

### 8.2.1 ASA Client

An ASA client includes a data retrieval mechanism for persistent data items, addressed with *PIDs* which lie in the P2P key space. Each data item is replicated four times when it is stored in ASA. The PID of each replica is specified by the *cross algorithm* (section 3.3). On a request for a specific PID a list of four server addresses is produced. Each of the servers identified with those addresses holds an identical data item, of which only one needs to be successfully fetched (see chapter 3).

The original ASA client was extended with a configurable DOC parameter. This allowed multiple replicas to be fetched concurrently during individual *get* requests. After the first replica was retrieved successfully, remaining *fetch* operations were terminated by killing the corresponding fetch threads.

### 8.2.2 Server Ranking

For an individual *get* request, the SRM ranked specific hosts, based on a prediction of how fast each server involved would transfer data to a client (fastest first, failed hosts at the end). The SRM's predictions were based on periodically gathered monitoring data. In order to gather this monitoring data[1] on individual participants, low level monitoring machinery ($LLMM$) was developed as part of the SRM and used in all clients and servers. To probe for LLMM monitoring data on certain known server hosts, further SRM monitoring machinery ($SR3M$) was developed, which was run only on a client. Data gathered by $SR3M$ was used by the SRM on a client to make the above-mentioned predictions.

#### LLMM

The LLMM was developed to periodically measure bandwidth and latency on a network link between a host which executes the LLMM (*LLMM-executor*) and an interconnection as introduced in the analytical model in section 5.4. Such a link was specified by the address of the specific LLMM-executor and a statically configured *known host* address in the interconnection which lies closest to the LLMM-executor. The monitoring of bandwidth and latency involved time measurements of three ping requests [33], each with a specific packet size, from the LLMM-executor to the specific known host. First a single ping with a negligible packet size was sent to avoid the measured times being skewed by a delay due to an ARP [72] request. No time was measured for the first ping. The second ping, of the same size as the first one, was carried out to derive the latency from its round trip time.

[1] bandwidth and latency on the links between a participant and the interconnection (figure 5.1)

The third ping was sent using a packet size ($S$) significantly bigger than the first two pings. The payload of the third ping and its round trip time were used to compute the bandwidth. The monitored bandwidths and latencies were stored locally by the LLMM-executor and provided for a remote caller on request.

**SR3M**

The client-specific monitoring component, SR3M, periodically contacted certain known servers to gather bandwidth and latency monitoring data. Gathered server-specific data was stored locally on the client along with client-specific bandwidth and latency measurements monitored by a local LLMM executed on the client.

**Ranking of Servers**

Monitoring data gathered from servers was used to compute an *expected data transfer time (EDTT)*, for each server, as proposed in section 5.5.2. Based on the assumption that variation originates from bandwidth and latency on the server side, only monitoring data from servers was considered when computing the EDTT. Which represented the expected time to transfer a data item with a specific size across a link with the monitored bandwidth and latency. Server-specific EDTTs were used to rank servers.

### 8.2.3 Autonomic Management Details

An event generator was added to the ASA client and to the SRM to provide the autonomic manager with information to compute metric values. It generated *network speed monitoring events* containing information about bandwidth, latency and the address on which host (the LLMM-executor) the data was monitored. Additionally *EDTT events* were generated for every computed EDTT. With respect to the ASA client's data retrieval mechanism *fetch failure events* and *initiated fetch operation events* were generated.

A metric extractor was used to compute the FFR, the FTV and the BN metric values as specified in section 5.5. Only events generated in a specific observation period were considered. The FFR metric was the ratio of failures to initiated fetch operations. If no operations (which also implies no errors) were monitored, the failure rate was defined to be $0$. The FTV was specified by the standard deviation normalised by the mean of all EDTTs. If no monitoring data was available to compute the FTV metric, the metric value was defined to be $0$. The BN metric was the ratio of the mean of all monitored client bandwidths to the mean of all server bandwidths. If no monitoring data was available to compute this metric, the metric value was defined as $1$. For providing consistent policy input a metric extractor was implemented to report the currently used DOC as a metric.

The policy was defined as follows:

- *If the DOC is smaller than the maximum DOC, and FFR and FTV are high, set the DOC to the highest possible value.*
- *Else if the DOC is smaller than the maximum DOC, and FFR or FTV is high, incrementally increase the DOC.*
- *Else if the DOC is greater than the minimum DOC, and FFR and FTV are low, decrease the DOC.*
  - *If BN denotes a bottleneck on the client side, set the DOC to the minimum DOC value (1).*
  - *Else incrementally decrease the DOC.*

A FFR or FTV metric value was specified to be high or low if it was greater or smaller than a specific threshold. A bottleneck at the client side was determined by BN values smaller than a specific threshold. Each policy evaluation triggered the metric extractor to compute the specific metrics and to apply a new DOC via an effector.

## 8.3 Experimental Parameters

Various policies were configured using the autonomic manager introduced in the previous section (8.2). The effects of policies on performance and resource consumption were experimentally evaluated in a local area ASA deployment, exposed to various conditions.

These conditions varied in churn pattern, workload, variation of available network speed, and data item size. In this section the conditions are defined in a general way and the machinery via which specific conditions were applied are explained. The configuration of specific conditions, for instance the particular data size or the particular churn pattern, are defined in a further section (8.4).

### 8.3.1 Data Item Size

The least complex experimental parameter was the data item size. It specified the uniform size of all data items retrieved from ASA servers during *get* requests. Data items of the specific size were uploaded before the experimental runs via standard ASA client and server interfaces and stored on the ASA servers' file system.

### 8.3.2 Workload

A workload specified a temporal pattern of *get* requests made against the distributed store client. The same principle as in the experiments reported in chapter 7 was used to specify *light weight workloads, heavy weight workloads* and *variable weight workloads*. The PIDs used to request specific data items were pseudo-randomly generated to allow the sequence of PIDs to be reproducible between experiment repetitions.

### 8.3.3 Churn Pattern

A churn pattern specified a temporal pattern with which storage servers were on-line or off-line, using the same principles as in the experiments reported in chapter 7 to support *low churn, high churn* and *temporally varying churn* patterns.

### 8.3.4 Network Speed

The network speed specified the bandwidth and latency between individual client and servers and an interconnection[2]. In the experimental setup used here the interconnection was represented by a central router via which all IP traffic was routed. Figure 8.1 shows the logical network topology used to represent a simplified distributed storage system.

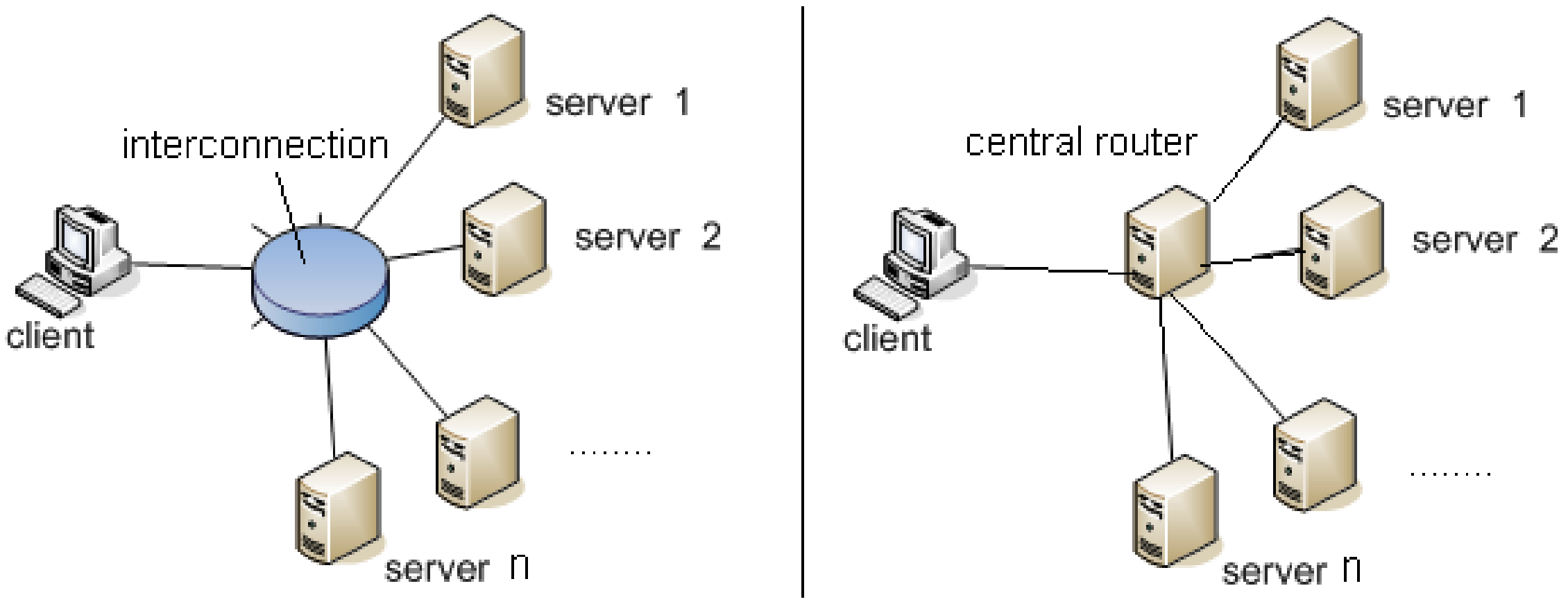

Figure 8.1: A simplified distributed storage system.

The network speed specified latency and bandwidth between client and servers and the central router. Bandwidth and latency applied to a specific link was the same in both direc-

[2]see chapter 5

tions.

Two categories of network speed configurations were defined, a uniform and a temporally varying one. In uniform network speed configurations, all servers exhibited the same network speed; they were used to simulate scenarios in which a network bottleneck was exhibited on the client or server side or on neither side.

**Simulated scenarios with uniform network speed configurations:**

- bottleneck on client side: The bandwidth between the client and the central router was significantly lower than between the individual servers and the central router. The latency between the client and the central router was significantly higher than between the individual servers and the central router.
- bottleneck on server side: The bandwidth between the client and the central router was significantly higher than between the individual servers and the central router. The latency between the client and the central router was significantly lower than between the individual servers and the central router.
- no bottleneck: All links between the individual participants and the central router exhibited the same bandwidth, no latency was configured.

These network speed configurations did not change during the course of an experiment.

**Simulated scenario with temporally varying network speed configuration:** The network speed varied between individual links to the central router and also varied over time.

The configured bandwidth and latency values were pseudo-randomly selected from ranges of bandwidth and latency values, defined by maximum and minimum values. This provided variation between the network speeds of the individual links during an experimental run, but allowed reproduction of the same network speed patters between repetitions of experimental runs. A high bandwidth value directly correlated with a low latency value. For instance, a configuration with a maximum bandwidth of 100 Mbps, a minimum bandwidth of 1 Mbps, a maximum latency of 100 ms and a minimum latency of 1 ms would result in a latency of 1 ms if a bandwidth of 100 Mbps were selected. At regular intervals a new network speed configuration was generated on every link.

**Implementation:** The network traffic shaping was implemented on the central router following guidelines from [38, 37, 35] using the Linux traffic control software $tc$, as available in *CentOS 5* (kernel 2.6.18-8.el5).

## 8.4 Experiment Setup

This section reports on the configuration values of the experimental parameters, including the churn pattern, workload, network speed, data item size and the autonomic manager configurations. Additional preliminary work is reported in appendix B.1, including the derivation of specific configurations and evaluations of the experiment harness.

### 8.4.1 The Test-Bed

The experiments were carried out in the same test-bed as used for P2P layer experiments, described in section 7.4.1. The network configuration was adapted so that one of the computers in the test-bed acted as the central router (see section 8.3.4). The client was configured with *global knowledge* of all storage server addresses and their key ranges.

### 8.4.2 Derivation of User-Level Metrics

These experiments were carried out to evaluate the effect on performance and resource consumption of specific policies which varied the DOC in response to specific conditions.

Performance was measured in terms of the time it took to complete *get* requests[3], the error rate and the time after which failing *get* requests reported an error. The network usage was obtained from the Linux traffic shaping software *tc*.

Above performance measurements were combined to give an *expected get time* using the same principles and motivations as for computing an *expected lookup time* in the experiments with ASA's P2P layer (see section 7.4.2, formula 7.1). To obtain the *expected get time*, the *get time*, the *get error rate* and the *get error time* were aggregated over observation periods of five minutes. The network usage measurement differed from earlier P2P layer-specific experiments due to a change in the version of the $RAFDA$ middleware. A new version in which the network traffic monitor was not implemented anymore was used here. Therefore *tc*-statistics were used to record the total amount of data sent from storage

[3] part of the executed workload

servers to the central router. Each experiment (specified by churn pattern, workload, data item size, network speed and policy) was repeated three times to verify the reproducibility of the observed effects. As in the P2P layer experiments, all expected *get time* values were aggregated over all three experiments and averaged in order to present a single performance measurement per experiment. The network usage was also specified by the mean of the three measurements (one measurement was available per experimental run).

### 8.4.3 Data Item Size Configurations

Two data item sizes were defined, one as *1024 KB* and another one as *100 KB*. These sizes were chosen after preliminary test runs showed that larger data item sizes caused *out of memory exceptions* in the experimental harness, while much smaller data item sizes did not give the manager enough time to show its full potential in situations with heavy weight workloads.

### 8.4.4 Workload Configurations

A *heavy weight workload*, a *light weight workload* and a *variable weight workload* were specified similar to the corresponding synthetic workloads in section 7.3.2. The workload determined the temporal pattern of *get* requests. The *heavy weight workload* contained *300 sequential get requests*. The *light weight workload* contained *10 sequential get requests*, with *120 seconds delay* between each of them. The *variable weight workload* was configured to execute *3 sequential get requests*, one after the other, with a *delay of 120 seconds* *after each sequence of 3 get requests*. The entire *variable weight workload* contained 30 get requests. The motivation for the specific configuration parameters of all workloads was to give the manager enough time to show its full potential and to stress the system to present a range of scenarios which starkly vary from each other.

### 8.4.5 Churn Pattern Configurations

Churn patterns were defined identically to the corresponding ones in section 7.3.1, representative of a storage server network with *low membership churn, high membership churn* and *temporally varying membership churn*. The low churn kept all storage nodes on-line for the duration of an entire experimental run. The high churn pattern had on-line durations of 37 (+/-5) seconds and off-line durations of 27 (+/-2) seconds. The temporally varying churn pattern was configured with about 5 minutes low churn followed by about 5 minutes high churn. In the high and temporally varying churn patterns an initial additional on-line phase of 20 (+/-5) seconds was included.

### 8.4.6 Network Speed Configurations

The static network speed configurations simulated network bottlenecks on either client, server or neither side. Each server and the client experienced a statically configured bandwidth and latency on their link to the central router for the entire experimental duration. The following values were used:

- A bottleneck at the server side:
  - Client bandwidth 78 Mbps & latency 0 ms
  - Uniform server bandwidth 3 Mbps & latency 20 ms
- A bottleneck at the client side:
  - Client bandwidth 3 Mbps & latency 20 ms
  - Uniform server bandwidth 22 Mbps & latency 0 ms
- No bottleneck:
  - Client bandwidth 18 Mbps & latency 0 ms
  - Uniform server bandwidth 18 Mbps & latency 0 ms

The temporally varying configuration changed the network speed configurations of the individual links every 10 seconds. Each individual link was configured with a pseudo-randomly selected latency from 0 ms to 20 ms and a bandwidth from 220 kbps to 22 Mbps.

In all of the configurations the maximum value for the latency of 20 ms was chosen after carrying out some network speed measurements on a common workstation connected via DSL to the internet. The sum of all configured bandwidths (on the client and server links to the central router) was chosen to not exceed the physical maximum of 94 Mbps[4]. 18 Mbps was used as the bandwidth for the individual links in configurations in which no bottleneck was exhibited, in order to equally share the network capacity amongst individual links.

[4]derived from measurements, part of the preliminary work, see appendix B.1

### 8.4.7 Policy Parameter Configurations

The behaviour of the policy introduced in section 8.2.3 was configurable via specific policy parameters, including metric specific thresholds and initial DOC values. The effects of five different DOC management policies on performance and resource consumption were evaluated. Two policies set a statically configured DOC; the other three performed autonomic adaptation of the DOC, as introduced in section 8.2. In order to make a fair comparison the autonomic manager was used to apply the static DOC configurations, but was configured not to react to any events.

All policies were evaluated once per $1$ minute; this also determined the monitoring interval over which events for individual metrics were extracted. The autonomic management policy determined a new DOC depending on metric values for the fetch failure rate (FFR), the fetch time variation (FTV), and the bottleneck (BN). Thresholds defined whether a specific metric value was (too) high or (too) low. The three autonomic management policies had in common that the threshold for the FTV ($T_{FTV}$) and the threshold for BN ($T_{BN}$) were kept at constant values[5]. The threshold for the FFR metric ($T_{FFR}$) was varied between the policies. Thus the behaviour of the autonomic management policies varied with respect to how eagerly they reacted to failures. As *policy 0* and as a baseline the static configured *DOC=1* was chosen. The motivation for that baseline was that this configuration, in combination with a SRM, is most comparable to existing systems as outlined in section 5.3. All policy parameter configurations are shown in table 8.1.

[5]Both thresholds were derived from preliminary experiments as outlined in appendix B.1.4.

| policy | $T_{FFR}$ | $T_{FTV}$ | $T_{BN}$ | initial DOC | policy type | description |
|---|---|---|---|---|---|---|
| 0 | - | - | - | 1 | static | $ST_{DOC=1}$ |
| 1 | - | - | - | 4 | static | $ST_{DOC=4}$ |
| 2 | 0.1 | 0.2 | 0.8 | 1 | autonomic | $AM_{T_{FFR}=0.1}$ |
| 3 | 0.3 | 0.2 | 0.8 | 1 | autonomic | $AM_{T_{FFR}=0.3}$ |
| 4 | 0.5 | 0.2 | 0.8 | 1 | autonomic | $AM_{T_{FFR}=0.5}$ |

Table 8.1: DOC Management Policy Configuration.

### 8.4.8 Server Ranking Mechanism Configuration

LLMM and SR3M components monitored and gathered data as described in section 8.2.2. Both were configured to execute monitoring and data gathering operations every 15 seconds. Additionally the address to which the LLMM sent pings in order to measure bandwidth and latency was configured at each individual server or client as its gateway address[6]. The data item size which was used to compute the EDTT was statically configured to be 1 MB at each client and server.

## 8.5 Experimental Results

### 8.5.1 Overview

The effects on the *expected get time (EGT)* and *network usage (NU)* of five DOC management policies were experimentally evaluated with three churn patterns, three workloads, four network speed configurations and two data sizes.

[6]represented by the central router in each case

In chapter 5 it was predicted that a statically configured low DOC would result in the best performance in cases in which a bottleneck exists on the client side or if no bottleneck is exhibited. Conversely in scenarios with a bottleneck on the client side or with temporally varying network speeds a static high DOC was predicted to result in the best performance, at the cost of additional resource usage. Autonomic management detected such situations and adapted the configuration accordingly without any prior knowledge of the existing conditions. This resulted in an overall improvement of performance at the cost of some additional resource usage when averaging *expected get time* and *network usage* over all the experiments. An analysis of the statistical significance of compared data sets was carried out following guidelines from [40]. It shows overlaps of the *90% confidence intervals of the means* of the compared measurement, as shown in figures 8.2 and 8.3.

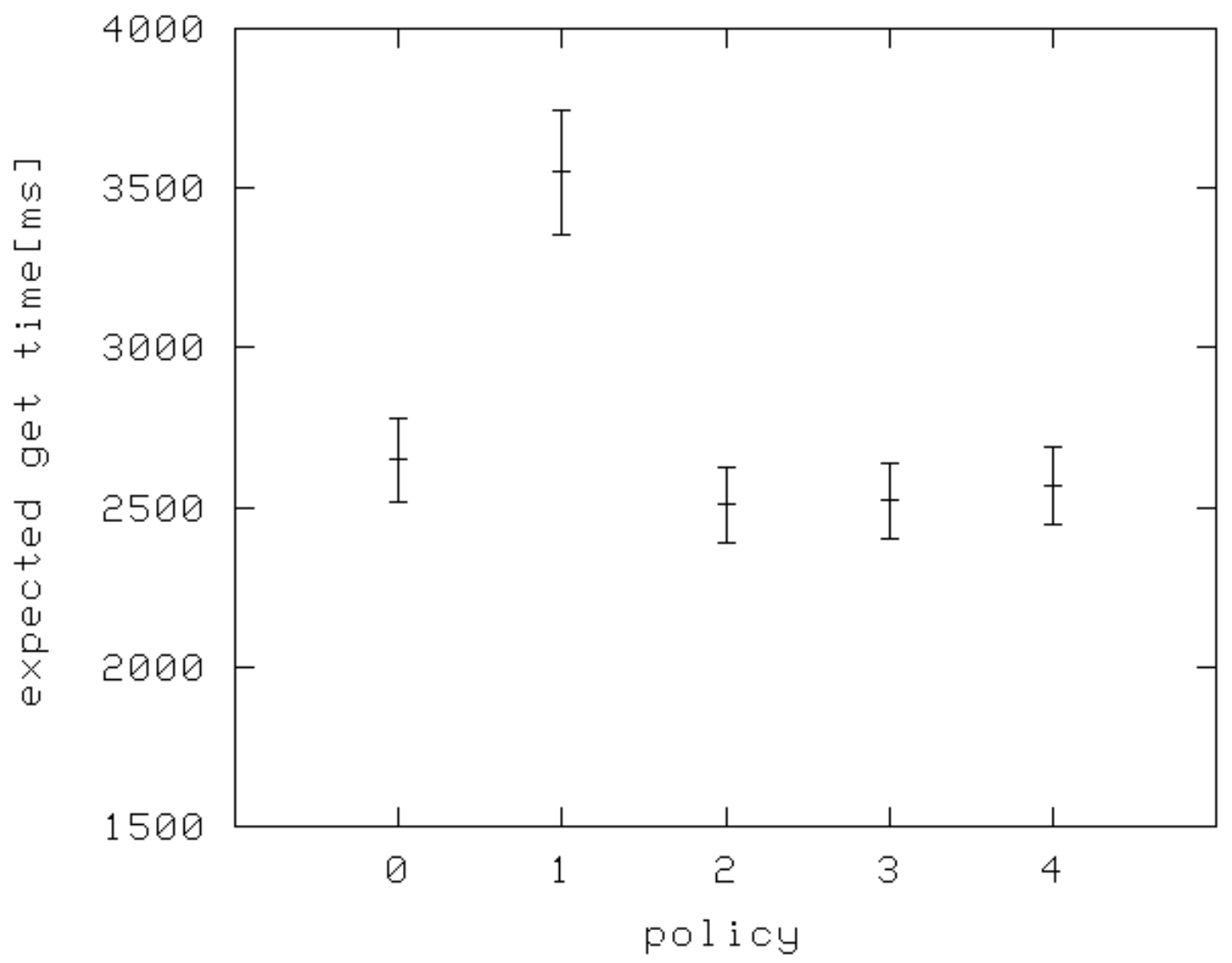

Figure 8.2: Visual Approximation for Statistical Significance of EGT measurements

In addition *t-tests* were carried out to evaluate the probabilities of the compared data sets being statistically significantly differently from each other. The results are presented in table 8.2 and 8.3. The *p-value* represents the probability that the compared measurements are not statistically significantly different.

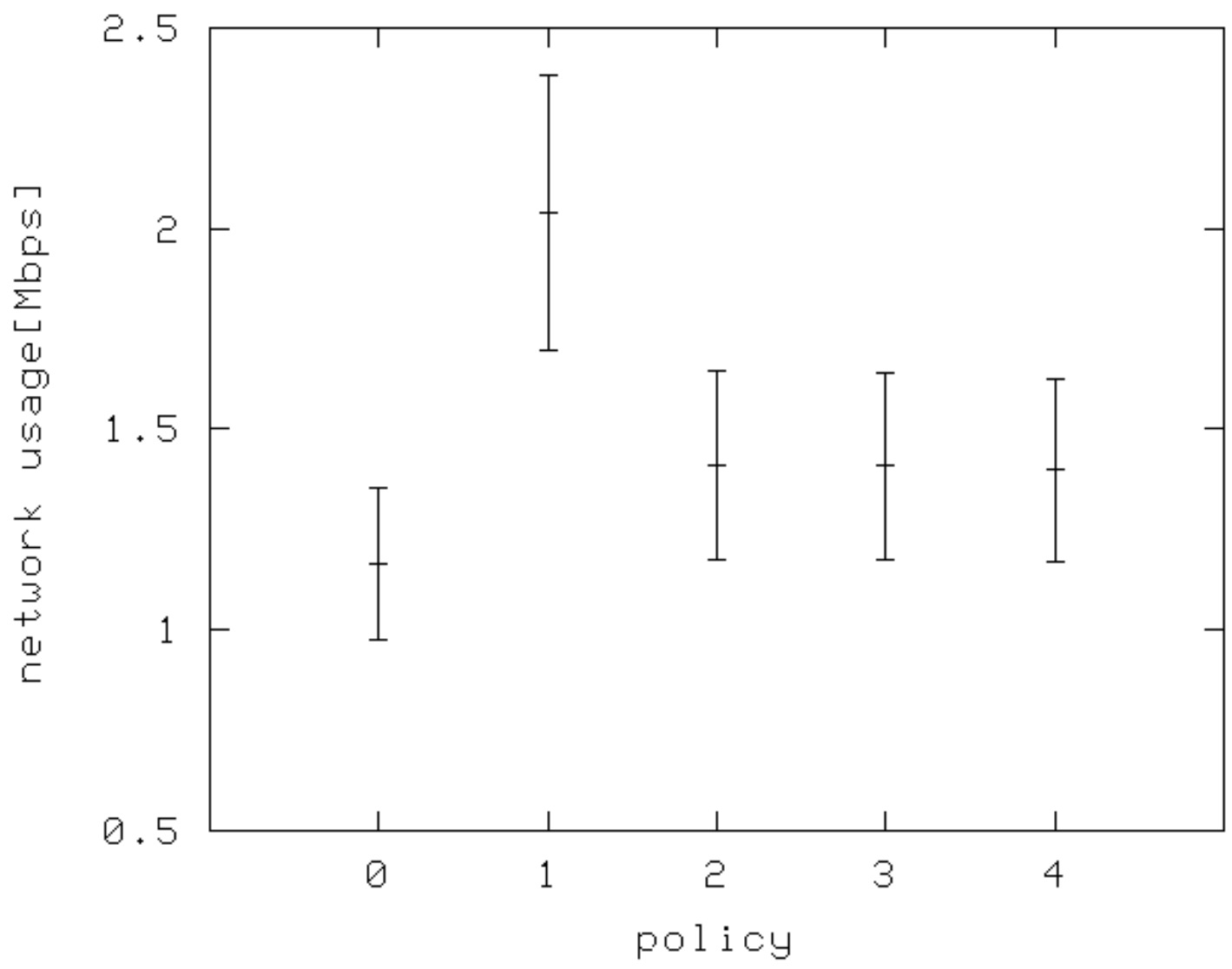

Figure 8.3: Visual Approximation for Statistical Significance of NU measurements

| policy 0 vs. | p-value |
|---|---|
| 4 | 0.44 |
| 3 | 0.23 |
| 2 | 0.19 |
| 1 | 0.00 |

Table 8.2: t-test Results for Expected Get Times

| policy 0 vs. | p-value |
|---|---|
| 4 | 0.20 |
| 3 | 0.18 |
| 2 | 0.18 |
| 1 | 0.00 |

Table 8.3: t-test Results for Network Usage

As in chapter 7 the *expected get time* was computed, based on measurements for *get time*, *get error time* and *get error rate* (see section 8.4.2). Figure 8.4 shows that, in contrast to the *expected get time*, the differences between *get times* resulting from autonomic management are significantly smaller than the ones resulting from any of the statically configured systems. Figure 8.5 shows that some of the confidence intervals for *get error time* overlap. The distributions of *get times* and *get error times* represent the raw data with sample sizes of about 20,000 and 5,000 individual measurements for each policy. The overall percentages of failed get requests are shown in table 8.4. The overall *get error rate* represents the *get error rate* per policy over all experiments.

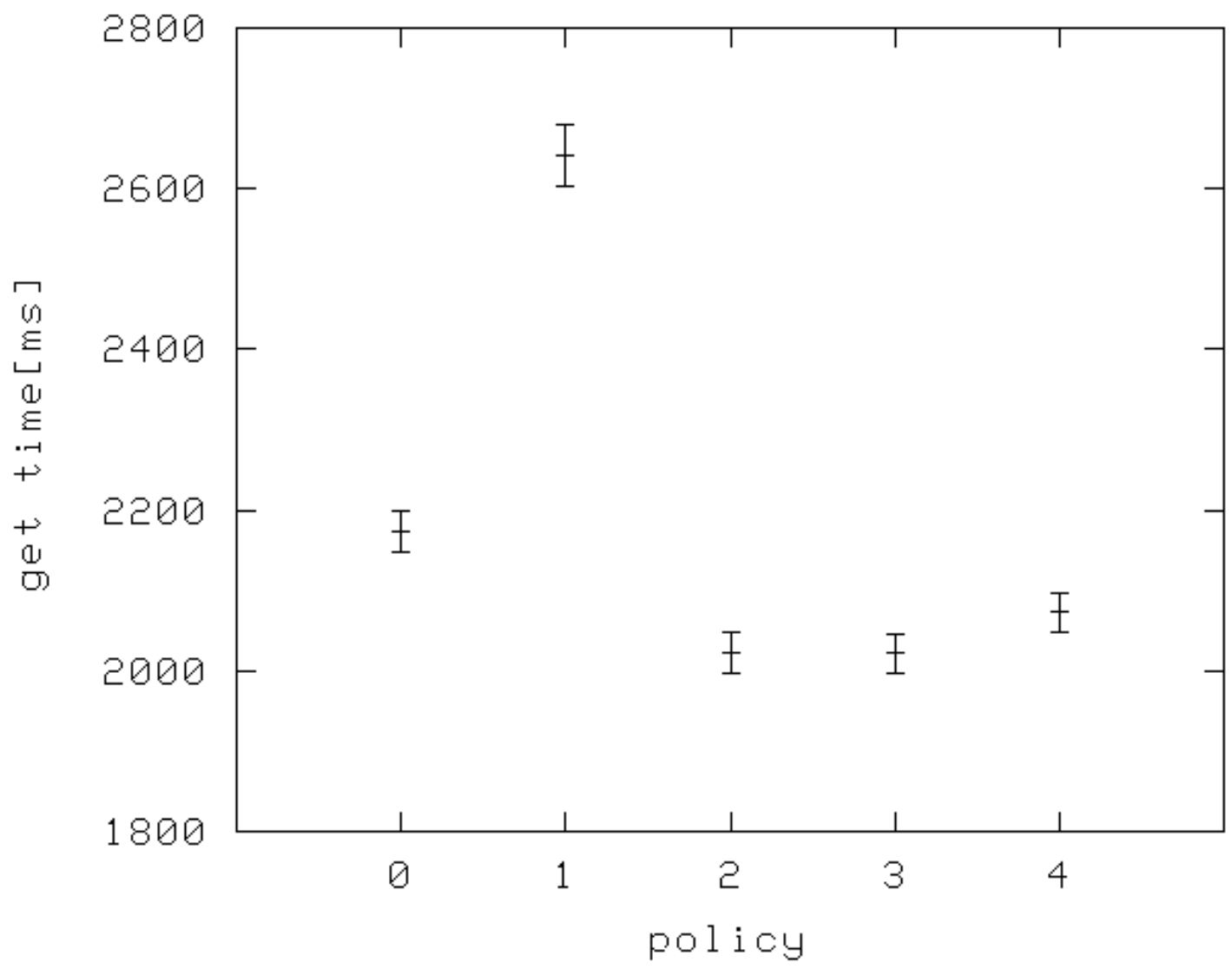

Figure 8.4: Visual Approximation for Statistical Significance of the Get Time

| policy | description | get error rate [%] |
|---|---|---|
| 4 | $AM_{T_{FFR}=0.5}$ | 17 |
| 3 | $AM_{T_{FFR}=0.3}$ | 19 |
| 2 | $AM_{T_{FFR}=0.1}$ | 23 |
| 1 | $ST_{DOC=4}$ | 19 |
| 0 | $ST_{DOC=1}$ | 17 |

Table 8.4: Overall Error Rate of Get Requests

The issue of the high probability that some data sets were not statistically significantly different is discussed in more detail in section 8.6. For simplicity the *expected get time* is used to analyse effects on performance in the following sections, rather than the individual secondary ULMs. To analyse effects of autonomic management on *expected get time* and *network usage* in more detail results are grouped in this section by the network speed configurations. Each of the following groups contains 18 experiments, specified by churn pattern, workload and data item size, in which each of the policies was tested. A combined analysis can be found in appendix B.2.

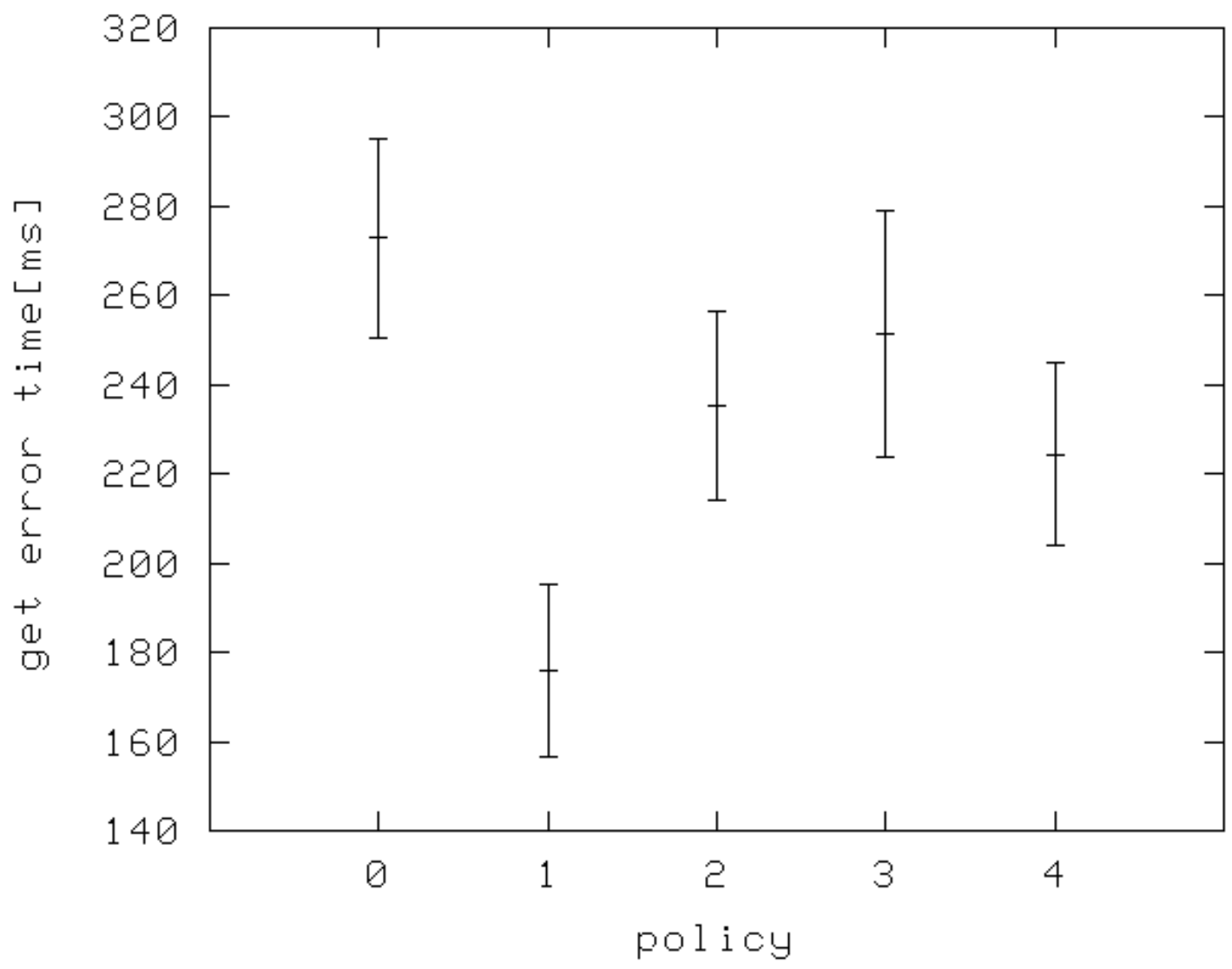

Figure 8.5: Visual Approximation for Statistical Significance of the Get Error Time

The results are organised in tables 8.5 to 8.8 showing the total number of experiments in which each policy yielded the shortest *expected get time (EGT)*, the *lowest network usage (NU)* and a combination of both (EGT & NU), for the individual network speed configurations. Tables B.3 to B.9 quantify the effects on EGT and NU individually (normalised to policy 0) and can be found in appendix B.2.

| policy | description | $EGT$ | $NU$ | $EGT$ & $NU$ |
|---|---|---|---|---|
| 4 | $AM_{T_{FFR}=0.5}$ | 2 | 2 | 1 |
| 3 | $AM_{T_{FFR}=0.3}$ | 5 | 3 | 1 |
| 2 | $AM_{T_{FFR}=0.1}$ | 4 | 7 | 1 |
| 1 | $ST_{DOC=4}$ | 1 | 0 | 0 |
| 0 | $ST_{DOC=1}$ | 6 | 6 | 3 |

Table 8.5: The number of experiments (out of a total of 18) in which the specific policies yielded the greatest benefits (bottleneck on the client side).

| policy | description | $EGT$ | $NU$ | $EGT$ & $NU$ |
|---|---|---|---|---|
| 4 | $AM_{T_{FFR}=0.5}$ | 3 | 5 | 1 |
| 3 | $AM_{T_{FFR}=0.3}$ | 3 | 2 | 2 |
| 2 | $AM_{T_{FFR}=0.1}$ | 3 | 3 | 0 |
| 1 | $ST_{DOC=4}$ | 4 | 0 | 0 |
| 0 | $ST_{DOC=1}$ | 5 | 8 | 3 |

Table 8.6: The number of experiments (out of a total of 18) in which the specific policies yielded the greatest benefits (no bottleneck).

| policy | description | $EGT$ | $NU$ | $EGT$ & $NU$ |
|---|---|---|---|---|
| 4 | $AM_{T_{FFR}=0.5}$ | 0 | 3 | 0 |
| 3 | $AM_{T_{FFR}=0.3}$ | 1 | 4 | 1 |
| 2 | $AM_{T_{FFR}=0.1}$ | 1 | 2 | 0 |
| 1 | $ST_{DOC=4}$ | 13 | 0 | 0 |
| 0 | $ST_{DOC=1}$ | 3 | 9 | 2 |

Table 8.7: The number of experiments (out of a total of 18) in which the specific policies yielded the greatest benefits (bottleneck on the server side).

| policy | description | $EGT$ | $NU$ | $EGT$ & $NU$ |
|---|---|---|---|---|
| 4 | $AM_{T_{FFR}=0.5}$ | 1 | 0 | 0 |
| 3 | $AM_{T_{FFR}=0.3}$ | 6 | 0 | 0 |
| 2 | $AM_{T_{FFR}=0.1}$ | 5 | 0 | 0 |
| 1 | $ST_{DOC=4}$ | 5 | 1 | 1 |
| 0 | $ST_{DOC=1}$ | 1 | 17 | 1 |

Table 8.8: The number of experiments (out of a total of 18) in which the specific policies yielded the greatest benefits (temporally varying network speed configuration).

### 8.5.2 Detailed Analysis

Some randomly chosen experiments for each network speed configuration are analysed in more detail in the following paragraphs, in order to explain the reasons for the effects of the individual policies on both ULMs, and how they evolved. Each analysis contains a plot of the *expected get time* over time. It shows the values averaged over five minute intervals and over three experiment repetitions. The individual *expected get time* values were also used for computing a single performance measurement for each experiment. Additionally the progression of the DOC is plotted, but with a smaller interval than the *expected get time* progression (1 minute), in order to allow more detail to be displayed. Progression plots of the same type use the same scale for comparability reasons. No progression of the network usage was plotted, as only a measurement of the total network usage was available for each experimental run due to the measurement technique (section 8.3.4).

**Static Client-Side Bottleneck**

For scenarios with a (static) network bottleneck on the client-side, policies which kept the DOC at a low value resulted in the shortest *expected get times* and the lowest network usage (table 8.5), as predicted. Such a situation was identified in an experiment with high churn, heavy weight workload, a bottleneck at client-side, and large data items. Table 8.9 shows the effect of the individual policies on the expected get time (EGT) and the network usage (NU). Both are normalised with respect to the corresponding measurements resulting from the baseline (policy 0, $ST_{DOC=1}$).

| policy | description | $EGT[\%]$ | $NU$ [%] |
|---|---|---|---|
| 4 | $AM_{T_{FFR}=0.5}$ | 106 | 103 |
| 3 | $AM_{T_{FFR}=0.3}$ | 116 | 104 |
| 2 | $AM_{T_{FFR}=0.1}$ | 116 | 104 |
| 1 | $ST_{DOC=4}$ | 122 | 112 |

Table 8.9: Effects of the individual policies on EGT and NU with respect to the baseline.

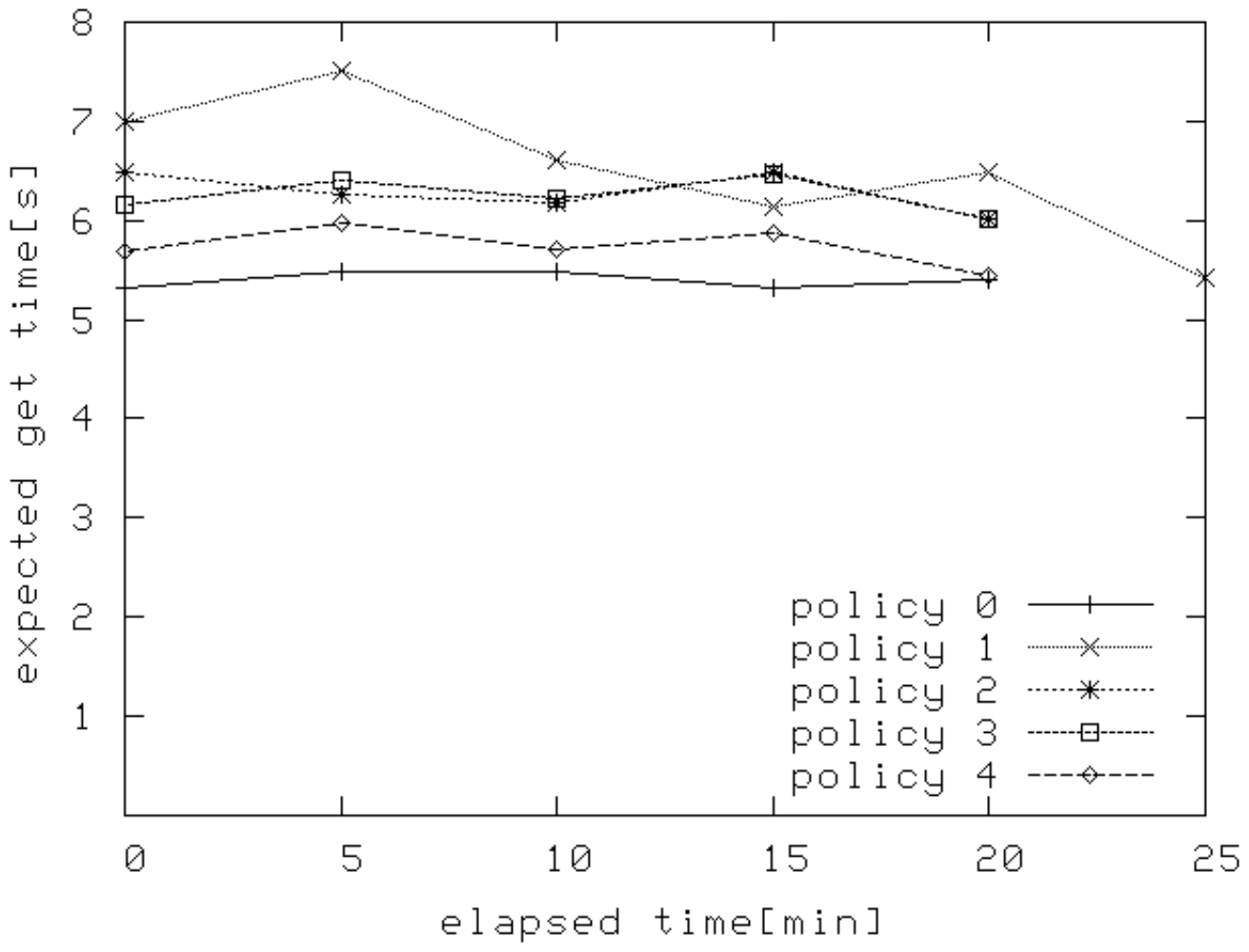

Figure 8.6: EGT progression for an experiment with a static client-side bottleneck.

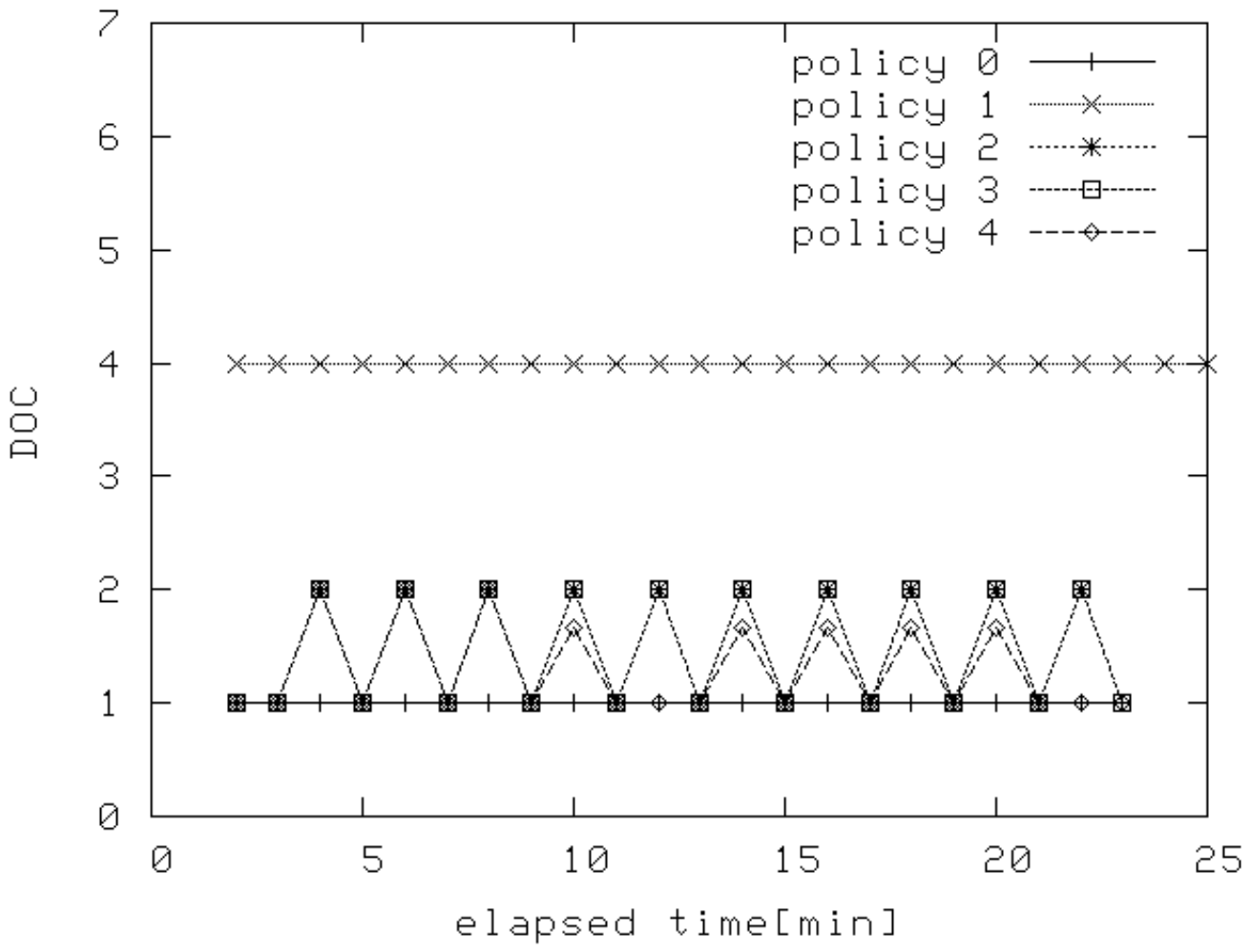

Figure 8.7: DOC progression for an experiment with a static client-side bottleneck.

The expected get time progression (figure 8.6) and the aggregated EGT values (table 8.9) shows that a low DOC corresponds here to the lowest expected get time. The policy that achieved the lowest expected get time was policy 0 ($ST_{DOC=1}$); it kept the DOC at a fixed, minimum value. This resulted in the same policy yielding the lowest network usage. An analysis of how the DOC values were adapted by the autonomic manager (figure 8.7) shows that autonomic management kept the DOC at small values which resulted in significantly better performance than policy 1 (in which a high DOC was statically configured). Any increase in the DOC in this case resulted in competition for resources between the individual fetch operations on the client link. A low DOC subsequently also resulted in a low network usage. The duration of the experiments in which policy 1 was tested was longer than that of any other experiment due to the increased expected get times resulting from the high DOC.

**Static Configuration with No Bottleneck**

For scenarios in which no (static) network bottleneck was configured, policies that kept the DOC at low values resulted in the shortest expected get times and lowest network usage (table 8.6), as predicted. Such a situation was identified in an experiment with low churn, heavy weight workload, no bottleneck and big data items. Table 8.10 shows the effects of the individual policies on the expected get time (EGT) and the network usage (NU). Both are normalised with respect to the corresponding measurements resulting from the baseline (policy 0, $ST_{DOC=1}$).

Similar reasons can be identified for the effects of the specific policies on the expected get times and the network usage as in experiments with a bottleneck at the client-side, analysed earlier. The network speed configuration in which no bottleneck was configured allowed equal bandwidths on the client and server links. This caused competition for bandwidth between individual fetch operations when the DOC was high, as was the case (to a higher degree though) in experiments in which a (static) bottleneck at the client side was con-figured. The duration of the experiments shown in figures 8.8 and 8.9 appears to be short in comparison to earlier (corresponding) plots. This is due to the network configuration which allowed data to be transferred at a faster rate than in experiments with other network speed configurations and thus caused the experiment to finish sooner. As in experiments with a (static) bottleneck at the client side, here policies with a high DOC resulted in a longer experimental run time than any other policies.

| policy | description | $EGT[\%]$ | $NU\ [\%]$ |
|---|---|---|---|
| 4 | $AM_{T_{FFR}=0.5}$ | 102 | 102 |
| 3 | $AM_{T_{FFR}=0.3}$ | 107 | 108 |
| 2 | $AM_{T_{FFR}=0.1}$ | 106 | 107 |
| 1 | $ST_{DOC=4}$ | 204 | 165 |

Table 8.10: Effects of the individual policies on EGT and NU with respect to the baseline.

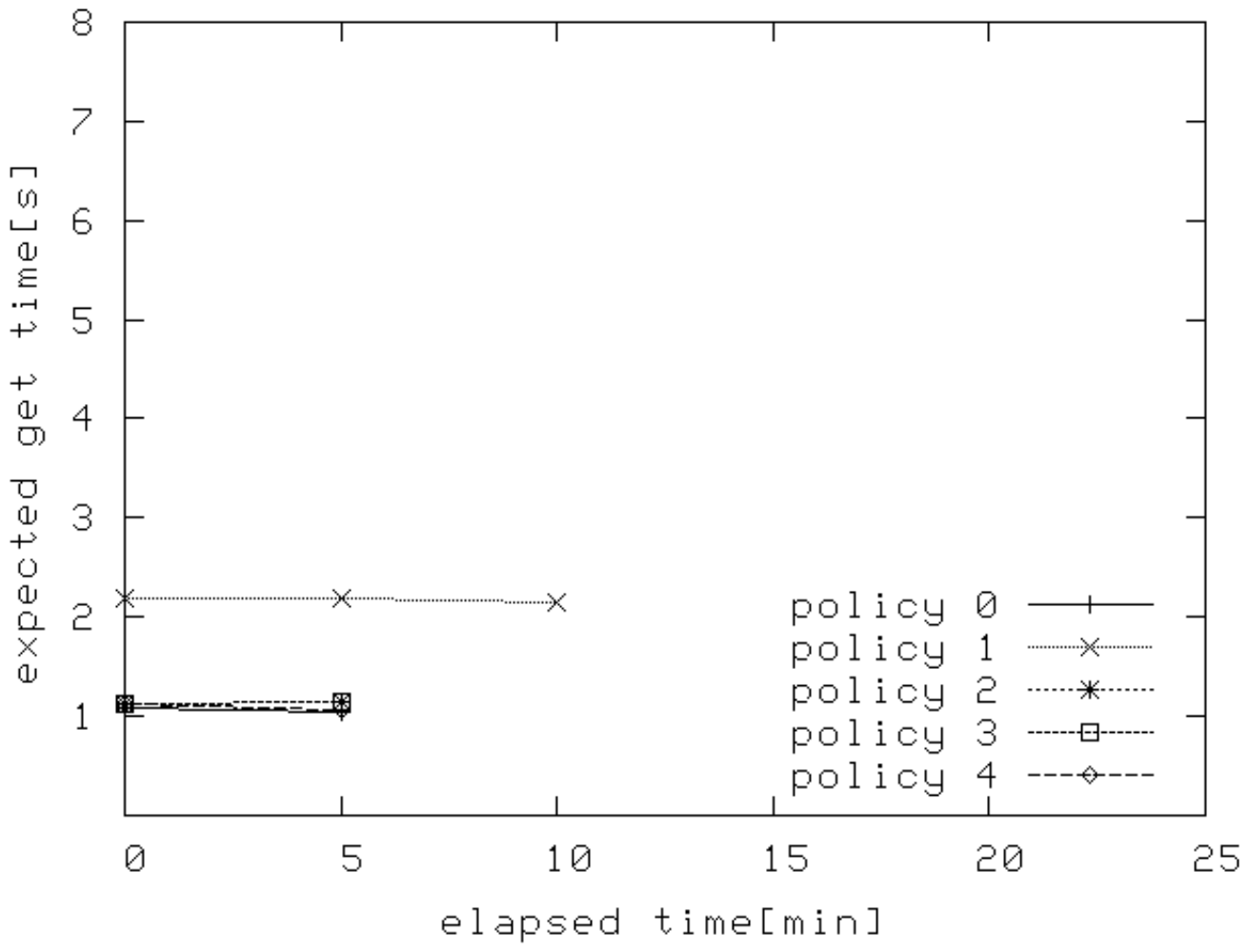

Figure 8.8: EGT progression for an experiment with no bottleneck.

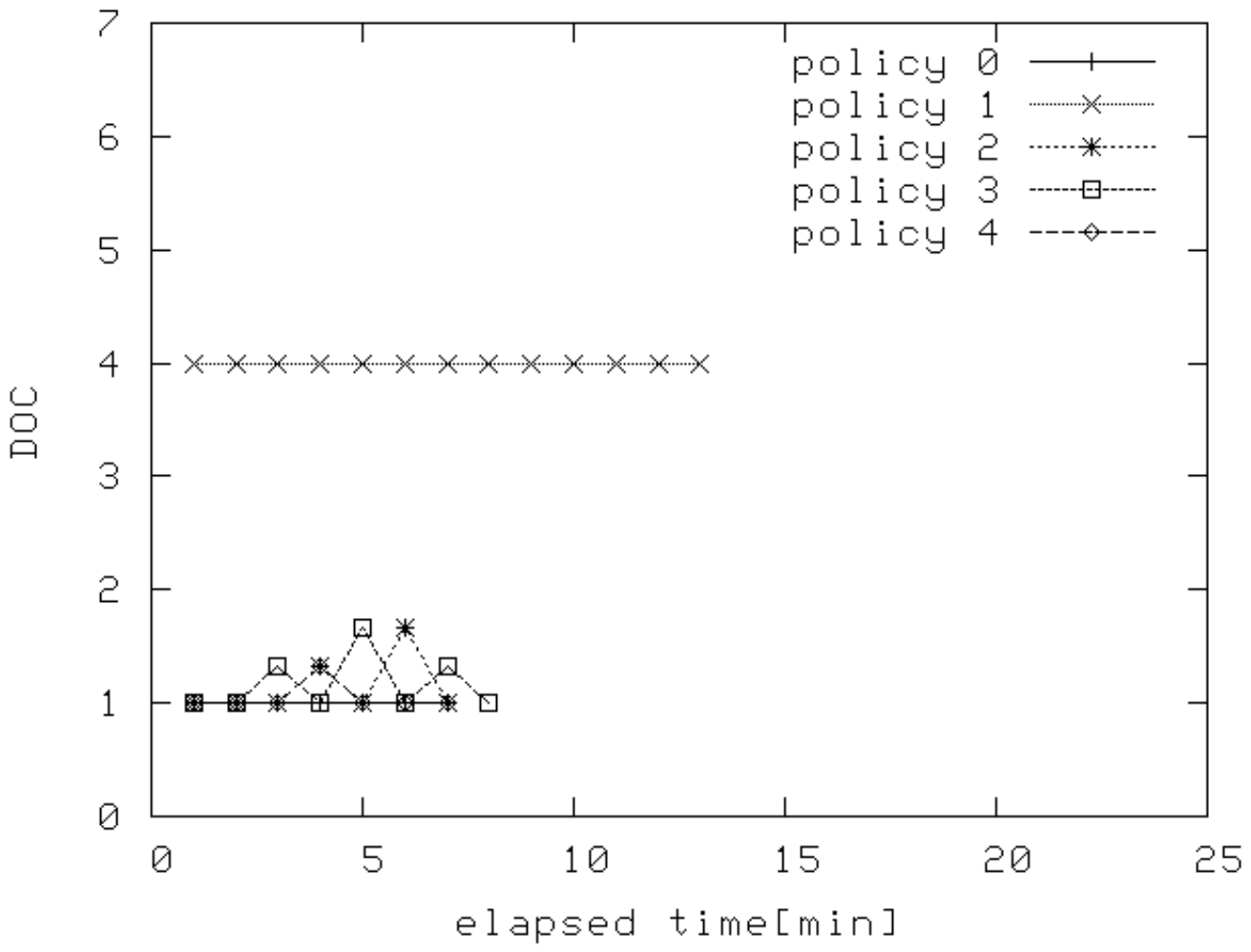

Figure 8.9: DOC progression for an experiment with no bottleneck.

### Static Server-Side Bottleneck

For scenarios in which a (static) network bottleneck on the server-side was experienced, policies which kept the DOC at a high value resulted in the shortest *expected get times* and lowest network usage (see table 8.7), as predicted. This is the analysis of an experiment with high churn, heavy weight workload, large data items and a bottleneck at the server-side. Table 8.11 shows the effects of the individual policies on the expected get time (EGT) and the network usage (NU). Both are normalised with respect to the corresponding measurements resulting from the baseline (policy 0, $ST_{DOC=1}$). The differences in the EGT values are insignificant. The effects of the individual policies on the network usage correspond with a configured and adapted high DOC.

| policy | description | $EGT$[%] | $NU$ [%] |
|---|---|---|---|
| 4 | $AM_{T_{FFR}=0.5}$ | 99 | 127 |
| 3 | $AM_{T_{FFR}=0.3}$ | 98 | 142 |
| 2 | $AM_{T_{FFR}=0.1}$ | 99 | 128 |
| 1 | $ST_{DOC=4}$ | 91 | 216 |

Table 8.11: Effects of the individual policies on EGT and NU relative to the baseline.

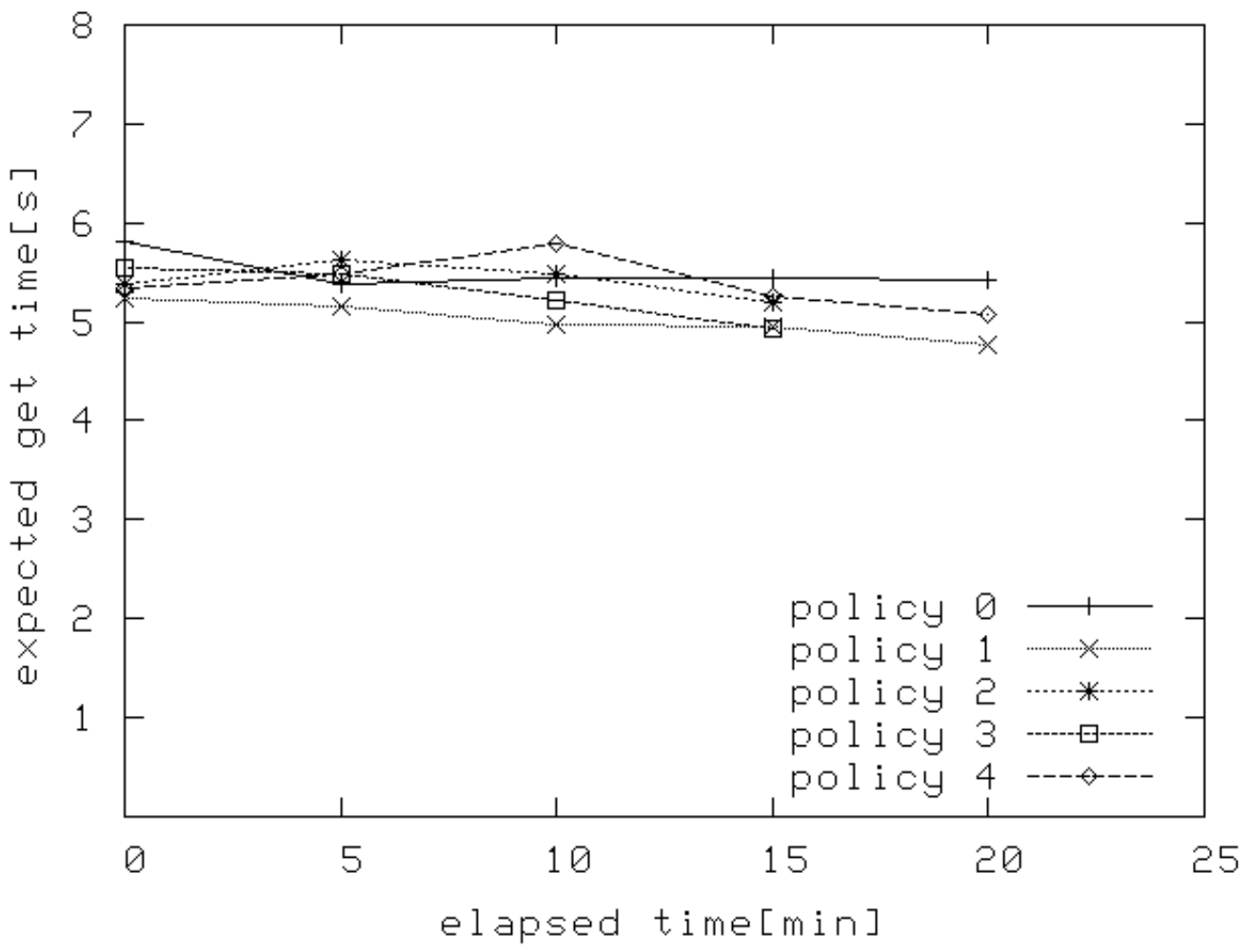

Figure 8.10: EGT progression for an experiment with a static server-side bottleneck.

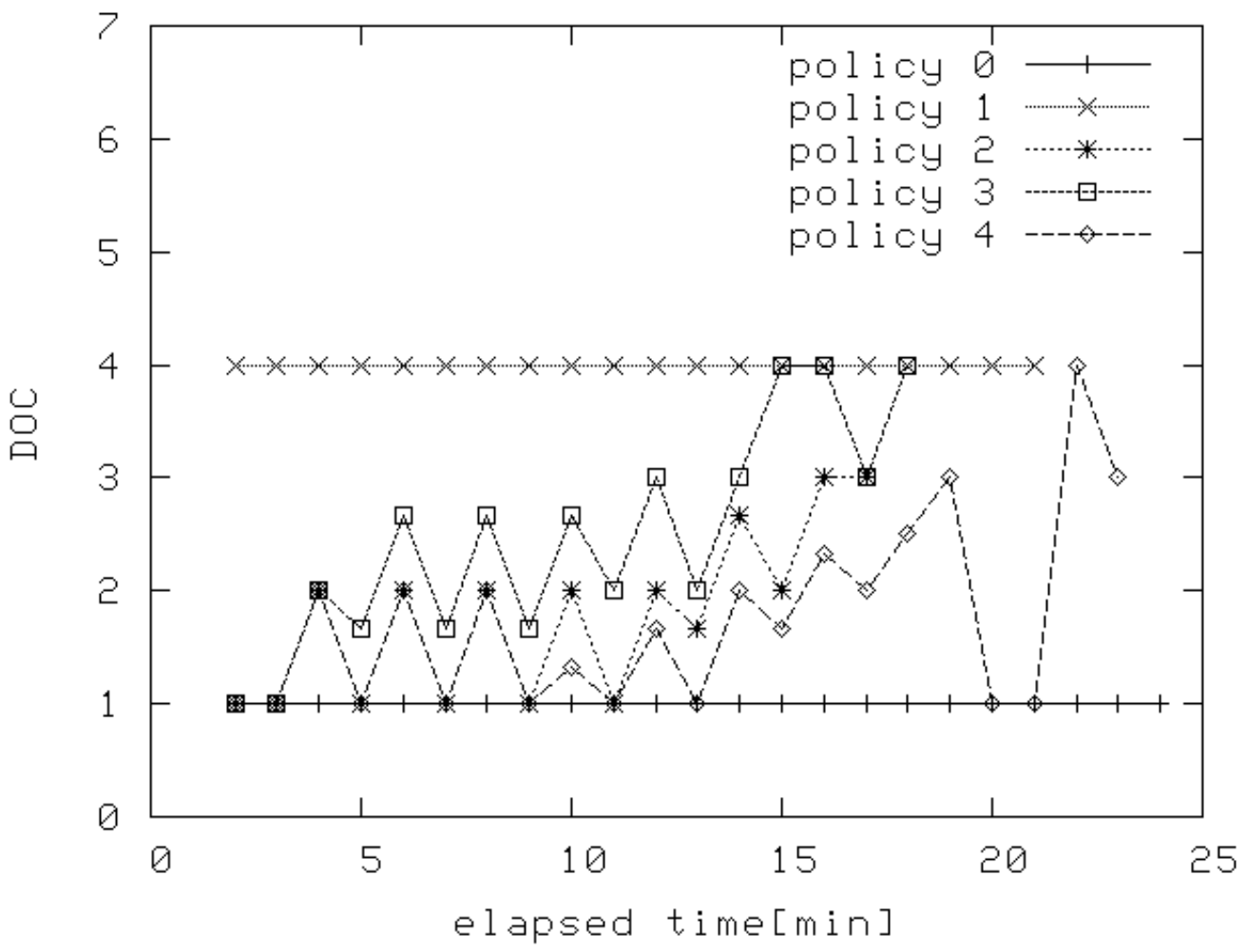

Figure 8.11: DOC progression for an experiment with a static server-side bottleneck.

The EGT progression (figure 8.10) and the aggregated EGT (table 8.11) identify policy 1 ($ST_{DOC=4}$) as the policy resulting in the lowest *expected get time*, as predicted for such a network speed configuration. This policy kept the DOC at a fixed, high value which was reflected by the network usage. Figure 8.11 shows that autonomic management slowly progressed from an initially low value to a high DOC. The increase in the DOC corresponds to the decrease in the expected get time. A further analysis reveals that the autonomic man-ager increased the DOC due to the increased *fetch time variation* metric (FTV) and *fetch failure rate* metric (FFR). This was a result of the congested links on the server-side which had a only a reduced bandwidth available, and of the high churn. The policy was designed to adapt the degree of modification of the DOC depending on the FFR and FTV. The value of the BN metric was used to determine the magnitude of the change of the DOC. A further analysis, which dealt with an other experiment with the same specifications (other then different workload), shows similar results. In that experiment the *get* requests were issued less frequently than in the experiment reported. Subsequently there was less congestion at the server links and fewer fetch operations failed. Hence the DOC was increased at a lower rate than in the experiment reported above.

**Temporally Varying Network Speed**

For scenarios in which a temporally varying network speed was configured, policies which kept the DOC at a high value resulted in the shortest *expected get times* and lowest network usage (table 8.8), as predicted. Here the variation in fetch times is too high for the SRM to make accurate predictions and thus removes any benefit from it. Such a situation was identified in an experiment with low churn, light weight workload, temporally varying network speed and big data items. Table 8.12 shows the effect of the individual policies on the expected get time (EGT) and the network usage (NU). Both are normalised with respect to the corresponding measurements resulting from the baseline (policy 0, $ST_{DOC=1}$).

The expected get time progression (figure 8.12) and aggregated EGT values (table 8.12) show that a high DOC corresponds here again with the lowest *expected get time*. The reason for this is that any prediction the SRM made was invalid due to the high frequency with which network speeds varied. The EGT peak of a system managed by policy 0 after about 5 minutes run time is due to the monitored *get time* and not due to get errors. The policy which achieved the lowest expected get time was policy 3 ($AM_{T_{FFR}=0.3}$). Here the difference between policies which kept the DOC at a high-level (which is the optimum in this case) was again quite small. However, after 15 minutes experimental run time a small, but visible, drop of the DOC by policy 3 can be identified despite any aggregation in figure 8.13. This allows the speculation that autonomic management would have set the DOC to a small value if, after a phase of temporally varying network speed, a phase with static network speed had been exhibited. Such an adaptation happened in experiments with static network speed as illustrated in figure 8.6. The initial DOC value used for autonomically managed systems was chosen to be 1. However, it appears in figure 8.13 as a higher value due to the aggregation and averaging of all DOC values for each interval (aggregation period).

| policy | description | $EGT$[%] | $NU$ [%] |
|---|---|---|---|
| 4 | $AM_{T_{FFR}=0.5}$ | 71 | 188 |
| 3 | $AM_{T_{FFR}=0.3}$ | 65 | 168 |
| 2 | $AM_{T_{FFR}=0.1}$ | 67 | 175 |
| 1 | $ST_{DOC=4}$ | 69 | 153 |

Table 8.12: Effects of the individual policies on EGT and NU with respect to the baseline.

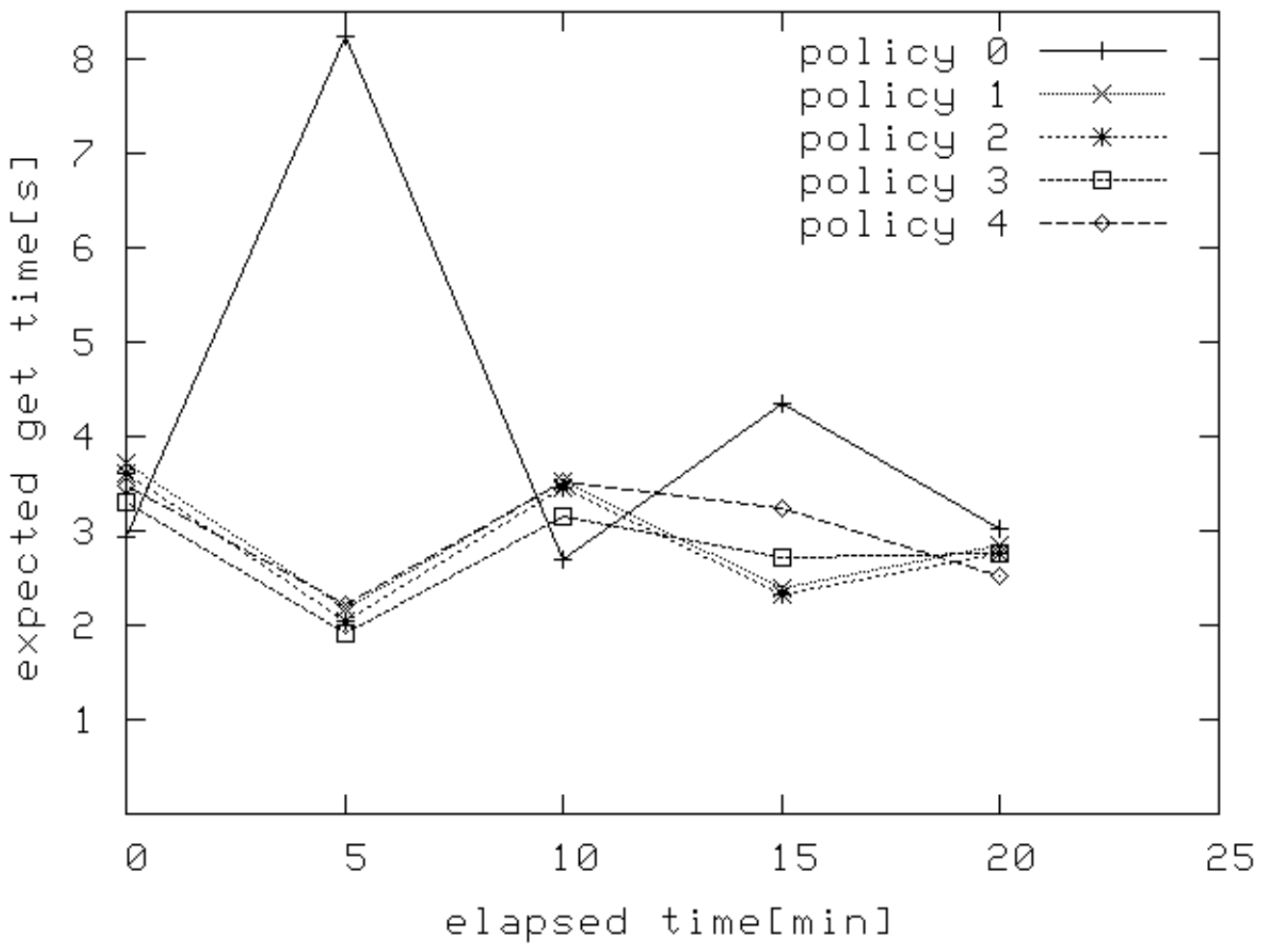

Figure 8.12: EGT progression for an experiment with temporally varying network speed.

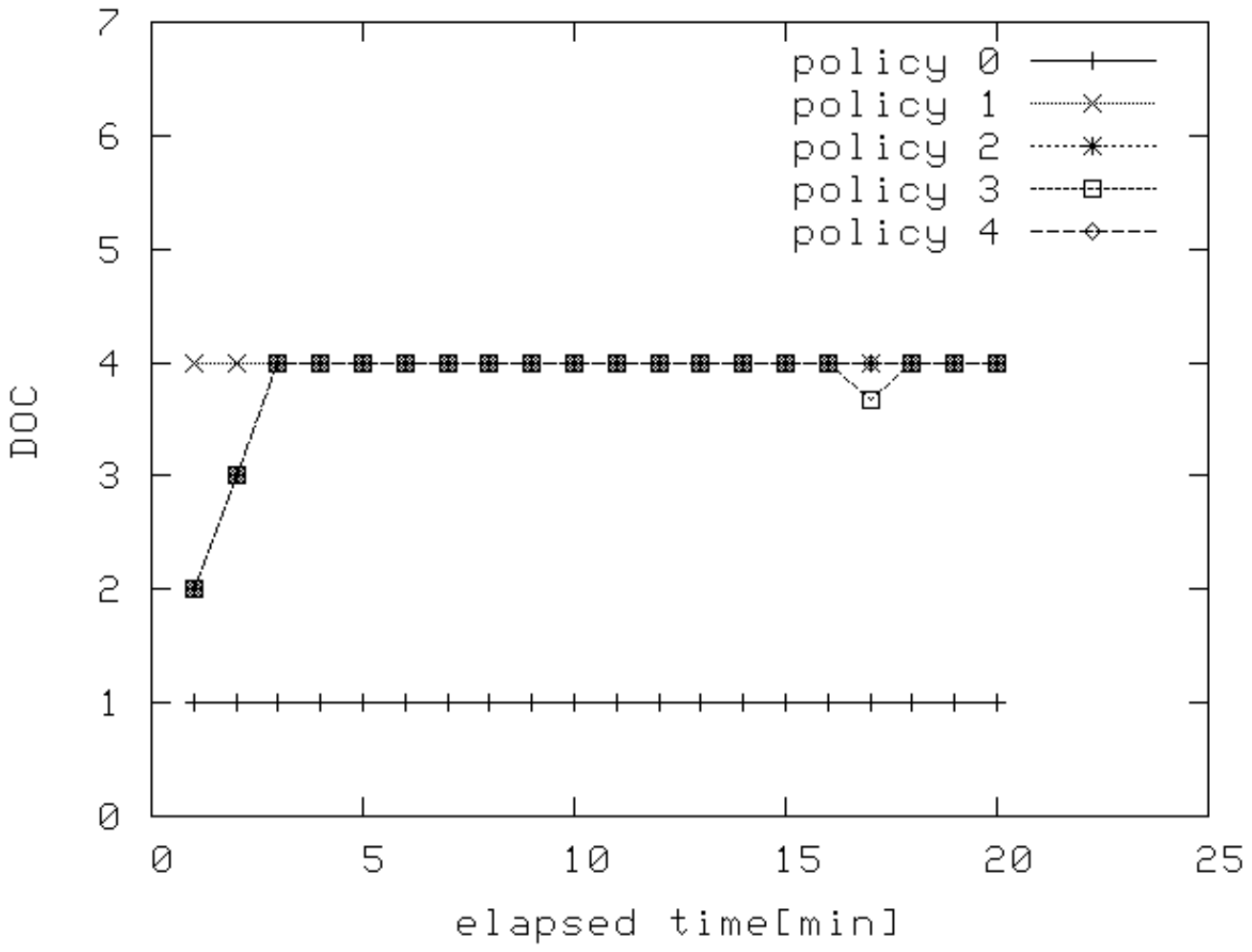

Figure 8.13: DOC progression for an experiment with temporally varying network speed.

### 8.5.3 Analysis of Reproducibility

Three repetitions (referred to as runs 1, 2 and 3) of each experiment were executed in order to verify reproducibility. To analyse this a *similarity metric* with respect to the *expected get time* was defined, in the same way as with the expected lookup time in chapter 7. The similarity metric was specified as the standard deviation of corresponding values from the three runs normalised by the mean ($NSD$). Figure 8.14 shows the cumulative frequency distribution of all similarity metric values, computed for all five policies in all seventy-two experiment groups. Ideally all of the experiments should exhibit a $NSD$ of $0.0$, which is represented by the straight line. The plot shows that the normalised standard deviation for the majority of the experiments is quite low, which suggests that there was little variation between the effects on the expected get time.

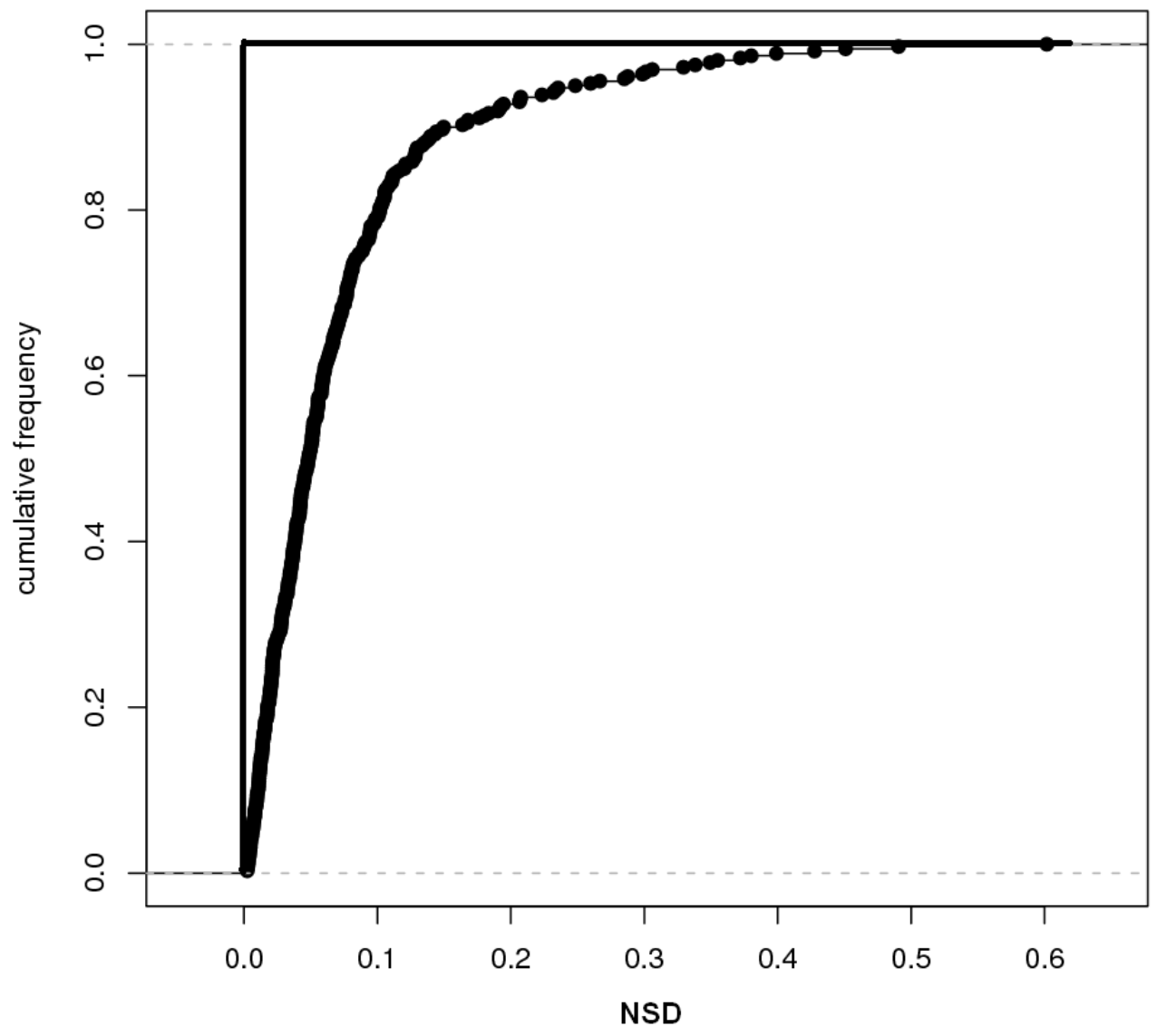

Figure 8.14: Cumulative frequency plot of all NSD - similarity metrics.

Some arbitrarily selected *expected get time progressions* from individual experimental runs are plotted to illustrate the relationship between the get times and specific NSD values. More than $75\%$ (the third quartile) of the experiments resulted in an $NSD < 0.13$ therefore the individual runs are considered as sufficiently reproducible.

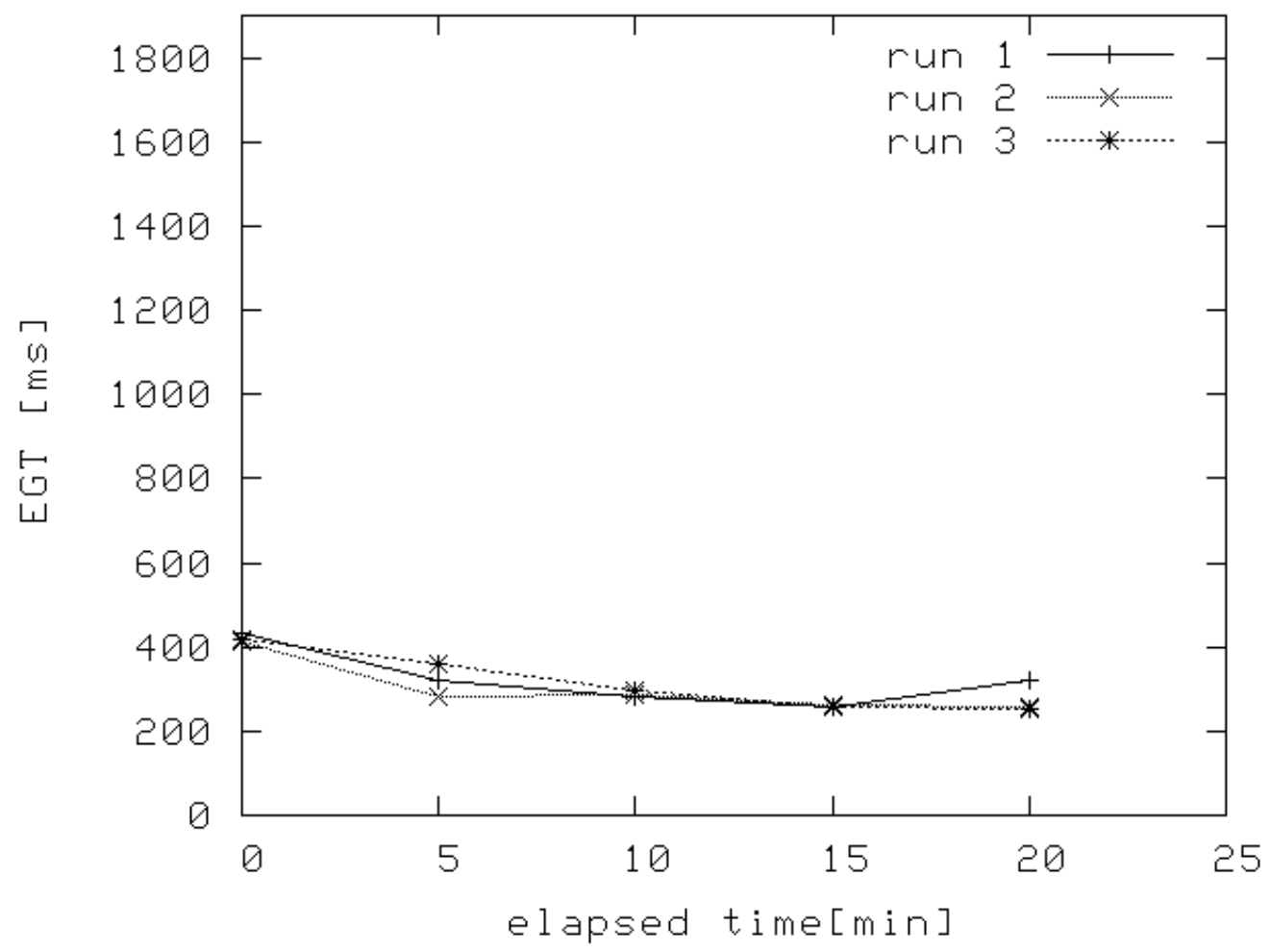

Figure 8.15: Expected get time progressions in experiment repetitions with synthetic light weight workload, high churn, no bottleneck, small data items, and managed by policy 1 ($NSD \approx 0.06$).

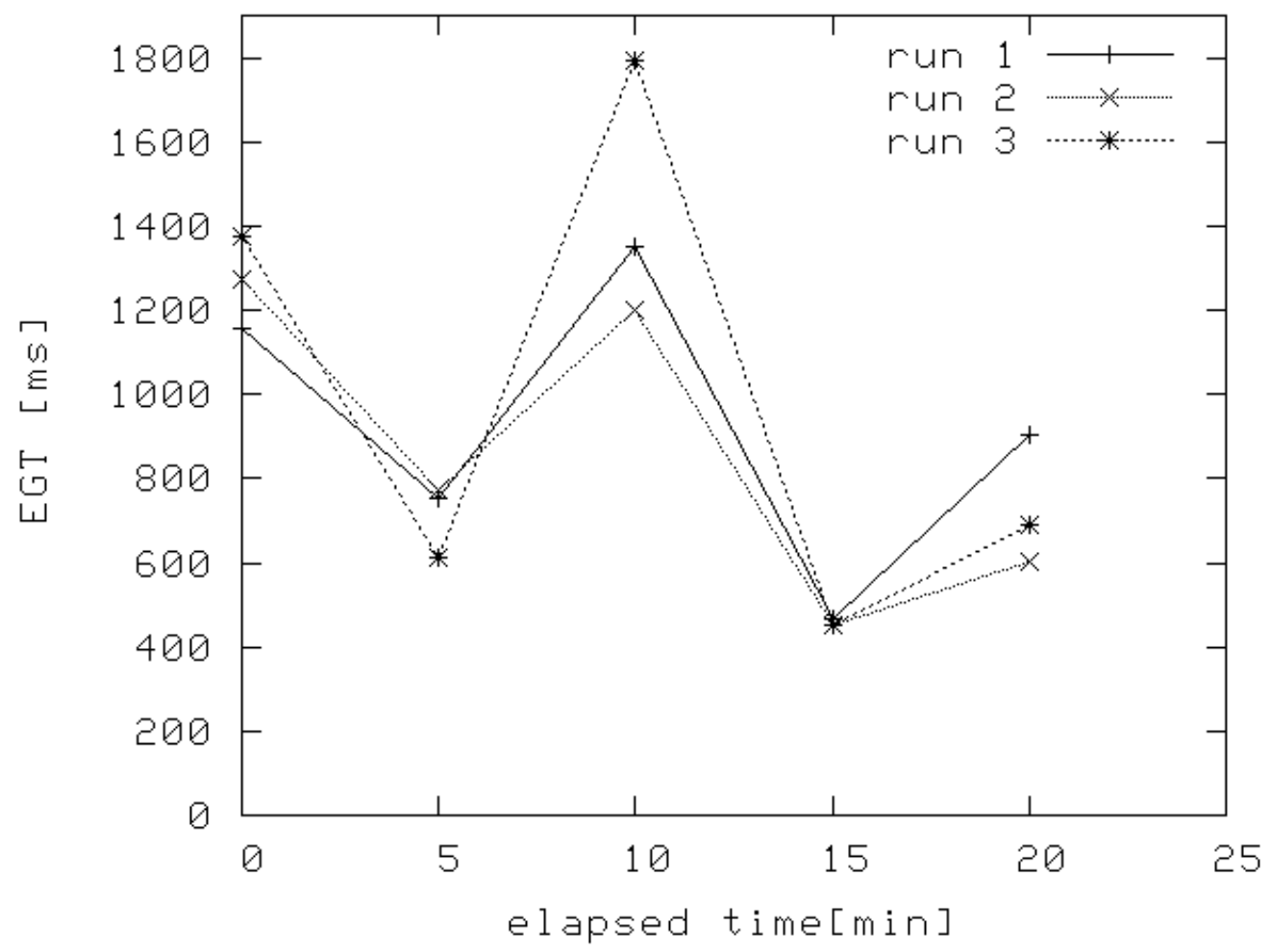

Figure 8.16: Expected get time progressions in experiment repetitions with light weight workload, low churn, temporally varied network speeds, and managed by policy 2 ($NSD \approx 0.13$).

## 8.6 Conclusions and Future Work

In the experiments reported in this chapter it was demonstrated that an autonomically managed distributed storage client is able to detect unsatisfactory situations with respect to the performance and resource consumption and to correct them by adapting the DOC accord-ingly. Autonomic management was compared here with two static configurations (policy 0, policy 1), each of which represented an optimal configuration for specific categories of scenarios. As predicted in chapter 5, in the category of experiments in which policy 0 represented an ideal configuration, policy 1 was disadvantageous and vice versa. Autonomic management adopted the DOC accordingly in each category appropriately and resulted thus in greater performance at the cost of some additional network usage. The differences in the effects of autonomic management on the user-level metrics *expected get time* and *network usage* were however not statistically significantly different with respect to the statically configured system (policy 0). The effects of autonomic management on the raw measurements of the *get time* however show a significant improvement. In the chosen experimental structure autonomic management had to compete against ideal static configurations in each category of experiments. The disadvantage of this setup was that it did not represent scenarios in which the dynamic adaptation of the DOC would have been an advantage. Such an experimental setup is proposed in section 8.6.2. Additionally situations have been identified in which this approach to autonomic management bears some potential for improvement (section 8.6.1).

### 8.6.1 Suggestions for Adaptations of the Managed ASA Client

In experiments with static server-side bottleneck configurations the DOC approached the predicted optimum (4) slowly. In the analysis reported in section 8.5.2, the DOC approached, but never reached, the optimum value over the course of the entire experimental run. A greater benefit could be achieved if autonomic management would set the DOC to an optimum value more quickly. This could be achieved by a more complex policy.

### 8.6.2 Alternative Management Approaches

The autonomic policy model used in the reported experiments was based on the idea that a single policy identifies both cases, one where the DOC is too high, and the other one where it is too low. A *sub-policy* approach as used in the P2P experiments could be adopted, which balances out requirements for an increase or a decrease in the DOC.

Such a policy would determine a new DOC depending on approximately the same metrics as used here. The policy would consist of an aggregation policy and sub-policies. A churn sub-policy would determine an increase in the DOC based on the observed FFR and FTV metrics in isolation. Instead of the current DOC, a metric would be used which represents unnecessary effort, or fetch operations (UFO). Out of a number $n$ of concurrently initiated fetch operations only 1 successfully finished operation is necessary. This means that, when a DOC of 3 is configured, at most 2 fetch operations are unnecessary. If, however, 1 of the 2 remaining fetch operations fails, only 1 operation is unnecessary. A high value for the UFO metric suggests that a lot of effort (and subsequently network resources) is wasted in fetching. The unnecessary work policy determines a decrease of the DOC, also in isolation. A bottleneck sub-policy determines the rate of change. All are combined by the aggregation policy which finally specifies a new DOC. Clearly there may be more issues to consider, and this would represent a promising future experiment.

### 8.6.3 Alternative Validation Approaches

In an experimental configuration in which phases with different requirements for an ideal DOC setting were exhibited, autonomic management would probably have been more beneficial, than the experiments reported here. Examples include experiments in which a phase during which a static bottleneck exists on the client side is followed by a phase during which varying network speeds are exhibited and so forth. During the first phase a low DOC would result in the greatest benefit. Conversely a high DOC would result in the greatest benefit in the second phase. Autonomic management may yield greater and more significant improvements than a statically managed system in such scenarios by adapting the DOC appropriately during the various different phases. Additionally experimental run times should be long enough to allow autonomic management to show its full potential. Situations where this was not the case were identified here in experiments with small data items and a heavy weight workload.

### 8.6.4 Wider Contribution

Even though this evaluation was carried out with an implementation of the distributed storage system ASA, it might be exploited in other distributed storage systems beyond this thesis. Any system discussed in chapter 5 could be simply adapted to use the autonomic manager introduced here. Even if the autonomic manager is not required in any of those systems, the server ranking mechanism can be used in isolation for any other system in which network speed measurements are important.

# Chapter 9

# Conclusions and Future Work

## Outline

This chapter concludes the thesis by summarising the work carried out and revisiting the hypotheses originally made, and goals stated. The contributions that this thesis makes to distributed storage systems architecture, autonomic, P2P and cloud computing are outlined and a synopsis of possible future work is provided.

## 9.1 Thesis Summary

In this thesis the hypothesis was evaluated that an autonomically managed distributed storage system yields better resource consumption and performance than a statically configured system. Autonomic management was applied to the peer-to-peer (P2P) and data retrieval components of ASA, a distributed storage system. The effects were measured experimentally under various conditions, including specific workloads and churn patterns. The management policies and mechanisms were implemented using a generic autonomic management framework developed during this work.

In the experiments with ASA's P2P component the *peer-set maintenance intervals* were autonomically adapted in response to various conditions, and the effects on user-perceived performance and network usage[1] compared with a P2P overlay in which the maintenance intervals were statically configured. The effects of a single static configuration were compared with two autonomic managers. The statically configured system resulted in the greatest performance in about 13% of the conducted experiments and in the lowest network usage in 6%. In all other experiments the two autonomically managed systems resulted in either the greatest performance or lowest network usage. In about 31% of the experiments one of the autonomic management policies resulted in the greatest performance and the lowest network usage in combination. The statically managed system never resulted in a combination of the greatest performance and lowest network usage. The majority of the improvements were statistically significant. More details about the results of the experimental evaluation are available in chapter 7.

[1]See section 7.4.2 for a detailed specification of performance and network usage.

In experiments with the data retrieval component, the *Degree of Concurrency (DOC)* in ASA's data retrieval mechanism was autonomically adapted. The effects on user-perceived performance and network usage[2] were compared with a static configuration. Here two statically configured systems were compared with three autonomically managed systems. The static configurations were optimised for specific scenarios. The static configurations resulted in the greatest benefits in the scenario for which they were optimised; autonomic management yielded an overall improvement with respect to the user-perceived performance ($P$) at the cost of some additional network usage ($NU$). The resulting $P$ and $NU$ measurements of the specific policy, normalised by the policy $static_1$ and averaged over all experiments, are shown in table 9.1. The effects shown in table 9.1 are however small and an analysis of statistical significance resulted in most of them not being significant. This is a result of the chosen experimental setup in which the autonomic management policies had to compete against static configurations which were specialised for each type of experiment. That means that none of the experiments required an dynamic adaptation due to autonomic management in order to improve performance or resource consumption. This issue is discussed in more detail in chapter 8. It is hypothesised that an experimental scenario with alternating phases, which differ with respect to the requirement for an ideal configuration, would allow autonomic management to yield more significant benefits than it did here.

| management | $P[\%]$ | $NU[\%]$ |
|---|---|---|
| $autonomic_3$ | 96 | 114 |
| $autonomic_2$ | 96 | 114 |
| $autonomic_1$ | 97 | 114 |
| $static_2$ | 115 | 153 |
| $static_1$ | 100 | 100 |

Table 9.1: Averaged Performance and Network Usage

[2]See section 8.4.2 for a detailed specification of performance and network usage.

## 9.2 Contributions

The work carried out for this thesis deals with aspects of various areas in computer science. These areas include distributed storage systems architecture, P2P computing, autonomic computing, and cloud computing. The findings and conclusions of this work could be exploited in any of the above-mentioned areas, and are outlined here as contributions to each specific area in turn.

### 9.2.1 Distributed Storage Systems Architecture

In chapters 4 and 7, the autonomic control of the scheduling of maintenance mechanisms was investigated. Even though the focus in those chapters lay on the maintenance mechanism of P2P overlays, the work carried out contributes to distributed storage systems in general. As outlined in chapters 3 and 7 various other distributed storage systems are built on Chord or another P2P overlay which carries out periodic peer-set maintenance operations. Thus all those distributed storage systems could be improved with a similar autonomic management mechanism to the one introduced in chapter 4.

A similar approach could also be used for centralised distributed storage systems like the Google File System (GFS) [27]. GFS does not utilise any P2P overlay for data to host map-pings, however it does carry out mapping-maintenance operations at static intervals. Here autonomic management could be used to adapt the intervals depending on the changing conditions in the same ways as were introduced in chapter 4.

In work on distributed storage systems like, for instance, PAST [21] and OceanStore [78], the ability to relocate data items to address membership churn, in order to maintain fault-tolerance, is referred to as a self-organising property. The investigation of autonomic management in the P2P layer has also an effect on this self-organising property of decentralised distributed storage systems. Decentralised distributed storage systems like, for instance, PAST [21] and OceanStore [78] claim to be self-organising. This self-organising capability relies on the underlying P2P layer to detect changes in the key to node mapping and to report them to a higher level storage layer. This storage layer then triggers data maintenance operations. An autonomically managed P2P overlay would improve the self-organising property of decentralised distributed storage systems because it evidently improved the capability of a P2P overlay to adapt its maintenance operations to various levels of churn.

When analysing the scope for optimisation with autonomic management of ASA's data retrieval component in chapter 5, an autonomic manager was introduced which can be used for other systems as well as for ASA. A wide range of distributed storage systems was identified in this chapter which are able to adapt the DOC. Additionally, the Server Ranking Mechanism (SRM) could be used in the same way in other similar distributed storage systems.

This work can be exploited for the design of any distributed storage system in which peri-odic maintenance operations are used or in which redundant data items can be fetched from multiple servers.

The reported experimental investigations included the simulation of scenarios which potentially occur when a distributed storage system is used to harness storage on user work stations. For instance, membership churn patterns (see sections 7.3 and 8.3) are representative of churn due to users in a corporate network. Another example is the varying network speed as used in experiments with ASA's data retrieval mechanism (section 8.3). This is representative of situations in which users execute network operations whilst a distributed storage system uses their workstations as storage providers. Here the distributed storage system has to share the available network speed with other applications. Autonomic management was beneficial in the majority of the experiments as it adapted the controlled parameters in response to such conditions. Thus it can be concluded that autonomic management, as used here, supports the usage of systems which harness storage on user workstations.

Such an application of distributed storage systems seems reasonable when considering the growing demand for storage [93] as well as the decrease in price and the increased availability of computing devices [75]. Studies carried out by Microsoft conclude that *the disk space of desktop (user) computers is mostly unused and is becoming less used over time* [16, 17, 18]. In other studies carried out by Microsoft, changes in the usage patterns of file systems over five years (2000 - 2004) were analysed [2]. One can extrapolate from the progression of file system capacity and usage that the total amount of unused disk capacity on individual user workstations has increased by about 90% within five years, up to tens of GB. Although the research projects cited above report on historical usage of storage, a potentially long term benefit could be obtained by harnessing unused storage on user machines. By extrapolating Moore's Law [64] and Bell's Law [6], it can be assumed that the capacity and performance of computing devices will improve over the next few years. Thus such an application of distributed storage systems may allow access to an already immensely big and still growing reservoir of storage.

### 9.2.2 P2P Computing

The investigation of the management of the scheduling of maintenance operations is applicable to a wide range of P2P overlays as reported in chapter 4. P2P overlays are not only used exclusively for distributed storage systems and are a research area in their own right. Thus the work reported in correlating chapters 4 and 7 can be not only considered as a contribution to distributed storage architecture but also to P2P computing.

### 9.2.3 Autonomic Management

This work makes a two-fold contribution to autonomic management. Firstly it allows the extraction of a set of guidelines on how to identify facets or components of an unmanaged target system which can be improved by applying autonomic management. Requirements for autonomically managed systems are often summarised with self-* (see chapter 2). This thesis describes how parameters are identified in ASA which would benefit from being adapted in an autonomic manner to changing conditions. This effectively makes ASA a self-* system. The same principles can be applied to any unmanaged target system and thus the corresponding chapters (4 and 5) provide representative examples of how to achieve this. Secondly, the generic autonomic management framework developed as part of this work (chapter 6) can be used to introduce autonomic management to systems outwith the scope of this thesis. This can be considered to be a contribution to the field of autonomic computing in its own right.

### 9.2.4 Cloud Computing

Recently the term cloud computing has become popular as a description of a network of computers through which services are provided. Such clouds are usually considered as huge internet scale distributed systems [26]. Clouds can however also be composed out of computers in local area networks such as a company's corporate network. Such clouds can then be referred to as *ad-hoc clouds* which can operate a cloud infrastructure on existing non-dedicated hardware like users' desktop computers or workstations. This includes benefits like very low additional cost and retaining control over data and processing. A distributed storage system which harnesses storage capacities on user workstations in a corporate network is an example of how such an *ad-hoc storage cloud* could be configured. Distributed cloud services other than storage may also benefit from autonomic management of specific distributed operations such as periodic maintenance or data accesses, as investigated here. In [61] the authors state that cloud systems can be described as *on-demand self-services* which can be acquired and used without the use of human interaction. With respect to *on-demand self-services* cloud systems can be considered as autonomic systems which share similar requirements to the autonomic distributed storage system investigated by this thesis. A first step towards the integration of autonomic management and cloud systems is reported in [87]. It introduces a *metascheduler* which dynamically adds and removes cloud nodes from a gateway service, used by a distributed application. The *metascheduler* makes its decisions based on the comparison of some performance heuristics with some statically configured thresholds. The use of the metascheduler is limited to a specific grid application. It however represents a use case for *nesting autonomic managers*, as discussed in this thesis (appendix C.3). The metascheduler could be implemented as an autonomic manager which evaluates the performance heuristic. The threshold this autonomic manager uses would then be controlled by another autonomic manager based on some high level policy. Guidelines for the application of autonomic management can be extracted from this thesis; it can thus be considered as a contribution to cloud computing.

## 9.3 Future Work

Some potential future work packages were already suggested in chapters 7 and 8. They include suggestions for improvements of the managed components in combination with the management; alternative management approaches; and different evaluation approaches. All of them are outlined in the corresponding chapters. An addition to these future work packages might be the evaluation of the effects of autonomic management on other facets of distributed storage systems. Such facets have been identified during the course of this work but not experimentally evaluated. Examples include the autonomic management of the scheduling of storage maintenance operations or the autonomic management of the degree of fragmentation. Both examples are briefly outlined in the following to motivate future research projects.

### 9.3.1 Autonomic Management of Storage Maintenance Operations

Data in decentralised distributed storage systems such as ASA is replicated on multiple host machines in order to improve its availability in the event of host machines failing. A certain number of replicas must always be available to maintain fault-tolerance. The detection of failed hosts and the copying of corresponding replicas to new hosts, which take over the key space of the failed host, is initiated, for instance, in ASA when P2P routing errors are observed. It is possible that replicas can be missed out which might have been agreed to be stored at such new hosts while the host was transiently unavailable or running slowly. Such issues need to be addressed by a maintenance operation which regularly searches for missing replicas. The responsible mechanism is referred to as a *storage maintenance mechanism* in ASA. In the original ASA design, every host runs a storage maintenance mechanism which periodically searches for and fetches missing replicas. Similar unsatisfactory situations can be identified here as for a periodic P2P peer-state maintenance operation, executed at statically configured intervals (see chapter 4). Autonomic management may be able to discover and correct unsatisfactory situations, similarly to the ones outlined in chapter 4, by autonomically adapting the corresponding interval.

### 9.3.2 Autonomic Management of Data Fragmentation

In storage systems in general, data items are often split up into smaller blocks. In a distributed storage system like ASA, the sizes of individual data blocks affect how data is distributed over the network; this specifies how evenly the storage hosts' storage capacities are utilised (*degree of data distribution*). The sizes of the data blocks also affect the time it takes to return a requested data item (*get time*). The amount of data transferred over the network is also affected, mainly due to maintenance and administrative overhead. As in any of the investigations carried out as part of this thesis, an optimal data block size depends on various conditions and cannot be predicted statically. Even if a data block size seems to be ideal initially, it may cease to be so as conditions vary. Clearly this issue needs to be analysed in more depth by, for instance, developing an analytic model as in chapter 5. However, autonomic management may be able to adapt the data block size in the presence of a changing environment in order to achieve better performance and resource consumption than a statically configured system would do.

### 9.3.3 Combination of Autonomic Elements

In chapter 2, the behaviour of an autonomically controlled system is described as being dependent on the autonomic behaviour of its constituent parts (autonomic elements). The autonomically managed scheduling mechanisms of ASA's P2P component and the managed DOC of the data retrieval mechanism correspond to individual autonomic elements. It was not within the scope of this work to evaluate the effect of autonomic management on performance and resource consumption if more than one facet was autonomically managed concurrently. It would however be an interesting future experiment to do so.

## 9.4 Conclusions

In chapter 1 the hypothesis was made that *autonomic management may be able to set a configuration which results in better performance and resource consumption than any that can be set a priori*. An autonomic manager was envisioned to *work without the need for a human operator*. One of ASA's design goals was stated in chapter 3 as: *A general autonomic tuning mechanism should be provided to allow low-level aspects of the system's operation to be managed automatically, controlled by policies that are driven by high-level user preferences*.

The hypothesis about the effects of autonomic management on resource consumption and performance was experimentally evaluated, analysed and reported in chapters 7 and 8. In both evaluations (a relatively small number of) cases have been identified in which the statically configured ASA components resulted in better performance and resource consumption than the autonomically managed ones. Following this, improvements of the specific manager (in combination with the managed component) were made in the corresponding chapters. However in the majority of the experiments the autonomically managed ASA components successfully identified and corrected unsatisfactory situations with respect to performance and resource consumption. This allows the conclusion that the hypothesis, that autonomically managed components may result in better performance and resource consumption than statically configured ones in various (changing) conditions, was successfully tested and found to hold.

The autonomic manager was envisioned to adapt any target system in response to various conditions without human interaction. The above outlined detection and correction of unsatisfactory situations happened autonomically by the manager without the need of a human administrator. Thus the corresponding design goal of the autonomic manager can be considered to have been met. An ASA design goal with respect to the autonomic manager stated that the manager's behaviour was envisioned to be governed by high-level policies. This is also considered as having been fully met as the autonomic management cycle and the developed framework use policy evaluators as fundamental building blocks.

## 9.5 Final Thoughts

This thesis reports on the benefits of a distributed storage system to which an autonomic manager is added. This allows the provision of storage via the utilisation of user workstations in an ad-hoc storage cloud. The differences between a dedicated data centre or storage server network and such an ad-hoc storage cloud were not analysed in this thesis. It can however be speculated that an ad-hoc storage cloud which provides comparable services like, for instance, a dedicated storage server will save the cost of building up a dedicated infrastructure in addition to an existing network of workstations. If users operating the latter should be provided with a storage system, then money, electrical power consumption and of course hardware can be saved by such a system, compared to a dedicated one. It can further be speculated that this work allows an economical and environmental improvement in the way that storage is provided in a corporate environment. Clearly more issues have to be addressed until this vision becomes reality, but this work may serve as a first step towards a new and better way of providing storage in corporate networks in comparison to contemporary approaches.

Finally, a last philosophical thought about the shared properties of the analysed effects of the autonomic management mechanisms introduced here. Both have in common that a better trade-off between resource consumption and performance than possible in statically configured systems was achieved. They showed that the more eagerly it was tried to improve a system's performance, the more likely the opposite effect was achieved in some situations. That means better performance would have been achieved with less effort. This might be applicable to a much wider area than just distributed storage systems, even to situations in our everyday life.

# Appendix A

# P2P Layer Experiments

## A.1 Preliminary Work

A number of experimental parameters used for the investigations reported in chapter 7 were derived from preliminary experiments. The purpose of these preliminary experiments was to gather data in order to make decisions as to how to configure the autonomic manager and experimental parameters.

Preliminary investigated experimental parameters were:

- The *static maintenance interval*, which determined the fixed interval with which unmanaged nodes were configured and with which autonomically managed nodes were initiated. A poorly chosen interval may have prevented a fair comparison between statically configured and autonomically managed nodes.

- The *workload length*, which determined the duration of a single experimental run and thus the time the autonomic manager had available to adapt the controlled system. In order to identify autonomic management as the cause for monitored changes in the system, the duration had to be long enough.
- The *policy parameters*, which determined the autonomic manager's behaviour. Poorly chosen parameters may have resulted in autonomic management either being harmful or showing only very little effect.

Additionally to the derivation of experimental parameters in appendix A.1.1, A.1.2 and A.1.3, preliminary work was carried out with respect to the derivation of P2P workload from FS traces as reported in A.1.4 and the selection of the experimental platform in A.1.5.

### A.1.1 Static Maintenance Intervals

This investigation was carried out to test various static maintenance intervals in order to select a suitable static interval for all succeeding experiments. The selected interval was then used to configure the manager's initial interval and the static interval in the unmanaged nodes. It was found that an interval of 2000 [ms] resulted in the greatest benefit. Benefit was derived from averaged and normalised measurements of: *network usage, lookup error rate* and *lookup time* for each tested interval[1]. The motivation for this experiment was to provide the means for a fair comparison of statically configured and autonomically managed nodes.

[1]In this preliminary experiments no *lookup error time* was monitored, thus no expected lookup time as in chapter 7 was computed.

Nodes were deployed in *networks with high membership churn* and in *networks with low membership churn* and a *synthetic heavy weight workload* was executed. Each experiment was repeated three times and all of the monitored measurements were averaged, in order to provide a single measurement for each interval. In order to allow comparison with other experiments an autonomic management process was executed but every sub-policy was configured with a threshold $t$ of $\infty$; this kept the nodes effectively unmanaged. In initial test runs, which were carried out prior to this investigation, 2000 [ms] yielded the biggest benefit, therefore an interval of 2000 [ms] was used here as the baseline.

Table A.1 shows all normalised and averaged measurements for the tested intervals. The smaller the normalised and averaged measurement, the greater was the benefit of the corresponding interval.

| interval [ms] | normalised and averaged measurements [%] |
|---|---|
| 10 | 128 |
| 100 | 190 |
| 1000 | 185 |
| 2000 | 100 |
| 10,000 | 114 |
| 100,000 | 26 |

Table A.1: Combined, averaged interval-specific ULM overview.

The results in table A.1 give the impression that an interval of 100,000 [ms] yielded the greatest benefit. However, when analysing the number of encountered nodes in the network it was found that no network was established in any experiment with an interval of 100,000 [ms]. Figure A.1 shows the number of encountered nodes averaged over 3 experiment repetitions and over 5 minutes in a network with low membership churn, as a representative example.

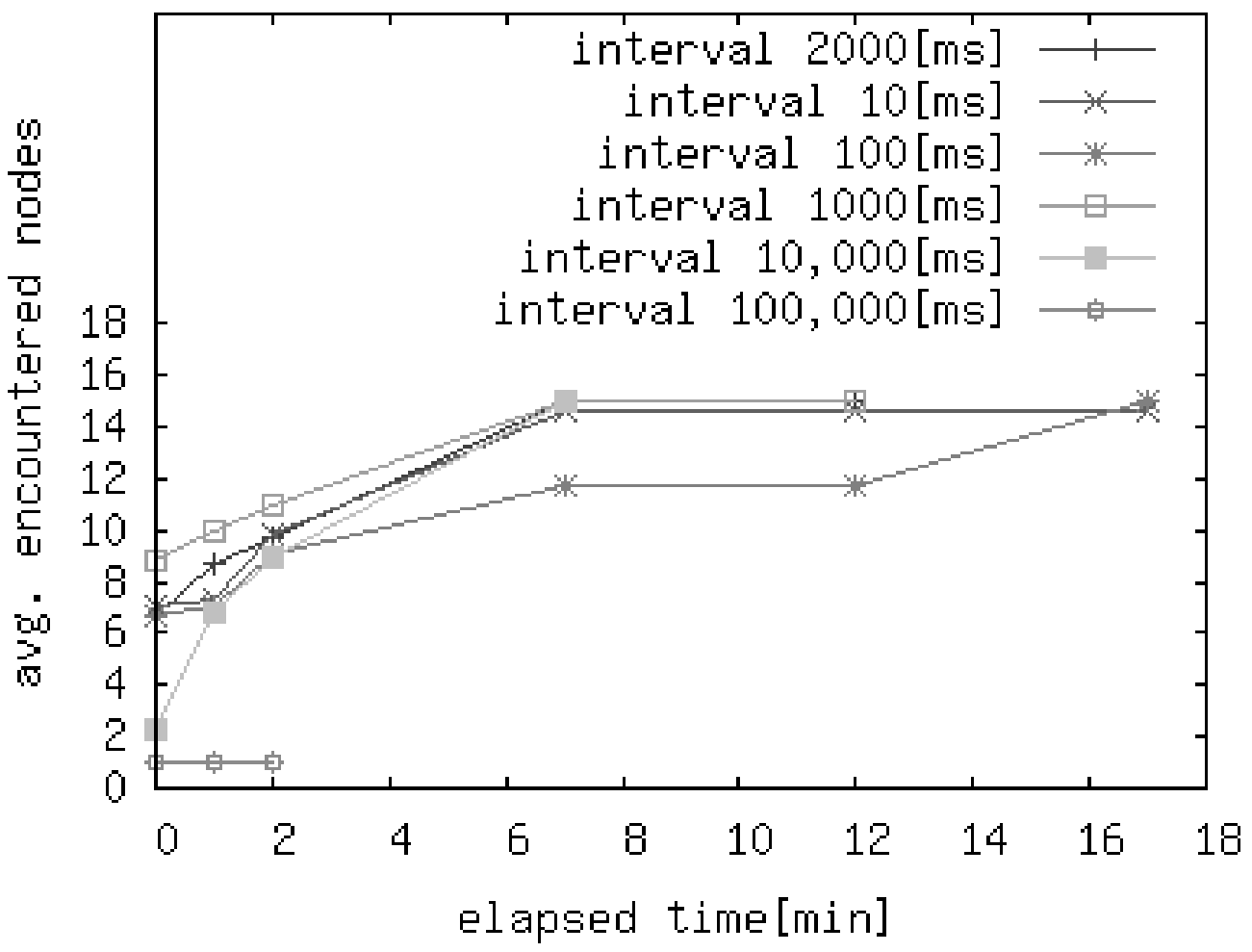

Figure A.1: Progression of the number of encountered network nodes.

In an empty network, lookups were locally resolved by the gateway. This resulted in shorter lookup times, less network usage and no lookup errors for nodes which were maintained every 100,000 [ms] in comparison to other intervals. Therefore it can be concluded that an interval of 2000 [ms] yielded the greatest overall benefit and was thus used in succeeding experiments.

### A.1.2 Workload Length

This preliminary investigation was carried out to establish the minimum time that an experiment should be allowed to run for in order to give autonomic management a chance for its full behaviour to be exhibited. The time it takes to issue all lookups of a workload determines the experimental runtime.

Experiments were carried out in which nodes were managed with different scheduling policies and deployed in a *network with high membership churn* under a *synthetic heavy* *weight workload*. Each policy, churn pattern and workload combination was repeated three times. The manager's configuration was derived from initial test runs which were carried out prior to this investigation. To identify if autonomic management resulted in a change of the controlled intervals the *finger table* maintenance interval progression on the gateway was plotted in figure A.2.

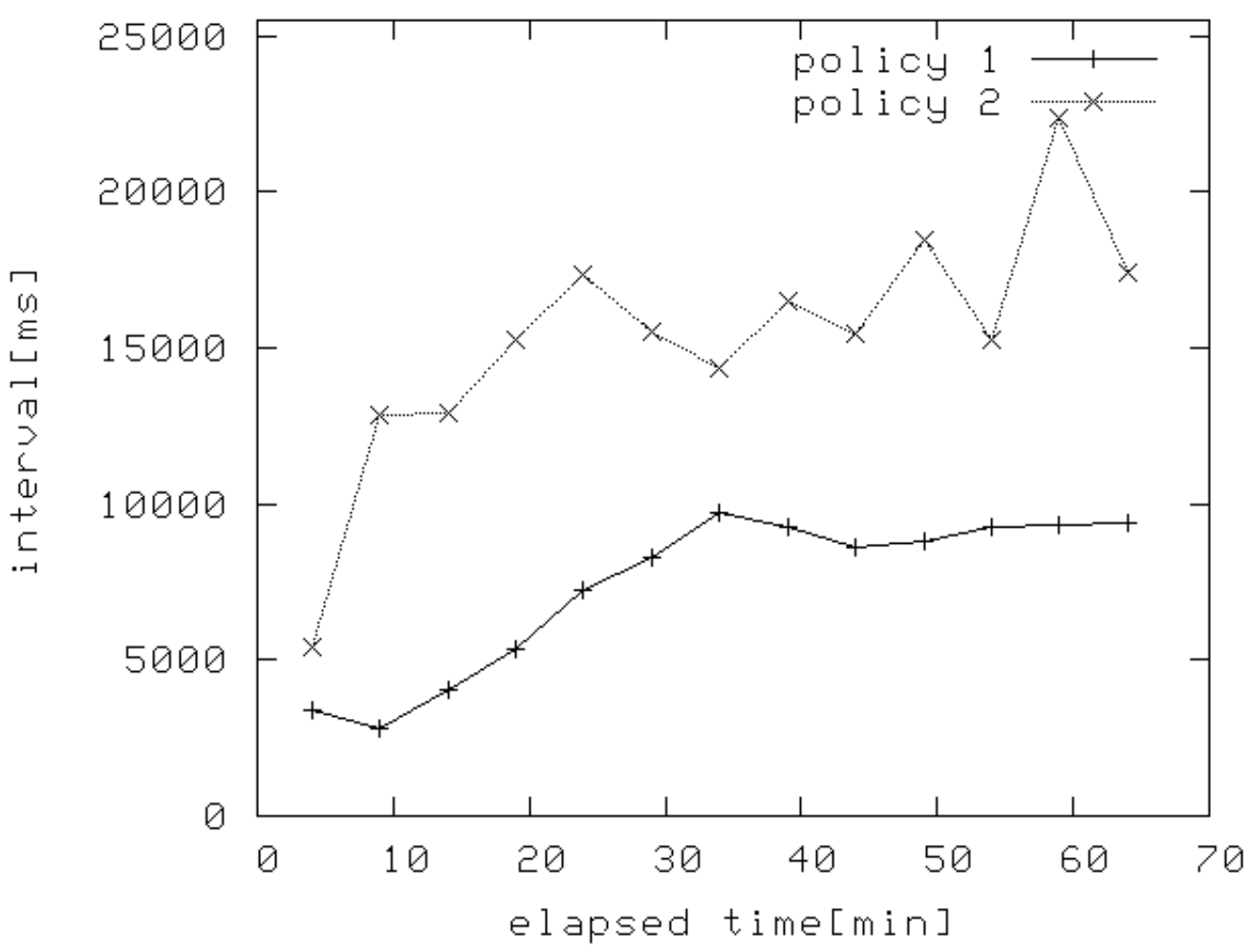

Figure A.2: Progression of the *finger table* maintenance interval on the gateway.

Figure A.2 shows that after an experiment runtime of about 40 minutes all maintenance progressions stabilise between some range. Therefore 40 minutes were considered as a long enough experimental duration to identify significant changes in the controlled system. In this investigation a workload of about 6000 sequential lookups correlated with 40 minutes run time. This information was used in succeeding experiments to define the workload length as due to the large number of experiments longer experiment durations would not have been feasible.

### A.1.3 Sampling the Policy Parameter Space

This preliminary evaluation was carried out to find a suitable set of parameters for configuring the manager's policy[2]. Various parameters were tested, those which resulted in the greatest benefit with respect to performance and network usage were used in succeeding experiments in order to make a fair comparison between unmanaged and managed systems. The same definition for benefit as in section A.1.1 was used.

Experiments were carried out in a *network with low membership churn* and a *network with high membership churn*. A *synthetic heavy weight workload* was executed during each experiment. An experiment was specified with a policy parameter set, workload and churn pattern. Each experiment was repeated three times. The tested policy parameters were:

- Threshold $t$ for the sub-policy which determined an interval with respect to *LILT, ($LILT_t$).*
- Factor $k$ for the sub-policy which determined an interval with respect to *LILT, ($LILT_k$).*
- Factor $k$ for the sub-policy which determined an interval with respect to *ER, ($ER_k$).*
- Factor $k$ for the sub-policy which determined an interval with respect to *NEMO, ($NEMO_k$).*

The thresholds for *ER* and *NEMO* related sub-policies were statically defined as $0$.

In four series of experiments various values for one parameter were tested, all other parameters were statically configured. The values were not varied during an individual ex-perimental run. The value of the varied parameter which yielded the greatest benefit was chosen as the winning parameter for an individual series. This resulted in a parameter set of optimum parameters, which were then tested in combination in a final series with a set of parameters from initial test runs. As baseline for each series statically configured nodes were used (see section A.1.1).

[2]See section 4.5.2 for the definition of the sub-policies and correlating parameters.

#### Policy Configuration

The autonomic adaptation of the intervals between *stabilize, fixNextFinger* and *checkPredecessor* operations was governed by policies which were of the same structure but were evaluated independently from each other. Each policy consisted of an aggregation-policy and of sub-policies (section 4.5.2). Each sup-policy determined an interval in reference to a single metric (*NEMO, ER* and *LILT*). For reasons of simplicity, each policy's sub-policies were configured with the same values for $t$ and $k$. This means that the $NEMO$ sub-policy used for tuning the interval between $stabilize$ operations was parameterised with the same $t$ and $k$ values as the $NEMO$ sub-policy used for tuning the intervals between $fixNextFinger$ and $checkPredecessor$. The same applies for the various $ER$ and $LILT$ sub-policies (with the caveat that $LILT$ is not considered in the $checkPredecessor$-interval adaptation).

**Tested Policy Parameters per Series and Results**

| | NEMO | | ER | | LILT | |
|---|---|---|---|---|---|---|
| policy parameter set | k | t | k | t | k | t |
| 1 | 8 | 0 | 8 | 0 | 800 | 100 |
| 2 | 8 | 0 | 8 | 0 | 800 | 200 |
| 3 | 8 | 0 | 8 | 0 | 800 | 400 |
| 4 | 8 | 0 | 8 | 0 | 800 | 800 |
| 5 | 8 | 0 | 8 | 0 | 800 | 1600 |
| 6 | 8 | 0 | 8 | 0 | 800 | 3200 |
| 7 | 8 | 0 | 8 | 0 | 800 | $\infty$ |

Table A.2: Management policy parameters in series A ($LILT_t$).

| | NEMO | | ER | | LILT | |
|---|---|---|---|---|---|---|
| policy parameter set | k | t | k | t | k | t |
| 1 | 8 | 0 | 8 | 0 | 100 | 800 |
| 2 | 8 | 0 | 8 | 0 | 200 | 800 |
| 3 | 8 | 0 | 8 | 0 | 400 | 800 |
| 4 | 8 | 0 | 8 | 0 | 800 | 800 |
| 5 | 8 | 0 | 8 | 0 | 1600 | 800 |
| 6 | 8 | 0 | 8 | 0 | 3200 | 800 |
| 7 | 8 | 0 | 8 | 0 | $\infty$ | 800 |

Table A.3: Management policy parameters in series B ($LILT_k$).

| | NEMO | | ER | | LILT | |
|---|---|---|---|---|---|---|
| policy parameter set | k | t | k | t | k | t |
| 1 | 8 | 0 | 1 | 0 | 800 | 800 |
| 2 | 8 | 0 | 2 | 0 | 800 | 800 |
| 3 | 8 | 0 | 4 | 0 | 800 | 800 |
| 4 | 8 | 0 | 8 | 0 | 800 | 800 |
| 5 | 8 | 0 | 16 | 0 | 800 | 800 |
| 6 | 8 | 0 | 32 | 0 | 800 | 800 |
| 7 | 8 | 0 | $\infty$ | 0 | 800 | 800 |

Table A.4: Management policy parameters in series C ($ER_k$).

| | NEMO | | ER | | LILT | |
|---|---|---|---|---|---|---|
| policy parameter set | k | t | k | t | k | t |
| 1 | 1 | 0 | 8 | 0 | 800 | 800 |
| 2 | 2 | 0 | 8 | 0 | 800 | 800 |
| 3 | 4 | 0 | 8 | 0 | 800 | 800 |
| 4 | 8 | 0 | 8 | 0 | 800 | 800 |
| 5 | 16 | 0 | 8 | 0 | 800 | 800 |
| 6 | 32 | 0 | 8 | 0 | 800 | 800 |
| 7 | $\infty$ | 0 | 8 | 0 | 800 | 800 |

Table A.5: Management policy parameters in series D ($NEMO_k$).

The parameter set which yielded the greatest benefit in series A was policy parameter set 5 ($LILT_t = 1600$); in series B, policy parameter set 7 ($LILT_k = \infty$); in series C, policy parameter set 6 ($ER_k = 32$); and in in series D, policy parameter set 4 ($NEMO_k = 8$). A combination of those were compared with a configuration derived from initial test runs in series E.

| | NEMO | | ER | | LILT | |
|---|---|---|---|---|---|---|
| policy parameter set | k | t | k | t | k | t |
| 1 | 8 | 0 | 32 | 0 | $\infty$ | 1600 |
| 2 | 5 | 0 | 10 | 1 | 800 | 1600 |

Table A.6: Management policy parameters in series E.

In series E, the winners from series A - D (set 1) resulted in the greatest benefit.

**Findings**

Policy parameter set 1 in series E (table A.6) resulted in the greatest overall benefit and was thus used in succeeding experiments. Additionally to that it was found that considering the metric $LILT$ (and subsequently the related sub-policy) is disadvantageous. In series A in networks with low membership churn, the lookup time worsened with respect to an unmanaged network with decreasing values for $LILT_t$. This had the highest (negative) effect on the ranking metric for benefit in that series. The reason for this was that low $LILT_t$ values decreased the managed interval and kept it at a low level, even though short maintenance intervals were not required in networks with low membership churns. This caused a P2P node to spend computing resources on maintenance instead of performing lookups.

### A.1.4 File System Specific Workload Generation

Some preliminary work was carried out with the objective to translate a file system workload into a corresponding temporal pattern of ASA lookup requests. This work involved the selection of an appropriate "real world" file system trace and the translation of this trace to a P2P workload.

#### File System Traces

Most of the identified available work on file system workloads [65, 4, 28, 97, 88, 80] does not include complete file system traces. The most recent available complete traces identified recorded various FS operations (including read and write) on workstations used by undergraduates at the University of Berkeley in 1996 [79]. The *Berkeley traces* have been made anonymous by translating the file names (including file paths) to numerical identifiers for privacy reasons. Of these *Berkeley traces* a file system trace of one individual work-station (host 30) from the $14^{th}$ December 1996 was randomly selected to be used for the experiments reported in chapter 7.

**Transformation of File System Traces**

To translate a file system trace into a pattern of P2P lookups read and write operations were extracted and transformed in corresponding ASA operations as described in chapter 3. Every directory along the path to a file is looked up separately; every data and meta-data item is replicated four times, in accordance with the *ASA cross-algorithm* (see chapter 3). A file path was represented by an unique file identifier. The length of the path was selected from a distribution of values based on work carried out by Microsoft Research Labs [2, 17]. Every element along this *virtual file path* was transformed into keys representative for data and meta-data. As data is self-verifying it was associated in a single lookup, preceded by a set of parallel lookups for meta-data.

### A.1.5 Choosing a Local Area Test-Bed

The motivation for choosing a local area test-bed in which nodes were deployed on multiple machines was that such an experimental platform is representative of envisioned ASA use cases. Additionally other options have do have limitations, with respect to this work, as briefly outlined in the following. Other options included P2P simulator such as p-sim [62], peersim [41], j-sim [29] or any other listed in [44, 67] as well as a wide area deployment like Planet Lab [71].

Existing P2P simulators simulate membership churn and workload specific to file sharing applications like napster [83] or gnutella [83]. Such scenarios are not relevant for this work as users in a distributed storage system behave differently from users that use P2P file sharing tools. Users of a file sharing tool may have no interest in staying on-line after they have completed their download, but users of a distributed storage system will not immediately disconnect after every operation. Workloads specific to file sharing applications [73, 84, 82] mainly consist of searches and downloads, whereas P2P infrastructures used for distributed storage systems experience workloads resulting from file system operations (see [31]).

Wide area deployments such as Planet Lab may experience high variation of the available network speeds because of being connected through multiple hops via the internet. This can skew the measurements of interest as, for this work, specific conditions are repeated in order to verify reproducibility of management actions. By not having full control over the infrastructure, variance in the network conditions between repetitions of the experiments which were supposed to be executed here under identical conditions may occur. Recent work [69] shows that a user of Planet Lab may observe a significant number of undesired or uncontrollable failures. To verify that autonomic management applies the same actions in identical, repeated, conditions experiments had to be reproducible. Therefore a local area test-bed was considered as more appropriate for the work reported here as a wide area test-bed.

## A.2 P2P Experimental Results

### A.2.1 Introduction

In the following sections detailed results from the experiments outlined in chapter 7 are reported.

Each report (section A.2.2 - A.2.17) shows the primary and secondary user-level metrics (ULM) values as specified in 7.4.2 for a specific policy, and information about their distribution. Policy 1 was configured with parameters derived from preliminary experiments and policy 2 was configured to exhibit an aggressive behaviour. Policy 0, the baseline, represented unmanaged nodes which were configured with a static interval.

The difference between the effects of the specific policies on the individual primary ULMs (*expected lookup time* and *network usage*) are illustrated as bars normalised with respect to the results from unmanaged nodes (policy 0). Each normalised user-level metric with a value of less than 100% therefore indicates an improvement of the specific ULM. Bars in the plot which were above a certain height have been cut off for readability reasons. Such a cut-off is indicated by an upwards arrow (↑) at the top of a cut-off bar and the information about the actual height of the bar. The raw data used for computing these normalised primary ULMs is listed below in a table. This table contains the mean values of the distribution of all primary ULMs combined from the specific experimental run repetitions. The distributions were generated by computing primary ULM values for five minute observation periods. The means of these distributions were used to quantify the effects of the specific policies in the plots. The table is followed by an individual table for each policy showing distribution characteristics of all primary and secondary user level metric distributions. The primary ULM *expected lookup time* is abbreviated as $ELT$, and the *network usage* as $NU$. The secondary ULM *lookup time* is abbreviated as $LT$, the *lookup error time* as $LET$, and the *lookup error rate* as $LER$. The distributions for $ELT$, $NU$ and $LER$ represent corresponding aggregated measurements over 5 minute intervals in individual experimental runs. $LT$ and $LET$ however represents distributions of individual measurements taken during the course of all experimental runs with the specific policy, churn and workload. Each distribution is characterised by the minimum value (min), the 90% confidence interval of the mean [40] given by ($ci_\mu$), the standard deviation ($s$), the first quartile ($Q_1$), the second quartile ($Q_2$), the third quartile ($Q_3$) and the maximum value (max). Unavailable values are abbreviated as *NA*; *lookup error time* values, for instance, were only monitored in experiments in which errors occurred.

All descriptive statistics in this work were computed using the *Commons-Math: The Apache Commons Mathematics Library*, commons-math 1.2, [25].

### A.2.2 Synthetic Light Weight Workload with Low Churn

#### Summarised Effects of Policies on Expected Lookup Time and Network Usage

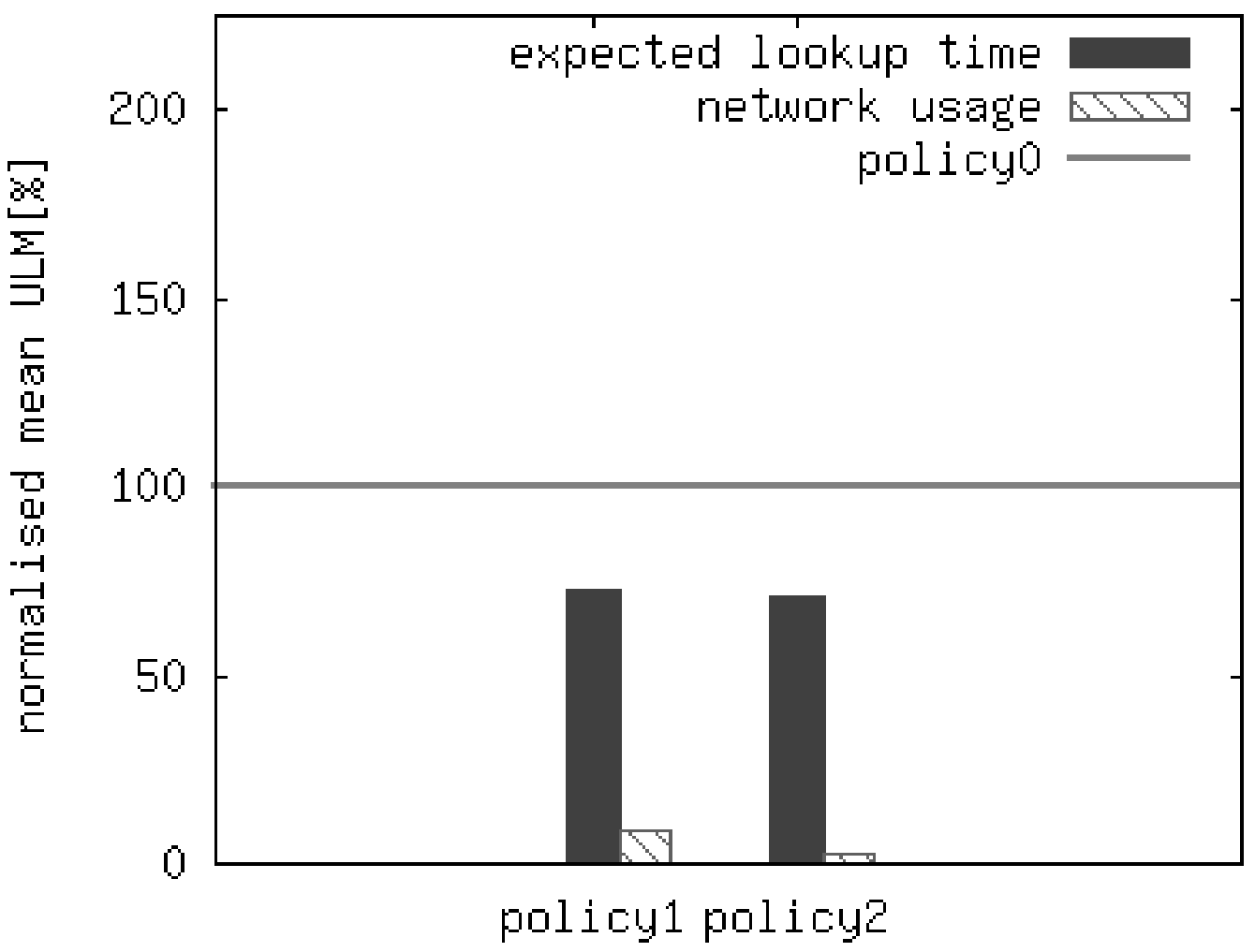

Figure A.3: normalised mean ULM values

| policy | mean expected lookup time [ms] | mean network usage [MB/5 min] |
|---|---|---|
| policy 0 | 720 | 60 |
| policy 1 | 522 | 5 |
| policy 2 | 507 | 2 |

Table A.7: mean ULM values

**Summary of ULM Distributions**

| ULM | unit | $ci_\mu$ | $s$ | min | $Q_1$ | $Q_2$ | $Q_3$ | max |
|---|---|---|---|---|---|---|---|---|
| ELT (primary) | [ms] | 720±62 | 207 | 357 | 516 | 777 | 857 | 1075 |
| NU (primary) | [MB/5 min] | 60±6 | 20 | 2 | 66 | 66 | 68 | 68 |
| LT (secondary) | [ms] | 720±62 | 207 | 357 | 516 | 777 | 857 | 1075 |
| LET (secondary) | [ms] | NA | NA | NA | NA | NA | NA | NA |
| LER (secondary) | [%] | 0±0 | 0 | 0 | 0 | 0 | 0 | 0 |

Table A.8: ULM distributions, measured with unmanaged nodes (policy 0).

| ULM | unit | $ci_\mu$ | $s$ | min | $Q_1$ | $Q_2$ | $Q_3$ | max |
|---|---|---|---|---|---|---|---|---|
| ELT (primary) | [ms] | 522±43 | 145 | 291 | 406 | 535 | 607 | 875 |
| NU (primary) | [MB/5 min] | 5±1 | 4 | 0 | 3 | 4 | 7 | 16 |
| LT (secondary) | [ms] | 522±43 | 145 | 291 | 406 | 535 | 607 | 875 |
| LET (secondary) | [ms] | NA | NA | NA | NA | NA | NA | NA |
| LER (secondary) | [%] | 0±0 | 0 | 0 | 0 | 0 | 0 | 0 |

Table A.9: ULM distributions, measured with managed nodes (policy 1).

| ULM | unit | $ci_\mu$ | $s$ | min | $Q_1$ | $Q_2$ | $Q_3$ | max |
|---|---|---|---|---|---|---|---|---|
| ELT (primary) | [ms] | 507±36 | 121 | 291 | 425 | 567 | 606 | 635 |
| NU (primary) | [MB/5 min] | 2±0 | 1 | 0 | 1 | 1 | 2 | 5 |
| LT (secondary) | [ms] | 507±36 | 121 | 291 | 425 | 567 | 606 | 635 |
| LET (secondary) | [ms] | NA | NA | NA | NA | NA | NA | NA |
| LER (secondary) | [%] | 0±0 | 0 | 0 | 0 | 0 | 0 | 0 |

Table A.10: ULM distributions, measured with managed nodes (policy 2).

### A.2.3 Synthetic Light Weight Workload with High Churn

#### Summarised Effects of Policies on Expected Lookup Time and Network Usage

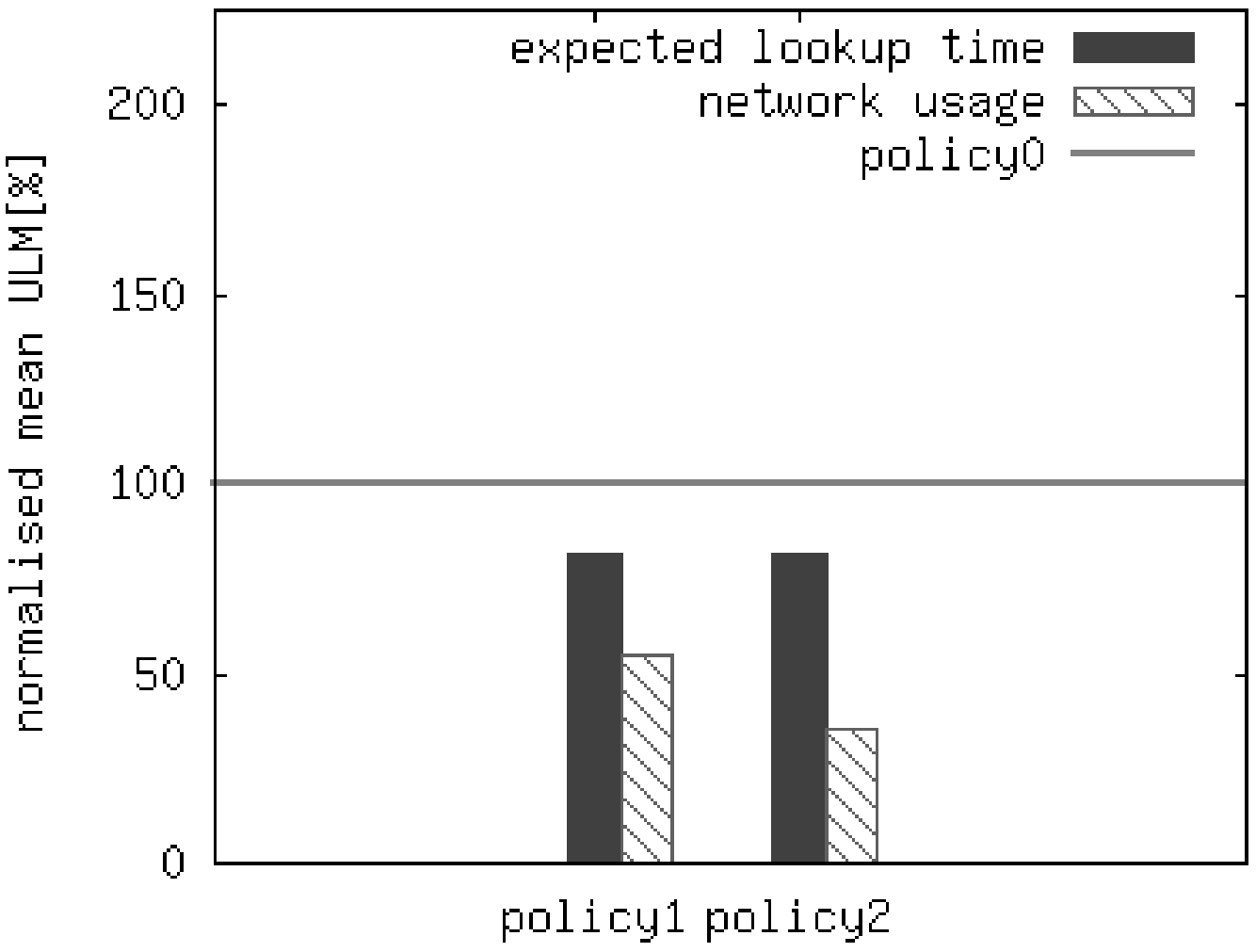

Figure A.4: normalised mean ULM values

| policy | mean expected lookup time [ms] | mean network usage [MB/5 min] |
|---|---|---|
| policy 0 | 562 | 33 |
| policy 1 | 457 | 18 |
| policy 2 | 459 | 12 |

Table A.11: mean ULM values

**Summary of ULM Distributions**

| ULM | unit | $ci_\mu$ | $s$ | min | $Q_1$ | $Q_2$ | $Q_3$ | max |
|---|---|---|---|---|---|---|---|---|
| ELT (primary) | [ms] | 562±92 | 296 | 176 | 398 | 551 | 637 | 1710 |
| NU (primary) | [MB/5 min] | 33±4 | 13 | 0 | 32 | 37 | 40 | 43 |
| LT (secondary) | [ms] | 562±92 | 296 | 176 | 398 | 551 | 637 | 1710 |
| LET (secondary) | [ms] | 11±4 | 4 | 8 | 8 | 11 | 13 | 13 |
| LER (secondary) | [%] | 7±8 | 25 | 0 | 0 | 0 | 0 | 100 |

Table A.12: ULM distributions, measured with unmanaged nodes (policy 0).

| ULM | unit | $ci_\mu$ | $s$ | min | $Q_1$ | $Q_2$ | $Q_3$ | max |
|---|---|---|---|---|---|---|---|---|
| ELT (primary) | [ms] | 457±60 | 166 | 128 | 386 | 482 | 544 | 779 |
| NU (primary) | [MB/5 min] | 18±2 | 7 | 0 | 17 | 21 | 22 | 23 |
| LT (secondary) | [ms] | 457±60 | 166 | 128 | 386 | 482 | 544 | 779 |
| LET (secondary) | [ms] | 16493±27105 | 49431 | 11 | 13 | 17 | 20 | 148308 |
| LER (secondary) | [%] | 30±14 | 47 | 0 | 0 | 0 | 100 | 100 |

Table A.13: ULM distributions, measured with managed nodes (policy 1).

| ULM | unit | $ci_\mu$ | $s$ | min | $Q_1$ | $Q_2$ | $Q_3$ | max |
|---|---|---|---|---|---|---|---|---|
| ELT (primary) | [ms] | 459±100 | 265 | 149 | 311 | 428 | 527 | 1434 |
| NU (primary) | [MB/5 min] | 12±1 | 4 | 0 | 9 | 12 | 15 | 17 |
| LT (secondary) | [ms] | 459±100 | 265 | 149 | 311 | 428 | 527 | 1434 |
| LET (secondary) | [ms] | 12555±20540 | 41412 | 9 | 13 | 47 | 49 | 137418 |
| LER (secondary) | [%] | 37±15 | 49 | 0 | 0 | 0 | 100 | 100 |

Table A.14: ULM distributions, measured with managed nodes (policy 2).

### A.2.4 Synthetic Light Weight Workload with Locally Varying Churn

#### Summarised Effects of Policies on Expected Lookup Time and Network Usage

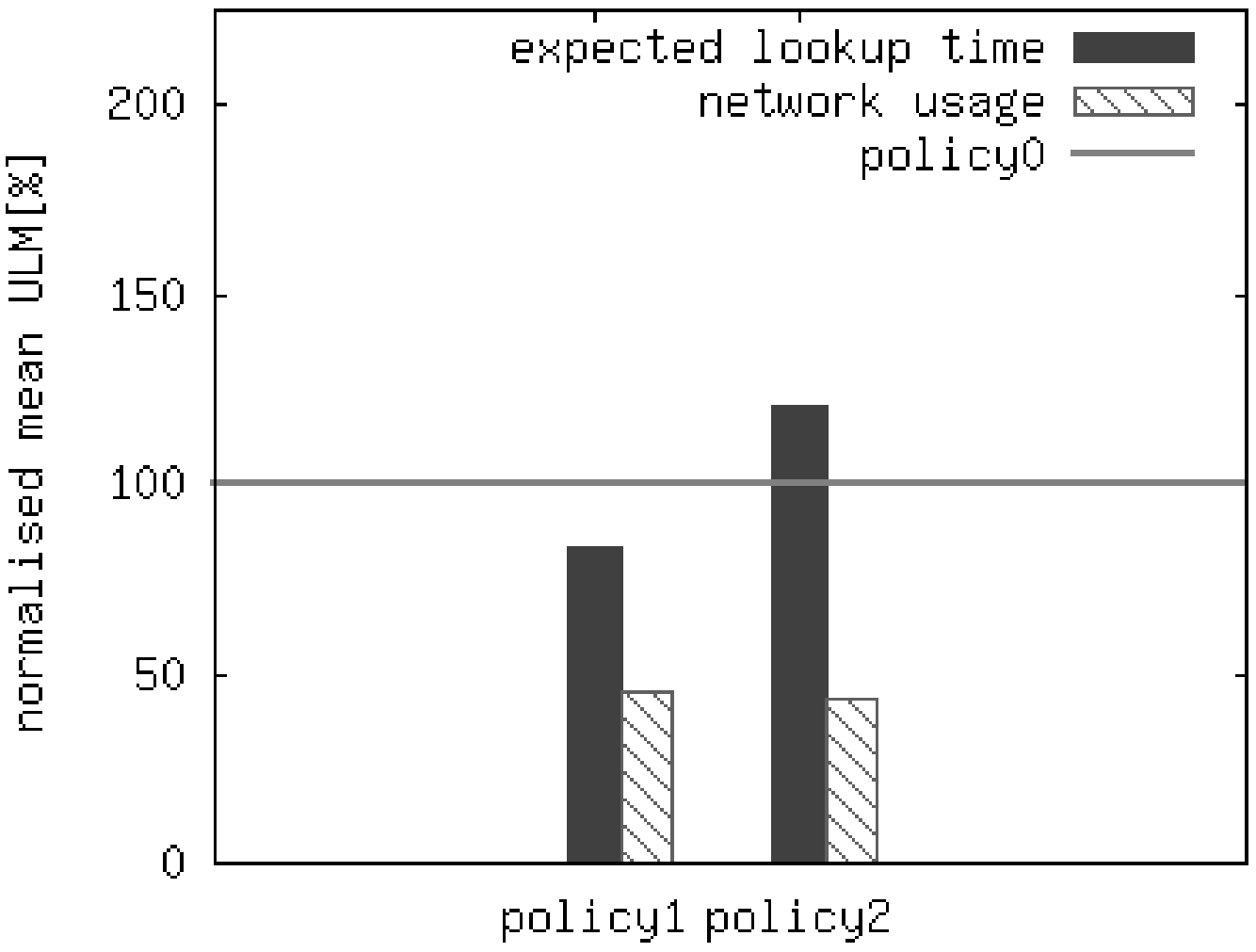

Figure A.5: normalised mean ULM values

| policy | mean expected lookup time [ms] | mean network usage [MB/5 min] |
|---|---|---|
| policy 0 | 689 | 40 |
| policy 1 | 575 | 18 |
| policy 2 | 831 | 17 |

Table A.15: mean ULM values

**Summary of ULM Distributions**

| ULM | unit | $ci_\mu$ | $s$ | min | $Q_1$ | $Q_2$ | $Q_3$ | max |
|---|---|---|---|---|---|---|---|---|
| ELT (primary) | [ms] | 689±75 | 248 | 137 | 550 | 708 | 830 | 1091 |
| NU (primary) | [MB/5 min] | 40±4 | 13 | 1 | 41 | 43 | 45 | 49 |
| LT (secondary) | [ms] | 689±75 | 248 | 137 | 550 | 708 | 830 | 1091 |
| LET (secondary) | [ms] | NA | NA | NA | NA | NA | NA | NA |
| LER (secondary) | [%] | 0±0 | 0 | 0 | 0 | 0 | 0 | 0 |

Table A.16: ULM distributions, measured with unmanaged nodes (policy 0).

| ULM | unit | $ci_\mu$ | $s$ | min | $Q_1$ | $Q_2$ | $Q_3$ | max |
|---|---|---|---|---|---|---|---|---|
| ELT (primary) | [ms] | 575±72 | 224 | 151 | 422 | 612 | 712 | 944 |
| NU (primary) | [MB/5 min] | 18±2 | 7 | 0 | 17 | 19 | 23 | 26 |
| LT (secondary) | [ms] | 575±72 | 224 | 151 | 422 | 612 | 712 | 944 |
| LET (secondary) | [ms] | 186±109 | 133 | 10 | 55 | 201 | 301 | 331 |
| LER (secondary) | [%] | 13±10 | 35 | 0 | 0 | 0 | 0 | 100 |

Table A.17: ULM distributions, measured with managed nodes (policy 1).

| ULM | unit | $ci_\mu$ | $s$ | min | $Q_1$ | $Q_2$ | $Q_3$ | max |
|---|---|---|---|---|---|---|---|---|
| ELT (primary) | [ms] | 831±185 | 574 | 136 | 448 | 701 | 947 | 2122 |
| NU (primary) | [MB/5 min] | 17±3 | 10 | 0 | 10 | 18 | 25 | 35 |
| LT (secondary) | [ms] | 831±185 | 574 | 136 | 448 | 701 | 947 | 2122 |
| LET (secondary) | [ms] | 56±76 | 93 | 9 | 9 | 11 | 149 | 195 |
| LER (secondary) | [%] | 13±10 | 35 | 0 | 0 | 0 | 0 | 100 |

Table A.18: ULM distributions, measured with managed nodes (policy 2).

### A.2.5 Synthetic Light Weight Workload with Temporally Varying Churn

#### Summarised Effects of Policies on Expected Lookup Time and Network Usage

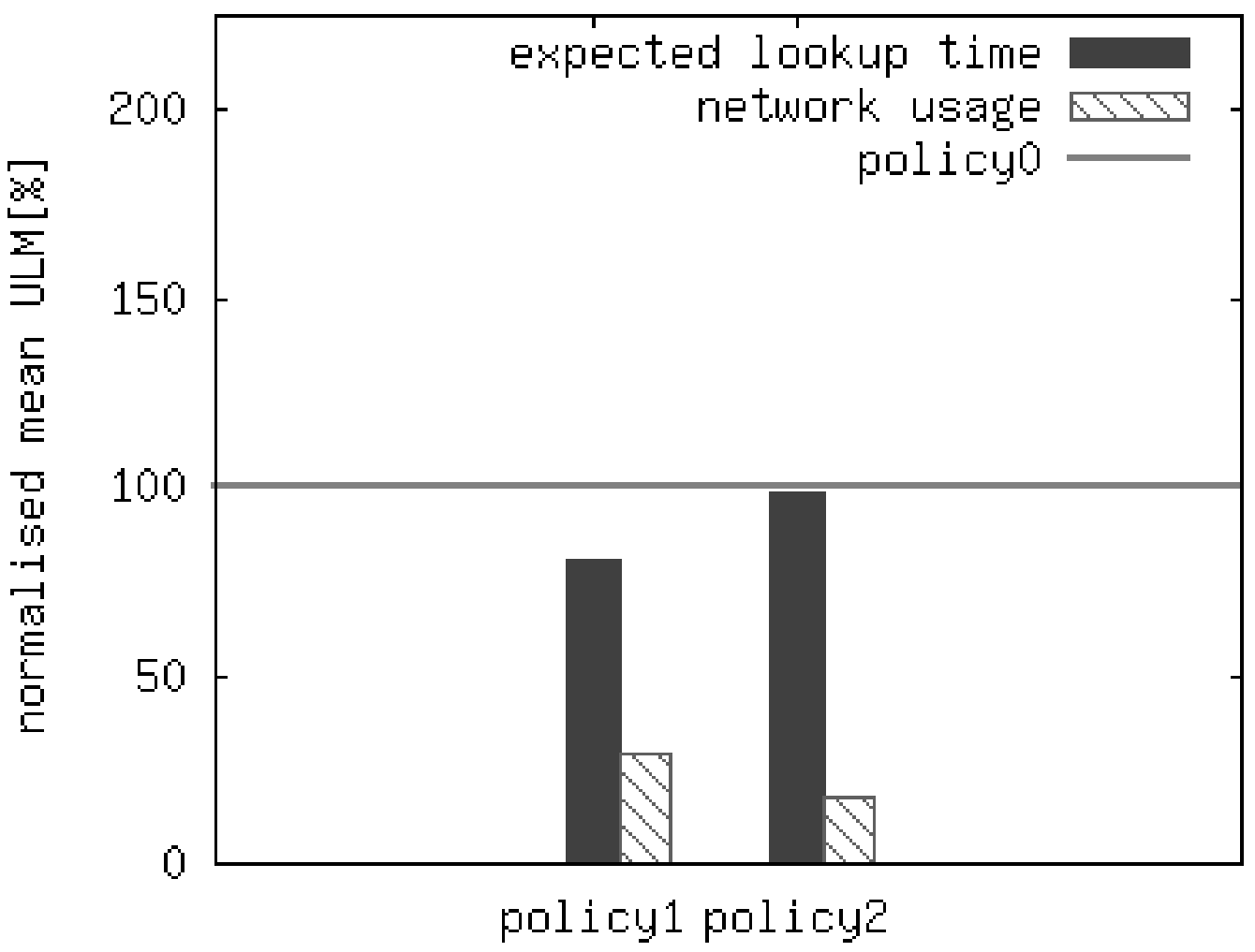

Figure A.6: normalised mean ULM values

| policy | mean expected lookup time [ms] | mean network usage [MB/5 min] |
|---|---|---|
| policy 0 | 643 | 44 |
| policy 1 | 518 | 13 |
| policy 2 | 629 | 8 |

Table A.19: mean ULM values

**Summary of ULM Distributions**

| ULM | unit | $ci_\mu$ | $s$ | min | $Q_1$ | $Q_2$ | $Q_3$ | max |
|---|---|---|---|---|---|---|---|---|
| ELT (primary) | [ms] | 643±75 | 232 | 139 | 463 | 602 | 871 | 1077 |
| NU (primary) | [MB/5 min] | 44±5 | 18 | 1 | 37 | 42 | 60 | 67 |
| LT (secondary) | [ms] | 643±75 | 232 | 139 | 463 | 602 | 871 | 1077 |
| LET (secondary) | [ms] | 397±631 | 767 | 13 | 13 | 14 | 1165 | 1548 |
| LER (secondary) | [%] | 13±10 | 35 | 0 | 0 | 0 | 0 | 100 |

Table A.20: ULM distributions, measured with unmanaged nodes (policy 0).

| ULM | unit | $ci_\mu$ | $s$ | min | $Q_1$ | $Q_2$ | $Q_3$ | max |
|---|---|---|---|---|---|---|---|---|
| ELT (primary) | [ms] | 518±48 | 152 | 308 | 441 | 466 | 604 | 935 |
| NU (primary) | [MB/5 min] | 13±2 | 6 | 0 | 9 | 14 | 17 | 22 |
| LT (secondary) | [ms] | 518±48 | 152 | 308 | 441 | 466 | 604 | 935 |
| LET (secondary) | [ms] | 12±0 | 0 | 12 | 12 | 12 | 12 | 12 |
| LER (secondary) | [%] | 10±9 | 31 | 0 | 0 | 0 | 0 | 100 |

Table A.21: ULM distributions, measured with managed nodes (policy 1).

| ULM | unit | $ci_\mu$ | $s$ | min | $Q_1$ | $Q_2$ | $Q_3$ | max |
|---|---|---|---|---|---|---|---|---|
| ELT (primary) | [ms] | 629±162 | 503 | 152 | 311 | 466 | 680 | 2039 |
| NU (primary) | [MB/5 min] | 8±2 | 5 | 0 | 4 | 7 | 12 | 18 |
| LT (secondary) | [ms] | 629±162 | 503 | 152 | 311 | 466 | 680 | 2039 |
| LET (secondary) | [ms] | 15855±15113 | 18374 | 8 | 9 | 14820 | 32735 | 33771 |
| LER (secondary) | [%] | 13±10 | 35 | 0 | 0 | 0 | 0 | 100 |

Table A.22: ULM distributions, measured with managed nodes (policy 2).

### A.2.6 Synthetic Heavy Weight Workload with Low Churn

#### Summarised Effects of Policies on Expected Lookup Time and Network Usage

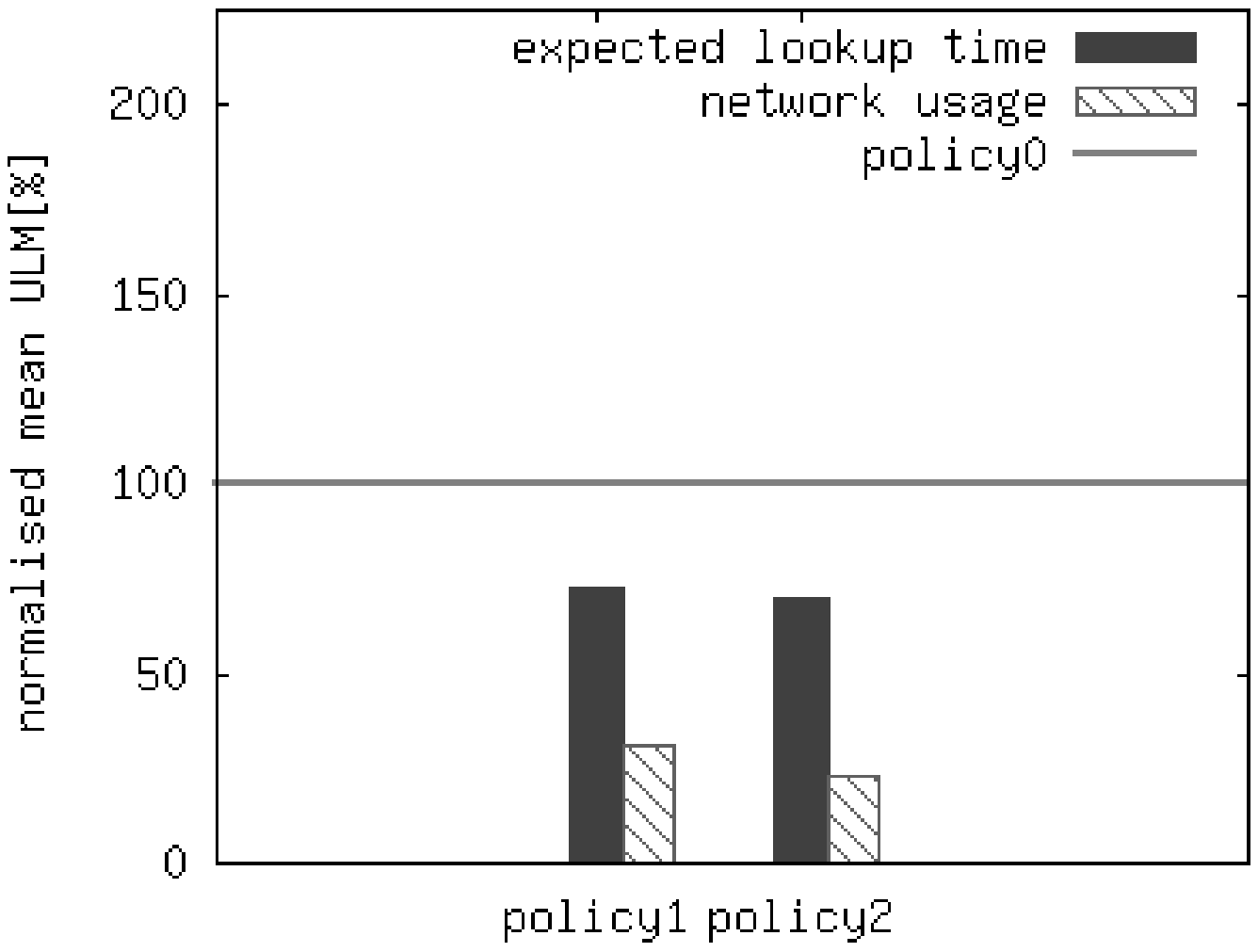

Figure A.7: normalised mean ULM values

| policy | mean expected lookup time [ms] | mean network usage [MB/5 min] |
|---|---|---|
| policy 0 | 613 | 69 |
| policy 1 | 445 | 22 |
| policy 2 | 428 | 16 |

Table A.23: mean ULM values

**Summary of ULM Distributions**

| ULM | unit | $ci_{\mu}$ | $s$ | min | $Q_1$ | $Q_2$ | $Q_3$ | max |
|---|---|---|---|---|---|---|---|---|
| ELT (primary) | [ms] | 613±7 | 25 | 538 | 611 | 619 | 627 | 640 |
| NU (primary) | [MB/5 min] | 69±4 | 15 | 17 | 73 | 74 | 74 | 75 |
| LT (secondary) | [ms] | 613±3 | 228 | 89 | 444 | 615 | 774 | 2505 |
| LET (secondary) | [ms] | NA | NA | NA | NA | NA | NA | NA |
| LER (secondary) | [%] | 0±0 | 0 | 0 | 0 | 0 | 0 | 0 |

Table A.24: ULM distributions, measured with unmanaged nodes (policy 0).

| ULM | unit | $ci_{\mu}$ | $s$ | min | $Q_1$ | $Q_2$ | $Q_3$ | max |
|---|---|---|---|---|---|---|---|---|
| ELT (primary) | [ms] | 445±6 | 19 | 425 | 431 | 438 | 458 | 501 |
| NU (primary) | [MB/5 min] | 22±3 | 8 | 13 | 17 | 18 | 22 | 41 |
| LT (secondary) | [ms] | 445±2 | 152 | 86 | 354 | 445 | 552 | 2234 |
| LET (secondary) | [ms] | NA | NA | NA | NA | NA | NA | NA |
| LER (secondary) | [%] | 0±0 | 0 | 0 | 0 | 0 | 0 | 0 |

Table A.25: ULM distributions, measured with managed nodes (policy 1).

| ULM | unit | $ci_{\mu}$ | $s$ | min | $Q_1$ | $Q_2$ | $Q_3$ | max |
|---|---|---|---|---|---|---|---|---|
| ELT (primary) | [ms] | 428±4 | 12 | 394 | 421 | 430 | 436 | 444 |
| NU (primary) | [MB/5 min] | 16±2 | 5 | 8 | 14 | 15 | 16 | 27 |
| LT (secondary) | [ms] | 428±2 | 142 | 87 | 322 | 419 | 549 | 1895 |
| LET (secondary) | [ms] | 206±0 | 0 | 206 | 206 | 206 | 206 | 206 |
| LER (secondary) | [%] | 0±0 | 0 | 0 | 0 | 0 | 0 | 0 |

Table A.26: ULM distributions, measured with managed nodes (policy 2).

### A.2.7 Synthetic Heavy Weight Workload with High Churn

#### Summarised Effects of Policies on Expected Lookup Time and Network Usage

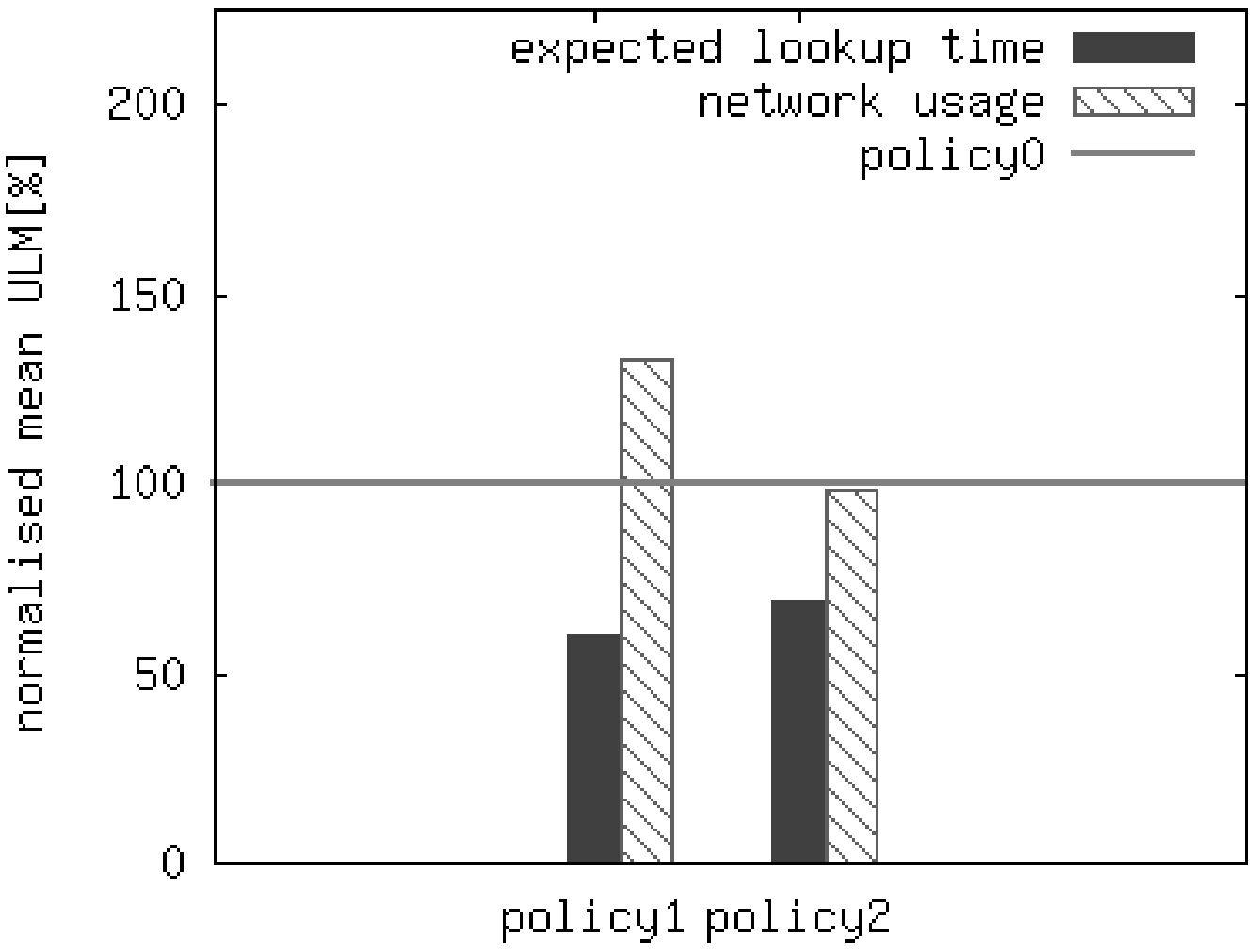

Figure A.8: normalised mean ULM values

| policy | mean expected lookup time [ms] | mean network usage [MB/5 min] |
|---|---|---|
| policy 0 | 727 | 25 |
| policy 1 | 440 | 33 |
| policy 2 | 504 | 24 |

Table A.27: mean ULM values

**Summary of ULM Distributions**

| ULM | unit | $ci_\mu$ | $s$ | min | $Q_1$ | $Q_2$ | $Q_3$ | max |
|---|---|---|---|---|---|---|---|---|
| ELT (primary) | [ms] | 727±71 | 229 | 476 | 509 | 713 | 927 | 1168 |
| NU (primary) | [MB/5 min] | 25±7 | 23 | 1 | 2 | 22 | 47 | 50 |
| LT (secondary) | [ms] | 590±4 | 256 | 97 | 406 | 623 | 750 | 4102 |
| LET (secondary) | [ms] | 152±25 | 1181 | 6 | 67 | 99 | 110 | 45045 |
| LER (secondary) | [%] | 26±9 | 29 | 1 | 3 | 18 | 37 | 86 |

Table A.28: ULM distributions, measured with unmanaged nodes (policy 0).

| ULM | unit | $ci_\mu$ | $s$ | min | $Q_1$ | $Q_2$ | $Q_3$ | max |
|---|---|---|---|---|---|---|---|---|
| ELT (primary) | [ms] | 440±9 | 30 | 372 | 435 | 449 | 455 | 501 |
| NU (primary) | [MB/5 min] | 33±2 | 6 | 13 | 33 | 34 | 36 | 38 |
| LT (secondary) | [ms] | 434±3 | 208 | 94 | 278 | 412 | 558 | 2336 |
| LET (secondary) | [ms] | 196±10 | 160 | 6 | 86 | 185 | 263 | 1506 |
| LER (secondary) | [%] | 4±0 | 2 | 1 | 3 | 3 | 4 | 7 |

Table A.29: ULM distributions, measured with managed nodes (policy 1).

| ULM | unit | $ci_\mu$ | $s$ | min | $Q_1$ | $Q_2$ | $Q_3$ | max |
|---|---|---|---|---|---|---|---|---|
| ELT (primary) | [ms] | 504±66 | 210 | 332 | 411 | 422 | 522 | 1071 |
| NU (primary) | [MB/5 min] | 24±2 | 7 | 4 | 24 | 26 | 28 | 33 |
| LT (secondary) | [ms] | 409±3 | 198 | 85 | 275 | 389 | 523 | 3555 |
| LET (secondary) | [ms] | 482±247 | 6545 | 6 | 142 | 185 | 251 | 165072 |
| LER (secondary) | [%] | 11±3 | 10 | 5 | 7 | 8 | 11 | 56 |

Table A.30: ULM distributions, measured with managed nodes (policy 2).

### A.2.8 Synthetic Heavy Weight Workload with Locally Varying Churn

#### Summarised Effects of Policies on Expected Lookup Time and Network Usage

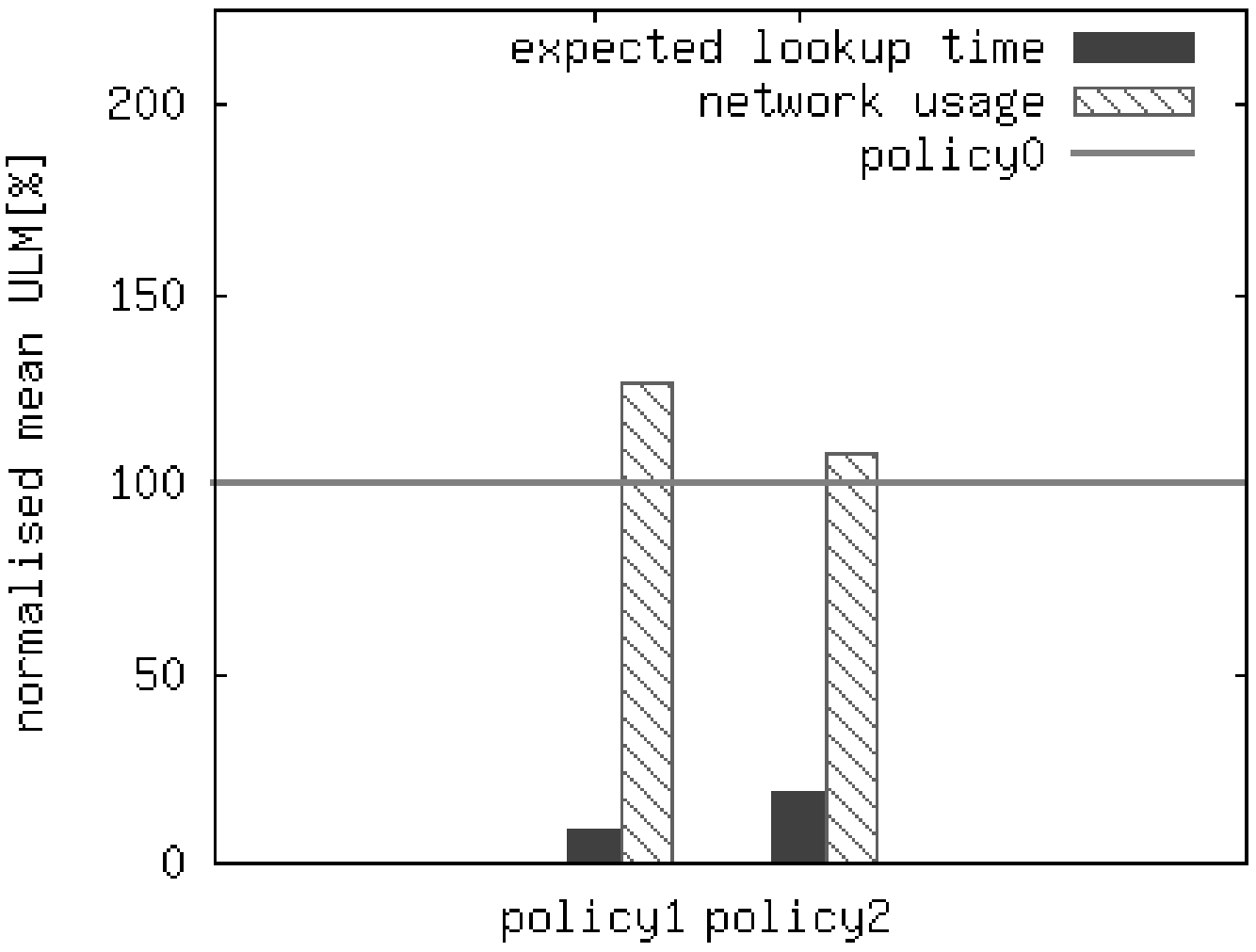

Figure A.9: normalised mean ULM values

| policy | mean expected lookup time [ms] | mean network usage [MB/5 min] |
|---|---|---|
| policy 0 | 6495 | 23 |
| policy 1 | 552 | 30 |
| policy 2 | 1190 | 25 |

Table A.31: mean ULM values

**Summary of ULM Distributions**

| ULM | unit | $ci_\mu$ | $s$ | min | $Q_1$ | $Q_2$ | $Q_3$ | max |
|---|---|---|---|---|---|---|---|---|
| ELT (primary) | [ms] | 6495±3107 | 15574 | 390 | 576 | 2120 | 8655 | 126505 |
| NU (primary) | [MB/5 min] | 23±4 | 21 | 0 | 6 | 13 | 51 | 57 |
| LT (secondary) | [ms] | 840±9 | 657 | 98 | 416 | 647 | 944 | 6366 |
| LET (secondary) | [ms] | 2447±531 | 17738 | 6 | 74 | 166 | 525 | 239475 |
| LER (secondary) | [%] | 33±5 | 24 | 0 | 2 | 43 | 53 | 69 |

Table A.32: ULM distributions, measured with unmanaged nodes (policy 0).

| ULM | unit | $ci_\mu$ | $s$ | min | $Q_1$ | $Q_2$ | $Q_3$ | max |
|---|---|---|---|---|---|---|---|---|
| ELT (primary) | [ms] | 552±117 | 389 | 375 | 438 | 460 | 474 | 2522 |
| NU (primary) | [MB/5 min] | 30±3 | 8 | 1 | 27 | 32 | 34 | 39 |
| LT (secondary) | [ms] | 444±3 | 204 | 84 | 279 | 421 | 589 | 2545 |
| LET (secondary) | [ms] | 871±639 | 12153 | 6 | 143 | 186 | 236 | 251373 |
| LER (secondary) | [%] | 5±1 | 3 | 0 | 3 | 6 | 7 | 13 |

Table A.33: ULM distributions, measured with managed nodes (policy 1).

| ULM | unit | $ci_\mu$ | $s$ | min | $Q_1$ | $Q_2$ | $Q_3$ | max |
|---|---|---|---|---|---|---|---|---|
| ELT (primary) | [ms] | 1190±325 | 1282 | 438 | 536 | 986 | 1268 | 7995 |
| NU (primary) | [MB/5 min] | 25±2 | 9 | 5 | 17 | 27 | 33 | 42 |
| LT (secondary) | [ms] | 541±6 | 438 | 93 | 283 | 444 | 662 | 4759 |
| LET (secondary) | [ms] | 1527±473 | 15181 | 6 | 141 | 186 | 243 | 239224 |
| LER (secondary) | [%] | 18±4 | 15 | 0 | 8 | 11 | 35 | 59 |

Table A.34: ULM distributions, measured with managed nodes (policy 2).

### A.2.9 Synthetic Heavy Weight Workload with Temporally Varying Churn

#### Summarised Effects of Policies on Expected Lookup Time and Network Usage

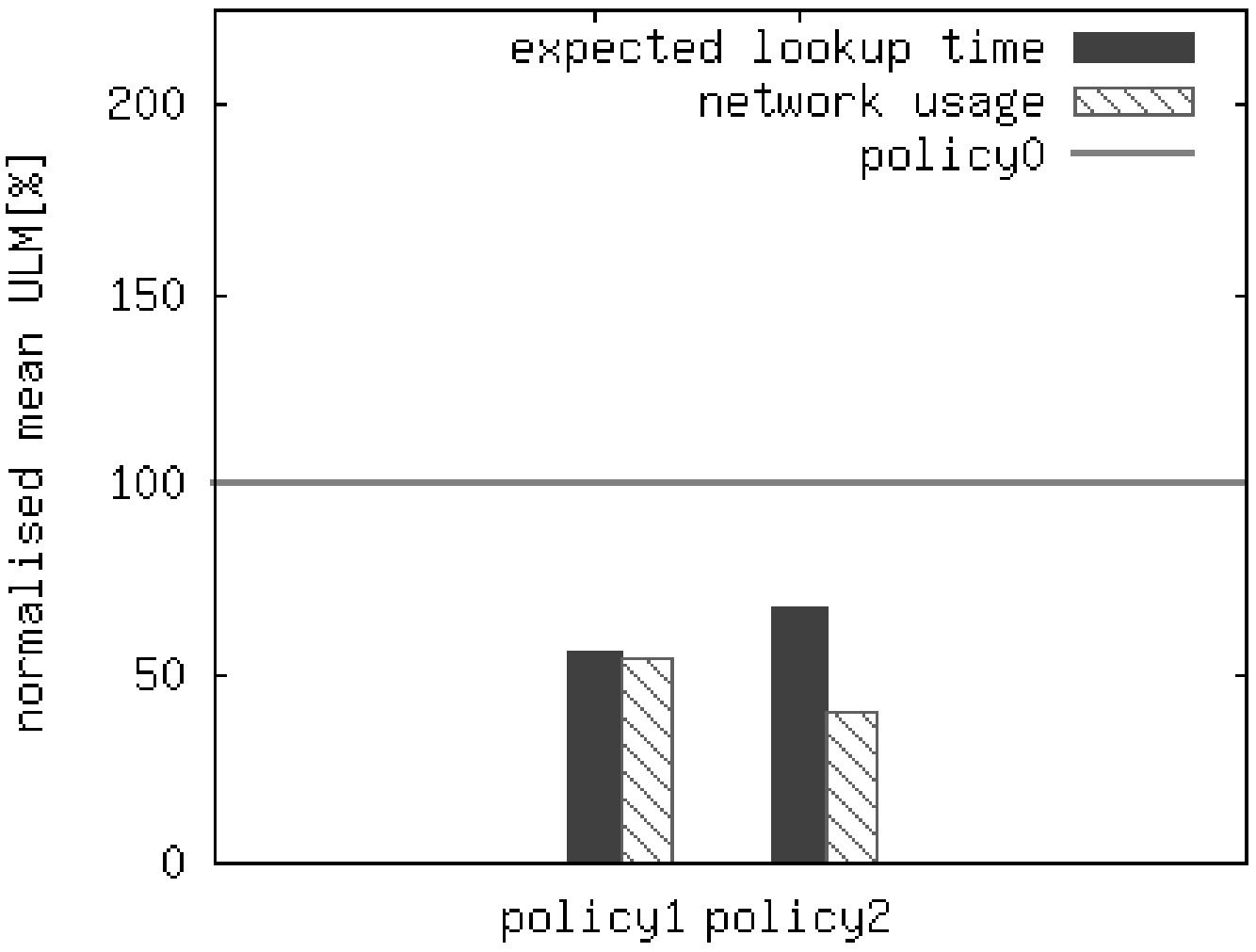

Figure A.10: normalised mean ULM values

| policy | mean expected lookup time [ms] | mean network usage [MB/5 min] |
|---|---|---|
| policy 0 | 804 | 51 |
| policy 1 | 452 | 28 |
| policy 2 | 540 | 20 |

Table A.35: mean ULM values

**Summary of ULM Distributions**

| ULM | unit | $ci_\mu$ | $s$ | min | $Q_1$ | $Q_2$ | $Q_3$ | max |
|---|---|---|---|---|---|---|---|---|
| ELT (primary) | [ms] | 804±172 | 629 | 519 | 545 | 579 | 628 | 2684 |
| NU (primary) | [MB/5 min] | 51±5 | 19 | 0 | 45 | 53 | 68 | 74 |
| LT (secondary) | [ms] | 622±5 | 408 | 98 | 409 | 570 | 739 | 6912 |
| LET (secondary) | [ms] | 196±27 | 745 | 7 | 97 | 105 | 128 | 30077 |
| LER (secondary) | [%] | 10±6 | 24 | 0 | 0 | 1 | 5 | 100 |

Table A.36: ULM distributions, measured with unmanaged nodes (policy 0).

| ULM | unit | $ci_\mu$ | $s$ | min | $Q_1$ | $Q_2$ | $Q_3$ | max |
|---|---|---|---|---|---|---|---|---|
| ELT (primary) | [ms] | 452±41 | 131 | 161 | 417 | 438 | 466 | 1030 |
| NU (primary) | [MB/5 min] | 28±3 | 9 | 8 | 22 | 29 | 35 | 40 |
| LT (secondary) | [ms] | 418±3 | 198 | 85 | 276 | 412 | 543 | 2662 |
| LET (secondary) | [ms] | 411±343 | 6669 | 6 | 177 | 184 | 231 | 213399 |
| LER (secondary) | [%] | 5±2 | 6 | 0 | 0 | 3 | 11 | 19 |

Table A.37: ULM distributions, measured with managed nodes (policy 1).

| ULM | unit | $ci_\mu$ | $s$ | min | $Q_1$ | $Q_2$ | $Q_3$ | max |
|---|---|---|---|---|---|---|---|---|
| ELT (primary) | [ms] | 540±58 | 195 | 321 | 435 | 445 | 578 | 1088 |
| NU (primary) | [MB/5 min] | 20±2 | 7 | 0 | 16 | 20 | 27 | 31 |
| LT (secondary) | [ms] | 444±2 | 185 | 85 | 304 | 425 | 560 | 2335 |
| LET (secondary) | [ms] | 916±571 | 12514 | 6 | 179 | 185 | 233 | 246854 |
| LER (secondary) | [%] | 7±3 | 11 | 0 | 0 | 1 | 7 | 32 |

Table A.38: ULM distributions, measured with managed nodes (policy 2).

### A.2.10 Synthetic Variabale Weight Workload with Low Churn

#### Summarised Effects of Policies on Expected Lookup Time and Network Usage

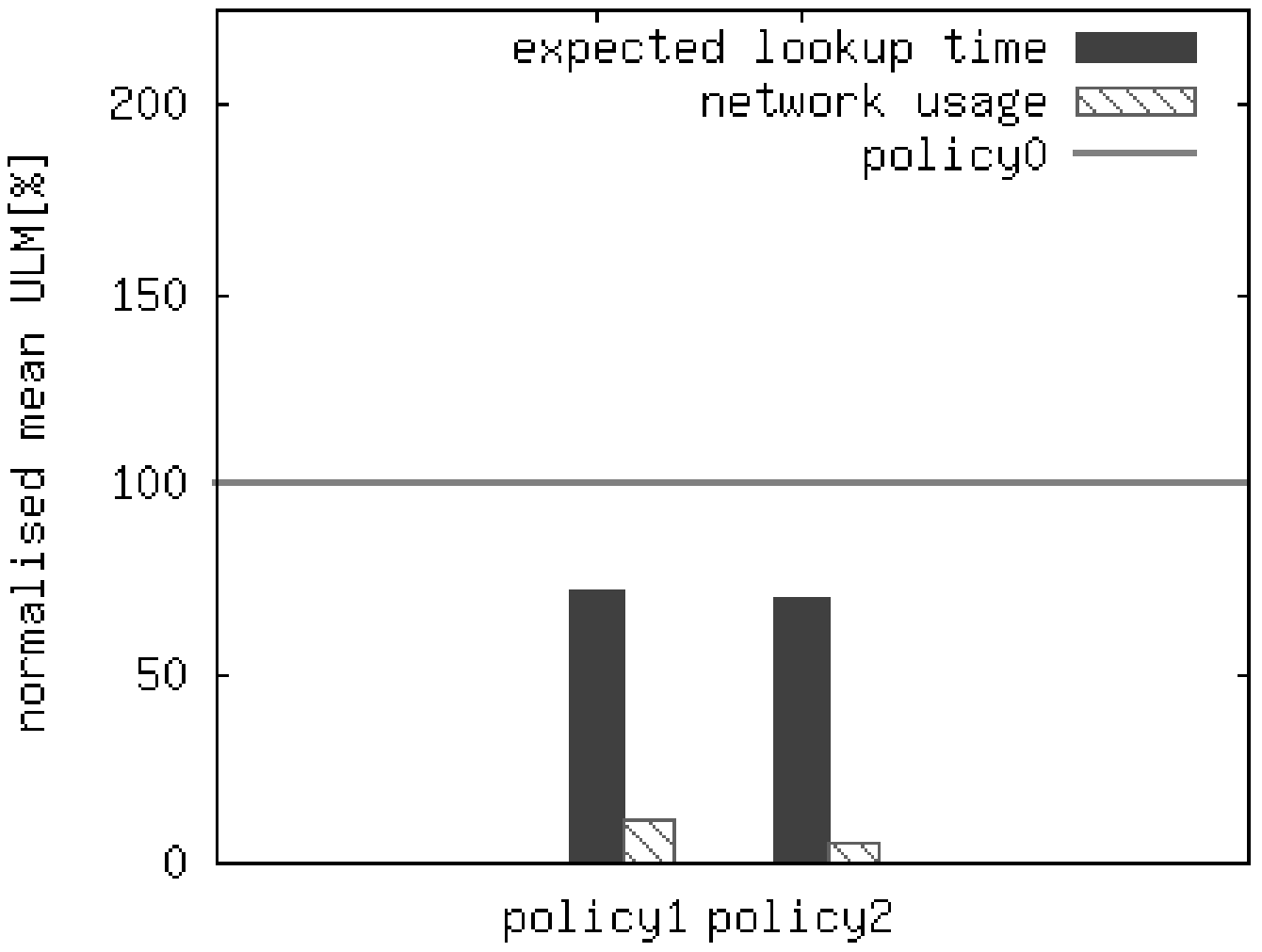

Figure A.11: normalised mean ULM values

| policy | mean expected lookup time [ms] | mean network usage [MB/5 min] |
|---|---|---|
| policy 0 | 636 | 63 |
| policy 1 | 454 | 7 |
| policy 2 | 445 | 3 |

Table A.39: mean ULM values

**Summary of ULM Distributions**

| ULM | unit | $ci_\mu$ | $s$ | min | $Q_1$ | $Q_2$ | $Q_3$ | max |
|---|---|---|---|---|---|---|---|---|
| ELT (primary) | [ms] | 636±6 | 23 | 593 | 618 | 633 | 654 | 699 |
| NU (primary) | [MB/5 min] | 63±4 | 16 | 8 | 67 | 68 | 68 | 69 |
| LT (secondary) | [ms] | 638±7 | 235 | 98 | 461 | 643 | 797 | 2315 |
| LET (secondary) | [ms] | NA | NA | NA | NA | NA | NA | NA |
| LER (secondary) | [%] | 0±0 | 0 | 0 | 0 | 0 | 0 | 0 |

Table A.40: ULM distributions, measured with unmanaged nodes (policy 0).

| ULM | unit | $ci_\mu$ | $s$ | min | $Q_1$ | $Q_2$ | $Q_3$ | max |
|---|---|---|---|---|---|---|---|---|
| ELT (primary) | [ms] | 454±7 | 24 | 426 | 437 | 451 | 462 | 525 |
| NU (primary) | [MB/5 min] | 7±1 | 4 | 3 | 4 | 6 | 8 | 18 |
| LT (secondary) | [ms] | 454±5 | 152 | 97 | 366 | 455 | 565 | 1826 |
| LET (secondary) | [ms] | NA | NA | NA | NA | NA | NA | NA |
| LER (secondary) | [%] | 0±0 | 0 | 0 | 0 | 0 | 0 | 0 |

Table A.41: ULM distributions, measured with managed nodes (policy 1).

| ULM | unit | $ci_\mu$ | $s$ | min | $Q_1$ | $Q_2$ | $Q_3$ | max |
|---|---|---|---|---|---|---|---|---|
| ELT (primary) | [ms] | 445±4 | 14 | 425 | 435 | 446 | 456 | 472 |
| NU (primary) | [MB/5 min] | 3±0 | 1 | 1 | 3 | 3 | 4 | 7 |
| LT (secondary) | [ms] | 445±4 | 142 | 95 | 363 | 452 | 554 | 1836 |
| LET (secondary) | [ms] | NA | NA | NA | NA | NA | NA | NA |
| LER (secondary) | [%] | 0±0 | 0 | 0 | 0 | 0 | 0 | 0 |

Table A.42: ULM distributions, measured with managed nodes (policy 2).

### A.2.11 Synthetic Variabale Weight Workload with High Churn

#### Summarised Effects of Policies on Expected Lookup Time and Network Usage

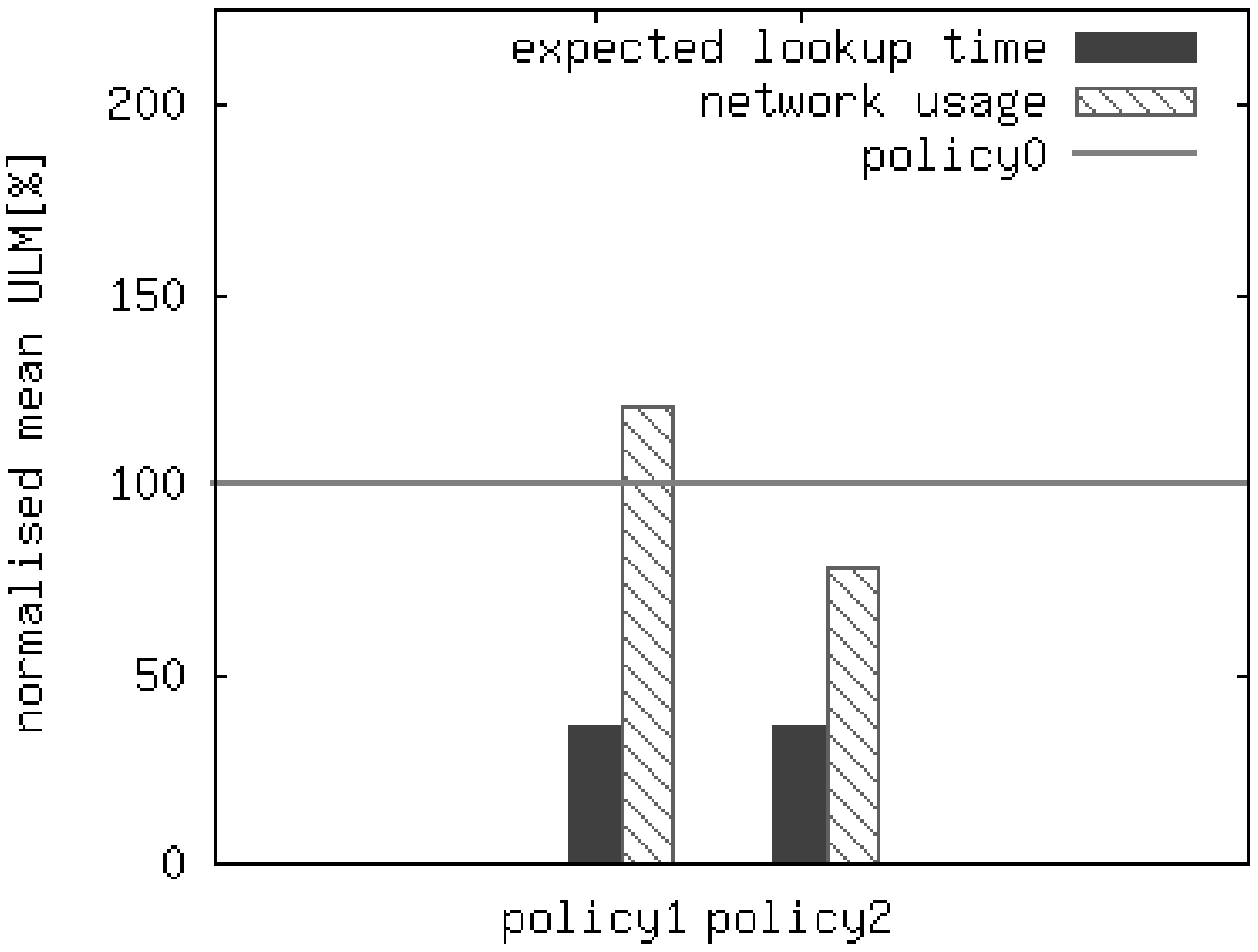

Figure A.12: normalised mean ULM values

| policy | mean expected lookup time [ms] | mean network usage [MB/5 min] |
|---|---|---|
| policy 0 | 1134 | 17 |
| policy 1 | 410 | 21 |
| policy 2 | 413 | 13 |

Table A.43: mean ULM values

**Summary of ULM Distributions**

| ULM | unit | $ci_\mu$ | $s$ | min | $Q_1$ | $Q_2$ | $Q_3$ | max |
|---|---|---|---|---|---|---|---|---|
| ELT (primary) | [ms] | 1134±469 | 1611 | 464 | 550 | 838 | 986 | 9665 |
| NU (primary) | [MB/5 min] | 17±6 | 19 | 0 | 0 | 4 | 38 | 45 |
| LT (secondary) | [ms] | 566±12 | 277 | 98 | 373 | 555 | 722 | 2535 |
| LET (secondary) | [ms] | 243±152 | 3572 | 8 | 97 | 102 | 107 | 132041 |
| LER (secondary) | [%] | 50±12 | 42 | 0 | 1 | 84 | 86 | 90 |

Table A.44: ULM distributions, measured with unmanaged nodes (policy 0).

| ULM | unit | $ci_\mu$ | $s$ | min | $Q_1$ | $Q_2$ | $Q_3$ | max |
|---|---|---|---|---|---|---|---|---|
| ELT (primary) | [ms] | 410±22 | 78 | 278 | 338 | 418 | 459 | 559 |
| NU (primary) | [MB/5 min] | 21±2 | 7 | 0 | 19 | 23 | 25 | 27 |
| LT (secondary) | [ms] | 405±6 | 194 | 97 | 274 | 371 | 519 | 1977 |
| LET (secondary) | [ms] | 122±10 | 105 | 6 | 85 | 87 | 144 | 633 |
| LER (secondary) | [%] | 9±6 | 20 | 0 | 1 | 2 | 8 | 86 |

Table A.45: ULM distributions, measured with managed nodes (policy 1).

| ULM | unit | $ci_\mu$ | $s$ | min | $Q_1$ | $Q_2$ | $Q_3$ | max |
|---|---|---|---|---|---|---|---|---|
| ELT (primary) | [ms] | 413±35 | 123 | 128 | 314 | 423 | 497 | 628 |
| NU (primary) | [MB/5 min] | 13±2 | 6 | 0 | 11 | 14 | 17 | 24 |
| LT (secondary) | [ms] | 389±7 | 202 | 89 | 244 | 354 | 506 | 2047 |
| LET (secondary) | [ms] | 174±9 | 106 | 6 | 86 | 183 | 233 | 576 |
| LER (secondary) | [%] | 13±6 | 20 | 0 | 1 | 6 | 15 | 86 |

Table A.46: ULM distributions, measured with managed nodes (policy 2).

### A.2.12 Synthetic Variabale Weight Workload with Locally Varying Churn

#### Summarised Effects of Policies on Expected Lookup Time and Network Usage

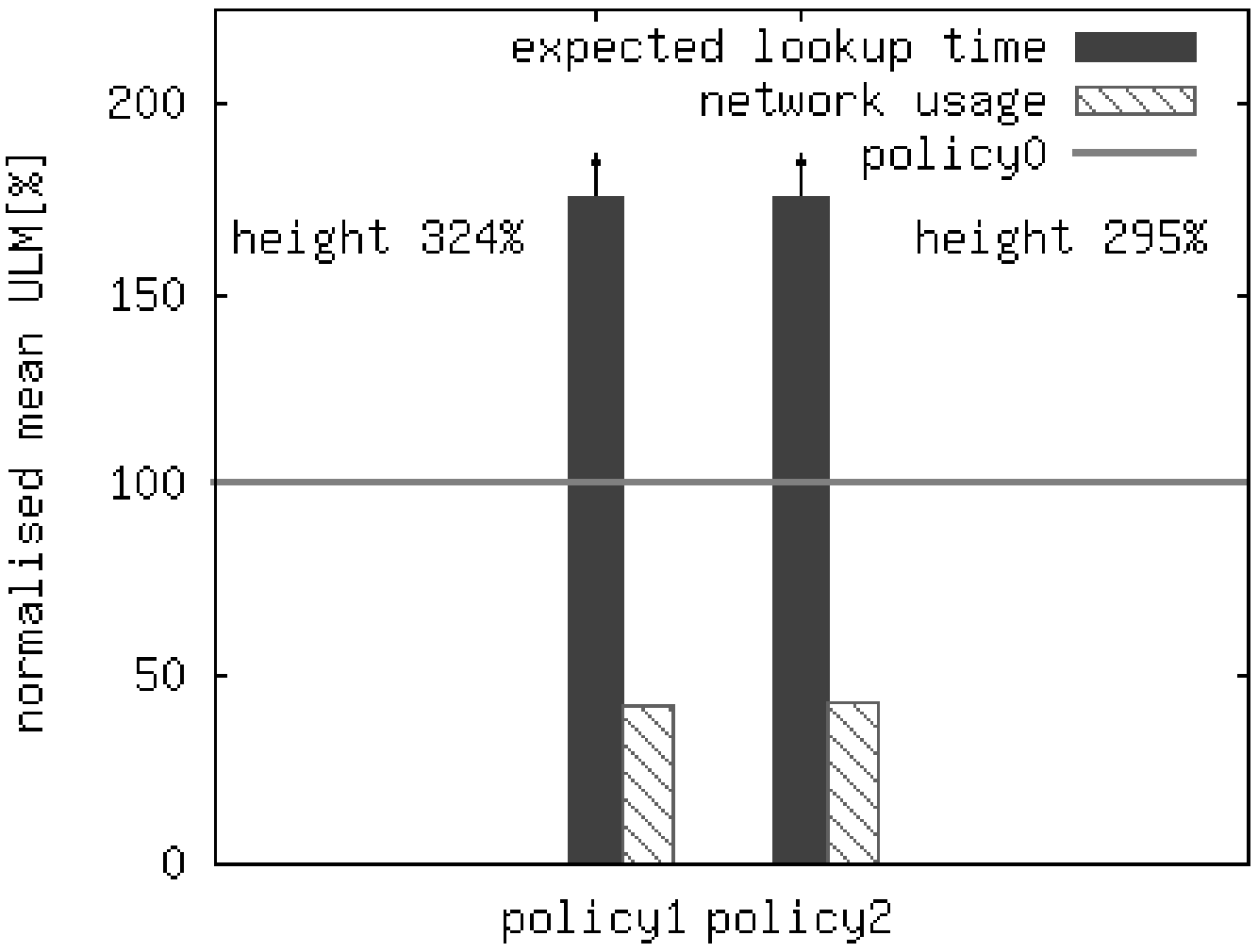

Figure A.13: normalised mean ULM values

| policy | mean expected lookup time [ms] | mean network usage [MB/5 min] |
|---|---|---|
| policy 0 | 574 | 45 |
| policy 1 | 1858 | 19 |
| policy 2 | 1695 | 19 |

Table A.47: mean ULM values

**Summary of ULM Distributions**

| ULM | unit | $ci_\mu$ | $s$ | min | $Q_1$ | $Q_2$ | $Q_3$ | max |
|---|---|---|---|---|---|---|---|---|
| ELT (primary) | [ms] | 574±28 | 97 | 367 | 535 | 574 | 601 | 1021 |
| NU (primary) | [MB/5 min] | 45±1 | 3 | 37 | 43 | 45 | 47 | 50 |
| LT (secondary) | [ms] | 570±8 | 267 | 98 | 367 | 551 | 758 | 2096 |
| LET (secondary) | [ms] | 966±262 | 1021 | 11 | 195 | 389 | 1726 | 4901 |
| LER (secondary) | [%] | 1±1 | 3 | 0 | 0 | 0 | 1 | 19 |

Table A.48: ULM distributions, measured with unmanaged nodes (policy 0).

| ULM | unit | $ci_\mu$ | $s$ | min | $Q_1$ | $Q_2$ | $Q_3$ | max |
|---|---|---|---|---|---|---|---|---|
| ELT (primary) | [ms] | 1858±1360 | 5031 | 356 | 431 | 460 | 512 | 29491 |
| NU (primary) | [MB/5 min] | 19±1 | 5 | 2 | 17 | 19 | 22 | 25 |
| LT (secondary) | [ms] | 444±6 | 200 | 98 | 282 | 427 | 589 | 2169 |
| LET (secondary) | [ms] | 6804±4396 | 34634 | 7 | 143 | 188 | 279 | 223616 |
| LER (secondary) | [%] | 7±3 | 11 | 0 | 0 | 2 | 9 | 42 |

Table A.49: ULM distributions, measured with managed nodes (policy 1).

| ULM | unit | $ci_\mu$ | $s$ | min | $Q_1$ | $Q_2$ | $Q_3$ | max |
|---|---|---|---|---|---|---|---|---|
| ELT (primary) | [ms] | 1695±967 | 3575 | 369 | 490 | 612 | 1434 | 21953 |
| NU (primary) | [MB/5 min] | 19±3 | 10 | 4 | 12 | 16 | 25 | 40 |
| LT (secondary) | [ms] | 572±16 | 477 | 99 | 279 | 470 | 673 | 3561 |
| LET (secondary) | [ms] | 2413±1427 | 18462 | 6 | 52 | 183 | 239 | 198501 |
| LER (secondary) | [%] | 18±5 | 18 | 0 | 1 | 12 | 33 | 71 |

Table A.50: ULM distributions, measured with managed nodes (policy 2).

### A.2.13 Synthetic Variabale Weight Workload with Temporally Varying Churn

#### Summarised Effects of Policies on Expected Lookup Time and Network Usage

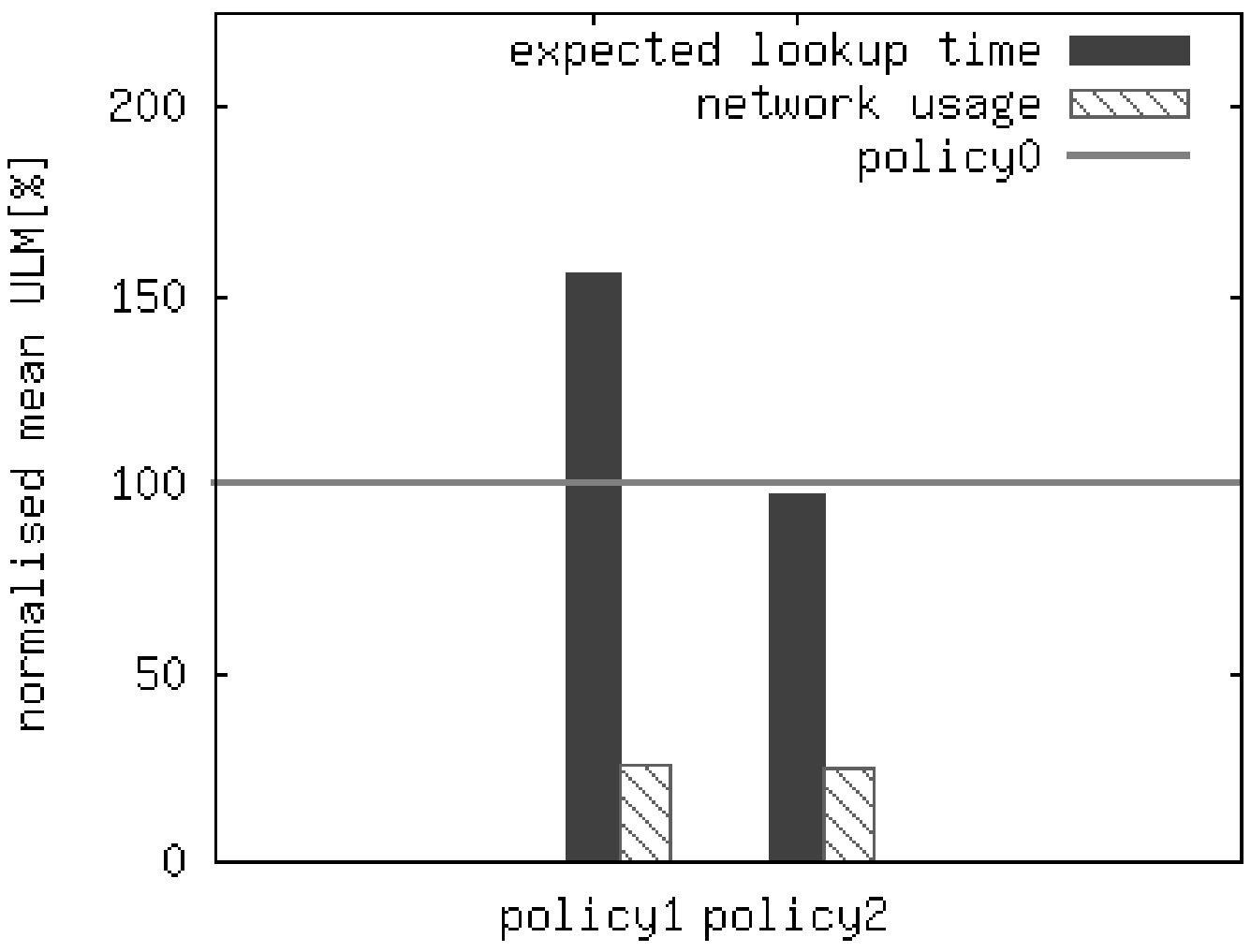

Figure A.14: normalised mean ULM values

| policy | mean expected lookup time [ms] | mean network usage [MB/5 min] |
|---|---|---|
| policy 0 | 570 | 52 |
| policy 1 | 889 | 13 |
| policy 2 | 555 | 13 |

Table A.51: mean ULM values

**Summary of ULM Distributions**

| ULM | unit | $ci_{\mu}$ | $s$ | min | $Q_1$ | $Q_2$ | $Q_3$ | max |
|---|---|---|---|---|---|---|---|---|
| ELT (primary) | [ms] | 570±21 | 72 | 335 | 525 | 575 | 626 | 689 |
| NU (primary) | [MB/5 min] | 52±3 | 10 | 40 | 45 | 47 | 63 | 68 |
| LT (secondary) | [ms] | 575±8 | 257 | 96 | 390 | 552 | 735 | 2111 |
| LET (secondary) | [ms] | 162±56 | 144 | 14 | 49 | 143 | 231 | 495 |
| LER (secondary) | [%] | 1±0 | 1 | 0 | 0 | 0 | 1 | 5 |

Table A.52: ULM distributions, measured with unmanaged nodes (policy 0).

| ULM | unit | $ci_{\mu}$ | $s$ | min | $Q_1$ | $Q_2$ | $Q_3$ | max |
|---|---|---|---|---|---|---|---|---|
| ELT (primary) | [ms] | 889±754 | 2633 | 149 | 386 | 444 | 489 | 15548 |
| NU (primary) | [MB/5 min] | 13±2 | 7 | 0 | 10 | 13 | 19 | 24 |
| LT (secondary) | [ms] | 435±5 | 168 | 95 | 312 | 418 | 540 | 2148 |
| LET (secondary) | [ms] | 8327±13486 | 103703 | 7 | 85 | 86 | 144 | 1311880 |
| LER (secondary) | [%] | 7±6 | 23 | 0 | 0 | 0 | 3 | 100 |

Table A.53: ULM distributions, measured with managed nodes (policy 1).

| ULM | unit | $ci_{\mu}$ | $s$ | min | $Q_1$ | $Q_2$ | $Q_3$ | max |
|---|---|---|---|---|---|---|---|---|
| ELT (primary) | [ms] | 555±49 | 172 | 253 | 450 | 518 | 615 | 1162 |
| NU (primary) | [MB/5 min] | 13±2 | 7 | 3 | 7 | 12 | 19 | 26 |
| LT (secondary) | [ms] | 479±6 | 186 | 96 | 362 | 461 | 588 | 2028 |
| LET (secondary) | [ms] | 252±54 | 725 | 7 | 181 | 224 | 267 | 15926 |
| LER (secondary) | [%] | 15±6 | 21 | 0 | 0 | 1 | 28 | 77 |

Table A.54: ULM distributions, measured with managed nodes (policy 2).

### A.2.14 File System Specific Workload with Low Churn

#### Summarised Effects of Policies on Expected Lookup Time and Network Usage

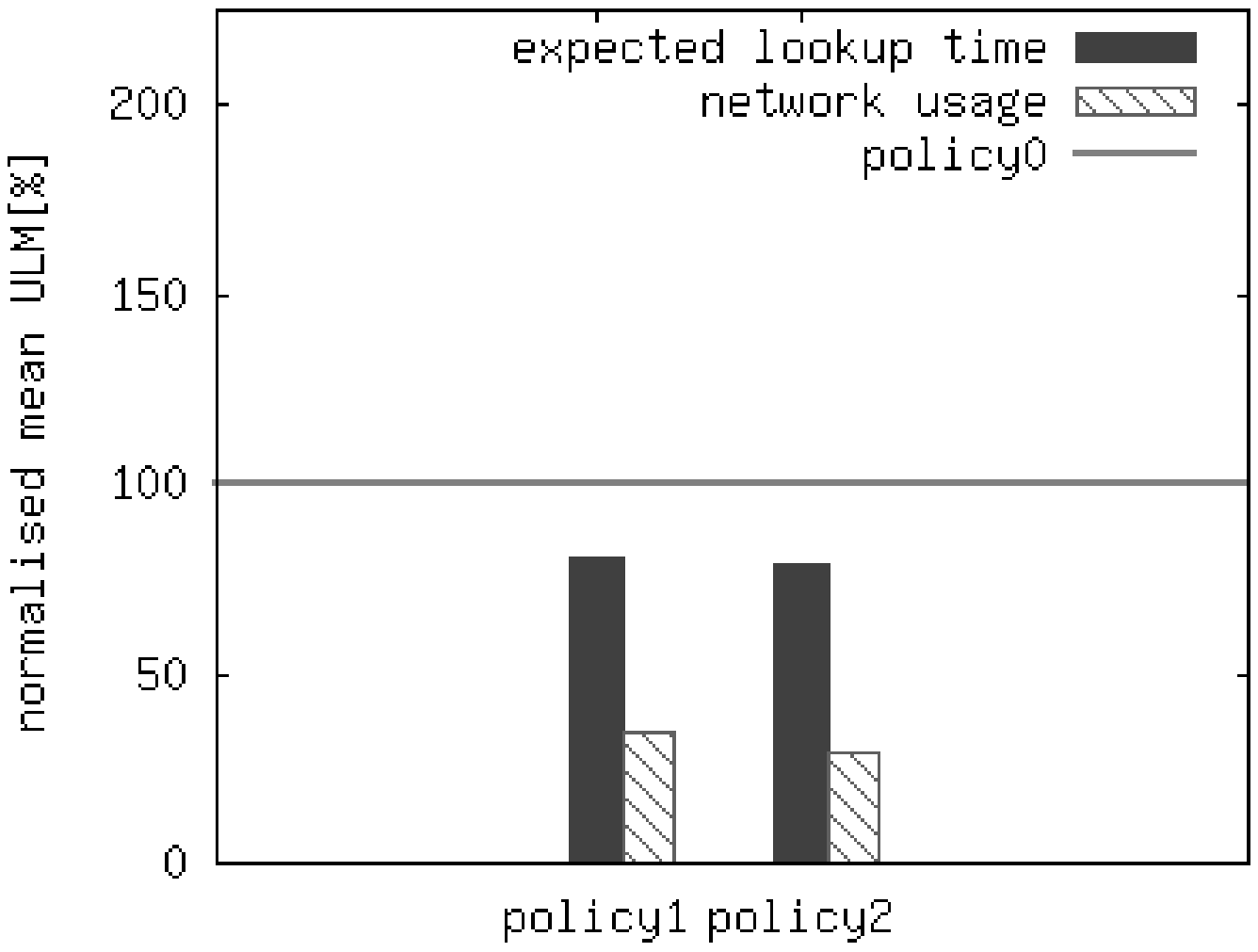

Figure A.15: normalised mean ULM values

| policy | mean expected lookup time [ms] | mean network usage [MB/5 min] |
|---|---|---|
| policy 0 | 877 | 76 |
| policy 1 | 705 | 26 |
| policy 2 | 690 | 22 |

Table A.55: mean ULM values

**Summary of ULM Distributions**

| ULM | unit | $ci_\mu$ | $s$ | min | $Q_1$ | $Q_2$ | $Q_3$ | max |
|---|---|---|---|---|---|---|---|---|
| ELT (primary) | [ms] | 877±6 | 26 | 787 | 868 | 884 | 890 | 911 |
| NU (primary) | [MB/5 min] | 76±2 | 11 | 21 | 79 | 79 | 80 | 81 |
| LT (secondary) | [ms] | 876±2 | 314 | 97 | 655 | 899 | 1110 | 2491 |
| LET (secondary) | [ms] | 78±0 | 0 | 78 | 78 | 78 | 78 | 78 |
| LER (secondary) | [%] | 0±0 | 0 | 0 | 0 | 0 | 0 | 0 |

Table A.56: ULM distributions, measured with unmanaged nodes (policy 0).

| ULM | unit | $ci_\mu$ | $s$ | min | $Q_1$ | $Q_2$ | $Q_3$ | max |
|---|---|---|---|---|---|---|---|---|
| ELT (primary) | [ms] | 705±3 | 14 | 684 | 694 | 699 | 714 | 732 |
| NU (primary) | [MB/5 min] | 26±2 | 8 | 7 | 23 | 24 | 27 | 48 |
| LT (secondary) | [ms] | 705±2 | 244 | 96 | 541 | 742 | 899 | 2546 |
| LET (secondary) | [ms] | NA | NA | NA | NA | NA | NA | NA |
| LER (secondary) | [%] | 0±0 | 0 | 0 | 0 | 0 | 0 | 0 |

Table A.57: ULM distributions, measured with managed nodes (policy 1).

| ULM | unit | $ci_\mu$ | $s$ | min | $Q_1$ | $Q_2$ | $Q_3$ | max |
|---|---|---|---|---|---|---|---|---|
| ELT (primary) | [ms] | 690±6 | 25 | 603 | 685 | 695 | 704 | 727 |
| NU (primary) | [MB/5 min] | 22±1 | 5 | 2 | 22 | 22 | 23 | 34 |
| LT (secondary) | [ms] | 690±2 | 238 | 93 | 530 | 726 | 887 | 2008 |
| LET (secondary) | [ms] | NA | NA | NA | NA | NA | NA | NA |
| LER (secondary) | [%] | 0±0 | 0 | 0 | 0 | 0 | 0 | 0 |

Table A.58: ULM distributions, measured with managed nodes (policy 2).

### A.2.15 File System Specific Workload with High Churn

#### Summarised Effects of Policies on Expected Lookup Time and Network Usage

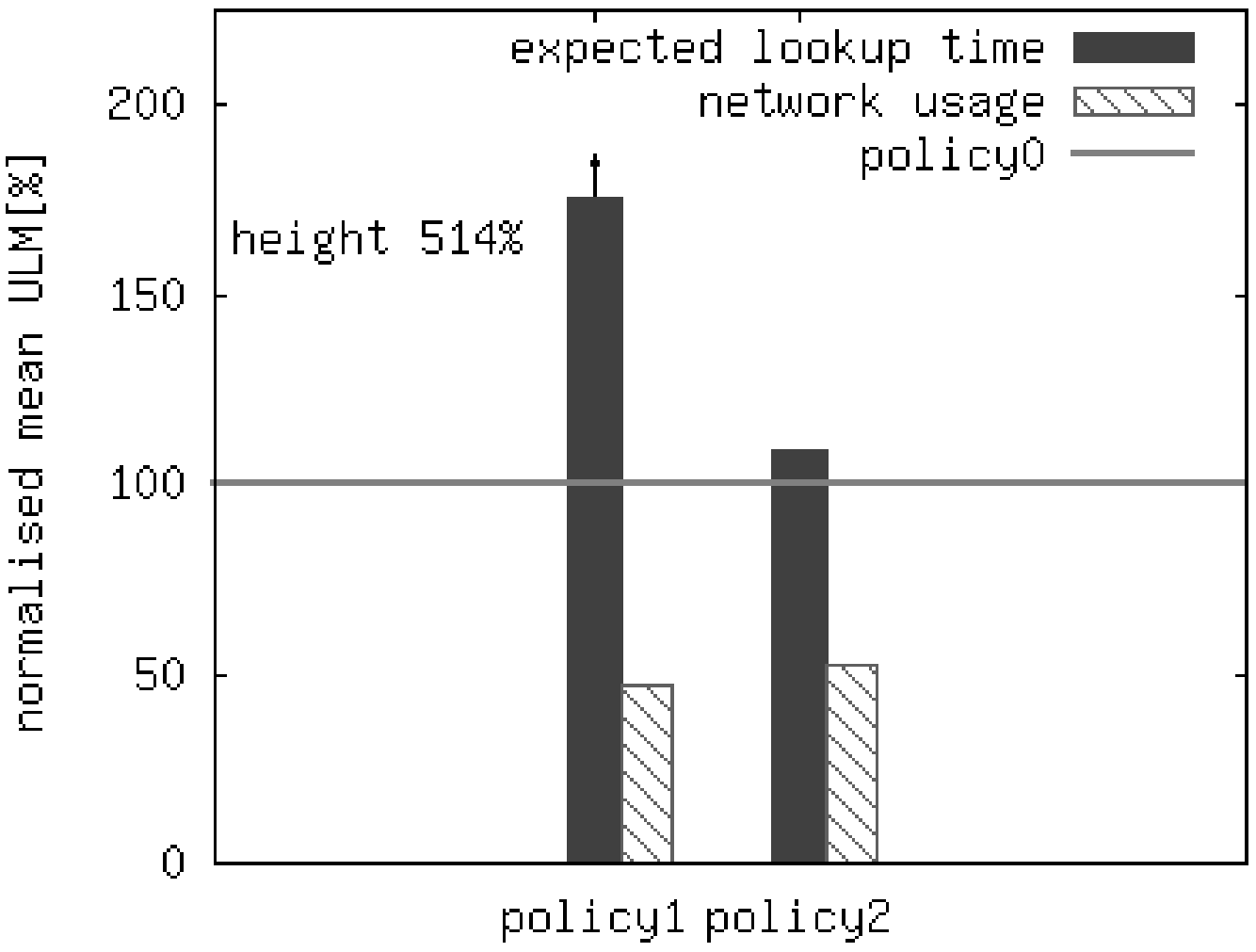

Figure A.16: normalised mean ULM values

| policy | mean expected lookup time [ms] | mean network usage [MB/5 min] |
|---|---|---|
| policy 0 | 720 | 45 |
| policy 1 | 3704 | 21 |
| policy 2 | 784 | 23 |

Table A.59: mean ULM values

**Summary of ULM Distributions**

| ULM | unit | $ci_\mu$ | $s$ | min | $Q_1$ | $Q_2$ | $Q_3$ | max |
|---|---|---|---|---|---|---|---|---|
| ELT (primary) | [ms] | 720±26 | 97 | 426 | 699 | 734 | 758 | 1003 |
| NU (primary) | [MB/5 min] | 45±4 | 15 | 2 | 45 | 50 | 52 | 58 |
| LT (secondary) | [ms] | 726±3 | 341 | 86 | 506 | 715 | 934 | 6065 |
| LET (secondary) | [ms] | 129±5 | 341 | 6 | 64 | 89 | 126 | 21026 |
| LER (secondary) | [%] | 10±7 | 25 | 0 | 2 | 3 | 4 | 97 |

Table A.60: ULM distributions, measured with unmanaged nodes (policy 0).

| ULM | unit | $ci_\mu$ | $s$ | min | $Q_1$ | $Q_2$ | $Q_3$ | max |
|---|---|---|---|---|---|---|---|---|
| ELT (primary) | [ms] | 3704±436 | 2528 | 455 | 688 | 4739 | 5327 | 14954 |
| NU (primary) | [MB/5 min] | 21±2 | 11 | 0 | 15 | 20 | 24 | 45 |
| LT (secondary) | [ms] | 1497±16 | 1970 | 97 | 506 | 732 | 1092 | 17915 |
| LET (secondary) | [ms] | 3736±4763 | 205918 | 6 | 80 | 142 | 375 | 14630025 |
| LER (secondary) | [%] | 8±2 | 17 | 0 | 0 | 3 | 7 | 92 |

Table A.61: ULM distributions, measured with managed nodes (policy 1).

| ULM | unit | $ci_\mu$ | $s$ | min | $Q_1$ | $Q_2$ | $Q_3$ | max |
|---|---|---|---|---|---|---|---|---|
| ELT (primary) | [ms] | 784±73 | 263 | 468 | 642 | 658 | 967 | 1688 |
| NU (primary) | [MB/5 min] | 23±3 | 12 | 0 | 16 | 30 | 32 | 35 |
| LT (secondary) | [ms] | 609±3 | 271 | 86 | 418 | 616 | 781 | 5744 |
| LET (secondary) | [ms] | 230±43 | 3153 | 6 | 72 | 118 | 183 | 244865 |
| LER (secondary) | [%] | 26±9 | 32 | 1 | 6 | 10 | 19 | 90 |

Table A.62: ULM distributions, measured with managed nodes (policy 2).

### A.2.16 File System Specific Workload with Locally Varying Churn

#### Summarised Effects of Policies on Expected Lookup Time and Network Usage

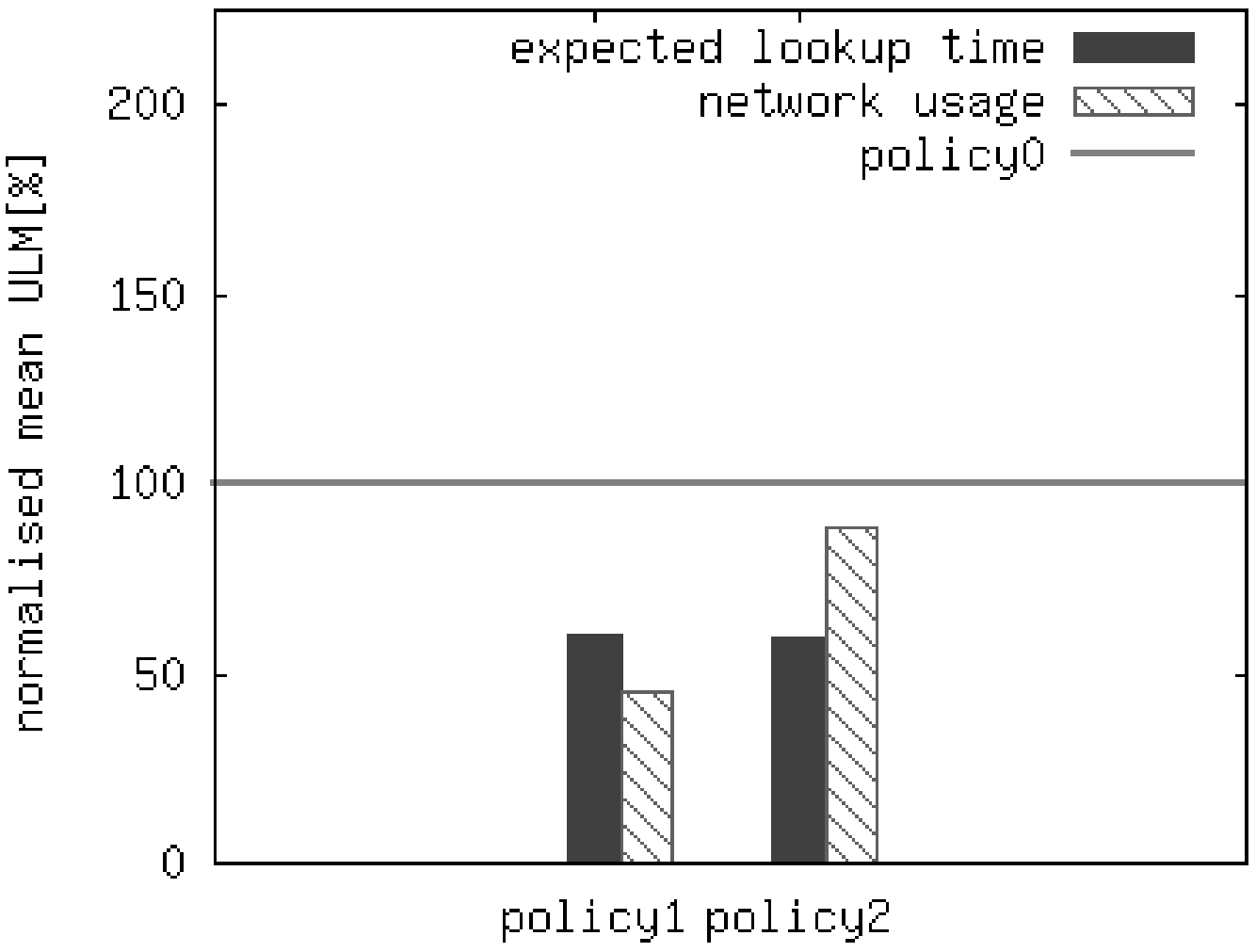

Figure A.17: normalised mean ULM values

| policy | mean expected lookup time [ms] | mean network usage [MB/5 min] |
|---|---|---|
| policy 0 | 6511 | 30 |
| policy 1 | 3908 | 13 |
| policy 2 | 3854 | 26 |

Table A.63: mean ULM values

**Summary of ULM Distributions**

| ULM | unit | $ci_\mu$ | $s$ | min | $Q_1$ | $Q_2$ | $Q_3$ | max |
|---|---|---|---|---|---|---|---|---|
| ELT (primary) | [ms] | 6511±1001 | 8254 | 694 | 1431 | 4700 | 6031 | 52998 |
| NU (primary) | [MB/5 min] | 30±2 | 13 | 5 | 18 | 30 | 38 | 61 |
| LT (secondary) | [ms] | 1821±16 | 1991 | 98 | 591 | 925 | 2479 | 22314 |
| LET (secondary) | [ms] | 6117±781 | 28939 | 6 | 191 | 449 | 993 | 241662 |
| LER (secondary) | [%] | 18±2 | 18 | 0 | 3 | 6 | 37 | 68 |

Table A.64: ULM distributions, measured with unmanaged nodes (policy 0).

| ULM | unit | $ci_\mu$ | $s$ | min | $Q_1$ | $Q_2$ | $Q_3$ | max |
|---|---|---|---|---|---|---|---|---|
| ELT (primary) | [ms] | 3908±398 | 2593 | 600 | 982 | 3866 | 5946 | 9414 |
| NU (primary) | [MB/5 min] | 13±2 | 15 | 1 | 2 | 3 | 28 | 49 |
| LT (secondary) | [ms] | 1053±8 | 946 | 97 | 529 | 759 | 1051 | 6816 |
| LET (secondary) | [ms] | 2027±260 | 15234 | 6 | 90 | 368 | 852 | 247382 |
| LER (secondary) | [%] | 37±3 | 22 | 0 | 10 | 51 | 55 | 60 |

Table A.65: ULM distributions, measured with managed nodes (policy 1).

| ULM | unit | $ci_\mu$ | $s$ | min | $Q_1$ | $Q_2$ | $Q_3$ | max |
|---|---|---|---|---|---|---|---|---|
| ELT (primary) | [ms] | 3854±660 | 3971 | 525 | 818 | 2401 | 5529 | 21939 |
| NU (primary) | [MB/5 min] | 26±2 | 10 | 10 | 18 | 24 | 35 | 49 |
| LT (secondary) | [ms] | 798±5 | 571 | 90 | 453 | 709 | 943 | 5477 |
| LET (secondary) | [ms] | 2716±421 | 21396 | 6 | 99 | 360 | 562 | 274519 |
| LER (secondary) | [%] | 30±4 | 23 | 1 | 8 | 28 | 52 | 68 |

Table A.66: ULM distributions, measured with managed nodes (policy 2).

### A.2.17 File System Specific Workload with Temporally Varying Churn

#### Summarised Effects of Policies on Expected Lookup Time and Network Usage

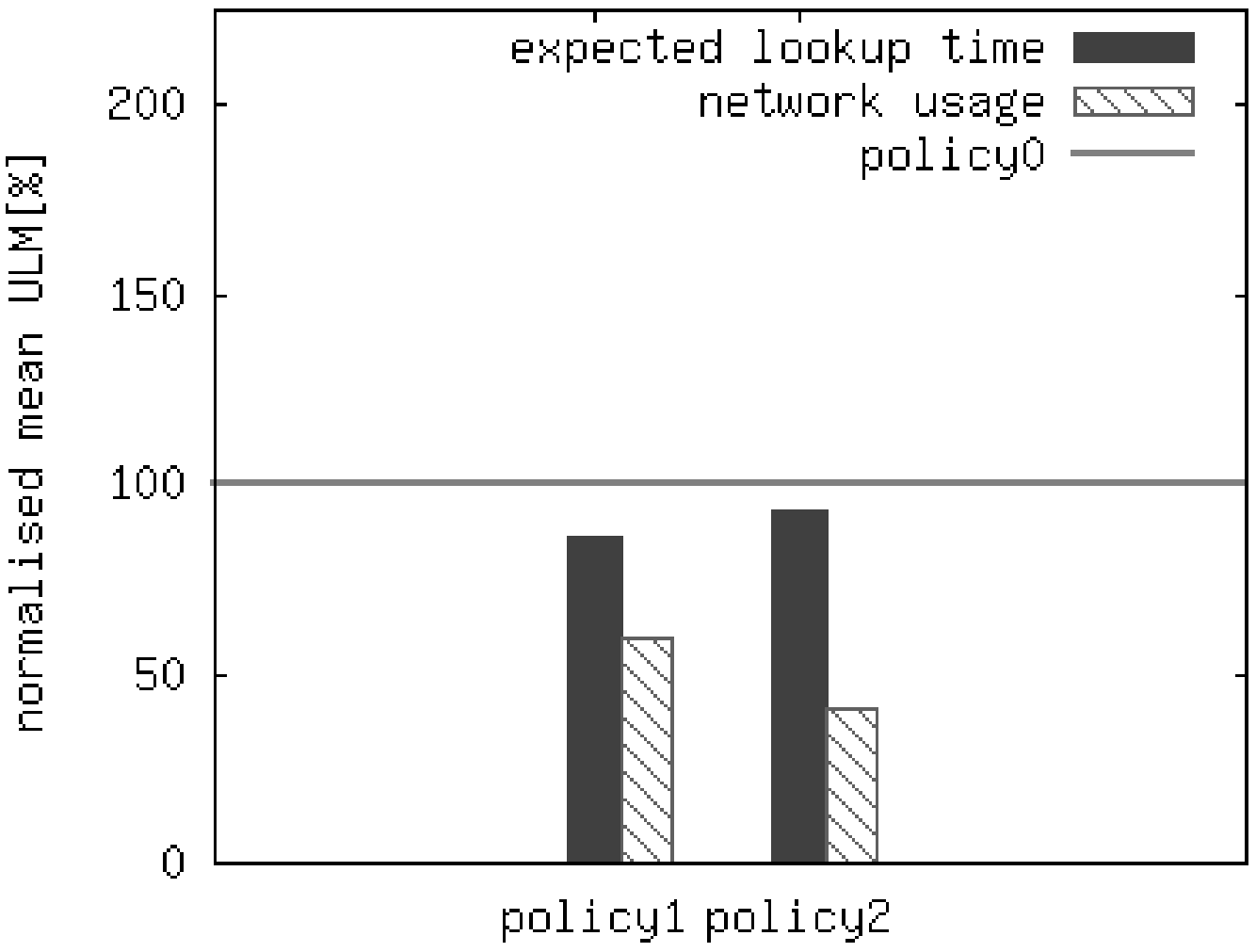

Figure A.18: normalised mean ULM values

| policy | mean expected lookup time [ms] | mean network usage [MB/5 min] |
|---|---|---|
| policy 0 | 783 | 58 |
| policy 1 | 675 | 34 |
| policy 2 | 730 | 24 |

Table A.67: mean ULM values

**Summary of ULM Distributions**

| ULM | unit | $ci_\mu$ | $s$ | min | $Q_1$ | $Q_2$ | $Q_3$ | max |
|---|---|---|---|---|---|---|---|---|
| ELT (primary) | [ms] | 783±17 | 69 | 467 | 743 | 777 | 841 | 876 |
| NU (primary) | [MB/5 min] | 58±4 | 15 | 3 | 54 | 58 | 63 | 80 |
| LT (secondary) | [ms] | 770±3 | 322 | 86 | 551 | 783 | 1004 | 4684 |
| LET (secondary) | [ms] | 114±6 | 232 | 5 | 47 | 54 | 73 | 8667 |
| LER (secondary) | [%] | 3±2 | 10 | 0 | 0 | 1 | 2 | 70 |

Table A.68: ULM distributions, measured with unmanaged nodes (policy 0).

| ULM | unit | $ci_\mu$ | $s$ | min | $Q_1$ | $Q_2$ | $Q_3$ | max |
|---|---|---|---|---|---|---|---|---|
| ELT (primary) | [ms] | 675±14 | 55 | 496 | 636 | 690 | 713 | 791 |
| NU (primary) | [MB/5 min] | 34±2 | 7 | 10 | 30 | 35 | 40 | 46 |
| LT (secondary) | [ms] | 654±2 | 263 | 86 | 471 | 669 | 842 | 2625 |
| LET (secondary) | [ms] | 330±9 | 258 | 6 | 211 | 327 | 414 | 9940 |
| LER (secondary) | [%] | 5±1 | 5 | 0 | 0 | 3 | 7 | 18 |

Table A.69: ULM distributions, measured with managed nodes (policy 1).

| ULM | unit | $ci_\mu$ | $s$ | min | $Q_1$ | $Q_2$ | $Q_3$ | max |
|---|---|---|---|---|---|---|---|---|
| ELT (primary) | [ms] | 730±85 | 335 | 272 | 570 | 681 | 754 | 2353 |
| NU (primary) | [MB/5 min] | 24±2 | 9 | 0 | 20 | 25 | 29 | 41 |
| LT (secondary) | [ms] | 608±2 | 273 | 85 | 388 | 617 | 817 | 2724 |
| LET (secondary) | [ms] | 1002±674 | 25285 | 6 | 145 | 282 | 381 | 1466637 |
| LER (secondary) | [%] | 9±3 | 14 | 0 | 0 | 2 | 10 | 60 |

Table A.70: ULM distributions, measured with managed nodes (policy 2).

### A.2.18 Analysis for Statistical Significance

An analysis of the statistical significance of the differences between the data sets was carried out following guidelines for a *t-test* from [40] and by using *Commons-Math: The Apache Commons Mathematics Library*, commons-math 1.2, [25]. The *p-values*, shown in table A.71 represent the probabilities that any differences in the monitored effects were only due to chance.

| experiment specification | | policy 1 | | policy 2 | |
|---|---|---|---|---|---|
| workload | churn pattern | expected lookup time | network usage | expected lookup time | network usage |
| synthetic light weight | low churn | 0.000 | 0.000 | 0.000 | 0.000 |
| | high churn | 0.123 | 0.000 | 0.222 | 0.000 |
| | locally varying churn | 0.078 | 0.000 | 0.249 | 0.000 |
| | temporally varying churn | 0.027 | 0.000 | 0.903 | 0.000 |
| synthetic heavy weight | low churn | 0.000 | 0.000 | 0.000 | 0.000 |
| | high churn | 0.000 | 0.074 | 0.000 | 0.925 |
| | locally varying churn | 0.002 | 0.039 | 0.007 | 0.517 |
| | temporally varying churn | 0.002 | 0.000 | 0.022 | 0.000 |
| synthetic variable weight | low churn | 0.000 | 0.000 | 0.000 | 0.000 |
| | high churn | 0.016 | 0.341 | 0.017 | 0.297 |
| | locally varying churn | 0.129 | 0.000 | 0.065 | 0.000 |
| | temporally varying churn | 0.492 | 0.000 | 0.647 | 0.000 |
| file system specific | low churn | 0.000 | 0.000 | 0.000 | 0.000 |
| | high churn | 0.000 | 0.000 | 0.181 | 0.000 |
| | locally varying churn | 0.000 | 0.000 | 0.000 | 0.012 |
| | temporally varying churn | 0.000 | 0.000 | 0.318 | 0.000 |

Table A.71: T-Test Results.

### A.2.19 Holistic Quantification of the Effects of Autonomic Management

Here a holistic approach towards the quantification of the effects of autonomic management is presented. The method for quantifying the effects of autonomic management on performance and resource consumption reported in chapter 7 is based on computing values for the expected lookup time for each observation time window. In a situation in which only successful lookups were executed for each observation period, the expected lookup time was equal to the monitored lookup time. In the case of errors it was computed as outlined in section 7.4.2, but only if monitoring data for at least one successful lookup was available. This approach provided a single set of values for quantifying the effects of autonomic management and showing how they changed over time. The shortcoming of this approach however was that observation periods in which only errors occurred were not considered when making an overall quantification of the effects.

It is also possible to quantify the effects of autonomic management in other ways which also detect the significant advantages of autonomic management. In this approach, the *expected lookup time* is computed by averaging all secondary ULMs over the entire experimental run time and over all repetitions. The amount of data sent to the network during the entire experimental run is used to quantify the *network usage*. This approach considers monitoring data in a more holistic way than by breaking it up into observation periods. However it does not provide a single set of values for showing ULM progressions and distributions and was therefore not used in chapter 7. In table A.72 the number of experiment groups[3] in which each policy yielded the greatest benefit, with respect to the expected lookup time ($ELT$) and the network usage ($NU$) individually and to both in combination. Policy 0 gave the best results in fewer experiments than policy 1 and 2.

| policy | description | $ELT$ | $NU$ | $ELT$ & $NU$ |
|---|---|---|---|---|
| 2 | $autonomic_2$ | 8 | 13 | 7 |
| 1 | $autonomic_1$ | 6 | 3 | 1 |
| 0 | $static$ | 2 | 0 | 0 |

Table A.72: The number of experiment groups (out of 16) in which the specific policies yielded the biggest (holistic) benefits.

[3]An experiment was specified by churn pattern, workload and policy; groups of experiments by churn pattern and workload.

Table A.73 shows how much autonomic management affected the observed ULMs in the various experiments that were conducted. It shows the specific mean ULMs in managed systems normalised with respect to the corresponding ones in an unmanaged system in the same experiment. Thus every normalised ULM $< 1$ represents a benefit of the specific autonomic management policy with respect to the unmanaged system. This provides an easy way to show that autonomic management also yielded significant benefits with a holistic analysis. Differences to the approach used in chapter 7 are outlined in section 7.7.3.

| experiment specification | | policy 1 | | policy 2 | |
|---|---|---|---|---|---|
| workload | churn pattern | ELT | NU | ELT | NU |
| synthetic light weight | low churn | 0.73 | 0.09 | 0.7 | 0.03 |
| | locally varying churn | 0.89 | 0.45 | 1.24 | 0.44 |
| | temporally varying churn | 0.73 | 0.29 | 4.21 | 0.18 |
| | high churn | 13.4 | 0.55 | 14.02 | 0.35 |
| synthetic heavy weight | low churn | 0.73 | 0.22 | 0.70 | 0.16 |
| | locally varying churn | 0.36 | 0.56 | 0.60 | 0.67 |
| | temporally varying churn | 0.68 | 0.39 | 0.79 | 0.34 |
| | high churn | 0.57 | 1.29 | 0.61 | 0.95 |
| synthetic varied weight | low churn | 0.71 | 0.10 | 0.70 | 0.05 |
| | locally varying churn | 1.52 | 0.47 | 1.75 | 0.47 |
| | temporally varying churn | 1.52 | 0.29 | 0.94 | 0.24 |
| | high churn | 0.39 | 1.24 | 0.39 | 0.81 |
| file system specific | low churn | 0.8 | 0.26 | 0.79 | 0.22 |
| | locally varying churn | 0.7 | 0.28 | 0.57 | 0.46 |
| | temporally varying churn | 0.86 | 0.51 | 0.91 | 0.4 |
| | high churn | 2.42 | 1.7 | 1 | 0.47 |

Table A.73: A summary of all normalised holistic ULMs.

# Appendix B

# Data Layer Experiments

## B.1 Preliminary Work

Preliminary work was carried out in order to make decisions as to how to configure the autonomic manager and to calibrate the experimental harness used in the experiments reported in chapter 8.

### B.1.1 Network Speed Configuration Verification

To verify that the network speed was applied correctly with *tc* (see section 8.3.4), sample configurations were tested with the third-party tool *iperf* [22] and with *traceroute* [58].

Measurements were carried out with a sample set of three hosts and the central router as introduced in section 8.3.4. The bandwidth between the hosts and the central router, and on the complete path between client and server was measured using *iperf*. The latency from one host to another via the central router was measured using *traceroute*. All measurements were carried out bi-directionally and repeated three times.

The results showed that the measurements with respect to bandwidth varied on average by a factor of $0.044$ in comparison to the configured bandwidth. The measurement with respect to latency varied by a factor of $0.045$ in comparison to the configured latency. Thus, and due to the absence of feasible alternatives, *tc* was considered to be accurate enough to be used for simulating various network conditions.

### B.1.2 Bandwidth Estimation

This preliminary work was carried out to test various SRM (see sections 5.5.2 and 8.2.2) configurations in order to select the one which provided a sufficiently accurate bandwidth estimation.

The SRM uses low level monitoring machinery (LLMM) which periodically estimates the bandwidth between two specific hosts. ICMP ping requests with different (configurable) packet sizes were used to measure the ping time and compute the available bandwidth from it. The biggest ICMP packet size used for this measurement determines the accuracy of the measurement. It was assumed that the biggest possible ICMP payload results in the most accurate measurement. However, this calibration was driven by the motivation to find the smallest payload size which is sufficiently accurate. The LLMM was tested in isolation and as part of the ASA components which were deployed here. It was exposed to a subset of the conditions in the experiments reported in chapter 8. In each case the LLMM was configured with four different ICMP payloads; *65507 bytes* is the maximum payload allowed by the ping utility.

The tested payloads were:

- 65507 bytes
- 32253 bytes
- 16126 bytes
- 8063 bytes

The results showed that *50%* of the maximum possible ICMP packet size (32253 bytes) provided sufficiently accurate bandwidth measurements. The bandwidth estimation results varied from the configured bandwidth by a factor between $0.03$ and $0.08$.

### B.1.3 Network Speed Configuration

This preliminary work was carried out to give an indication of the range of values for latency that can be used to simulate small network speeds.

The latency was measured from a typical user's workstation connection via a private *BT Total Broadband* internet connection to a host in the same network in which the ASA storage cloud[1] was located. Measurements were conducted at various times of the day. The latency was measured on the first hop between the workstation and the above-mentioned host. The local area network (LAN) connection between the workstation and the modem was ignored[2]. The reason for ignoring it was that such a LAN connection would not represent a connection with limited network speed. The average measured round trip time was *42 ± 1.8 [ms]*; a latency of *20 [ms]* was thus configured in the experiments in which limited network speed was simulated.

[1] as used for the experiments reported in chapters 7 and 8

[2] This link was specified with RFC 1918 [76] addresses.

### B.1.4 Autonomic Manager Configuration

These preliminary experiments were carried out to get an intuition of how to configure the autonomic manager used in the experiments reported in chapter 8. The autonomic manager determines a new DOC depending on an observed fetch failure rate (FFR) metric, a bottleneck (BN) metric and a fetch time variation (FTV) metric. A threshold can be configured for each metric to determine whether this specific metric value is high or low. All metrics were measured in preliminary experiments with high and low churn, a heavy and a light weight workload, small and big data item sizes, bottlenecks on client, server and on neither side as well as with a statically high and low configured DOC. The metrics were averaged over all experiments, which included, all together, three repetitions of each experiment.

| metric | min | mean | standard deviation | median | $Q_3$ | max |
|---|---|---|---|---|---|---|
| FFR | 0 | 0.143 | 0.26 | 0 | 0.188 | 0.75 |
| FTV | 0.021 | 0.034 | 0.004 | 0.033 | 0.036 | 0.046 |
| BN | 0.956 | 1.004 | 0.021 | 1.003 | 1.016 | 1.065 |

Table B.1: Observed and averaged metrics values.

A FFR threshold determines the failure rate below which no responses (increase of the DOC) are initiated by the autonomic manager. The FFR metric values shown in table B.1 exhibited a stark variation as it was aggregated over high and low churn. Therefore a range of thresholds (0.1, 0.3 and 0.5), was used in the experiments reported in chapter 8. Like the FFR threshold, the FTV threshold also determines a metric value below which no responses are initiated by the DOC. The variation of fetch times as shown by FTV metrics values in table B.1 are monitored even though no network speed variation was configured. Therefore a FTV threshold, below which no policy response is triggered was chosen to be 0.2, significantly above the maximum observed variation. The BN threshold determines that a bottleneck is present at the client side, if any monitored BN metric is smaller than the corresponding threshold. Only BN metric values monitored in experiments in which no bottlenecks were exhibited are shown in table B.1. A BN threshold of 0.8 was chosen for the experiments reported in chapter 8.

## B.2 Experimental Results

### B.2.1 Introduction

The following sections provide additional detailed results to those reported in chapter 8. This section explains how they are organised and what information each report contains.

After an overview of the policies and of how often individual policies yielded the greatest in section B.2.2, details about the effects of individual policies on the expected get time ($EGT$) and the network usage ($NU$) are provided in sections B.2.3 to B.2.6. The detailed results in sections B.2.3 to B.2.6 are grouped by network speed configuration. All experiments in which a static bottleneck at the server was exhibited are reported in section B.2.3, all in which a bottleneck on the client side was exhibited are reported in section B.2.4 and so forth. Each of these sections contains two tables, one with the effects of the individual policy on the specific ULM, another one explaining the abbreviations used to specify the individual experiment. The effects are normalised by the baseline in the corresponding experiment. As baseline (*policy 0*) a static DOC configuration with no concurrency was used (in combination with a SRM).

### B.2.2 Policy Effects Overview

Table B.2 indicates how often each policy resulted either in the shortest expected get time ($EGT$), the lowest network usage ($NU$) and in both in combination (out of 72 different

experiment groups).

| policy | description | $EGT$ | $NU$ | $EGT$ & $NU$ |
|---|---|---|---|---|
| 4 | $AM_{T_{FFR}=0.5}$ | 6 | 10 | 2 |
| 3 | $AM_{T_{FFR}=0.3}$ | 15 | 9 | 4 |
| 2 | $AM_{T_{FFR}=0.1}$ | 13 | 12 | 1 |
| 1 | $ST_{DOC=4}$ | 23 | 1 | 1 |
| 0 | $ST_{DOC=1}$ | 15 | 40 | 9 |

Table B.2: Benefits summary.

Table B.2 contains a brief description of each policy. Policy 0, the baseline, has a statically configured low DOC (1). Policy 1 has a statically configured high DOC (4). Policies 2 - 4 are policies which specify an autonomic adaptation of the DOC. The autonomic policies vary in terms of the threshold for the number of failed fetch operations per observation period.

### B.2.3 Static Bottleneck at Server Side

| | policy 1 | | policy 2 | | policy 3 | | policy 4 | |
|---|---|---|---|---|---|---|---|---|
| experiment specification | EGT | NU | EGT | NU | EGT | NU | EGT | NU |
| CP1 WL1 DS1 | 1.041 | 3.580 | 1.003 | 1.027 | 1.002 | 1.052 | 1.003 | 1.054 |
| CP1 WL1 DS2 | 0.913 | 3.675 | 1.009 | 0.994 | 0.991 | 1.055 | 1.005 | 0.994 |
| CP1 WL2 DS1 | 1.005 | 1.809 | 0.994 | 0.996 | 1.000 | 1.014 | 1.006 | 0.995 |
| CP1 WL2 DS2 | 0.908 | 1.088 | 1.004 | 0.985 | 0.985 | 1.000 | 1.027 | 1.000 |
| CP1 WL3 DS1 | 1.024 | 2.497 | 1.011 | 0.989 | 1.003 | 0.982 | 1.004 | 0.993 |
| CP1 WL3 DS2 | 0.895 | 1.362 | 0.992 | 1.041 | 0.992 | 1.004 | 1.005 | 0.998 |
| CP2 WL1 DS1 | 0.912 | 2.157 | 0.989 | 1.278 | 0.976 | 1.419 | 0.988 | 1.267 |
| CP2 WL1 DS2 | 0.828 | 2.124 | 0.967 | 1.157 | 0.973 | 1.149 | 0.985 | 1.125 |
| CP2 WL2 DS1 | 0.976 | 1.639 | 1.006 | 1.000 | 0.965 | 0.990 | 1.023 | 1.007 |
| CP2 WL2 DS2 | 0.897 | 1.057 | 1.013 | 1.000 | 1.017 | 1.000 | 0.962 | 1.000 |
| CP2 WL3 DS1 | 0.975 | 1.308 | 1.019 | 1.024 | 1.015 | 1.026 | 1.002 | 1.015 |
| CP2 WL3 DS2 | 0.937 | 1.161 | 0.984 | 1.002 | 1.020 | 1.002 | 0.957 | 1.001 |
| CP3 WL1 DS1 | 0.981 | 2.688 | 1.043 | 1.366 | 1.027 | 1.341 | 1.037 | 1.401 |
| CP3 WL1 DS2 | 0.836 | 2.123 | 1.012 | 1.153 | 0.979 | 1.144 | 0.990 | 1.138 |
| CP3 WL2 DS1 | 1.017 | 1.782 | 1.045 | 1.010 | 1.010 | 1.025 | 1.055 | 1.030 |
| CP3 WL2 DS2 | 0.849 | 1.082 | 0.960 | 0.992 | 0.976 | 0.999 | 0.972 | 0.999 |
| CP3 WL3 DS1 | 0.980 | 1.989 | 1.013 | 1.009 | 0.996 | 0.999 | 1.013 | 1.005 |
| CP3 WL3 DS2 | 0.841 | 1.228 | 0.998 | 1.006 | 0.995 | 1.006 | 0.978 | 1.002 |

Table B.3: Effects on ULM when bottleneck is on the server side.

| abbreviation | description |
|---|---|
| CP1 | low churn |
| CP2 | high churn |
| CP3 | temporally varying churn |
| WL1 | heavy weight workload |
| WL2 | light weight workload |
| WL3 | temporally varying workload |
| DS1 | big data item size |
| DS2 | small data item size |

Table B.4: Abbreviations

### B.2.4 Static Bottleneck at Client Side

| | policy 1 | | policy 2 | | policy 3 | | policy 4 | |
|---|---|---|---|---|---|---|---|---|
| experiment specification | EGT | NU | EGT | NU | EGT | NU | EGT | NU |
| CP1 WL1 DS1 | 2.363 | 1.161 | 1.001 | 0.998 | 0.999 | 1.002 | 1.000 | 0.998 |
| CP1 WL1 DS2 | 1.562 | 1.502 | 0.990 | 1.009 | 1.003 | 0.999 | 1.001 | 1.001 |
| CP1 WL2 DS1 | 2.674 | 1.637 | 1.009 | 1.001 | 0.999 | 1.004 | 1.006 | 0.988 |
| CP1 WL2 DS2 | 1.492 | 1.081 | 0.992 | 0.998 | 1.001 | 1.000 | 0.991 | 0.963 |
| CP1 WL3 DS1 | 2.432 | 1.734 | 0.998 | 0.994 | 1.004 | 0.998 | 1.003 | 0.995 |
| CP1 WL3 DS2 | 1.728 | 1.224 | 1.003 | 0.998 | 0.997 | 1.017 | 1.000 | 1.025 |
| CP2 WL1 DS1 | 1.220 | 1.122 | 1.164 | 1.037 | 1.158 | 1.040 | 1.062 | 1.026 |
| CP2 WL1 DS2 | 1.007 | 1.348 | 0.991 | 1.079 | 1.026 | 1.084 | 1.011 | 1.070 |
| CP2 WL2 DS1 | 1.707 | 1.335 | 1.007 | 1.003 | 0.954 | 0.993 | 1.008 | 1.003 |
| CP2 WL2 DS2 | 1.209 | 1.044 | 1.043 | 0.999 | 1.020 | 1.000 | 1.022 | 1.000 |
| CP2 WL3 DS1 | 1.255 | 1.165 | 1.040 | 1.016 | 1.069 | 1.035 | 1.011 | 1.005 |
| CP2 WL3 DS2 | 1.094 | 1.077 | 0.938 | 1.000 | 0.947 | 1.002 | 0.989 | 1.001 |
| CP3 WL1 DS1 | 1.589 | 1.154 | 1.088 | 1.022 | 1.092 | 1.018 | 1.048 | 1.022 |
| CP3 WL1 DS2 | 0.978 | 1.327 | 0.988 | 1.072 | 1.016 | 0.997 | 0.982 | 1.082 |
| CP3 WL2 DS1 | 2.115 | 1.426 | 1.005 | 0.999 | 1.009 | 1.012 | 1.002 | 1.011 |
| CP3 WL2 DS2 | 1.125 | 1.053 | 1.022 | 0.983 | 1.002 | 0.996 | 1.001 | 1.004 |
| CP3 WL3 DS1 | 1.760 | 1.531 | 0.964 | 1.066 | 0.984 | 1.070 | 0.957 | 1.059 |
| CP3 WL3 DS2 | 1.192 | 1.115 | 1.015 | 0.997 | 0.994 | 1.000 | 1.007 | 1.000 |

Table B.5: Effects on ULM when bottleneck is on the client side.

| abbreviation | description |
|---|---|
| CP1 | low churn |
| CP2 | high churn |
| CP3 | temporally varying churn |
| WL1 | heavy weight workload |
| WL2 | light weight workload |
| WL3 | temporally varying workload |
| DS1 | big data item size |
| DS2 | small data item size |

Table B.6: Abbreviations

### B.2.5 No Bottleneck

| | policy 1 | | policy 2 | | policy 3 | | policy 4 | |
|---|---|---|---|---|---|---|---|---|
| experiment specification | EGT | NU | EGT | NU | EGT | NU | EGT | NU |
| CP1 WL1 DS1 | 2.042 | 1.651 | 1.060 | 1.074 | 1.069 | 1.085 | 1.024 | 1.023 |
| CP1 WL1 DS2 | 1.523 | 1.985 | 1.001 | 1.006 | 0.990 | 1.010 | 0.998 | 1.006 |
| CP1 WL2 DS1 | 1.649 | 1.598 | 0.987 | 1.028 | 1.005 | 1.015 | 1.008 | 1.009 |
| CP1 WL2 DS2 | 0.955 | 1.032 | 1.095 | 1.002 | 1.015 | 1.009 | 0.970 | 1.001 |
| CP1 WL3 DS1 | 1.786 | 2.015 | 0.980 | 1.004 | 0.971 | 0.992 | 0.978 | 0.993 |
| CP1 WL3 DS2 | 1.055 | 1.160 | 0.989 | 0.987 | 1.021 | 0.991 | 0.974 | 1.002 |
| CP2 WL1 DS1 | 1.290 | 1.428 | 1.193 | 1.167 | 1.115 | 1.127 | 1.097 | 1.077 |
| CP2 WL1 DS2 | 0.894 | 1.243 | 1.008 | 0.986 | 1.038 | 0.963 | 1.042 | 0.955 |
| CP2 WL2 DS1 | 1.239 | 1.325 | 0.995 | 1.000 | 0.999 | 1.001 | 0.997 | 0.999 |
| CP2 WL2 DS2 | 1.084 | 1.041 | 1.005 | 1.000 | 1.083 | 1.001 | 1.085 | 0.999 |
| CP2 WL3 DS1 | 1.381 | 1.669 | 0.988 | 0.987 | 0.965 | 0.985 | 0.995 | 0.997 |
| CP2 WL3 DS2 | 1.075 | 1.132 | 1.050 | 1.003 | 1.012 | 1.000 | 0.987 | 0.999 |
| CP3 WL1 DS1 | 1.440 | 1.623 | 1.100 | 1.134 | 1.151 | 1.077 | 1.049 | 1.051 |
| CP3 WL1 DS2 | 0.860 | 1.300 | 0.991 | 1.078 | 0.960 | 1.040 | 1.011 | 0.987 |
| CP3 WL2 DS1 | 1.478 | 1.413 | 1.017 | 0.998 | 1.028 | 1.003 | 1.014 | 1.002 |
| CP3 WL2 DS2 | 1.052 | 1.063 | 0.993 | 1.007 | 1.064 | 1.008 | 1.006 | 1.008 |
| CP3 WL3 DS1 | 1.573 | 1.881 | 1.012 | 0.988 | 1.009 | 1.003 | 0.997 | 1.007 |
| CP3 WL3 DS2 | 0.976 | 1.131 | 1.040 | 1.005 | 1.004 | 1.004 | 0.994 | 1.006 |

Table B.7: Effects on ULM when there is no bottleneck on either side.

| abbreviation | description |
|---|---|
| CP1 | low churn |
| CP2 | high churn |
| CP3 | temporally varying churn |
| WL1 | heavy weight workload |
| WL2 | light weight workload |
| WL3 | temporally varying workload |
| DS1 | big data item size |
| DS2 | small data item size |

Table B.8: Abbreviations

### B.2.6 Temporally Varying Network Speed

| | policy 1 | | policy 2 | | policy 3 | | policy 4 | |
|---|---|---|---|---|---|---|---|---|
| experiment specification | EGT | NU | EGT | NU | EGT | NU | EGT | NU |
| CP1 WL1 DS1 | 0.947 | 2.336 | 0.930 | 2.320 | 0.931 | 2.343 | 0.945 | 2.340 |
| CP1 WL1 DS2 | 0.441 | 2.927 | 0.788 | 1.464 | 0.841 | 1.425 | 0.776 | 1.481 |
| CP1 WL2 DS1 | 0.690 | 1.526 | 0.668 | 1.747 | 0.652 | 1.680 | 0.705 | 1.883 |
| CP1 WL2 DS2 | 0.991 | 1.088 | 0.948 | 1.114 | 0.956 | 1.144 | 1.225 | 1.170 |
| CP1 WL3 DS1 | 0.827 | 1.988 | 0.850 | 1.960 | 0.742 | 1.996 | 0.875 | 2.059 |
| CP1 WL3 DS2 | 0.801 | 1.153 | 0.827 | 1.087 | 0.752 | 1.101 | 0.826 | 1.157 |
| CP2 WL1 DS1 | 0.757 | 1.745 | 0.732 | 1.848 | 0.736 | 1.863 | 0.757 | 1.860 |
| CP2 WL1 DS2 | 0.760 | 1.272 | 0.862 | 1.380 | 0.916 | 1.588 | 0.937 | 1.522 |
| CP2 WL2 DS1 | 1.214 | 1.394 | 0.923 | 1.316 | 1.095 | 1.461 | 1.366 | 1.399 |
| CP2 WL2 DS2 | 0.807 | 1.019 | 0.789 | 1.030 | 0.738 | 1.028 | 0.763 | 1.034 |
| CP2 WL3 DS1 | 0.692 | 1.399 | 0.675 | 1.395 | 0.719 | 1.400 | 0.768 | 1.352 |
| CP2 WL3 DS2 | 0.801 | 1.099 | 0.811 | 1.091 | 0.750 | 1.070 | 0.829 | 1.094 |
| CP3 WL1 DS1 | 0.869 | 1.941 | 0.884 | 1.857 | 0.815 | 2.049 | 0.903 | 2.038 |
| CP3 WL1 DS2 | 0.665 | 1.378 | 0.861 | 1.611 | 0.705 | 1.082 | 0.796 | 1.667 |
| CP3 WL2 DS1 | 1.103 | 1.315 | 1.224 | 1.249 | 1.264 | 1.354 | 1.264 | 1.386 |
| CP3 WL2 DS2 | 0.531 | 0.994 | 0.693 | 1.004 | 0.562 | 1.057 | 0.662 | 1.037 |
| CP3 WL3 DS1 | 0.499 | 1.704 | 0.522 | 1.539 | 0.617 | 1.580 | 0.506 | 1.675 |
| CP3 WL3 DS2 | 0.742 | 1.095 | 0.753 | 1.101 | 0.767 | 1.102 | 0.687 | 1.106 |

Table B.9: Effects on ULM when network speed varies temporally.

| abbreviation | description |
|---|---|
| CP1 | low churn |
| CP2 | high churn |
| CP3 | temporally varying churn |
| WL1 | heavy weight workload |
| WL2 | light weight workload |
| WL3 | temporally varying workload |
| DS1 | big data item size |
| DS2 | small data item size |

Table B.10: Abbreviations

# Appendix C

# GAMF

In chapter 6 the generic autonomic management framework *GAMF* was introduced. The following sections briefly outline how the GAMF can be used for applying autonomic management to various target systems outwith the scope of this thesis.

## C.1 Customising Triggers

The internal GAMF building blocks which are used to initiate the execution of metric extractors or policy evaluators are referred to as triggers. Triggers for periodically triggering or for triggering at the arrival of a specific event are provided by the GAMF. If needed, triggers can be customised by the system adapter developer. In such cases the provided trigger-interface and system-adapter-interface need to be implemented. This can be useful if a triggering mechanism is required for handling complex scheduling, for instance, the evaluation of a policy every $100^{th}$ time a specified event type is recorded; or to trigger an evaluation if, after a given time, no event of a given event type arrives.

## C.2 Triggering By Combining System Adapters

If a policy evaluation is very expensive, with respect to compute resource usage, it can be configured that it is, for instance, only evaluated in specific conditions. Such conditions can be represented by specific metric values. This can be achieved without customising triggers by combining a metric extractor with an event generator. In such a case the metric extractor can generate an event when a metric value is above a certain threshold. A policy can be configured to be triggered at the arrival of a specific event type (generated by the metric).

## C.3 Nesting Autonomic Managers

It may be required that an autonomically managed system is controlled by a higher level manager, via provision of *effectors*. This might be useful if the behaviour of a manager can be configured with some constant parameters (as it is the case in chapter 4). If this behaviour should be again controlled autonomically, nested autonomic managers can be used. In such a case an autonomically managed system is considered as the target system by a higher level manager (figure C.1). The lower level manager may be controlled by a set of parameters directly influencing its policy objectives. GAMF can provide events through which a higher level manager can gather information.

## C.4 Supporting Team Development

In a large scale development project with more than one system adapter developer, working on multiple system adapters, it may become very complex to keep track of the system adapters used and how they interact with each other. It may even be more complex and therefore error prone if various individual system adapters are reused for the management of multiple facets. GAMF's registration mechanism prevents accidental removal of system adapters during a reassembly of the management components if GAMF is configured to do so.

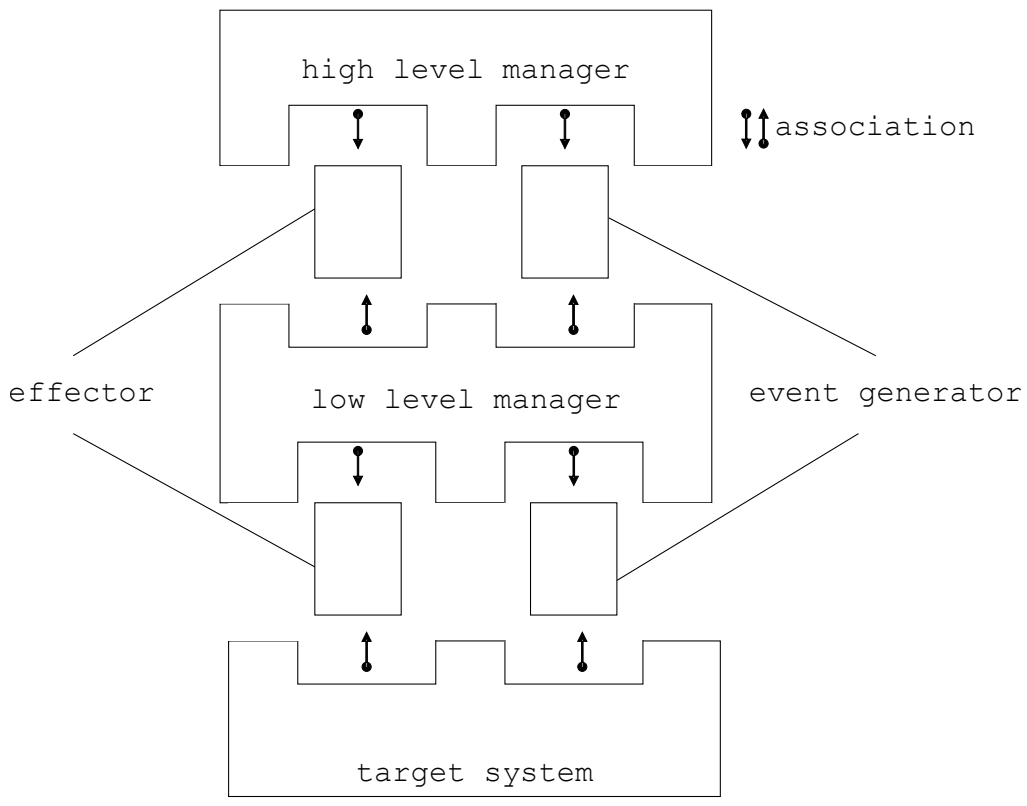

Figure C.1: Nested Manager

## C.5 A Target System Without Access to Source

In the experimental work reported in this thesis, an autonomic manager was added to a system whose source was available. Here it is briefly outlined how GAMF can be used to apply autonomic behaviour to a system whose source is not available. The focus of this use case lies on the interaction between the manager and the target system as provided by the event generator and effector. The objective of the manager is to control the resource allocation of a specific web site autonomically in order to maintain free resources for other websites on the same host in the case that the website under consideration exhibits a high access rate.

The target system in this case is a web server, whose source code is not available. The web server's configuration files and the log files allow read and write access. GAMF runs in a separate address space but on the same physical machine as the web server. The website consists of an application which carries out resource consuming computations. If the number of accesses to the application per observation period exceeds a specific threshold, the configuration setting for the number of web server child processes is reduced in order to allow other services on the web server to use enough CPU.

A *log file watcher* which implements an event generator is developed. This log file watcher periodically searches for events in the web server's access log file which indicate that the website was accessed. The event generator sends the number of accesses as an event to the GAMF which may or may not reside in the same address space as the log file watcher. An effector is implemented which changes the maximum number of child processes for the website under consideration in the web server configuration. After a successful change of the number of child processes, the effector restarts the web server so that the configuration change has an effect.

This shows how an autonomic manager can be implemented with GAMF in order to dynamically adapt a target system of which the source code is not available. It is hoped that the GAMF can thus be used by other people in various other projects besides this thesis. The GAMF is implemented in *Java* its source code and additional information, including an API, can be obtained from http://www-systems.cs.st-andrews.ac.uk/gamf.

# Bibliography

[1] M. Agarwal, V. Vhat, H. Liu, V. Matossian, V. Putty, C. Schmidt, G. Zhang, L. Zhen, and M. Parashar. Automate: Enabling autonomic applications on the grid. CAIP TR-269, Department of Electrical and Computer Engineering, Rutgers University, Seattle, WA, 2003.

[2] N. Agrawal, W. J. Bolosky, J. R. Douceur, and J. R. Lorch. A five-year study of file-system metadata. *ACM Transactions on Storage (TOS)*, 3(3), October 2007.

[3] S. Ajmani, D. E. Clarke, C. Moh, and S. Richman. ConChord: Cooperative SDSI Certificate Storage and Name Resolution. In *First International Workshop on Peer-to-Peer Systems*, pages 141–154, 2002.

[4] M. G. Baker, J. H. Hartman, M. D. Kupfer, K. W. Shirriff, and J. K. Ousterhout. Measurements of a distributed file system. In *Proceedings of 13th ACM Symposium on Operating Systems Principles*, pages 198–212. Association for Computing Machinery SIGOPS, 1991.

[5] S. Balasubramaniam, D. Botvich, B. Jennings, S. Davy, W. Donnelly, and J. Strassner. Policy-constrained bio-inspired processes for autonomic route management. *Computer Networks*, 53:1666–1682, 2009.

[6] G. Bell. Bell's law for the birth and death of computer classes. *Commun. ACM*, 51(1):86–94, 2008.

[7] A. Binzenhöfer and K. Leibnitz. Estimating Churn in Structured P2P Networks. Technical Report 404, University of Würzburg, 2007. Also appears in "Managing Traffic Performance in Converged Networks", http://www.springerlink.com/content/d613316022674h0k/.

[8] A. Binzenhöfer and H. Schnabel. Improving the Performance and Robustness of Kademlia-based Overlay Networks. Technical report, University of Würzburg, 2007. Also appears in the book "Informatik aktuell": http://www.springerlink.com/content/v06258j277t27066/.

[9] A. Binzenhöfer, D. Staehle, and R. Henjes. On the Stability of Chord-based P2P Systems. Technical report, University of Wuerzburg, 2004.

[10] W. J. Bolosky, J. R. Douceur, and J. Howell. The Farsite project: a retrospective. *SIGOPS Oper. Syst. Rev.*, 41(2):17–26, 2007.

[11] N. Chase. *Understand the Autonomic Management Engine*. IBM, June 2004. http://ibm.com/developerworks/autonomic (login required).

[12] B. Chun, F. Dabek, A. Haeberlen, E. Sit, H. Weatherspoon, M. Kaashoek, J. Kubiatowicz, and R. Morris. Efficient replica maintenance for distributed storage systems, 2006.

[13] F. Dabek, M. F. Kaashoek, D. Karger, R. Morris, and I. Stoica. Wide-area cooperative storage with CFS. In *SOSP '01: Proceedings of the eighteenth ACM Symposium on Operating Systems Principles*, pages 202–215, New York, NY, USA, 2001. ACM Press.

[14] F. Dabek, B. Zhao, P. Druschel, J. Kubiatowicz, and I. Stoica. Towards a Common API for Structured Peer-to-Peer Overlays. In IPTPS '03, Berkeley, CA, February 2003.

[15] A. Dearle, G. Kirby, and S. Norcross. Hosting Byzantine Fault Tolerant Services on a Chord Ring. Technical Report CS/07/1, University of St Andrews, 2007.

[16] J. R. Douceur, A. Adya, W. J. Bolosky, D. Simon, and M. Theimer. Reclaiming space from duplicate files in a serverless distributed file system. In *Proceedings of 22nd International Conference on Distributed Computing Systems (ICDCS)*, 2002.

[17] J. R. Douceur and W. J. Bolosky. A large-scale study of file-system contents. In *Proceedings of the 1999 ACM SIGMETRICS international conference on Measurement and modeling of computer systems*, pages 59–70, Atlanta, Georgia, USA, 1999.

[18] J.R. Douceur and R.P. Wattenhofer. Optimizing File Availability in a Secure Serverless Distributed File System. In *20th IEEE Symposium on Reliable Distributed Systems*, 2001.

[19] J. Dowling. *The Decentralised Coordination of Self-Adaptive Components for Autonomic Distributed Systems*. PhD thesis, University of Dublin, Trinity College, 2004.

[20] J. Dowling and V. Cahill. The k-component architecture meta-model for self-adaptive software. In *In Akinori Yonezawa and Satoshi Matsuoka, editors, Proceedings of 3rd International Conference on Metalevel Architectures and Separation of Crosscutting Concerns (Reflection2001), LNCS 2192*, pages 81–88. Springer-Verlag, 2001.

[21] P. Druschel and A. Rowstron. PAST: A large-scale, persistent peer-to-peer storage utility. In *Proceedings HotOS VIII*, May 2001.

[22] J. Dugan and M. Kutzko. Iperf. http://sourceforge.net/projects/iperf/, 2009.

[23] A. Faller. *Der Körper des Menschen*, chapter Vegetatives Nervensystem, pages 425–435. Georg Thieme Verlag, 1995.

[24] M.J. Farabee. The nervous system. Webpage, 2001. http://www.emc.maricopa.edu/faculty/farabee/BIOBK/BioBookNERV.html.

[25] The Apache Software Foundation. *Common [Maths], Statistics*, 2008. Online userguide, http://commons.apache.org/math/userguide/stat.html.

[26] G. Galen and E. Knorr. What cloud computing really means, April 2008. http://www.infoworld.com/article/08/04/07/15FE-cloud-computing-reality_1.html.

[27] S. Ghemawat, H. Gobioff, and S. Leung. The Google File System. In *SOSP '03: Proceedings of the nineteenth ACM symposium on operating systems principles*, pages 29–43, New York, NY, USA, October 2003. ACM Press.

[28] D. S. Gill, S. Zhou, and H. S. Sandhu. A Case Study of File System Workload in a Large-Scale Distributed Environment. In *Measurement and Modeling of Computer Systems*, pages 276–277, 1994.

[29] H. Tyan. *Tutorial: Working With J-Sim.* Ohio State University, Electrical Engineering, http://www.j-sim.org/tutorial/jsim_tutorial.html, 2002.

[30] W. Haager. *Regelungstechnik.* Hoelder-Pichler-Tempsky, 1997.

[31] A. Haberlen, A. Mislove, A. Post, and P. Druschel. Fallacies in evaluating decentralized systems. In *Proceedings of the 5th International Workshop on Peer-to-Peer Systems (IPTPS'06)*, Santa Barbara, CA, February 2006.

[32] J. E. Hanson, I. Whalley, D. M. Chess, and J. O. Kephart. An Architectural Approach to Autonomic Computing. In *Proceedings of the First International Conference on Autonomic Computing (ICAC'04)*, pages 2–9. IEEE Computer Society, 2004.

[33] S. K. Hares and C. J. Wittbrodt. RFC 1574 - Essential Tools for the OSI Internet. http://www.faqs.org/rfcs/rfc1574.html, February 1994.

[34] S. Hariri, B. Khargharia, H. Chen, J. Yang, Y. Zhang, M. Parashar, and H. Liu. The Autonomic Computing Paradigm. *Cluster Computing*, 9(1):5–17, 2006.

[35] S. Hemminger. *Net:Netem*, Jannuary 2008. http://www.linuxfoundation.org/en/Net:Netem.

[36] P. Horn. Autonomic computing: IBM's perspective on the state of information technology, October 2001.

[37] B. Hubert. *tc(8) - Linux man page*. die.net. Obtained online via. http://linux.die.net/man/8/tc on 14/10/2009.

[38] B. Hubert. *Linux Advanced Routing & Traffic Control HOWTO*. Netherlabs BV, 1.43 edition, October 2003.

[39] IBM and autonomic computing (no author provided). *An Architectural Blueprint For Autonomic Computing*. IBM, April 2006. http://www-03.ibm.com/autonomic/library.html.

[40] R. Jain. *The art of computer systems performance analysis, techniques for experimental design, measurement, simulation and modeling*. digital equipment corporation, 1991.

[41] G. P. Jesi. *PeerSim HOWTO: Build a new protocol for the PeerSim 1.0 simulator*. http://peersim.sourceforge.net/tutorial1/tutorial1.html, 2002.

[42] M. F. Kaashoek and D. R. Karger. Koorde: A Simple Degree-optimal Hash Table. In *Proceedings of the 2nd International Workshop on Peer-To-Peer Systems (IPTPS '03)* , pages 98–107, 2003.

[43] X. Kaiping, H. Peilin, and L. Jinsheng. FS-Chord: A New P2P Model with Fractional Steps Joining. In *Proceedings: Advanced International Conference on Telecommunications and International Conference on Internet and Web Applications and Services (AICT-ICIW'06)*, page 98, 2006.

[44] J. Kangasharju, U. Schmidt, D. Bradler, and J. Schräbernhardi. Chunksim: Simulating peer-to-peer content distribution. Bachelor Thesis, 2007.

[45] J. O. Kephart and D. M. Chess. The vision of autonomic computing. *IEEE Computer*, 36(1):41–50, 2003.

[46] B. Khargharia, S. Hariri, M. Parshar, L. Ntaimo, and B. Kim. vgrid: A framework for building autonomic applications. In *Proceedings of the International Workshop on Challenges of Large Applications in Distributed Enviroments (CLADE'03)*, 2003.

[47] G. Kirby, A. Dearle, R. Morrison, and S. Norcross. Secure Location-Independent Autonomic Storage Architectures. Poster at EPSRC Computer Science for e-Science Meeting, National e-Science Centre, March 2004. http://asa.cs.st-andrews.ac.uk/.

[48] G. Kirby, A. Dearle, R. Morrison, S. Norcross, M. Tauber, and R. MacInnis. Report on GR/S44501/01: Secure Location-Independent Autonomic Storage Architectures. Technical report, University of St Andrews, School of Computer Science, 2008.

[49] G. Kirby, S. Norcross, A. Dearle, R. Morrison, M. Tauber, and R. MacInnis. ASA Infrastructure Overview. Handout at the EPSRC e-Science Projects All Hands Meeting, March 2007. http://asa.cs.st-andrews.ac.uk/.

[50] J. Kubiatowicz, D. Bindel, Y. Chen, P. Eaton, D. Geels, R. Gummadi, S. Rhea, H. Weatherspoon, W. Weimer, C. Wells, and B. Zhao. OceanStore: An Architecture for Global-scale Persistent Storage. In *Proceedings of ACM ASPLOS*. ACM, November 2000.

[51] G. Kunzmann and A. Binzenhöfer. Autonomically Improving the Security and Robustness of Structured P2P Overlays. In *International Conference on Systems and Networks Communications. ICSNC*, 2006.

[52] G. Kunzmann, A. Binzenhöfer, and R. Henjes. Analyzing and Modifying Chord's Stabilization Algorithm to Handle High Churn Rates. In *MICC& ICON*, 2005.

[53] J. Ledlie, J. M. Taylor, L. Serban, and M. Seltzer. Self-organization in peer-to-peer systems. In *EW10: Proceedings of the 10th workshop on ACM SIGOPS European workshop*, pages 125–132, New York, NY, USA, 2002. ACM.

[54] X. Li, J. Misra, and C. G. Plaxton. Concurrent Maintenance of Rings. *Distributed Computing*, 2006.

[55] D. Liben-Nowell, H. Balakrishnan, and D. Karger. Analysis of the evolution of peer-to-peer systems. In *PODC '02: Proceedings of the twenty-first annual symposium on Principles of distributed computing*, pages 233–242, New York, NY, USA, 2002. ACM.

[56] H. Liu and M. Parashar. A component based programming framework for autonomic applications. In *the International Conference on Autonomic Computing*, New York, NY, USA, 2004.

[57] R. Mahajan, M. Castro, and A. Rowstron. Controlling the Cost of Reliability in Peer-to-Peer Overlays. In *Proceedings of the 2nd International Workshop on Peer-To-Peer Systems (IPTPS '03)* , pages 21–32, 2003.

[58] G. Malkin. Traceroute Using an IP Option. RFC1393, Jannuary 1993.

[59] P. Maymounkov and D. Mazieres. Kademlia: A Peer-to-peer Information System Based on the XOR Metric. In *Proceedings of the 1st International Workshop on Peer-To-Peer Systems (IPTPS '02)* , 2002.

[60] B. Melcher and B. Mitchell. Towards an Autonomic Framework: Self-Configuring Network Services and Developing Autonomic Applications. *Intel Techology Journal*, 8(4):279–290, November 2004.

[61] P. Mell and T. Grance. The NIST Definition of Cloud Computing, July 2009. http://csrc.nist.gov/groups/SNS/cloud-computing/index.html.

[62] S. Merugu, S. Srinivasan, and E. Zegura. p-sim: A Simulator for Peer-to-Peer Networks. In *IEEE/ACM International Symposium on Modeling, Analysis and Simulation of Computer and Telecommunication Systems (MASCOTS)*, 2003.

[63] D. L. Mills. *The Network Time Protocol (NTP) Distribution*, March 2008. http://www.eecis.udel.edu/ mills/ntp/html/index.html.

[64] G. E. Moore. Cramming more components onto integrated circuits. *Electronics*, 38(8):55–60, 1965.

[65] L. Mummert and M. Satyanarayanan. Long term distributed file reference tracing: Implementation and experience. Technical Report CMU-CS-94-213, Carnegie Mellon School of Computer Science, 1994.

[66] A. Muthitacharoen, R. Morris, T. M. Gil, and B. Chen. Ivy: A Read/Write Peer-to-peer File System. In *The Fifth Symposium on Operating Systems Design and Implementation (OSDI)*, Boston, MA, December 2002.

[67] S. Naicken, B. Livingston, A. Basu, S. Rodhetbhai, I. Wakeman, and D. Chalmers. The state of peer-to-peer simulators and simulations. *SIGCOMM Comput. Commun. Rev.*, 37(2):95–98, 2007.

[68] N. F. Neimark. Mind/Body Education Center - The Fight or Flight Response, November 2007. Webpage: http://www.thebodysoulconnection.com/EducationCenter/fight.html.

[69] I. Norros, V. Pehkonen, H. Reittu, A. Binzenhöfer, and K. Tutschku. Relying on Randomness - PlanetLab Experiments with Distributed File-sharing Protocols. Technical Report 407, University of Würzburg, 2007. Also appears in proceedings of the 3rd EURO-NGI Conference on Next Generation Internet Networks (NGI 2007).

[70] D. E. Perry and A. L. Wolf. Foundations for the study of software architecture. *SIGSOFT Softw. Eng. Notes*, 17(4):40–52, 1992.

[71] L. Peterson, A. Bavier, M. E. Fiuczynski, and S. Muir. Experiences building planetlab. In *Proceedings of the 7th USENIX Symp. on Operating Systems Design and Implementation (OSDI)*, 2006.

[72] D. C. Plummer. An ethernet address resolution protocol. http://www.faqs.org/rfcs/rfc826.html, November 1982.

[73] Y. Qiao and F. E. Bustamante. Structured and Unstructured Overlays Under the Microscope - A Measurement-based View of Two P2P Systems That People Use. In *In Proc. of the 2006 USENIX Annual Technical Confrence*, 2006.

[74] S. Ratnasamy, P. Francis, M. Handley, R. Karp, and S. Shenker. A Scalable Content Addressable Network. In *SIGCOMM '01: Proceedings of the 2001 conference on Applications, technologies, architectures, and protocols for computer communications*, pages 161–172, New York, NY, USA, 2001. ACM.

[75] J. Reimer. Total share: 30 years of personal computer market share figures. *ars technica*, page 10, 2005. http://arstechnica.com/articles/culture/total-share.ars.

[76] Y. Rekhter, B. Moskowitz, D. Karrenberg, G. J. de Groot, and E. Lear. Address Allocation for Private Internets. ftp://ftp.ripe.net/rfc/rfc1918.txt, February 1996.

[77] S. Rhea, P. Eaton, D. Geels, H. Weatherspoon, B. Zhao, and J. Kubiatowicz. Pond: The Oceanstore prototype. In *Proceedings of the Conference on File and Storage Technologies*. USENIX, 2003.

[78] S. Rhea, C. Wells, P. Eaton, D. Geels, B. Zhao, H. Weatherspoon, and J. Kubiatowicz. Maintenance-Free Global Data Storage. *IEEE Internet Computing*, 5(5):40–49, 2001.

[79] D. Roselli. Characteristics of File System Workloads. Technical Report CSD-98-1029, University of California at Berkeley, 1998.

[80] D. Roselli, J. R. Lorch, and T. E. Anderson. A Comparison of File System Workloads. In *2000 USENIX Annual Technical Conference*, pages 41–54, 2000.

[81] A. Rowstron and P. Druschel. Pastry: Scalable, distributed object location and routing for large-scale peer-to-peer systems. In *IFIP/ACM International Conference on Distributed Systems Platforms (Middleware)*, pages 329–350, Heidelberg, Germany, November 2001.

[82] O. Saleh and M. Hefeeda. Modeling and Caching of Peer-to-Peer Traffic. In *The 14th IEEE International Conference on Network Protocols*, pages 249–257. IEEE, November 2006.

[83] S. Saroiu, K. Gummadi, and S. Gribble. Measuring and analyzing the characteristics of napster and gnutella hosts, 2003.

[84] S. Sen and J. Wang. Analyzing peer-to-peer traffic across large networks. In *Second Annual ACM Internet Measurement Workshop*, November 2002.

[85] I. Stoica, R. Morris, D. Liben-Nowell, D. R. Karger, M. F. Kaashoek, F. Dabek, and H. Balakrishnan. Chord: A Scalable Peer-to-peer Lookup Protocol for Internet Applications. In *The Proceedings of ACM SIGCOMM*, San Diego, CA, Augsut 2001.

[86] S. van der Meer. Architectural artefacts for autonomic distributed systems- contract language. In *EASE '09: Proceedings of the 2009 Sixth IEEE Conference and Workshops on Engineering of Autonomic and Autonomous Systems*, pages 99–108, Washington, DC, USA, 2009. IEEE Computer Society.

[87] C. Vazquez, E. Huedo, R.S. Montero, and I.M. Llorente. Dynamic provision of computing resources from grid infrastructures and cloud providers. In *Grid and Pervasive Computing Conference, 2009. GPC '09*, 2009.

[88] W. Vogels. File system usage in Windows NT 4.0. In *Symposium on Operating Systems Principles*, pages 93–109, 1999.

[89] S. Walker. *A Flexible, Policy-Aware Middleware System.* Phd, University of St Andrews, 2005.

[90] S. Walker, A. Dearle, S. Norcross, G. Kirby, and A. McCarthy. RAFDA: A Policy-Aware Middleware Supporting the Flexible Separation of Application Logic from Distribution. Technical Report CS/06/2, University of St Andrews, 2006.

[91] C. Wells. The OceanStore Archive: Goals, Structure, and Self-Repair. Master's thesis, U.C. Berkeley, 2000.

[92] M. Wolsk, C. Mazurek, P. Spychaa, and A. Sumowski. *Software Engineering Techniques: Design for Quality*, volume 227/2007, chapter The architecture of distributed systems driven by autonomic patterns, pages 49–60. Springer Boston, 2007.

[93] P. N. Yianilos and S. Sobti. The evolving field of distributed storage. *IEEE Internet Computing*, pages 35 – 39, September 2001.

[94] J. Yuh-Jzer and W. Jiaw-Chang. Chord2: A two-layer Chord for reducing maintenance overhead via heterogeneity. *Computer Networks*, 51(3):712–731, 2007.

[95] Y. Zhang, A. Liu, and W. Qu. Software Architecture Design of an Autonomic System. In *Fifth Australasian Workshop on Software and System Architectures*, April 2004. In conjunction with Australian Software Engineering Conference (ASWEC 2004) Melbourne, Australia April 13 and 14, 2004.

[96] B. Zhao, L. Huang, J. Stribling, S. Rhea, A. Joseph, and J. Kubiatowicz. Tapestry: A resilient global-scale overlay for service deployment. IEEE Journal on Selected Areas in Communications, Vol 22, No. 1, January 2004.

[97] N. Zhu, J. Chen, T. Chiueh, and D. Ellard. An NFS Trace Player for File System Evaluation. Technical report, Harvard Computer Science, December 2003. Harvard Computer Science Technical Report TR-16-03.